\documentclass[aps,prx,reprint,superscriptaddress]{revtex4-2}
\usepackage{graphicx}
\usepackage{float}
\usepackage{placeins}
\usepackage{epstopdf}
\usepackage{tocloft}
\usepackage{amsmath,amssymb,amsthm,amsbsy,amsfonts}
\usepackage{latexsym}
\usepackage{esint,physics}
\usepackage{longtable,booktabs,array}

\usepackage{array,tabularx,booktabs,makecell}
\newcolumntype{P}[1]{>{\centering\arraybackslash}p{#1}}
\newcolumntype{M}[1]{>{\centering\arraybackslash}m{#1}}
\newcolumntype{Y}{>{\centering\arraybackslash}X}
\usepackage{mathtools}
\usepackage{xcolor}
\usepackage{tikz}
\usetikzlibrary{positioning,arrows.meta}
\usepackage{adjustbox}

\usepackage[english]{babel}
\usepackage{enumitem}
\usepackage{CJKutf8} 
 
\definecolor{myred}{RGB}{210, 70, 90}     
\definecolor{myblue}{RGB}{70, 120, 200}   
\usepackage{hyperref}
\hypersetup{
    pdfinfo={
		Title={},
		Author={},
	},
    colorlinks=true,
    linkcolor=myred,
    citecolor=myblue,
    filecolor=magenta,
    urlcolor=myblue,
}
\usepackage[capitalize,nameinlink]{cleveref}
\crefname{smsection}{SM Sec.}{SM Secs.}
\Crefname{smsection}{SM Sec.}{SM Secs.}

\renewcommand{\thesection}{\arabic{section}}

\definecolor{profred}{HTML}{8B0000}
\makeatletter
\newcommand{\runinsection}[1]{%
  \par\addvspace{0.8ex}
  \refstepcounter{section}
  \def\@currentlabelname{#1}
  \noindent
  {\normalfont\normalsize\bfseries\color{profred}
  (\thesection) \itshape #1.}
}
\makeatother

\usepackage{appendix}
\usepackage{minitoc}
\usepackage{titletoc}

\titlecontents{section}
  [1.6em]
  {\normalsize\bfseries}
  {\contentslabel{1.4em}}
  {}
  {\hfill\contentspage}
  [\addvspace{3pt}]

\titlecontents{subsection}
  [3.2em]
  {\normalsize}
  {\contentslabel{3.0em}}
  {}
  {\titlerule*[0.6pc]{.}\contentspage}
  [\addvspace{1.5pt}]

\newcommand\DoToC{%
  \startcontents
  \hrule height 0.6pt depth 0.6pt width \linewidth
  \printcontents{}{1}[2]{}
  \hrule height 0.6pt depth 0.6pt width \linewidth
}

\makeatletter
\newcommand{\AppendixSectionPrefixNumbers}{%
  \renewcommand{\thesubsection}{\thesection.\arabic{subsection}}%
  \renewcommand{\thesubsubsection}{\thesubsection.\alph{subsubsection}}%
  \renewcommand{\p@subsection}{}%
  \renewcommand{\p@subsubsection}{}%
}
\makeatother

\newcommand{\knotlinewidth}{1.5pt}
\tikzset{
  knot/.style   ={line width=\knotlinewidth, baseline=-.5ex},
  move/.style   ={line width=\knotlinewidth},
  rest/.style   ={line width=\knotlinewidth,dashed},
  vertex/.style ={circle,fill=black,inner sep=1.6pt},
  overcross/.style={double, line width=1.5, white, double=#1, double distance=\knotlinewidth},
  overcross/.default=black,
}

\newcommand{\CrossPic}[1][]{%
  \tikz[knot,#1]{%
    \draw[move,blue] (-.48, .48) -- (.48,-.48);
    \draw[overcross=red] (-.48,-.48) -- (.48, .48);
  }%
}

\newcommand{\SidePic}[1][]{\tikz[knot, #1]{\draw[red, looseness=1.4] (-.5,-.5) to[out=0, in=0] (-.5,.5);\draw[blue, looseness=1.4] (.5,.5) to[out=180,in=180] (.5,-.5);}}

\newcommand{\TopPic}[1][]{%
  \tikz[knot, rotate=90, #1]{%
    \draw[red , looseness=1.4] (-.5,-.5) to[out=0  ,in=0  ] (-.5, .5);
    \draw[blue, looseness=1.4] ( .5, .5) to[out=180,in=180] ( .5,-.5);
  }%
}

\newcommand{\VertexPic}[1][]{%
  \tikz[knot,#1]{%
    \node[vertex] (v) at (0,0){};
    \draw[move,red  ] (v) -- ++( 45:0.55);
    \draw[move,green] (v) -- ++(135:0.55);
    \draw[move,black] (v) -- ++(225:0.55);
    \draw[move,blue ] (v) -- ++(315:0.55);
  }%
}

\newcommand{\edgeline}{1.5pt}

\newcommand{\ThreeGraphZeroEdge}[1][]{%
  \tikz[scale=0.7,
        every node/.style={circle,draw=black,fill=red,
                           inner sep=2pt,font=\bfseries\small,text=white},
        edge/.style={
          preaction={draw=black,line width=3pt},
          draw=blue,line width=\edgeline
        },
        #1
       ]{%
    \node (C) at (-1.1,-2.5) {C};
    \node (D) at ( 1.1,-2.5) {D};
  }%
}

\newcommand{\ThreeGraphOneEdge}[1][]{%
  \tikz[scale=0.7,
        every node/.style={circle,draw=black,fill=red,
                           inner sep=2pt,font=\bfseries\small,text=white},
        edge/.style={
          preaction={draw=black,line width=3pt},
          draw=blue,line width=\edgeline
        },
        #1
       ]{%
    \node (C) at (-1.1,-2.5) {C};
    \node (D) at ( 1.1,-2.5) {D};
    \draw[edge] (C) -- (D);
  }%
}

\newcommand{\ThreeGraphTwoEdge}[1][]{%
  \tikz[scale=0.7,
        every node/.style={circle,draw=black,fill=red,
                           inner sep=2pt,font=\bfseries\small,text=white},
        edge/.style={
          preaction={draw=black,line width=3pt},
          draw=blue,line width=\edgeline
        },
        #1
       ]{%
    \useasboundingbox (-1.4,-1.47) rectangle (1.42,-2.85);
    \node (C) at (-1.1,-2.5) {C};
    \node (D) at ( 1.1,-2.5) {D};
    \draw[edge] (C) -- (D);
    \draw[edge] (C) to[out=90,in=90,looseness=1] (D);
  }%
}

\newcommand{\ThreeGraphThreeEdgeCD}[1][]{%
  \tikz[scale=0.7,
        every node/.style={circle,draw=black,fill=red,
                           inner sep=2pt,font=\bfseries\small,text=white},
        edge/.style={
          preaction={draw=black,line width=3pt},
          draw=blue,line width=\edgeline
        },
        #1
       ]{%
    \useasboundingbox (-1.4,-1.47) rectangle (1.42,-3.51);
    \node (C) at (-1.1,-2.5) {C};
    \node (D) at ( 1.1,-2.5) {D};
    \draw[edge] (C) -- (D);
    \draw[edge] (C) to[out=270,in=270,looseness=1] (D);
    \draw[edge] (C) to[out=90, in=90, looseness=1] (D);
  }%
}

\newcommand{\ThreeGraphThreeEdgeAB}[1][]{%
  \tikz[scale=0.7,
        every node/.style={circle,draw=black,fill=red,
                           inner sep=2pt,font=\bfseries\small,text=white},
        edge/.style={
          preaction={draw=black,line width=3pt},
          draw=blue,line width=\edgeline
        },
        #1
       ]{%
    \useasboundingbox (-1.4,-1.47) rectangle (1.42,-3.51);
    \node (C) at (-1.1,-2.5) {A};
    \node (D) at ( 1.1,-2.5) {B};
    \draw[edge] (C) -- (D);
    \draw[edge] (C) to[out=270,in=270,looseness=1] (D);
    \draw[edge] (C) to[out=90, in=90, looseness=1] (D);
  }%
}

\newcommand{\Overallgraph}[1][]{%
  \tikz[%
      scale=0.9,
      edge/.style={
        preaction={draw=black,line width=3pt},
        draw=blue,line width=\edgeline
      },
      overcross/.style={
        preaction={draw=white,line width=\edgeline+8pt},
        preaction={draw=black,line width=3pt},
        draw=blue,line width=\edgeline
      },
      every node/.style={
        circle,draw=black,fill=red,
        inner sep=2pt,
        font=\bfseries\small,
        text=white
      },
      #1
    ]{%
    \useasboundingbox (-3.3,  1.6) rectangle ( 3.3, -3.5);
    \node (A) at (-1.1,  1.0) {A};
    \node (B) at ( 1.1,  1.0) {B};
    \node (C) at (-1.1, -1.5) {C};
    \node (D) at ( 1.1, -1.5) {D};
    \draw[edge]      (A) -- (B);
    \draw[edge]      (C) -- (D);
    
    \draw[edge]      (A) -- (D);
    \draw[overcross] (B) -- (C);
    
    \draw[edge]      (C) .. controls ( 4, -6) and ( 4,  2.5) .. (B);
    \draw[overcross] (A) .. controls (-4,  2.5) and (-4, -6) .. (D);
  }%
}

\usepackage[most]{tcolorbox}
\tcbuselibrary{breakable,skins,theorems}
\definecolor{statementgreen}{RGB}{55,125,85}
\definecolor{statementorange}{RGB}{180,95,30}

\tcbset{
  statementbox/.style={
    enhanced,
    breakable,
    boxrule=0.6pt,
    arc=2pt,
    left=7pt,
    right=7pt,
    top=7pt,
    bottom=7pt,
    before skip=0.8em,
    after skip=0.8em,
    fonttitle=\bfseries,
    coltitle=white,
    segmentation style={draw=black!45,line width=0.4pt},
    before upper={
      \setlength{\parindent}{0pt}
      \setlength{\parskip}{0.75em}
    }
  }
}

\newtcbtheorem[
  number within=section,
  crefname={definition}{definitions},
  Crefname={Definition}{Definitions}
]{boxeddefinition}{Definition}{
  statementbox,
  colback=myblue!4,
  colframe=myblue!70!black,
  colbacktitle=myblue!70!black
}{def}

\newtcbtheorem[
  number within=section,
  crefname={lemma}{lemmas},
  Crefname={Lemma}{Lemmas}
]{boxedlemma}{Lemma}{
  statementbox,
  colback=statementgreen!4,
  colframe=statementgreen!75!black,
  colbacktitle=statementgreen!75!black
}{lem}

\newtcbtheorem[
  number within=section,
  crefname={theorem}{theorems},
  Crefname={Theorem}{Theorems}
]{boxedtheorem}{Theorem}{
  statementbox,
  colback=myred!4,
  colframe=myred!80!black,
  colbacktitle=myred!80!black
}{thm}

\newtcbtheorem[
  number within=section,
  crefname={proposition}{propositions},
  Crefname={Proposition}{Propositions}
]{boxedproposition}{Proposition}{
  statementbox,
  colback=statementorange!4,
  colframe=statementorange!80!black,
  colbacktitle=statementorange!80!black
}{prop}

\newtcbtheorem[
  number within=section,
  crefname={remark}{remarks},
  Crefname={Remark}{Remarks}
]{boxedremark}{Remark}{
  statementbox,
  colback=black!2,
  colframe=black!55,
  colbacktitle=black!55
}{rem}

\newtcolorbox{proofbox}{
  enhanced,
  breakable,
  colback=white,
  colframe=black!65,
  boxrule=0.6pt,
  arc=2pt,
  left=7pt,
  right=7pt,
  top=7pt,
  bottom=7pt,
  before skip=0.8em,
  after skip=0.8em,
  before upper={
    \setlength{\parindent}{0pt}
    \setlength{\parskip}{0.75em}
  }
}

\graphicspath{{./figs/}{./figs/appx/}{./figs/ExperimentFigures/}}

\begin{document}

\begin{CJK}{UTF8}{gbsn}
\title{
Topological classification through knotted graphs: Fermi surface dispersions and Lifshitz transitions
}

\author{Hakan Akg\"un}
\altaffiliation{These authors contributed equally, and are ordered alphabetically}
\email{hakan.akgun@u.nus.edu}
\affiliation{Department of Physics, National University of Singapore, Singapore 117551}

\author{Xianquan Yan}
\altaffiliation{These authors contributed equally, and are ordered alphabetically}
\email{yanx@u.nus.edu}
\affiliation{Department of Physics, National University of Singapore, Singapore 117551}
\affiliation{Department of Computer Science, National University of Singapore, Singapore 117417}

\author{Ching Hua Lee}
\email{phylch@nus.edu.sg}
\affiliation{Department of Physics, National University of Singapore, Singapore 117551}

\begin{abstract}
Knot theory has provided a rich topological taxonomy for band structures, but its reach is fundamentally limited: knot invariants classify only 1D nodal lines at gap closure, and cannot encode the full dispersion or rich Fermi surface structure of realistic materials. Here we show that \textbf{knotted graphs} (knots that admit graph-like intersections in 3D space) -- which have so far been elusive in condensed matter literature --- provide a unified topological language for classifying the entire band dispersion, and even the eigenstate topology in some contexts. We propose a new framework beyond the existing Yamada polynomials that can  topologically characterize the intricacies of realistic Fermi surfaces completely,  crucially including how their multiple disconnected pieces are nested. This yields the \textbf{Yamada set}, a boundary-resolved extension which organizes the full topological evolution across energy into a \textbf{Yamada sequence}: a compact dispersion-level fingerprint directly tied to experimental signatures of Lifshitz transitions. Our framework is demonstrated with DFT-based band structures of real materials. Beyond dispersion-level classifications, this framework can be extended to non-Hermitian exceptional surfaces, where Berry-curvature flux further equips the knotted-graph skeleton with a directed \textit{Abelian edge flow} that also captures the eigenstate topology.
\end{abstract}

\maketitle
\end{CJK}

\section{Introduction\label{sec:intro}}
Topological classification has stood as a dominant theme in contemporary physics, particularly condensed matter. It is currently achieved through two dichotomous approaches, powered by knot theory and graph theory respectively: The first is based on how non-degenerate quantities wind around each other, as in band structure windings~\cite{Ryu_2010,Schnyder2008,Chiu2016,vanderbilt2018berry}  and, more recently, the knot topology of nodal or exceptional bands~\cite{Bi_2017,Yan_2017,li2018realistic,St_lhammar_2019,Li2019,Carlstrom2019,Lee_2020b, Patil_2022}. The second approach, focusing on the graph topology abstracted from state space connectivity, has been extensively used for classifying non-Hermitian spectra~\cite{Tai2023,Lin_2023,qin2024kinked,yang2020non,li2025phase,gu2026long,yang2026nonlocality,Xiong2024,Pi_2025,yan2026hsg12mlargescalebenchmarkspatial} and Hilbert space fragmentation~\cite{Sala2020,Khemani2020}. %
Yet, it is clear that these paradigms, taken alone, often cannot fully capture the underlying mathematical structure -- knot topology becomes ill-defined when the strands are allowed to intersect%
, while graph topology is agnostic to the global winding structure.

These limitations are particularly evident when the focus is on the governing \emph{laws} behind a sequence of geometric objects in 3D, rather than their individual topological character. 
Examples in this broad context include Fermi seas ~\cite{lifshitz1960anomalies,blanter1994theory,Varlamov_2021}, evolving equipotentials~\cite{Franzosi_2000,BIASOTTI20085}, cell growth~\cite{gibson2006emergence,GIBSON200987}, galactic structures~\cite{MELOTT19901,copi2026topologyuniverse} and other families of extended physical manifolds controlled by an external parameter. Hidden behind their sequences of configurational intricacies such as junctions and twists are underlying rules that govern the evolution.    
Characterizing these rules requires a global framework that encodes both connectivity (\textit{isomorphism}) and spatial entanglement (\textit{ambient isotopy}), manifestly beyond the reach of knot theory and graph theory, as well as the Betti numbers from algebraic topology~\cite{hatcher2005algebraic}\footnote{The genus only captures the topological connectivity of the surface of the 3D structure, not how its 3D ``bulk'' is embedded}.

In this work, we surmounted these challenges by developing a general topological classification language for the \emph{evolution pattern} of generic smooth 3D manifolds, not just their static forms.  Inspired by a distant analogy with persistent homology from topological data analysis~\cite{edelsbrunner2008persistent,carlsson2009topology}, our core idea is to abstract out a \emph{sequence} of topological invariants representing how the essential manifold structure evolves. We encode this structural essence by ``shrinking''---i.e., morphologically skeletonizing---the manifold into a \emph{knotted graph}, such that changes in the manifold's shape are encoded in the form of new braids or vertices.  %

For concreteness, we apply our framework to generic electronic materials, specifically to uncover the topological origin behind the Lifshitz transitions~\cite{lifshitz1960anomalies} induced by tuning the Fermi energy $E$. 
These transitions in the shape of the Fermi surface fundamentally govern various condensed matter observables, including signatures in transport~\cite{transport_ref,Zhang_2017}, thermodynamic response~\cite{thermo_superconductivityref,Slizovskiy2015}, superconductivity~\cite{Kang_2015,Shi_2017} and beyond~\cite{Varlamov_2021, superconduct_transition_temp, blanter1994theory, lin2017line}. 
While the genus encodes basic topological information about the Fermi surface at a particular energy,
understanding what governs the onset of Lifshitz transitions across different energies requires a more nuanced topological characterization.

To motivate the construction of our new framework, we take conventional knots as the starting point. Evidently, they can only represent a narrow variety of Fermi surfaces, specifically tube-like scenarios associated with nodal or exceptional lines at small Fermi energy. Other possibilities in this limit, such as nodal chains, cages, and nets found in real materials~\cite{Chang2017,Feng2018,Yi2018,Xie_2019,ding2022ideal} and metamaterials ~\cite{Yan_2018,Yang_2020} are already nontrivially connected, and inherit a graph structure too. In almost all materials, this nontrivial connectivity also sets in inevitably as the Fermi energy is increased, since Fermi surfaces generically inflate and develop additional crossings and connections, %
as we will demonstrate for Ti$_3$Al~\cite{Zhang2018}, TiB$_2$~\cite{Feng2018,Liu2018,Yi2018}, YH$_3$~\cite{Shao_2018,huiberts1996synthesis,wang1995structural}, and Co$_2$MnGa~\cite{Chang2017,Ilya2019,Guin_2019,Markou_2019}
explicitly in \cref{sec:materials_main}.

As such, we shall topologically encode generic Fermi surfaces with a little-studied class of mathematical objects known as \emph{knotted graphs}%
~\cite{conway1983,kauffmanInvariantsGraphsThreeSpace1989,flapan2017}%
\footnote{In the mathematics literature, knotted graphs are usually called \textit{spatial graphs}.}. Through our mapping, they form skeletonized representations of Fermi surfaces that preserve spatial embedding and connectivity. Defined as knots with allowed intersections, or equivalently as graphs that are not agnostic to their spatial knotting/winding structure, knotted graphs possess extremely rich topological structure, as detailed in \cref{sec:hopf_main,sec:yamada_main}. Although conventional knots are already systematically classified through extensive knot tables and well-developed polynomial invariants~\cite{alexander1928topological,jones1985polynomial,hoste1998first,rolfsen2003knots} , research on the rich landscape of knotted graphs remains largely in its infancy. Investigating these objects across diverse physical settings will likely uncover new organizing principles and reveal previously inaccessible distinctions between complex three-dimensional structures. To illustrate this broader applicability, we extend our framework beyond Fermi surfaces to exceptional surfaces in non-Hermitian systems~\cite{bergholtz2021rmp,wang2024berry}, where Berry flux naturally orients the graph edges and gives rise to the novel notion of \textit{flowed} knotted graphs, thereby jointly characterizing the spatial topology of the dispersion and the eigenstate geometry within a unified framework.

Existing Yamada polynomials can classify knotted graphs associated with single-boundary handlebodies, but the geometric structures encountered in most realistic Fermi surfaces present an even broader challenge: owing to their multiple connected components and nested inner boundaries, the corresponding Fermi volumes are generally compression bodies rather than handlebodies. This connects to a recently articulated open problem in topology---determining which invariants of knotted graphs can be transformed into invariants of their associated handlebody-links \cite{bardakov2025invariantshandlebodylinksspatialgraphs}. Driven by these practical demands~\cite{Zhang2018,Feng2018,Liu2018,Yi2018,Shao_2018,huiberts1996synthesis,wang1995structural,Chang2017,Ilya2019,Guin_2019,Markou_2019}, we extend the existing handlebody-based framework to compression bodies and introduce the \textit{Yamada set}, which collectively records the boundary-wise knotted graphs and Yamada invariants associated with all connected components and nested void boundaries (\cref{appx:compression-body,appx:materials}). Through these sets, we show that the full energy dispersion---sequence of Fermi surfaces across different energies---can be identified with an ordered sequence of \emph{topological invariants}, which we refer to as the \textit{Yamada sequence}.
  
Tracking the Yamada sequence as the Hamiltonian is continuously deformed by experiment control parameters extends this framework from a fixed dispersion to a continuous family of band structures. Combining the deformation parameter with the Fermi energy yields the two-dimensional topological phase space developed in \cref{sec:deform_main}, which resolves finite-energy Lifshitz transitions alongside changes in nodal topology. We also construct parameter-resolved topology phase diagrams for the effective Hamiltonians of the experimentally studied materials TiB$_2$~\cite{Feng2018,Yi2018,Liu2018} and Co$_2$MnGa~\cite{Chang2017,Ilya2019}. In all, this provides a novel knotted-graph-powered \emph{dispersion-level} fingerprint for generic electronic materials, far beyond characterization at a single Fermi energy.

\begin{figure*}[t]
    \centering
    \includegraphics[width=1\textwidth]{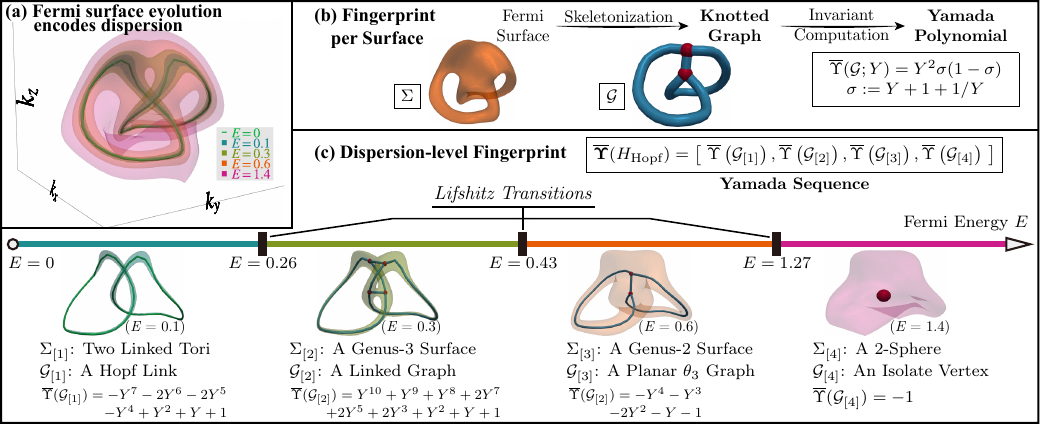}
    \caption{\small
    \textbf{Anatomy of Fermi-sea dispersion through knotted graphs.}
    \textbf{(a)} Nested constant-energy (Fermi) surfaces $\{\mathbf{k}\in \text{BZ}: |f(\mathbf{k})|=E\}$ of the Hopf link Hamiltonian $H_{\text{Hopf}}$ [\cref{eq:two_band_hamiltonian,eq:H_Hopf}] at various Fermi energies $E$.  
    \textbf{(b)} A Fermi surface $\Sigma$ is characterized by a knotted graph $\mathcal{G}$ obtained by morphological skeletonization. Such a graph admits a robust topological invariant, the Yamada polynomial $\Upsilon(\mathcal{G})$, which we extensively generalize in \cref{sec:yamada_main} to encode the intricacies of realistic material Fermi surfaces.
    \textbf{(c)} For a given Hamiltonian, energy windows separated by Lifshitz transitions correspond to distinct Fermi-surface topologies $\Sigma_{[i]}$ (translucent) represented by  knotted graphs $\mathcal{G}_{[i]}$ (blue edges; red vertices) topologically classified by  normalized Yamada polynomials $\overline{\Upsilon}(\mathcal{G}_{[i]})$, where $[i]$ indexes the energy window. Representative Fermi-surface snapshots are shown at selected energies (marked in parentheses), accompanied by their $\Sigma_{[i]}$, $\mathcal{G}_{[i]}$, and $\overline{\Upsilon}(\mathcal{G}_{[i]})$ displayed below. As $E$ increases, the surface evolves from two linked tori to higher-genus surfaces and finally to a simply connected pocket, while the skeleton evolves from a Hopf link to genuine knotted graphs and ultimately to an isolated vertex. The full \textit{Yamada sequence}, which we substantially extend to admit the description of realistic nested Fermi surface structures (see \cref{appx:compression-body}), across energy serves as a \textit{dispersion-level} topological fingerprint.
    }
    \label{fig:hopfsequence}
\end{figure*}

\begin{figure*}[t]
    \centering
    \includegraphics[width=1\linewidth]{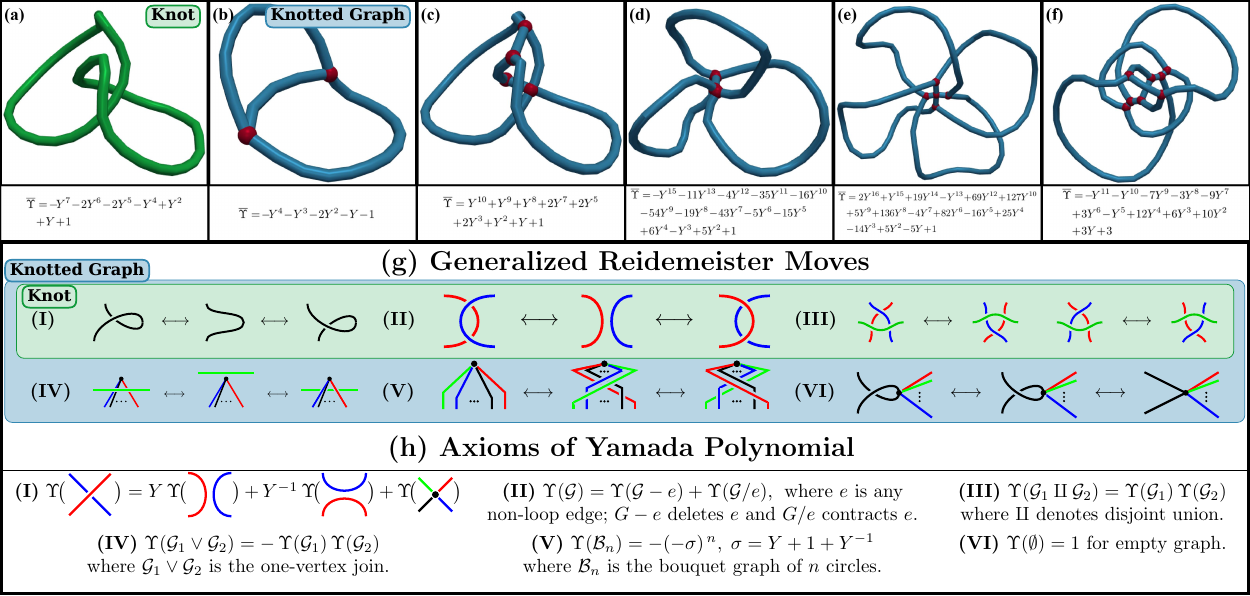}
    \caption{\small
    \textbf{Knotted graphs and their Yamada polynomial classification.}
    \textbf{(a)} A Hopf link (i.e. a two-component knotted graph with no vertices) amenable to conventional nodal-knot classification.
    \textbf{(b--f)} Illustrative knotted-graph skeletons (blue edges; red vertices) extracted from Fermi surfaces of the ansatz Hamiltonians in \cref{appx:ansatz}, with their normalized Yamada polynomials $\overline{\Upsilon}$ shown below.
    \textbf{(g)} Generalized Reidemeister moves for knotted graphs: the classical moves \textbf{(I)--(III)} for all knots/links and the vertex moves \textbf{(IV)--(VI)} specific to knotted graphs.
    \textbf{(h)} Defining relations (axioms) for the Yamada polynomial $\Upsilon(G;Y)$, which enable recursive evaluation of graph diagrams; $\overline{\Upsilon}$ denotes normalized polynomial (details in \cref{appx:Yamada}). See \cref{appx:compression-body} for further details on encoding the boundary hierarchy, which is an integral yet hitherto unexplored aspect that is crucial for their application in fingerprinting the dispersions of realistic materials.
    }
    \label{fig:gallery}
\end{figure*}

\section{Results\label{sec:results}}
\subsection{Dispersion-Level Topological Fingerprint Based on Knotted Graphs\label{sec:hopf_main}}
We first outline our framework, giving an overview of the key steps that culminate in a topological fingerprint for a given electronic material that encodes its properties across \emph{all} energies, not just a particular Fermi surface slice. 

The key observation [\cref{fig:hopfsequence}(a)] is that with increasing Fermi energy $E$, a generic Fermi surface goes through a succession of distinct surface topologies i.e. undergoes a series of \textit{Lifshitz transitions}. %
Physically, such transitions are captured by distinct experimental signatures such as singularities in the density of states~\cite{dos_ref,transport_ref}, Hall effect reversal~\cite{transport_ref}, resistivity anomalies~\cite{Zhang_2017}, thermodynamic and superconducting effects~\cite{Slizovskiy2015,thermo_superconductivityref,Kang_2015,Shi_2017}, and more~\cite{Varlamov_2021,superconduct_transition_temp, blanter1994theory}. %

How and when these Lifshitz transitions occur is intimately dictated by the energy dispersion $E(\bold k)$. For systematic characterization, we introduce the procedure outlined in \cref{fig:hopfsequence}(b). First, starting at the level of an individual Fermi surface $\Sigma$, we extract its key structure through a \textit{morphological skeletonization} procedure detailed in \cref{appx:skeletonization}. The result is a knotted graph that captures its global connectivity structure, which reduces to an ordinary knot in the case of a nodal band structure. Its graph topology encodes the history of coalescence in various parts of the Fermi surface.  
Next, one assigns a topological invariant to this knotted graph, such as but not limited to the normalized Yamada polynomial $\overline{\Upsilon}$%
, which we will elaborate on later. 

For concrete illustration, we define an ansatz two-component Hamiltonian that allows for convenient implementation of any desired knotted graph structure:
\begin{equation}
H(\mathbf{k})=[\mathrm{Re}\,f(\mathbf{k})]\,\sigma_x+[\mathrm{Im}\,f(\mathbf{k})]\,\sigma_z,
\label{eq:two_band_hamiltonian}
\end{equation}
where $f(\mathbf{k}): \text{BZ}\mapsto\mathbb{C}$, BZ$=\mathbb{T}^3$ determines the dispersion relation $E_{\pm}(\mathbf{k})=\pm |f(\mathbf{k})|$~\footnote{The nodal set at $E=0$ is just the locus $\{\mathbf{k}\in \text{BZ}: f(\mathbf{k})=0\}$.}. To prescribe a particular Fermi surface structure, it is helpful to decompose $f(\bold k)$ into $f(z(\bold k),w(\bold k)):\text{BZ}\mapsto\mathbb{C}^2\mapsto\mathbb{C}$~\cite{bode2016constructingpolynomialnodalset,Bode_2017}, such that $f(z,w)$ specifies the knotted structure, while the map $z(\bold k),w(\bold k)$ specifies the non-universal embedding from physical $\bold k$, designed based on specific symmetry constraints on the physical ingredients~\cite{Lee_2020b,lee2021tidal}. Here we fix~\cite{Lee_2020b}
\begin{align}
z(\mathbf{k})&=\cos(2k_z)+\tfrac12+i\bigl(\cos k_x+\cos k_y+\cos k_z-2\bigr),\notag\\
w(\mathbf{k})&=\sin k_x+i\,\sin k_y, \label{eq:z_w}
\end{align}
even though other valid choices merely distort the Fermi surface. We consider the form $f(z,w)=z^p-w^q$, which provides vivid examples of Fermi surface topology by generating a broad family of $(p,q)$-torus knots at $E=0$ (\cref{appx:ansatz}). i.e. setting $(p,q)$ = (1,1) and (2,2) defines an unknot and a Hopf-link nodal Hamiltonian:
\begin{align}
H_{\mathrm{Unknot}}(\mathbf{k}) &: \qquad f(\mathbf{k}) = z(\mathbf{k}) - w(\mathbf{k}),
\label{eq:H_Unknot}\\
H_{\mathrm{Hopf}}(\mathbf{k}) &: \qquad f(\mathbf{k}) = z(\mathbf{k})^2 - w(\mathbf{k})^2.
\label{eq:H_Hopf} 
\end{align}

As featured in \cref{fig:hopfsequence}(a), $H_{\mathrm{Hopf}}$ exhibits topologically distinct Fermi surfaces at different Fermi energies. We hence partition the Fermi energy scaling into various windows, each corresponding to a particular Fermi surface topology. 
For the $i$th energy window, we denote the corresponding Fermi-surface topology by $\Sigma_{[i]}$, its skeletonized knotted graph by $\mathcal{G}_{[i]}$, and the associated normalized Yamada invariant by $\overline{\Upsilon}\!\left(\mathcal{G}_{[i]}\right)$. Accordingly, \cref{fig:hopfsequence}(c) displays, for each window, the Fermi surface, its knotted-graph skeleton, and the invariant that distinguishes the resulting topological sector.

For this example, the four distinct windows correspond to: a pair of linked tori, whose
skeleton is a Hopf link, for $E<0.26$; a genus-$3$ surface with a
linked-graph skeleton for $0.26<E<0.43$; a genus-$2$ surface with a
$\theta_3$-graph (two vertices joined by three edges) skeleton for $0.43<E<1.27$; and a simply connected
pocket, whose skeleton reduces to an isolated vertex, for $E>1.27$. 
The ordered \emph{sequence} of these topologies hence encodes the entire energy dispersion from $E=0$ to the large $E$ limit, where all Fermi surfaces become topologically trivial. Its dispersion-level fingerprint is given by its Yamada sequence $\boldsymbol{\overline{\Upsilon}}(H_{\mathrm{Hopf}})$, which is ordered list of the corresponding normalized Yamada invariants (elaborated in \cref{sec:yamada_main}). The whole workflow remains unchanged even if the concerned Hamiltonian does not exhibit nodal-type degeneracies at low energies.

The same methodology extends beyond this example to the generic multiband dispersions of real materials, which can possess far more complicated Fermi surface topologies. A central advance in this work is the development of the tools for handling these intricacies of practical importance, as discussed in \cref{sec:yamada_main} and further applied to representative material Fermi surfaces in  \cref{sec:materials_main} -- see \cref{appx:materials,appx:compression-body} for elaborations on the boundary-wise treatment required for multicomponent surfaces and nested inner boundaries encountered in these materials. %

\subsection{Topological Characterization of Knotted Graphs\label{sec:yamada_main}}

Having explained the general workflow, we now develop the specific topological characterization tools (see \cref{appx:glossary} for a glossary of technical terms). As described previously, what needs to be characterized is the sequence of knotted graphs that represent the embedded skeletons corresponding to the Fermi-surfaces at various $E$.
Formally, a \textit{knotted graph} is an embedding of an abstract finite graph $\mathcal{G}$ in a three-manifold, here the Brillouin zone $\cong T^3$, considered up to ambient isotopy i.e. spatial entanglement. Ordinary knots and links (shown in green) constitute the vertex-free subclass of generic knotted graphs (shown in blue) [\cref{fig:gallery}(a)]. Allowing the knots/links to intersect at vertices (red balls) produces the considerably broader range of structures in \cref{fig:gallery}(b--f), in which distinct vertex configurations give rise to dissimilar graph connectivity structures. %
Whether two knotted graphs are topologically equivalent or not is governed by the
generalized Reidemeister moves [\cref{fig:gallery}(g)]: the classical
 moves \textbf{(I)--(III)} discern whether the conventional knotting structures are equivalent, while the additional vertex moves \textbf{(IV)--(VI)} allow for the presence of vertices %
(\cref{appx:Yamada}).

Ordinary knot invariants cannot characterize knotted graphs because %
they contain vertices, and are hence not embedded
1D manifolds. Going beyond knot invariants, a starting point is the \textit{Yamada polynomial} invariant $\Upsilon(\mathcal{G};Y)$
\cite{yamada1989invariant,
kauffmanInvariantsGraphsThreeSpace1989,
mellorInvariantsSpatialGraphs2018}. While existing Yamada polynomials are insufficient for representing many real material dispersions, as explained later below, we set the stage by first explaining their construction. 

We begin with the planar projection of a given knotted graph $\mathcal{G}$. Its Yamada polynomial can be obtained by recursively decomposing $\mathcal{G}$ via Axiom \textbf{(I)} in
\cref{fig:gallery}(h): at each iteration, each crossing is resolved into two ordinary
crossings and an additional vertex, producing three ``descendent" knotted graphs with one fewer crossing or vertex. The Yamada polynomial of $\mathcal{G}$ can then be expressed as a linear superposition of the Yamada polynomials for these three descendents, weighted by powers of the argument $Y$ as indicated (see \cref{fig:YamadaResolutions} for a worked example). The constant-weight vertex term is precisely what is absent in ordinary knot-polynomial constructions~\cite{yamada1989invariant,kauffmanInvariantsGraphsThreeSpace1989,mellorInvariantsSpatialGraphs2018}.
Recursively repeating this process produces a number of ever-simpler knotted graphs, until all crossings and vertices have been resolved. The other axioms serve to ``clean up" the Yamada polynomial: Axiom \textbf{(II)} handles non-loop edges through deletion--contraction, while Axioms \textbf{(III)--(VI)} encode residual disjoint unions, one-vertex joins, bouquets, and empty graphs.

Below each knotted graph in \cref{fig:gallery}, we display $\overline{\Upsilon}(\mathcal{G};Y)$, the normalized form of this Yamada polynomial~\footnote{The unnormalized Yamada polynomial $\Upsilon$ is invariant under moves \textbf{(II)--(IV)}, corresponding to regular rigid-vertex isotopy. Its normalized form $\overline{\Upsilon}$ removes the monomial factors associated with moves \textbf{(I)} and \textbf{(V)} and is therefore invariant under moves \textbf{(I)--(V)}, corresponding to rigid-vertex isotopy. Ambient isotopy additionally permits edge reordering at vertices through move \textbf{(VI)}. If all vertices have degree $\le 3$---as is the case for every example in \cref{fig:gallery}---$\overline{\Upsilon}$ is consequently a full ambient-isotopy invariant. For graphs containing higher-degree vertices, which do arise for the real materials in \cref{appx:materials}, we use the minimum-crossing planar-projection convention detailed in \cref{appx:Yamada}.}.
The displayed polynomials correspond respectively to the skeletonized Fermi surfaces of $H_{\text{Hopf}}(E=0)$, $H_{\text{Unknot}}(E=0.9)$, $H_{\text{Hopf}}(E=0.2)$, $H_{\text{Trefoil}}(E=0.3)$, $H_{\text{Solomon}}(E=0.3)$, and $H_{\text{Three-link}}(E=0.1)$; their explicit expressions are given in \cref{appx:ansatz}. A larger number of vertices and crossings generally enlarges the recursive state sum and produces higher-degree polynomials.
Note that for a single knotted closed loop without vertices, this construction reduces to a specialization of the Jones polynomial for ordinary knots; for \cref{fig:gallery}(b--f), the vertex-resolution term retains the branching information that ordinary knot invariants necessarily discard.

\subsection{Fingerprinting Real Materials\label{sec:materials_main}}
\begin{figure}[t]
    \centering
    \includegraphics[width=\linewidth]{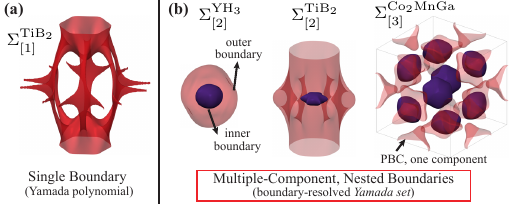}
    \caption{\small
    \textbf{Inadequacy of Yamada polynomials in topologically characterizing realistic material Fermi surfaces. }
    \textbf{(a)} Existing Yamada polynomials can only characterize one embedded knotted graph, which corresponds to a single-boundary (handlebody) Fermi volume, for instance that of $\Sigma_{[1]}^{\mathrm{TiB}_{2}}$ shown.
    \textbf{(b)} Realistic materials with multiple components and nested boundaries (compression bodies) cannot be properly encoded with a single knotted graph. Topologically characterizing them requires our new boundary-resolved Yamada set description, as elaborated in \cref{appx:compression-body}.
    Here, $\Sigma_{[2]}^{\mathrm{YH}_{3}}$ contains an enclosed inner pocket; $\Sigma_{[2]}^{\mathrm{TiB}_{2}}$ contains a more intricate inner boundary inside the outer shell; and $\Sigma_{[3]}^{\mathrm{Co}_{2}\mathrm{MnGa}}$ contains a globally spherical outer surface enclosing eight disconnected internal gaps.
    }
    \label{fig:handlebody_compression_dichotomy}
\end{figure}

The above Yamada polynomial characterization is sufficient only for the restricted class of Fermi volumes that are \textit{handlebodies}---solid regions obtained from a ball by attaching handles. The key simplifying condition is that the Fermi volume has a single boundary component, such that it retracts onto one embedded knotted graph $\mathcal{G}$, %
as is the case for $\Sigma_{[1]}^{\mathrm{TiB}_{2}}$ in \cref{fig:handlebody_compression_dichotomy}(a). For such cases, this knotted graph can be straightforwardly extracted using the algorithmic skeletonization procedure elaborated in \cref{appx:skeletonization}. For example, the orange Fermi volume bounded by $\Sigma$ in \cref{fig:hopfsequence}(b) retracts onto the knotted graph $\mathcal{G}$ shown within it. The graphs $\mathcal{G}_{[i]}$ embedded in the surfaces $\Sigma_{[i]}$ in \cref{fig:hopfsequence}(c), together with the examples in \cref{fig:gallery}(a--f), are obtained in the same way.

Realistic Fermi surfaces, however, generally go beyond the single-boundary handlebody setting, and \emph{cannot} be topologically classified with the existing Yamada polynomials formalism described. The obstruction is directly visible in the material examples in \cref{fig:handlebody_compression_dichotomy}(b), with full Fermi surface sequences shown in \cref{fig:YH3,fig:TiB2_c,fig:Co2mnGa2cellsize}. In $\Sigma_{[2]}^{\mathrm{YH}_{3}}$, the gaps separating the outer lobes close as energy $E$ increases while a new enclosed pocket remains, producing a closed outer boundary together with a distinct inner boundary. In $\Sigma_{[2]}^{\mathrm{TiB}_{2}}$, the outer shell encloses a more intricate, flower-shaped inner boundary, while $\Sigma_{[3]}^{\mathrm{Co}_{2}\mathrm{MnGa}}$ contains a globally spherical outer surface surrounding eight disconnected internal gaps.
These multi-boundary Fermi surfaces are common in realistic materials and are no longer handlebodies that can be characterized by Yamada polynomials -- mathematically, they are termed \emph{compression bodies}, which can contain additional inner boundary components that represent the enclosed voids of the Fermi volume. 

To overcome this key limitation, we develop a boundary-resolved extension of the Yamada classification from handlebodies to compression bodies. Our boundary-resolved Yamada set can encode the boundary hierarchy that records the outer and inner boundary components, including their nesting. 
Its complete mathematical construction and equivalence relations, as well as its implementation for our material examples, are mathematically subtle and elaborated at length in \cref{appx:compression-body,appx:materials}. Here in the main text, we simply summarize the key steps: For every connected component of every outer or inner boundary, our construction extracts a knotted graph $\mathcal{G}_{\alpha}$ and retains its normalized Yamada polynomial separately. We collect these boundary-wise invariants into the new \textit{Yamada set}
\begin{equation}
\overline{\Upsilon}_{\partial\mathcal{F}}
=
\left\{
\overline{\Upsilon}
\!\left(\mathcal{G}_{\alpha};Y\right)
\right\}_{\alpha\in\pi_0(\partial\mathcal{F})},
\label{eq:yamada-set-main}
\end{equation}
where $\mathcal{F}$ denotes the Fermi volume, $\partial\mathcal{F}$ its boundary, and $\alpha$ runs over the connected components of $\partial\mathcal{F}$. The nesting relations among these components are retained together with their individual invariants; the full definition is given in \cref{eq:YamadaSET,eq:yamada_Set_seq_def}. This construction records the outer boundary, the boundaries of enclosed cavities, and their rearrangements across Lifshitz transitions. It reduces to a single Yamada polynomial when $\mathcal{F}$ is a handlebody without inner boundaries.

\begin{figure*}[t]
    \centering
    \includegraphics[width=1\linewidth]{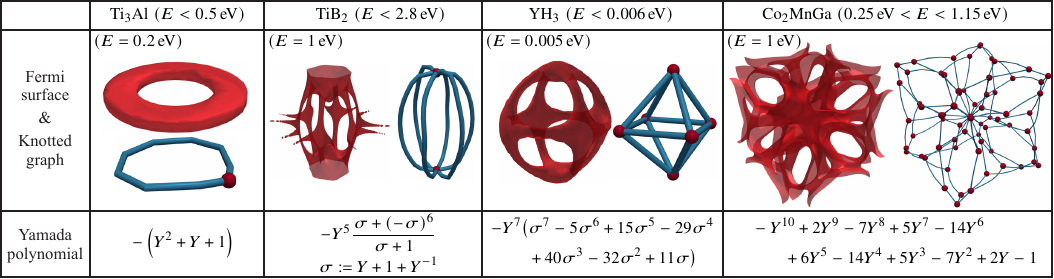}
    \caption{\small
    \textbf{Knotted-graph fingerprints of illustrative Fermi surfaces in real materials.}
    \textbf{(Top Row)} Fermi surfaces of Ti$_3$Al~\cite{Zhang2018}, TiB$_2$~\cite{Feng2018,Liu2018,Yi2018}, YH$_3$~\cite{Shao_2018,huiberts1996synthesis,wang1995structural}, and Co$_2$MnGa~\cite{Chang2017,Ilya2019,Guin_2019,Markou_2019} at selected energies (indicated in parentheses), computed from DFT-fitted effective Hamiltonians~\cite{Kohn1996,kormanyos2015k} (see \cref{appx:materials} for details), together with their skeletonized knotted graphs;
    \textbf{(Bottom Row)} Corresponding normalized Yamada polynomials $\overline{\Upsilon}$, where $\sigma := Y + 1 + Y^{-1}$.
    For graphical clarity, only representative single-boundary (handlebody) windows are shown; the full sequences, including windows that require our extended Yamada set formulation, are given in \cref{appx:materials}.
    }
    \label{fig:materials}
\end{figure*}

With this boundary-resolved framework established, we now compare the resulting dispersion-level fingerprints across a broader set of real materials. For graphical clarity, \cref{fig:materials} displays representative single-component windows from each material, while the full fingerprints, including multicomponent windows that require the Yamada-set formulation, are detailed in \cref{appx:materials}.
The first row shows their Fermi surfaces together with the extracted knotted graphs at the indicated Fermi energies, while the second row shows the corresponding Yamada polynomials. From left to right, these are a single torus-like pocket (Ti$_3$Al~\cite{Zhang2018}), symmetry-organized multi-tube surfaces (TiB$_2$~\cite{Feng2018,Liu2018,Yi2018}), an octahedral skeleton motif (YH$_3$~\cite{Shao_2018,huiberts1996synthesis,wang1995structural}), and a highly connected multi-component structure (Co$_2$MnGa~\cite{Chang2017,Ilya2019,Guin_2019,Markou_2019}). The Yamada sequences of these material Hamiltonians (\cref{appx:materials}) provide a systematic theoretical map of the Lifshitz transitions expected under Fermi-level tuning. Consequently, they furnish a compact topological guide to the associated changes in the density of states, transport properties, and other experimentally accessible observables. %
This connection opens new avenues for understanding how the intrinsic connectivity of the Fermi surface governs measurable electronic signatures. Beyond these individual Fermi surface fingerprints, we next show how the same framework resolves continuous topological evolution under microscopic deformations of both model and realistic material Hamiltonians.

\subsection{Phase Diagram of Knotted Graph Topology\label{sec:deform_main}}
We next ask how an energy-resolved topological fingerprint evolves when the Hamiltonian itself is deformed, which gives a broader view of the intricate topological structures revealed by our knotted graph framework. As a minimal example, we interpolate between the unknot and Hopf-link nodal Hamiltonians:
\begin{equation}
H_\lambda
=
(1-\lambda)\,H_{\mathrm{Unknot}}
+
\lambda\,H_{\mathrm{Hopf}},
\qquad
\lambda\in[0,1],
\label{eq:lambda_continous_transform}
\end{equation}
where $H_{\mathrm{Unknot}}$ and $H_{\mathrm{Hopf}}$ are given by
\cref{eq:H_Unknot,eq:H_Hopf}. %

\begin{figure}[t]
    \centering
    \includegraphics[width=\linewidth]
    {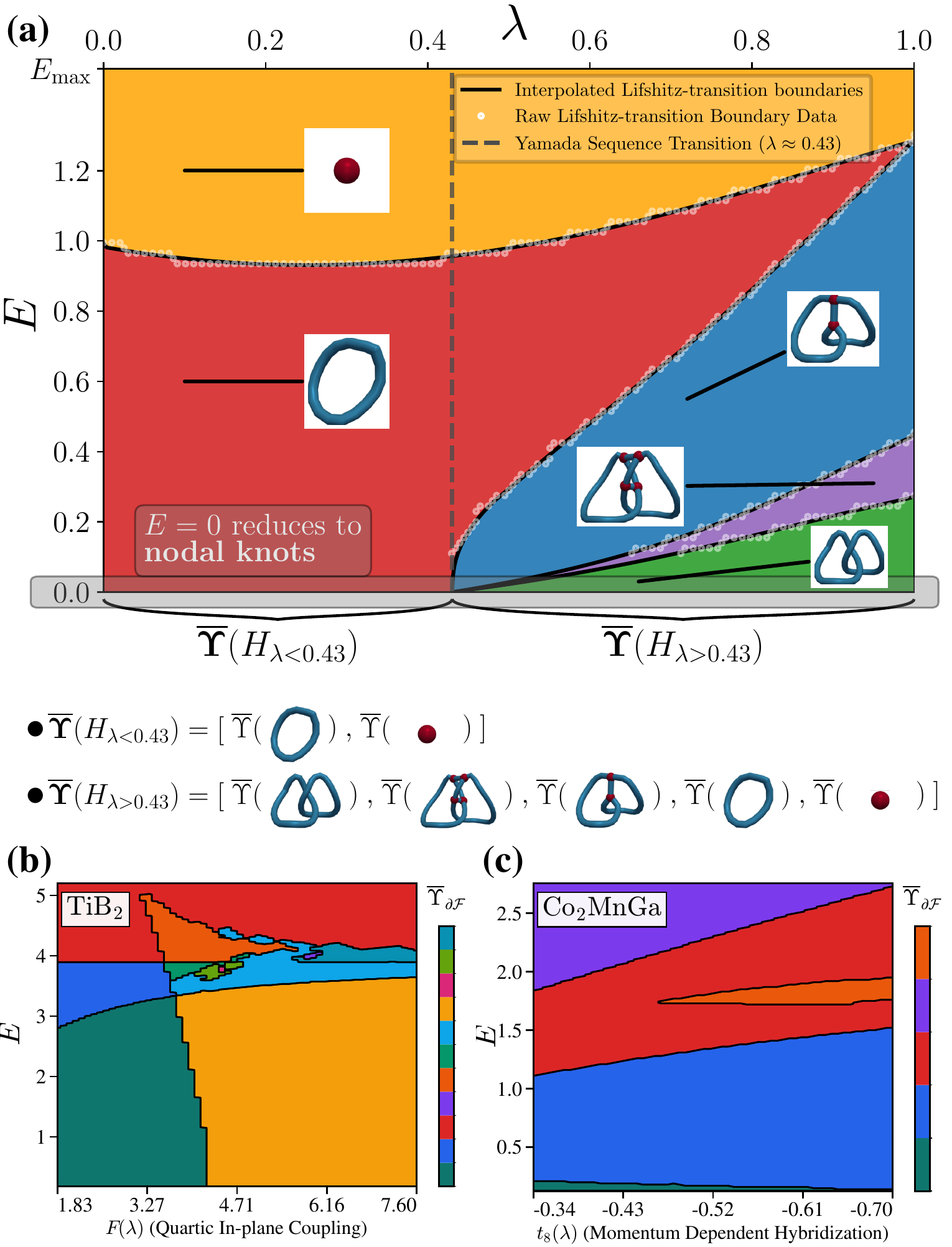}
     \caption{\small \textbf{Topological phase diagrams and dispersion-level knotted-graph fingerprints.} \textbf{(a)} Phase diagram of the interpolating Hamiltonian $H(\lambda)$ from \cref{eq:lambda_continous_transform}, resolved as a function of Fermi energy $E$ and interpolation parameter $\lambda$. Colored regions denote distinct knotted-graph ambient-isotopy classes of the constant-energy surfaces, separated by numerically detected Lifshitz-transition boundaries (faint white points). For each fixed $\lambda$, a vertical cut defines the Yamada sequence \cref{eq:yamadaseq_def} encountered as $E$ is increased. The dashed vertical line at $\lambda\simeq0.43$ marks the Yamada-sequence transition, coincident with the conventional nodal-knot transition at $E=0$. \textbf{(b,c)} Material topology phase diagrams obtained by varying one microscopic parameter while holding all others fixed: \textbf{(b)} the quartic in-plane coupling $F$, from $1.83$ to $7.60\,\mathrm{eV}\,\text{\AA}^{4}$, in the $\Gamma$-centered three-band $k\!\cdot\!p$ model of TiB$_2$~\cite{Feng2018}; \textbf{(c)} the momentum-dependent $d$--$p$ hybridization coefficient $t_8$, from $-0.34$ to $-0.70$, in the six-band model of Co$_2$MnGa~\cite{Chang2017}. The parameter paths and the energy coordinate $E$ are specified in \cref{appx:material_parameter_maps}. Colors label different topological sectors that are characterized by Yamada set $\overline{\Upsilon}_{\partial\mathcal{F}}$ [\cref{eq:yamada-set-main}]. Representative surfaces, knotted-graphs, and numerical details are in \cref{appx:material_parameter_maps}.}
    \label{fig:topophasespace}
\end{figure}

Such interpolations model the continuous band-structure changes driven by experimental tuning parameters under which the evolving Hamiltonian takes the form $H_{\mathrm{mat}}(\mathbf{k};\lambda)=\sum_a c_a\!\left[\lambda\right]O_a(\mathbf{k})$, with fixed operators $O_a(\mathbf{k})$ encoding the momentum, orbital, or hopping structure and the coefficients $c_a[\lambda]$ representing parameter-dependent microscopic quantities such as hopping amplitudes, orbital energies, and spin--orbit couplings~\cite{Shi_2017,dos_ref,transport_ref,temp_induced_lifshitz,Zhang_2017}. We first establish the resulting parameter-resolved topology through this minimal unknot--Hopf interpolation in \cref{fig:topophasespace}(a), and subsequently demonstrate the same construction for the material Hamiltonians of TiB$_2$ and Co$_2$MnGa in \cref{fig:topophasespace}(b,c).

Since the unknot and Hopf link are topologically inequivalent, the $E\rightarrow 0$ Fermi surface of $H_\lambda$ must already possess a topological transition at some intermediate $\lambda$. What is more interesting and unexplored is the prospect of new topological transitions across \emph{all} Fermi energies $E$, which would collectively define the topological fingerprint of its entire dispersion. At every fixed $\lambda$, recall [see \cref{fig:hopfsequence}(c)] that we can characterize this fingerprint by the ordered \textit{Yamada sequence}
\begin{equation}
\boldsymbol{\overline{\Upsilon}}(H):=
\bigl[\,
\overline{\Upsilon}\left(\mathcal{G}_{[1]}\right),
\overline{\Upsilon}\left(\mathcal{G}_{[2]}\right),
\ldots,
\overline{\Upsilon}\left(\mathcal{G}_{[m]}\right)
\,\bigr],
\label{eq:yamadaseq_def}
\end{equation}
where the $\overline{\Upsilon}\!\left(\mathcal{G}_{[i]}\right)$ are the normalized Yamada polynomials for all the distinct knotted graph topologies encountered as the Fermi surface varies across the full range of $E$. Note that $\boldsymbol{\overline{\Upsilon}}$ (boldface) is a property of the full Hamiltonian $H$, not of any single Fermi surface or knotted graph; each entry $\overline{\Upsilon}(\mathcal{G}_{[i]})$ (non-bold) is the invariant of an individual knotted graph. For volumes with multiple boundary components, each entry is replaced by the boundary-resolved Yamada set $\overline{\Upsilon}_{\partial\mathcal F_{[i]}}$ in \cref{eq:yamada-set-main}.

As shown in \cref{fig:topophasespace}(a), the Fermi energy $E$ and the deformation parameter $\lambda$ define a two-dimensional topological phase space. The expected unknot to Hopf-link transition (red to green) at $\lambda\simeq0.43$ (dashed) lives in the one-dimensional $E=0$ subspace, which bounds a much richer extended $(E,\lambda)$ phase diagram. The Lifshitz-transition boundaries marked by black curves partition it into several distinct finite-energy Fermi-surface configurations that cannot be inferred from the $E=0$ nodal slice alone.

This phase diagram can be systematically understood by examining vertical cuts at fixed $\lambda$, starting at $E=0$ and ending at $E_\text{max}> 1.2$ where all Fermi surfaces become topologically trivial balls.
For $\lambda<0.43$, there are only two topological configurations: a toroidal Fermi surface corresponding to the unknot (red), and a simply-connected pocket whose skeleton reduces to the isolated vertex (yellow).
For $\lambda>0.43$, the Fermi surface instead undergoes five distinct knotted graph configurations with varying $E$: a low-energy phase with two linked toroidal components corresponding to a Hopf-link skeleton (green), which merge to form a higher-genus Fermi surface characterized by a linked-graph skeleton (purple), and subsequently a connected Fermi surface with $\theta_3$-type skeleton (blue). At still higher energies, the system enters the same red unknot and yellow isolated-vertex phases encountered universally; the latter is inevitable because the Fermi surface thickens and eventually congeals into a ball as $E$ increases.

These energy-resolved topological regions provide an organizing framework for material behavior: within each energy window, the Fermi-surface topology remains fixed, whereas the boundaries between windows identify Lifshitz transitions at which Fermi-surface-sensitive observables---such as the density of states, transport coefficients, thermodynamic response, and superconducting behavior---can exhibit singularities or qualitative changes~\cite{dos_ref,transport_ref,Zhang_2017,Slizovskiy2015,Shi_2017,Varlamov_2021}. The Yamada sequences [\cref{eq:yamadaseq_def}] listed in \cref{fig:topophasespace}(a) provide compact labels for these distinct paths through the topological phase space. For $\lambda > 0.43$, the sequence is already given in \cref{fig:hopfsequence}(c); while for $\lambda<0.43$ it is given by
\begin{equation}
\boldsymbol{\overline{\Upsilon}}(H_{\lambda<0.43})
=
\bigl[-(Y^2+Y+1),-1\bigr].
\end{equation}
as follows from Axioms \textbf{(V)} from \cref{fig:gallery}. Horizontal cuts at constant $E$ instead reveal how a fixed surface topology persists under Hamiltonian deformation. At $E=0.8$, for example, the red region extends beyond $\lambda=0.7$, indicating that the toroidal (thickened nodal ring) topology is far more stable than what the conventional $E=0$ knot invariant suggests: while the $E\approx 0$ Fermi surface transforms from an unknot to a Hopf link at $\lambda =0.43$, its thickened larger-$E$ counterpart remains an unknot much longer.

This parameter-resolved construction extends directly to realistic multiband Hamiltonians. In TiB$_2$, we vary the quartic in-plane coupling $F$ entering $h_{12}=Ck_-^2+Fk_+^4$ [\cref{eq:tib2_h12_parameter_map}], which controls the competition between two in-plane hybridization channels; in Co$_2$MnGa, we vary the momentum-dependent $d$--$p$ hybridization coefficient $t_8$ entering the hybridization terms in \cref{eq:co2mnga_t8_terms}. For the material phase diagrams in \cref{fig:topophasespace}(b,c), $E$
parametrizes the interband-gap scale, as detailed in \cref{appx:materials}. The resulting topology phase diagrams in \cref{fig:topophasespace}(b,c) reveal multiple intermediate topological sectors for TiB$_2$ and five distinct topological sectors for Co$_2$MnGa, resolving how the momentum-space topology reorganizes as microscopic Hamiltonian parameters are continuously varied. Such microscopic deformations can be facilitated experimentally through tuning parameters such as strain~\cite{Sunko2019}, pressure~\cite{Xiang2015}, or composition~\cite{Dziawa2012}; once these controls are related to the microscopic Hamiltonian coefficients, the resulting phase maps can guide searches for Lifshitz transitions and allow experimentalists to determine where in parameter space their characteristic signatures should emerge in the density of states~\cite{dos_ref}, transport~\cite{transport_ref,Zhang_2017,Guin_2019,Markou_2019}, thermodynamic response~\cite{Slizovskiy2015}, and superconductivity~\cite{Shi_2017}. The parameter paths, representative surfaces and knotted-graph skeletons, and numerical details are provided in \cref{appx:material_parameter_maps}.

\begin{figure}[t]
    \centering
    \includegraphics[width=1\linewidth]{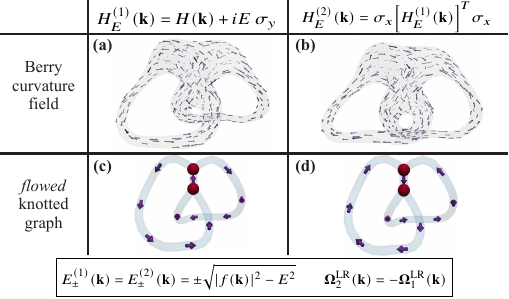}
   \caption{\small
    \textbf{Flowed knotted graphs capture eigenstate geometry beyond dispersion.}
    For the non-Hermitian Hopf-link model $H_E^{(1)}(\mathbf{k})=H_{\mathrm{Hopf}}(\mathbf{k})+iE\sigma_y$ [\cref{eq:H_Hopf}],
    \textbf{(a)} the biorthogonal Berry-curvature field inside the exceptional surface and
    \textbf{(c)} the corresponding Berry flux orient the skeleton edges, producing a \emph{flowed knotted graph} (\cref{appx:ES,appx:oriantability_of_knottedgraphs}).
    This extends knotted-graph topology to surfaces endowed with an internal vector field.
    \textbf{(b,\,d)} The flow carries information beyond the dispersion. We construct the isospectral partner $H_E^{(2)}=\sigma_x[H_E^{(1)}]^T\sigma_x$ [\cref{eq:isospectral_pair}], which shares the same complex spectrum, exceptional surface, and unflowed skeleton. Its Berry curvature and resulting flow are everywhere opposite.
    This distinction is invisible to the spectrum or to any invariant of the unflowed graph; it is captured only by the flow, which fingerprints the eigenstate geometry beyond merely dispersion (\cref{appx:isospectral-opposite-flow}).
    }
    \label{fig:directed}
\end{figure}

\subsection{Flowed Knotted Graphs beyond Dispersion Topology: Exceptional Surfaces with Berry Curvature Flow\label{sec:nh_main}}
Having established a rigorous fermiology based on knotted graphs, we show that the framework is not limited to this setting, but adapts naturally to generic topological classifications of 2D surfaces in 3D beyond the genus. As one example, it applies to non-Hermitian band structures, specifically in symmetry-protected settings where band degeneracies are constrained to form \textit{exceptional surfaces}~\cite{yang2026exceptional,stalhammar2021prb,kawabata2019prl,wang2024berry}. In the two-band case, the exceptional surface is the locus of momentum points that satisfy the band-closing condition of
\begin{equation}\label{eq:NonhermitianThickening}
H_E(\mathbf{k})=H(\mathbf{k})+iE\,\sigma_y,
\end{equation}
which is the Hamiltonian $H(k)$ of \cref{eq:two_band_hamiltonian} subject to a gain/loss term proportional to $E$~\footnote{Here, the non-reciprocal $\sigma_y$ term is also the non-Hermitian contribution, but basis-equivalent results can be obtained even if the Pauli matrices are permuted. }. Its exceptional surface is given by precisely the same condition $|f(\mathbf{k})|=E$. Thus, our procedure given in \cref{fig:hopfsequence}(b) and the Yamada fingerprints naturally classify exceptional surfaces as well. But, as described below, exceptional surfaces are accompanied by additional topological information beyond their spatial geometry: the Berry-curvature field within them equips the corresponding knotted-graph skeleton with an intrinsic flow. Our framework can therefore be extended from classifying the exceptional surface alone to jointly characterizing its topology and the vector field supported within it.

While the full Hamiltonian remains $\mathcal{PT}$ symmetric, its eigenstates spontaneously break this symmetry inside the exceptional surface, where the eigenvalues form a complex-conjugate pair. For a chosen nondegenerate band of $H_E$, let $|u^R\rangle$ and $\langle u^L|$ be the right and left eigenstates, normalized by $\langle u^L|u^R\rangle=1$. Their biorthogonal Berry curvature is $\boldsymbol{\Omega}=i\nabla_{\mathbf{k}}\times\langle u^L|\nabla_{\mathbf{k}}u^R\rangle$, which becomes singular on approaching the exceptional surface~\cite{wang2024berry} (\cref{appx:ES}). Its net flux through a cross section of each tubular branch defines a natural edge orientation, which we later prove to be an \textit{Abelian edge flow}, thus producing a \emph{flowed} knotted graph as illustrated in \cref{fig:directed}.

Flux conservation within the enclosed region then imposes the Kirchhoff-type vertex condition $\sum_{i=1}^{N}\Phi_i=0$, where $\Phi_i:=\iint_{D_i}\boldsymbol{\Omega}(\mathbf{k})\cdot d\mathbf{S}_i$ is the biorthogonal Berry-curvature flux through a transverse section $D_i$ of the $i$-th incident tube, with $d\mathbf{S}_i$ oriented outward from the vertex [see \cref{eq:edge_flux_def,eq:kirchhoff_stationary}]. These fluxes carry information about the \emph{eigenstate} geometry beyond the eigenenergy dispersions. Accordingly, the edge orientations are not arbitrary labels: the vertex condition restricts the admissible flow patterns and determines how the Berry-curvature flux is distributed among branches meeting at a vertex. Whereas the ordinary Yamada fingerprint characterizes the underlying undirected knotted graph, the flowed graph additionally retains how the Berry-curvature flux is routed through its branches.

To demonstrate concretely that the flow carries information beyond the dispersion, we construct an isospectral pair of Hamiltonians $H_E^{(1)}$ and $H_E^{(2)}=\sigma_x[H_E^{(1)}]^T\sigma_x$ that share the same complex spectrum, exceptional surface, and unflowed skeleton, yet have everywhere opposite biorthogonal Berry curvatures $\boldsymbol{\Omega}_1^\text{LR}=-\boldsymbol{\Omega}_2^\text{LR}$ (\cref{fig:directed}). Their flowed knotted graphs are $(\mathcal{G},\varphi)$ and $(\mathcal{G},-\varphi)$---a distinction invisible to any dispersion-based or unflowed-graph invariant. The flow therefore constitutes a \textit{joint spectral--eigenstate fingerprint}: the unflowed graph $\mathcal{G}$ captures the exceptional-volume embedding determined by the eigenvalues, while the signed flow $\varphi$ captures the Berry-curvature geometry of the biorthogonal eigenstates (see \cref{appx:isospectral-opposite-flow} for details).

More broadly, the flowed knotted graphs connect to the notion of Abelian $A$-flows~\cite{ishii2012quandle}, and further flowed-graph invariants (e.g., quandle/cocycle constructions) may refine exceptional-surface fingerprints with flux information invisible to undirected invariants, with potential applications to knotted vortex tubes~\cite{ricca1996topological,Kleckner2013,kleckner2016superfluid} and magnetic flux ropes~\cite{Liu_2020,Bellan_2003}.

\section{Discussion}
We have developed a knotted-graph framework that provides a rigorous, dispersion-level topological fingerprint for generic electronic materials and beyond. Introducing knotted graphs and their Yamada polynomial invariants into fermiology already establishes a classification language far richer than conventional knot-theoretic or genus-based approaches. On top of this, applying the framework to realistic Fermi surfaces required going beyond the existing mathematical literature: the Fermi volumes of real materials generically possess multiple connected components and nested inner boundaries---compression bodies, not handlebodies---which existing Yamada polynomials cannot encode. To overcome this, we proved a new compression-body equivalence theorem, formulated boundary-wise knotted graph data, and introduced the Yamada set, enabling the first systematic Yamada-polynomial fingerprinting of real materials. The framework even further extends to surfaces endowed with vector fields: in non-Hermitian systems, the biorthogonal Berry curvature equips the knotted-graph skeleton with an Abelian edge flow, and this flow captures eigenstate geometry that is entirely invisible to the complex spectrum alone or to any unflowed-graph invariant.

Looking ahead, a natural mathematical direction is to determine whether other embedding-sensitive spatial-graph invariants can be extended to compression bodies through representative sets analogous to the Yamada set, and whether such constructions provide more discriminating or computationally efficient descriptions of their spatial topology. In particular, the same boundary-resolved construction suggests analogous Yokota- and Thompson-polynomial sets, providing alternative compression-body fingerprints with different invariance and computational properties. The cycle structure of flowed spatial graphs may further organize inequivalent Berry-flux integration paths with distinct topological roles in higher-dimensional parameter spaces where the corresponding fluxes can become quantized. Moreover our construction naturally motivates the broader notion of \textit{flowed compression bodies}, in which a vector field or directional structure carried by the embedded region is represented through corresponding sets of oriented or flow-carrying spatial graphs, enabling the joint characterization of spatial embedding and physically relevant flow structure. At a still more global level, the evolving 1D knotted graph generated as the Fermi energy is swept traces a 2D knotted surface in the extended $(3+1)$-dimensional momentum--energy space. Describing the full dispersion as a single embedded surface therefore offers a complementary formulation that may connect the present framework to surface-knot theory, including 2-knots in appropriate closed cases, and capture global topological information across the complete energy evolution. This perspective also raises an inverse problem: given a Yamada sequence, or more generally an energy-resolved sequence of spatial-topological invariants, what constraints does it impose on the class of compatible dispersions, and can candidate band structures or Hamiltonians be reconstructed directly from such data? 

A broader application-oriented direction is to develop this framework into a scalable methodology for systematically characterizing realistic electronic structures across materials, energies and experimentally accessible control parameters. Further optimization of graph extraction, invariant evaluation and phase-space tracking could enable high-throughput integration with DFT and Wannier-based workflows, allowing Lifshitz transitions and changes in Fermi-surface connectivity and spatial entanglement to be identified across large material and parameter spaces. At the materials-database scale, Yamada sequences and related spatial-topological descriptors could provide compact fingerprints for comparing materials, identifying robust and recurring topological structures, and screening for candidate systems with targeted Fermi-surface connectivity, Berry-flux structure or Lifshitz-transition behavior. More broadly, such a high-throughput topological description could enable systematic investigation of how momentum-space topology correlates with material properties across large material families, providing a route toward identifying topology–property relationships and using them to guide materials discovery.

\paragraph*{Acknowledgments}
This research is supported by the Singapore Ministry of Education (MOE) through the Academic Research Fund Tier-II Grant MOE-T2EP50224-0007.
During the preparation of this manuscript, the authors used ChatGPT and Claude to assist with language refinement and the revision of selected passages originally drafted by the authors. All AI-assisted revisions were reviewed and edited by the authors, who take full responsibility for the final content.

\paragraph*{Data and Code Availability}
All data and code associated with this work are available from the corresponding authors upon reasonable request.

\bibliographystyle{apsrev4-2}
\bibliography{references}

\begin{thebibliography}{129}%
\makeatletter
\providecommand \@ifxundefined [1]{%
 \@ifx{#1\undefined}
}%
\providecommand \@ifnum [1]{%
 \ifnum #1\expandafter \@firstoftwo
 \else \expandafter \@secondoftwo
 \fi
}%
\providecommand \@ifx [1]{%
 \ifx #1\expandafter \@firstoftwo
 \else \expandafter \@secondoftwo
 \fi
}%
\providecommand \natexlab [1]{#1}%
\providecommand \enquote  [1]{``#1''}%
\providecommand \bibnamefont  [1]{#1}%
\providecommand \bibfnamefont [1]{#1}%
\providecommand \citenamefont [1]{#1}%
\providecommand \href@noop [0]{\@secondoftwo}%
\providecommand \href [0]{\begingroup \@sanitize@url \@href}%
\providecommand \@href[1]{\@@startlink{#1}\@@href}%
\providecommand \@@href[1]{\endgroup#1\@@endlink}%
\providecommand \@sanitize@url [0]{\catcode `\\12\catcode `\$12\catcode `\&12\catcode `\#12\catcode `\^12\catcode `\_12\catcode `\%12\relax}%
\providecommand \@@startlink[1]{}%
\providecommand \@@endlink[0]{}%
\providecommand \url  [0]{\begingroup\@sanitize@url \@url }%
\providecommand \@url [1]{\endgroup\@href {#1}{\urlprefix }}%
\providecommand \urlprefix  [0]{URL }%
\providecommand \Eprint [0]{\href }%
\providecommand \doibase [0]{https://doi.org/}%
\providecommand \selectlanguage [0]{\@gobble}%
\providecommand \bibinfo  [0]{\@secondoftwo}%
\providecommand \bibfield  [0]{\@secondoftwo}%
\providecommand \translation [1]{[#1]}%
\providecommand \BibitemOpen [0]{}%
\providecommand \bibitemStop [0]{}%
\providecommand \bibitemNoStop [0]{.\EOS\space}%
\providecommand \EOS [0]{\spacefactor3000\relax}%
\providecommand \BibitemShut  [1]{\csname bibitem#1\endcsname}%
\let\auto@bib@innerbib\@empty
\bibitem [{\citenamefont {Ryu}\ \emph {et~al.}(2010)\citenamefont {Ryu}, \citenamefont {Schnyder}, \citenamefont {Furusaki},\ and\ \citenamefont {Ludwig}}]{Ryu_2010}%
  \BibitemOpen
  \bibfield  {author} {\bibinfo {author} {\bibfnamefont {S.}~\bibnamefont {Ryu}}, \bibinfo {author} {\bibfnamefont {A.~P.}\ \bibnamefont {Schnyder}}, \bibinfo {author} {\bibfnamefont {A.}~\bibnamefont {Furusaki}},\ and\ \bibinfo {author} {\bibfnamefont {A.~W.~W.}\ \bibnamefont {Ludwig}},\ }\href {https://doi.org/10.1088/1367-2630/12/6/065010} {\bibfield  {journal} {\bibinfo  {journal} {New Journal of Physics}\ }\textbf {\bibinfo {volume} {12}},\ \bibinfo {pages} {065010} (\bibinfo {year} {2010})}\BibitemShut {NoStop}%
\bibitem [{\citenamefont {Schnyder}\ \emph {et~al.}(2008)\citenamefont {Schnyder}, \citenamefont {Ryu}, \citenamefont {Furusaki},\ and\ \citenamefont {Ludwig}}]{Schnyder2008}%
  \BibitemOpen
  \bibfield  {author} {\bibinfo {author} {\bibfnamefont {A.~P.}\ \bibnamefont {Schnyder}}, \bibinfo {author} {\bibfnamefont {S.}~\bibnamefont {Ryu}}, \bibinfo {author} {\bibfnamefont {A.}~\bibnamefont {Furusaki}},\ and\ \bibinfo {author} {\bibfnamefont {A.~W.~W.}\ \bibnamefont {Ludwig}},\ }\href {https://doi.org/10.1103/PhysRevB.78.195125} {\bibfield  {journal} {\bibinfo  {journal} {Phys. Rev. B}\ }\textbf {\bibinfo {volume} {78}},\ \bibinfo {pages} {195125} (\bibinfo {year} {2008})}\BibitemShut {NoStop}%
\bibitem [{\citenamefont {Chiu}\ \emph {et~al.}(2016)\citenamefont {Chiu}, \citenamefont {Teo}, \citenamefont {Schnyder},\ and\ \citenamefont {Ryu}}]{Chiu2016}%
  \BibitemOpen
  \bibfield  {author} {\bibinfo {author} {\bibfnamefont {C.-K.}\ \bibnamefont {Chiu}}, \bibinfo {author} {\bibfnamefont {J.~C.~Y.}\ \bibnamefont {Teo}}, \bibinfo {author} {\bibfnamefont {A.~P.}\ \bibnamefont {Schnyder}},\ and\ \bibinfo {author} {\bibfnamefont {S.}~\bibnamefont {Ryu}},\ }\href {https://doi.org/10.1103/RevModPhys.88.035005} {\bibfield  {journal} {\bibinfo  {journal} {Rev. Mod. Phys.}\ }\textbf {\bibinfo {volume} {88}},\ \bibinfo {pages} {035005} (\bibinfo {year} {2016})}\BibitemShut {NoStop}%
\bibitem [{\citenamefont {Vanderbilt}(2018)}]{vanderbilt2018berry}%
  \BibitemOpen
  \bibfield  {author} {\bibinfo {author} {\bibfnamefont {D.}~\bibnamefont {Vanderbilt}},\ }\href@noop {} {\emph {\bibinfo {title} {Berry phases in electronic structure theory: electric polarization, orbital magnetization and topological insulators}}}\ (\bibinfo  {publisher} {Cambridge University Press},\ \bibinfo {year} {2018})\BibitemShut {NoStop}%
\bibitem [{\citenamefont {Bi}\ \emph {et~al.}(2017)\citenamefont {Bi}, \citenamefont {Yan}, \citenamefont {Lu},\ and\ \citenamefont {Wang}}]{Bi_2017}%
  \BibitemOpen
  \bibfield  {author} {\bibinfo {author} {\bibfnamefont {R.}~\bibnamefont {Bi}}, \bibinfo {author} {\bibfnamefont {Z.}~\bibnamefont {Yan}}, \bibinfo {author} {\bibfnamefont {L.}~\bibnamefont {Lu}},\ and\ \bibinfo {author} {\bibfnamefont {Z.}~\bibnamefont {Wang}},\ }\bibfield  {journal} {\bibinfo  {journal} {Physical Review B}\ }\textbf {\bibinfo {volume} {96}},\ \href {https://doi.org/10.1103/physrevb.96.201305} {10.1103/physrevb.96.201305} (\bibinfo {year} {2017})\BibitemShut {NoStop}%
\bibitem [{\citenamefont {Yan}\ \emph {et~al.}(2017)\citenamefont {Yan}, \citenamefont {Bi}, \citenamefont {Shen}, \citenamefont {Lu}, \citenamefont {Zhang},\ and\ \citenamefont {Wang}}]{Yan_2017}%
  \BibitemOpen
  \bibfield  {author} {\bibinfo {author} {\bibfnamefont {Z.}~\bibnamefont {Yan}}, \bibinfo {author} {\bibfnamefont {R.}~\bibnamefont {Bi}}, \bibinfo {author} {\bibfnamefont {H.}~\bibnamefont {Shen}}, \bibinfo {author} {\bibfnamefont {L.}~\bibnamefont {Lu}}, \bibinfo {author} {\bibfnamefont {S.-C.}\ \bibnamefont {Zhang}},\ and\ \bibinfo {author} {\bibfnamefont {Z.}~\bibnamefont {Wang}},\ }\bibfield  {journal} {\bibinfo  {journal} {Physical Review B}\ }\textbf {\bibinfo {volume} {96}},\ \href {https://doi.org/10.1103/physrevb.96.041103} {10.1103/physrevb.96.041103} (\bibinfo {year} {2017})\BibitemShut {NoStop}%
\bibitem [{\citenamefont {Li}\ \emph {et~al.}(2018)\citenamefont {Li}, \citenamefont {Lee},\ and\ \citenamefont {Gong}}]{li2018realistic}%
  \BibitemOpen
  \bibfield  {author} {\bibinfo {author} {\bibfnamefont {L.}~\bibnamefont {Li}}, \bibinfo {author} {\bibfnamefont {C.~H.}\ \bibnamefont {Lee}},\ and\ \bibinfo {author} {\bibfnamefont {J.}~\bibnamefont {Gong}},\ }\href@noop {} {\bibfield  {journal} {\bibinfo  {journal} {Physical review letters}\ }\textbf {\bibinfo {volume} {121}},\ \bibinfo {pages} {036401} (\bibinfo {year} {2018})}\BibitemShut {NoStop}%
\bibitem [{\citenamefont {Stålhammar}\ \emph {et~al.}(2019)\citenamefont {Stålhammar}, \citenamefont {Rødland}, \citenamefont {Arone}, \citenamefont {Budich},\ and\ \citenamefont {Bergholtz}}]{St_lhammar_2019}%
  \BibitemOpen
  \bibfield  {author} {\bibinfo {author} {\bibfnamefont {M.}~\bibnamefont {Stålhammar}}, \bibinfo {author} {\bibfnamefont {L.}~\bibnamefont {Rødland}}, \bibinfo {author} {\bibfnamefont {G.}~\bibnamefont {Arone}}, \bibinfo {author} {\bibfnamefont {J.~C.}\ \bibnamefont {Budich}},\ and\ \bibinfo {author} {\bibfnamefont {E.}~\bibnamefont {Bergholtz}},\ }\bibfield  {journal} {\bibinfo  {journal} {SciPost Physics}\ }\textbf {\bibinfo {volume} {7}},\ \href {https://doi.org/10.21468/scipostphys.7.2.019} {10.21468/scipostphys.7.2.019} (\bibinfo {year} {2019})\BibitemShut {NoStop}%
\bibitem [{\citenamefont {Li}\ \emph {et~al.}(2019)\citenamefont {Li}, \citenamefont {Lee},\ and\ \citenamefont {Gong}}]{Li2019}%
  \BibitemOpen
  \bibfield  {author} {\bibinfo {author} {\bibfnamefont {L.}~\bibnamefont {Li}}, \bibinfo {author} {\bibfnamefont {C.~H.}\ \bibnamefont {Lee}},\ and\ \bibinfo {author} {\bibfnamefont {J.}~\bibnamefont {Gong}},\ }\href {https://doi.org/10.1038/s42005-019-0235-4} {\bibfield  {journal} {\bibinfo  {journal} {Communications Physics}\ }\textbf {\bibinfo {volume} {2}},\ \bibinfo {pages} {135} (\bibinfo {year} {2019})}\BibitemShut {NoStop}%
\bibitem [{\citenamefont {Carlstr\"om}\ \emph {et~al.}(2019)\citenamefont {Carlstr\"om}, \citenamefont {St\aa{}lhammar}, \citenamefont {Budich},\ and\ \citenamefont {Bergholtz}}]{Carlstrom2019}%
  \BibitemOpen
  \bibfield  {author} {\bibinfo {author} {\bibfnamefont {J.}~\bibnamefont {Carlstr\"om}}, \bibinfo {author} {\bibfnamefont {M.}~\bibnamefont {St\aa{}lhammar}}, \bibinfo {author} {\bibfnamefont {J.~C.}\ \bibnamefont {Budich}},\ and\ \bibinfo {author} {\bibfnamefont {E.~J.}\ \bibnamefont {Bergholtz}},\ }\href {https://doi.org/10.1103/PhysRevB.99.161115} {\bibfield  {journal} {\bibinfo  {journal} {Phys. Rev. B}\ }\textbf {\bibinfo {volume} {99}},\ \bibinfo {pages} {161115(R)} (\bibinfo {year} {2019})}\BibitemShut {NoStop}%
\bibitem [{\citenamefont {Lee}\ \emph {et~al.}(2020{\natexlab{a}})\citenamefont {Lee}, \citenamefont {Sutrisno}, \citenamefont {Hofmann}, \citenamefont {Helbig}, \citenamefont {Liu}, \citenamefont {Ang}, \citenamefont {Ang}, \citenamefont {Zhang}, \citenamefont {Greiter},\ and\ \citenamefont {Thomale}}]{Lee_2020b}%
  \BibitemOpen
  \bibfield  {author} {\bibinfo {author} {\bibfnamefont {C.~H.}\ \bibnamefont {Lee}}, \bibinfo {author} {\bibfnamefont {A.}~\bibnamefont {Sutrisno}}, \bibinfo {author} {\bibfnamefont {T.}~\bibnamefont {Hofmann}}, \bibinfo {author} {\bibfnamefont {T.}~\bibnamefont {Helbig}}, \bibinfo {author} {\bibfnamefont {Y.}~\bibnamefont {Liu}}, \bibinfo {author} {\bibfnamefont {Y.~S.}\ \bibnamefont {Ang}}, \bibinfo {author} {\bibfnamefont {L.~K.}\ \bibnamefont {Ang}}, \bibinfo {author} {\bibfnamefont {X.}~\bibnamefont {Zhang}}, \bibinfo {author} {\bibfnamefont {M.}~\bibnamefont {Greiter}},\ and\ \bibinfo {author} {\bibfnamefont {R.}~\bibnamefont {Thomale}},\ }\bibfield  {journal} {\bibinfo  {journal} {Nature Communications}\ }\textbf {\bibinfo {volume} {11}},\ \href {https://doi.org/10.1038/s41467-020-17716-1} {10.1038/s41467-020-17716-1} (\bibinfo {year} {2020}{\natexlab{a}})\BibitemShut {NoStop}%
\bibitem [{\citenamefont {Patil}\ \emph {et~al.}(2022)\citenamefont {Patil}, \citenamefont {Höller}, \citenamefont {Henry}, \citenamefont {Guria}, \citenamefont {Zhang}, \citenamefont {Jiang}, \citenamefont {Kralj}, \citenamefont {Read},\ and\ \citenamefont {Harris}}]{Patil_2022}%
  \BibitemOpen
  \bibfield  {author} {\bibinfo {author} {\bibfnamefont {Y.~S.~S.}\ \bibnamefont {Patil}}, \bibinfo {author} {\bibfnamefont {J.}~\bibnamefont {Höller}}, \bibinfo {author} {\bibfnamefont {P.~A.}\ \bibnamefont {Henry}}, \bibinfo {author} {\bibfnamefont {C.}~\bibnamefont {Guria}}, \bibinfo {author} {\bibfnamefont {Y.}~\bibnamefont {Zhang}}, \bibinfo {author} {\bibfnamefont {L.}~\bibnamefont {Jiang}}, \bibinfo {author} {\bibfnamefont {N.}~\bibnamefont {Kralj}}, \bibinfo {author} {\bibfnamefont {N.}~\bibnamefont {Read}},\ and\ \bibinfo {author} {\bibfnamefont {J.~G.~E.}\ \bibnamefont {Harris}},\ }\href {https://doi.org/10.1038/s41586-022-04796-w} {\bibfield  {journal} {\bibinfo  {journal} {Nature}\ }\textbf {\bibinfo {volume} {607}},\ \bibinfo {pages} {271–275} (\bibinfo {year} {2022})}\BibitemShut {NoStop}%
\bibitem [{\citenamefont {Tai}\ and\ \citenamefont {Lee}(2023)}]{Tai2023}%
  \BibitemOpen
  \bibfield  {author} {\bibinfo {author} {\bibfnamefont {T.}~\bibnamefont {Tai}}\ and\ \bibinfo {author} {\bibfnamefont {C.~H.}\ \bibnamefont {Lee}},\ }\href {https://doi.org/10.1103/PhysRevB.107.L220301} {\bibfield  {journal} {\bibinfo  {journal} {Phys. Rev. B}\ }\textbf {\bibinfo {volume} {107}},\ \bibinfo {pages} {L220301} (\bibinfo {year} {2023})}\BibitemShut {NoStop}%
\bibitem [{\citenamefont {Lin}\ \emph {et~al.}(2023)\citenamefont {Lin}, \citenamefont {Tai}, \citenamefont {Li},\ and\ \citenamefont {Lee}}]{Lin_2023}%
  \BibitemOpen
  \bibfield  {author} {\bibinfo {author} {\bibfnamefont {R.}~\bibnamefont {Lin}}, \bibinfo {author} {\bibfnamefont {T.}~\bibnamefont {Tai}}, \bibinfo {author} {\bibfnamefont {L.}~\bibnamefont {Li}},\ and\ \bibinfo {author} {\bibfnamefont {C.~H.}\ \bibnamefont {Lee}},\ }\bibfield  {journal} {\bibinfo  {journal} {Frontiers of Physics}\ }\textbf {\bibinfo {volume} {18}},\ \href {https://doi.org/10.1007/s11467-023-1309-z} {10.1007/s11467-023-1309-z} (\bibinfo {year} {2023})\BibitemShut {NoStop}%
\bibitem [{\citenamefont {Qin}\ \emph {et~al.}(2024)\citenamefont {Qin}, \citenamefont {Shen}, \citenamefont {Li},\ and\ \citenamefont {Lee}}]{qin2024kinked}%
  \BibitemOpen
  \bibfield  {author} {\bibinfo {author} {\bibfnamefont {F.}~\bibnamefont {Qin}}, \bibinfo {author} {\bibfnamefont {R.}~\bibnamefont {Shen}}, \bibinfo {author} {\bibfnamefont {L.}~\bibnamefont {Li}},\ and\ \bibinfo {author} {\bibfnamefont {C.~H.}\ \bibnamefont {Lee}},\ }\href@noop {} {\bibfield  {journal} {\bibinfo  {journal} {Physical Review A}\ }\textbf {\bibinfo {volume} {109}},\ \bibinfo {pages} {053311} (\bibinfo {year} {2024})}\BibitemShut {NoStop}%
\bibitem [{\citenamefont {Yang}\ \emph {et~al.}(2020{\natexlab{a}})\citenamefont {Yang}, \citenamefont {Zhang}, \citenamefont {Fang},\ and\ \citenamefont {Hu}}]{yang2020non}%
  \BibitemOpen
  \bibfield  {author} {\bibinfo {author} {\bibfnamefont {Z.}~\bibnamefont {Yang}}, \bibinfo {author} {\bibfnamefont {K.}~\bibnamefont {Zhang}}, \bibinfo {author} {\bibfnamefont {C.}~\bibnamefont {Fang}},\ and\ \bibinfo {author} {\bibfnamefont {J.}~\bibnamefont {Hu}},\ }\href@noop {} {\bibfield  {journal} {\bibinfo  {journal} {Physical Review Letters}\ }\textbf {\bibinfo {volume} {125}},\ \bibinfo {pages} {226402} (\bibinfo {year} {2020}{\natexlab{a}})}\BibitemShut {NoStop}%
\bibitem [{\citenamefont {Li}\ \emph {et~al.}(2025)\citenamefont {Li}, \citenamefont {Jiang},\ and\ \citenamefont {Lee}}]{li2025phase}%
  \BibitemOpen
  \bibfield  {author} {\bibinfo {author} {\bibfnamefont {Q.}~\bibnamefont {Li}}, \bibinfo {author} {\bibfnamefont {H.}~\bibnamefont {Jiang}},\ and\ \bibinfo {author} {\bibfnamefont {C.~H.}\ \bibnamefont {Lee}},\ }\href@noop {} {\bibfield  {journal} {\bibinfo  {journal} {Advanced Science}\ }\textbf {\bibinfo {volume} {12}},\ \bibinfo {pages} {e08047} (\bibinfo {year} {2025})}\BibitemShut {NoStop}%
\bibitem [{\citenamefont {Gu}\ \emph {et~al.}(2026)\citenamefont {Gu}, \citenamefont {Fu}, \citenamefont {Hu},\ and\ \citenamefont {Wang}}]{gu2026long}%
  \BibitemOpen
  \bibfield  {author} {\bibinfo {author} {\bibfnamefont {D.}~\bibnamefont {Gu}}, \bibinfo {author} {\bibfnamefont {Z.}~\bibnamefont {Fu}}, \bibinfo {author} {\bibfnamefont {Y.-M.}\ \bibnamefont {Hu}},\ and\ \bibinfo {author} {\bibfnamefont {Z.}~\bibnamefont {Wang}},\ }\href@noop {} {\bibfield  {journal} {\bibinfo  {journal} {arXiv preprint arXiv:2608.28577}\ } (\bibinfo {year} {2026})}\BibitemShut {NoStop}%
\bibitem [{\citenamefont {Yang}\ \emph {et~al.}(2026{\natexlab{a}})\citenamefont {Yang}, \citenamefont {Poddubny},\ and\ \citenamefont {Lee}}]{yang2026nonlocality}%
  \BibitemOpen
  \bibfield  {author} {\bibinfo {author} {\bibfnamefont {M.}~\bibnamefont {Yang}}, \bibinfo {author} {\bibfnamefont {A.~N.}\ \bibnamefont {Poddubny}},\ and\ \bibinfo {author} {\bibfnamefont {C.~H.}\ \bibnamefont {Lee}},\ }\href@noop {} {\bibfield  {journal} {\bibinfo  {journal} {arXiv preprint arXiv:2608.02746}\ } (\bibinfo {year} {2026}{\natexlab{a}})}\BibitemShut {NoStop}%
\bibitem [{\citenamefont {Xiong}\ and\ \citenamefont {Hu}(2024)}]{Xiong2024}%
  \BibitemOpen
  \bibfield  {author} {\bibinfo {author} {\bibfnamefont {Y.}~\bibnamefont {Xiong}}\ and\ \bibinfo {author} {\bibfnamefont {H.}~\bibnamefont {Hu}},\ }\href {https://doi.org/10.1103/PhysRevB.109.L100301} {\bibfield  {journal} {\bibinfo  {journal} {Phys. Rev. B}\ }\textbf {\bibinfo {volume} {109}},\ \bibinfo {pages} {L100301} (\bibinfo {year} {2024})}\BibitemShut {NoStop}%
\bibitem [{\citenamefont {Pi}\ \emph {et~al.}(2025)\citenamefont {Pi}, \citenamefont {Wang}, \citenamefont {Liu},\ and\ \citenamefont {Yan}}]{Pi_2025}%
  \BibitemOpen
  \bibfield  {author} {\bibinfo {author} {\bibfnamefont {J.}~\bibnamefont {Pi}}, \bibinfo {author} {\bibfnamefont {C.}~\bibnamefont {Wang}}, \bibinfo {author} {\bibfnamefont {Y.-C.}\ \bibnamefont {Liu}},\ and\ \bibinfo {author} {\bibfnamefont {Y.}~\bibnamefont {Yan}},\ }\bibfield  {journal} {\bibinfo  {journal} {Physical Review B}\ }\textbf {\bibinfo {volume} {111}},\ \href {https://doi.org/10.1103/physrevb.111.165407} {10.1103/physrevb.111.165407} (\bibinfo {year} {2025})\BibitemShut {NoStop}%
\bibitem [{\citenamefont {Yan}\ \emph {et~al.}(2025)\citenamefont {Yan}, \citenamefont {Akgün}, \citenamefont {Kawaguchi}, \citenamefont {Loh},\ and\ \citenamefont {Lee}}]{yan2026hsg12mlargescalebenchmarkspatial}%
  \BibitemOpen
  \bibfield  {author} {\bibinfo {author} {\bibfnamefont {X.}~\bibnamefont {Yan}}, \bibinfo {author} {\bibfnamefont {H.}~\bibnamefont {Akgün}}, \bibinfo {author} {\bibfnamefont {K.}~\bibnamefont {Kawaguchi}}, \bibinfo {author} {\bibfnamefont {N.~D.}\ \bibnamefont {Loh}},\ and\ \bibinfo {author} {\bibfnamefont {C.~H.}\ \bibnamefont {Lee}},\ }\href {https://arxiv.org/abs/2506.08618} {\bibinfo {title} {Hsg-12m: A large-scale spatial multigraph dataset}} (\bibinfo {year} {2025}),\ \Eprint {https://arxiv.org/abs/2506.08618} {arXiv:2506.08618 [cs.LG]} \BibitemShut {NoStop}%
\bibitem [{\citenamefont {Sala}\ \emph {et~al.}(2020)\citenamefont {Sala}, \citenamefont {Rakovszky}, \citenamefont {Verresen}, \citenamefont {Knap},\ and\ \citenamefont {Pollmann}}]{Sala2020}%
  \BibitemOpen
  \bibfield  {author} {\bibinfo {author} {\bibfnamefont {P.}~\bibnamefont {Sala}}, \bibinfo {author} {\bibfnamefont {T.}~\bibnamefont {Rakovszky}}, \bibinfo {author} {\bibfnamefont {R.}~\bibnamefont {Verresen}}, \bibinfo {author} {\bibfnamefont {M.}~\bibnamefont {Knap}},\ and\ \bibinfo {author} {\bibfnamefont {F.}~\bibnamefont {Pollmann}},\ }\href {https://doi.org/10.1103/PhysRevX.10.011047} {\bibfield  {journal} {\bibinfo  {journal} {Phys. Rev. X}\ }\textbf {\bibinfo {volume} {10}},\ \bibinfo {pages} {011047} (\bibinfo {year} {2020})}\BibitemShut {NoStop}%
\bibitem [{\citenamefont {Khemani}\ \emph {et~al.}(2020)\citenamefont {Khemani}, \citenamefont {Hermele},\ and\ \citenamefont {Nandkishore}}]{Khemani2020}%
  \BibitemOpen
  \bibfield  {author} {\bibinfo {author} {\bibfnamefont {V.}~\bibnamefont {Khemani}}, \bibinfo {author} {\bibfnamefont {M.}~\bibnamefont {Hermele}},\ and\ \bibinfo {author} {\bibfnamefont {R.}~\bibnamefont {Nandkishore}},\ }\href {https://doi.org/10.1103/PhysRevB.101.174204} {\bibfield  {journal} {\bibinfo  {journal} {Phys. Rev. B}\ }\textbf {\bibinfo {volume} {101}},\ \bibinfo {pages} {174204} (\bibinfo {year} {2020})}\BibitemShut {NoStop}%
\bibitem [{\citenamefont {Lifshitz}\ \emph {et~al.}(1960)\citenamefont {Lifshitz} \emph {et~al.}}]{lifshitz1960anomalies}%
  \BibitemOpen
  \bibfield  {author} {\bibinfo {author} {\bibfnamefont {I.}~\bibnamefont {Lifshitz}} \emph {et~al.},\ }\href@noop {} {\bibfield  {journal} {\bibinfo  {journal} {Sov. Phys. JETP}\ }\textbf {\bibinfo {volume} {11}},\ \bibinfo {pages} {1130} (\bibinfo {year} {1960})}\BibitemShut {NoStop}%
\bibitem [{\citenamefont {Blanter}\ \emph {et~al.}(1994)\citenamefont {Blanter}, \citenamefont {Kaganov}, \citenamefont {Pantsulaya},\ and\ \citenamefont {Varlamov}}]{blanter1994theory}%
  \BibitemOpen
  \bibfield  {author} {\bibinfo {author} {\bibfnamefont {Y.~M.}\ \bibnamefont {Blanter}}, \bibinfo {author} {\bibfnamefont {M.}~\bibnamefont {Kaganov}}, \bibinfo {author} {\bibfnamefont {A.}~\bibnamefont {Pantsulaya}},\ and\ \bibinfo {author} {\bibfnamefont {A.}~\bibnamefont {Varlamov}},\ }\href@noop {} {\bibfield  {journal} {\bibinfo  {journal} {Physics Reports}\ }\textbf {\bibinfo {volume} {245}},\ \bibinfo {pages} {159} (\bibinfo {year} {1994})}\BibitemShut {NoStop}%
\bibitem [{\citenamefont {Varlamov}\ \emph {et~al.}(2021)\citenamefont {Varlamov}, \citenamefont {Galperin}, \citenamefont {Sharapov},\ and\ \citenamefont {Yerin}}]{Varlamov_2021}%
  \BibitemOpen
  \bibfield  {author} {\bibinfo {author} {\bibfnamefont {A.~A.}\ \bibnamefont {Varlamov}}, \bibinfo {author} {\bibfnamefont {Y.~M.}\ \bibnamefont {Galperin}}, \bibinfo {author} {\bibfnamefont {S.~G.}\ \bibnamefont {Sharapov}},\ and\ \bibinfo {author} {\bibfnamefont {Y.}~\bibnamefont {Yerin}},\ }\href {https://doi.org/10.1063/10.0005556} {\bibfield  {journal} {\bibinfo  {journal} {Low Temperature Physics}\ }\textbf {\bibinfo {volume} {47}},\ \bibinfo {pages} {672–683} (\bibinfo {year} {2021})}\BibitemShut {NoStop}%
\bibitem [{\citenamefont {Franzosi}\ \emph {et~al.}(2000)\citenamefont {Franzosi}, \citenamefont {Pettini},\ and\ \citenamefont {Spinelli}}]{Franzosi_2000}%
  \BibitemOpen
  \bibfield  {author} {\bibinfo {author} {\bibfnamefont {R.}~\bibnamefont {Franzosi}}, \bibinfo {author} {\bibfnamefont {M.}~\bibnamefont {Pettini}},\ and\ \bibinfo {author} {\bibfnamefont {L.}~\bibnamefont {Spinelli}},\ }\href {https://doi.org/10.1103/physrevlett.84.2774} {\bibfield  {journal} {\bibinfo  {journal} {Physical Review Letters}\ }\textbf {\bibinfo {volume} {84}},\ \bibinfo {pages} {2774–2777} (\bibinfo {year} {2000})}\BibitemShut {NoStop}%
\bibitem [{\citenamefont {Biasotti}\ \emph {et~al.}(2008)\citenamefont {Biasotti}, \citenamefont {Giorgi}, \citenamefont {Spagnuolo},\ and\ \citenamefont {Falcidieno}}]{BIASOTTI20085}%
  \BibitemOpen
  \bibfield  {author} {\bibinfo {author} {\bibfnamefont {S.}~\bibnamefont {Biasotti}}, \bibinfo {author} {\bibfnamefont {D.}~\bibnamefont {Giorgi}}, \bibinfo {author} {\bibfnamefont {M.}~\bibnamefont {Spagnuolo}},\ and\ \bibinfo {author} {\bibfnamefont {B.}~\bibnamefont {Falcidieno}},\ }\href {https://doi.org/10.1016/j.tcs.2007.10.018} {\bibfield  {journal} {\bibinfo  {journal} {Theoretical Computer Science}\ }\textbf {\bibinfo {volume} {392}},\ \bibinfo {pages} {5} (\bibinfo {year} {2008})},\ \bibinfo {note} {computational Algebraic Geometry and Applications}\BibitemShut {NoStop}%
\bibitem [{\citenamefont {Gibson}\ \emph {et~al.}(2006)\citenamefont {Gibson}, \citenamefont {Patel}, \citenamefont {Nagpal},\ and\ \citenamefont {Perrimon}}]{gibson2006emergence}%
  \BibitemOpen
  \bibfield  {author} {\bibinfo {author} {\bibfnamefont {M.~C.}\ \bibnamefont {Gibson}}, \bibinfo {author} {\bibfnamefont {A.~B.}\ \bibnamefont {Patel}}, \bibinfo {author} {\bibfnamefont {R.}~\bibnamefont {Nagpal}},\ and\ \bibinfo {author} {\bibfnamefont {N.}~\bibnamefont {Perrimon}},\ }\href@noop {} {\bibfield  {journal} {\bibinfo  {journal} {Nature}\ }\textbf {\bibinfo {volume} {442}},\ \bibinfo {pages} {1038} (\bibinfo {year} {2006})}\BibitemShut {NoStop}%
\bibitem [{\citenamefont {Gibson}\ and\ \citenamefont {Gibson}(2009)}]{GIBSON200987}%
  \BibitemOpen
  \bibfield  {author} {\bibinfo {author} {\bibfnamefont {W.~T.}\ \bibnamefont {Gibson}}\ and\ \bibinfo {author} {\bibfnamefont {M.~C.}\ \bibnamefont {Gibson}},\ }in\ \href {https://doi.org/10.1016/S0070-2153(09)89004-2} {\emph {\bibinfo {booktitle} {Current Topics in Developmental Biology}}},\ \bibinfo {series} {Current Topics in Developmental Biology}, Vol.~\bibinfo {volume} {89}\ (\bibinfo  {publisher} {Academic Press},\ \bibinfo {year} {2009})\ pp.\ \bibinfo {pages} {87--114}\BibitemShut {NoStop}%
\bibitem [{\citenamefont {Melott}(1990)}]{MELOTT19901}%
  \BibitemOpen
  \bibfield  {author} {\bibinfo {author} {\bibfnamefont {A.~L.}\ \bibnamefont {Melott}},\ }\href {https://doi.org/10.1016/0370-1573(90)90162-U} {\bibfield  {journal} {\bibinfo  {journal} {Physics Reports}\ }\textbf {\bibinfo {volume} {193}},\ \bibinfo {pages} {1} (\bibinfo {year} {1990})}\BibitemShut {NoStop}%
\bibitem [{\citenamefont {Copi}\ \emph {et~al.}(2026)\citenamefont {Copi}, \citenamefont {Mihaylov}, \citenamefont {Negro}, \citenamefont {Samandar}, \citenamefont {Starkman}, \citenamefont {Akrami}, \citenamefont {Alestas}, \citenamefont {Anselmi}, \citenamefont {Duque}, \citenamefont {Cornet-Gomez}, \citenamefont {Lu}, \citenamefont {Jaffe}, \citenamefont {Kosowsky}, \citenamefont {Barandiaran}, \citenamefont {Pereira}, \citenamefont {Petretti},\ and\ \citenamefont {Tamosiunas}}]{copi2026topologyuniverse}%
  \BibitemOpen
  \bibfield  {author} {\bibinfo {author} {\bibfnamefont {C.~J.}\ \bibnamefont {Copi}}, \bibinfo {author} {\bibfnamefont {D.~P.}\ \bibnamefont {Mihaylov}}, \bibinfo {author} {\bibfnamefont {A.}~\bibnamefont {Negro}}, \bibinfo {author} {\bibfnamefont {A.}~\bibnamefont {Samandar}}, \bibinfo {author} {\bibfnamefont {G.~D.}\ \bibnamefont {Starkman}}, \bibinfo {author} {\bibfnamefont {Y.}~\bibnamefont {Akrami}}, \bibinfo {author} {\bibfnamefont {G.}~\bibnamefont {Alestas}}, \bibinfo {author} {\bibfnamefont {S.}~\bibnamefont {Anselmi}}, \bibinfo {author} {\bibfnamefont {J.~C.}\ \bibnamefont {Duque}}, \bibinfo {author} {\bibfnamefont {F.}~\bibnamefont {Cornet-Gomez}}, \bibinfo {author} {\bibfnamefont {L.~H.}\ \bibnamefont {Lu}}, \bibinfo {author} {\bibfnamefont {A.~H.}\ \bibnamefont {Jaffe}}, \bibinfo {author} {\bibfnamefont {A.}~\bibnamefont {Kosowsky}}, \bibinfo {author} {\bibfnamefont {M.~M.}\ \bibnamefont {Barandiaran}}, \bibinfo {author} {\bibfnamefont {T.~S.}\ \bibnamefont {Pereira}}, \bibinfo {author} {\bibfnamefont {C.}~\bibnamefont {Petretti}},\ and\ \bibinfo {author} {\bibfnamefont {A.}~\bibnamefont {Tamosiunas}},\ }\href {https://arxiv.org/abs/2606.24886} {\bibinfo {title} {The topology of the universe}} (\bibinfo {year} {2026}),\ \Eprint {https://arxiv.org/abs/2606.24886} {arXiv:2606.24886 [astro-ph.CO]} \BibitemShut {NoStop}%
\bibitem [{\citenamefont {Hatcher}(2002)}]{hatcher2005algebraic}%
  \BibitemOpen
  \bibfield  {author} {\bibinfo {author} {\bibfnamefont {A.}~\bibnamefont {Hatcher}},\ }\href@noop {} {\emph {\bibinfo {title} {Algebraic topology}}}\ (\bibinfo  {publisher} {Cambridge University Press},\ \bibinfo {year} {2002})\BibitemShut {NoStop}%
\bibitem [{Note1()}]{Note1}%
  \BibitemOpen
  \bibinfo {note} {The genus only captures the topological connectivity of the surface of the 3D structure, not how its 3D ``bulk'' is embedded}\BibitemShut {NoStop}%
\bibitem [{\citenamefont {Edelsbrunner}\ and\ \citenamefont {Harer}(2008)}]{edelsbrunner2008persistent}%
  \BibitemOpen
  \bibfield  {author} {\bibinfo {author} {\bibfnamefont {H.}~\bibnamefont {Edelsbrunner}}\ and\ \bibinfo {author} {\bibfnamefont {J.}~\bibnamefont {Harer}},\ }in\ \href {https://doi.org/10.1090/conm/453/08802} {\emph {\bibinfo {booktitle} {Surveys on Discrete and Computational Geometry: Twenty Years Later}}},\ \bibinfo {series} {Contemporary Mathematics}, Vol.\ \bibinfo {volume} {453}\ (\bibinfo  {publisher} {American Mathematical Society},\ \bibinfo {year} {2008})\ pp.\ \bibinfo {pages} {257--282}\BibitemShut {NoStop}%
\bibitem [{\citenamefont {Carlsson}(2009)}]{carlsson2009topology}%
  \BibitemOpen
  \bibfield  {author} {\bibinfo {author} {\bibfnamefont {G.}~\bibnamefont {Carlsson}},\ }\href@noop {} {\bibfield  {journal} {\bibinfo  {journal} {Bulletin of the American mathematical society}\ }\textbf {\bibinfo {volume} {46}},\ \bibinfo {pages} {255} (\bibinfo {year} {2009})}\BibitemShut {NoStop}%
\bibitem [{\citenamefont {Galeski}\ \emph {et~al.}(2022)\citenamefont {Galeski}, \citenamefont {Legg}, \citenamefont {Wawrzyńczak}, \citenamefont {Förster}, \citenamefont {Zherlitsyn}, \citenamefont {Gorbunov}, \citenamefont {Uhlarz}, \citenamefont {Lozano}, \citenamefont {Li}, \citenamefont {Gu}, \citenamefont {Felser}, \citenamefont {Wosnitza}, \citenamefont {Meng},\ and\ \citenamefont {Gooth}}]{transport_ref}%
  \BibitemOpen
  \bibfield  {author} {\bibinfo {author} {\bibfnamefont {S.}~\bibnamefont {Galeski}}, \bibinfo {author} {\bibfnamefont {H.~F.}\ \bibnamefont {Legg}}, \bibinfo {author} {\bibfnamefont {R.}~\bibnamefont {Wawrzyńczak}}, \bibinfo {author} {\bibfnamefont {T.}~\bibnamefont {Förster}}, \bibinfo {author} {\bibfnamefont {S.}~\bibnamefont {Zherlitsyn}}, \bibinfo {author} {\bibfnamefont {D.}~\bibnamefont {Gorbunov}}, \bibinfo {author} {\bibfnamefont {M.}~\bibnamefont {Uhlarz}}, \bibinfo {author} {\bibfnamefont {P.~M.}\ \bibnamefont {Lozano}}, \bibinfo {author} {\bibfnamefont {Q.}~\bibnamefont {Li}}, \bibinfo {author} {\bibfnamefont {G.~D.}\ \bibnamefont {Gu}}, \bibinfo {author} {\bibfnamefont {C.}~\bibnamefont {Felser}}, \bibinfo {author} {\bibfnamefont {J.}~\bibnamefont {Wosnitza}}, \bibinfo {author} {\bibfnamefont {T.}~\bibnamefont {Meng}},\ and\ \bibinfo {author} {\bibfnamefont {J.}~\bibnamefont {Gooth}},\ }\bibfield  {journal} {\bibinfo  {journal} {Nature Communications}\ }\textbf {\bibinfo {volume} {13}},\ \href {https://doi.org/10.1038/s41467-022-35106-7} {10.1038/s41467-022-35106-7} (\bibinfo {year} {2022})\BibitemShut {NoStop}%
\bibitem [{\citenamefont {Zhang}\ \emph {et~al.}(2017)\citenamefont {Zhang}, \citenamefont {Wang}, \citenamefont {Yu}, \citenamefont {Liu}, \citenamefont {Liang}, \citenamefont {Huang}, \citenamefont {Nie}, \citenamefont {Sun}, \citenamefont {Zhang}, \citenamefont {Shen}, \citenamefont {Liu}, \citenamefont {Weng}, \citenamefont {Zhao}, \citenamefont {Chen}, \citenamefont {Jia}, \citenamefont {Hu}, \citenamefont {Ding}, \citenamefont {Zhao}, \citenamefont {Gao}, \citenamefont {Li}, \citenamefont {He}, \citenamefont {Zhao}, \citenamefont {Zhang}, \citenamefont {Zhang}, \citenamefont {Yang}, \citenamefont {Wang}, \citenamefont {Peng}, \citenamefont {Dai}, \citenamefont {Fang}, \citenamefont {Xu}, \citenamefont {Chen},\ and\ \citenamefont {Zhou}}]{Zhang_2017}%
  \BibitemOpen
  \bibfield  {author} {\bibinfo {author} {\bibfnamefont {Y.}~\bibnamefont {Zhang}}, \bibinfo {author} {\bibfnamefont {C.}~\bibnamefont {Wang}}, \bibinfo {author} {\bibfnamefont {L.}~\bibnamefont {Yu}}, \bibinfo {author} {\bibfnamefont {G.}~\bibnamefont {Liu}}, \bibinfo {author} {\bibfnamefont {A.}~\bibnamefont {Liang}}, \bibinfo {author} {\bibfnamefont {J.}~\bibnamefont {Huang}}, \bibinfo {author} {\bibfnamefont {S.}~\bibnamefont {Nie}}, \bibinfo {author} {\bibfnamefont {X.}~\bibnamefont {Sun}}, \bibinfo {author} {\bibfnamefont {Y.}~\bibnamefont {Zhang}}, \bibinfo {author} {\bibfnamefont {B.}~\bibnamefont {Shen}}, \bibinfo {author} {\bibfnamefont {J.}~\bibnamefont {Liu}}, \bibinfo {author} {\bibfnamefont {H.}~\bibnamefont {Weng}}, \bibinfo {author} {\bibfnamefont {L.}~\bibnamefont {Zhao}}, \bibinfo {author} {\bibfnamefont {G.}~\bibnamefont {Chen}}, \bibinfo {author} {\bibfnamefont {X.}~\bibnamefont {Jia}}, \bibinfo {author} {\bibfnamefont {C.}~\bibnamefont {Hu}}, \bibinfo {author} {\bibfnamefont {Y.}~\bibnamefont {Ding}}, \bibinfo {author} {\bibfnamefont {W.}~\bibnamefont {Zhao}}, \bibinfo {author} {\bibfnamefont {Q.}~\bibnamefont {Gao}}, \bibinfo {author} {\bibfnamefont {C.}~\bibnamefont {Li}}, \bibinfo {author} {\bibfnamefont {S.}~\bibnamefont {He}}, \bibinfo {author} {\bibfnamefont {L.}~\bibnamefont {Zhao}}, \bibinfo {author} {\bibfnamefont {F.}~\bibnamefont {Zhang}}, \bibinfo {author} {\bibfnamefont {S.}~\bibnamefont {Zhang}}, \bibinfo {author} {\bibfnamefont {F.}~\bibnamefont {Yang}}, \bibinfo {author} {\bibfnamefont {Z.}~\bibnamefont {Wang}}, \bibinfo {author} {\bibfnamefont {Q.}~\bibnamefont {Peng}}, \bibinfo {author} {\bibfnamefont {X.}~\bibnamefont {Dai}}, \bibinfo {author} {\bibfnamefont {Z.}~\bibnamefont {Fang}}, \bibinfo {author} {\bibfnamefont {Z.}~\bibnamefont {Xu}}, \bibinfo {author} {\bibfnamefont {C.}~\bibnamefont {Chen}},\ and\ \bibinfo {author} {\bibfnamefont {X.~J.}\ \bibnamefont {Zhou}},\ }\bibfield  {journal} {\bibinfo  {journal} {Nature Communications}\ }\textbf {\bibinfo {volume} {8}},\ \href {https://doi.org/10.1038/ncomms15512} {10.1038/ncomms15512} (\bibinfo {year} {2017})\BibitemShut {NoStop}%
\bibitem [{\citenamefont {Zhong}\ \emph {et~al.}(2022)\citenamefont {Zhong}, \citenamefont {Chen}, \citenamefont {Chen}, \citenamefont {Xu}, \citenamefont {Hashimoto}, \citenamefont {He}, \citenamefont {ichi Uchida}, \citenamefont {Lu}, \citenamefont {Mo},\ and\ \citenamefont {Shen}}]{thermo_superconductivityref}%
  \BibitemOpen
  \bibfield  {author} {\bibinfo {author} {\bibfnamefont {Y.}~\bibnamefont {Zhong}}, \bibinfo {author} {\bibfnamefont {Z.}~\bibnamefont {Chen}}, \bibinfo {author} {\bibfnamefont {S.-D.}\ \bibnamefont {Chen}}, \bibinfo {author} {\bibfnamefont {K.-J.}\ \bibnamefont {Xu}}, \bibinfo {author} {\bibfnamefont {M.}~\bibnamefont {Hashimoto}}, \bibinfo {author} {\bibfnamefont {Y.}~\bibnamefont {He}}, \bibinfo {author} {\bibfnamefont {S.}~\bibnamefont {ichi Uchida}}, \bibinfo {author} {\bibfnamefont {D.}~\bibnamefont {Lu}}, \bibinfo {author} {\bibfnamefont {S.-K.}\ \bibnamefont {Mo}},\ and\ \bibinfo {author} {\bibfnamefont {Z.-X.}\ \bibnamefont {Shen}},\ }\href {https://doi.org/10.1073/pnas.2204630119} {\bibfield  {journal} {\bibinfo  {journal} {Proceedings of the National Academy of Sciences}\ }\textbf {\bibinfo {volume} {119}},\ \bibinfo {pages} {e2204630119} (\bibinfo {year} {2022})},\ \Eprint {https://arxiv.org/abs/https://www.pnas.org/doi/pdf/10.1073/pnas.2204630119} {https://www.pnas.org/doi/pdf/10.1073/pnas.2204630119} \BibitemShut {NoStop}%
\bibitem [{\citenamefont {Slizovskiy}\ \emph {et~al.}(2015)\citenamefont {Slizovskiy}, \citenamefont {Chubukov},\ and\ \citenamefont {Betouras}}]{Slizovskiy2015}%
  \BibitemOpen
  \bibfield  {author} {\bibinfo {author} {\bibfnamefont {S.}~\bibnamefont {Slizovskiy}}, \bibinfo {author} {\bibfnamefont {A.~V.}\ \bibnamefont {Chubukov}},\ and\ \bibinfo {author} {\bibfnamefont {J.~J.}\ \bibnamefont {Betouras}},\ }\href {https://doi.org/10.1103/PhysRevLett.114.066403} {\bibfield  {journal} {\bibinfo  {journal} {Phys. Rev. Lett.}\ }\textbf {\bibinfo {volume} {114}},\ \bibinfo {pages} {066403} (\bibinfo {year} {2015})}\BibitemShut {NoStop}%
\bibitem [{\citenamefont {Kang}\ \emph {et~al.}(2015)\citenamefont {Kang}, \citenamefont {Zhou}, \citenamefont {Yi}, \citenamefont {Yang}, \citenamefont {Guo}, \citenamefont {Shi}, \citenamefont {Zhang}, \citenamefont {Wang}, \citenamefont {Zhang}, \citenamefont {Jiang}, \citenamefont {Li}, \citenamefont {Yang}, \citenamefont {Wu}, \citenamefont {Zhang}, \citenamefont {Sun},\ and\ \citenamefont {Zhao}}]{Kang_2015}%
  \BibitemOpen
  \bibfield  {author} {\bibinfo {author} {\bibfnamefont {D.}~\bibnamefont {Kang}}, \bibinfo {author} {\bibfnamefont {Y.}~\bibnamefont {Zhou}}, \bibinfo {author} {\bibfnamefont {W.}~\bibnamefont {Yi}}, \bibinfo {author} {\bibfnamefont {C.}~\bibnamefont {Yang}}, \bibinfo {author} {\bibfnamefont {J.}~\bibnamefont {Guo}}, \bibinfo {author} {\bibfnamefont {Y.}~\bibnamefont {Shi}}, \bibinfo {author} {\bibfnamefont {S.}~\bibnamefont {Zhang}}, \bibinfo {author} {\bibfnamefont {Z.}~\bibnamefont {Wang}}, \bibinfo {author} {\bibfnamefont {C.}~\bibnamefont {Zhang}}, \bibinfo {author} {\bibfnamefont {S.}~\bibnamefont {Jiang}}, \bibinfo {author} {\bibfnamefont {A.}~\bibnamefont {Li}}, \bibinfo {author} {\bibfnamefont {K.}~\bibnamefont {Yang}}, \bibinfo {author} {\bibfnamefont {Q.}~\bibnamefont {Wu}}, \bibinfo {author} {\bibfnamefont {G.}~\bibnamefont {Zhang}}, \bibinfo {author} {\bibfnamefont {L.}~\bibnamefont {Sun}},\ and\ \bibinfo {author} {\bibfnamefont {Z.}~\bibnamefont {Zhao}},\ }\bibfield  {journal} {\bibinfo  {journal} {Nature Communications}\ }\textbf {\bibinfo {volume} {6}},\ \href {https://doi.org/10.1038/ncomms8804} {10.1038/ncomms8804} (\bibinfo {year} {2015})\BibitemShut {NoStop}%
\bibitem [{\citenamefont {Shi}\ \emph {et~al.}(2017)\citenamefont {Shi}, \citenamefont {Han}, \citenamefont {Peng}, \citenamefont {Richard}, \citenamefont {Qian}, \citenamefont {Wu}, \citenamefont {Qiu}, \citenamefont {Wang}, \citenamefont {Hu}, \citenamefont {Sun},\ and\ \citenamefont {Ding}}]{Shi_2017}%
  \BibitemOpen
  \bibfield  {author} {\bibinfo {author} {\bibfnamefont {X.}~\bibnamefont {Shi}}, \bibinfo {author} {\bibfnamefont {Z.-Q.}\ \bibnamefont {Han}}, \bibinfo {author} {\bibfnamefont {X.-L.}\ \bibnamefont {Peng}}, \bibinfo {author} {\bibfnamefont {P.}~\bibnamefont {Richard}}, \bibinfo {author} {\bibfnamefont {T.}~\bibnamefont {Qian}}, \bibinfo {author} {\bibfnamefont {X.-X.}\ \bibnamefont {Wu}}, \bibinfo {author} {\bibfnamefont {M.-W.}\ \bibnamefont {Qiu}}, \bibinfo {author} {\bibfnamefont {S.~C.}\ \bibnamefont {Wang}}, \bibinfo {author} {\bibfnamefont {J.~P.}\ \bibnamefont {Hu}}, \bibinfo {author} {\bibfnamefont {Y.-J.}\ \bibnamefont {Sun}},\ and\ \bibinfo {author} {\bibfnamefont {H.}~\bibnamefont {Ding}},\ }\bibfield  {journal} {\bibinfo  {journal} {Nature Communications}\ }\textbf {\bibinfo {volume} {8}},\ \href {https://doi.org/10.1038/ncomms14988} {10.1038/ncomms14988} (\bibinfo {year} {2017})\BibitemShut {NoStop}%
\bibitem [{\citenamefont {Volovik}(2018)}]{superconduct_transition_temp}%
  \BibitemOpen
  \bibfield  {author} {\bibinfo {author} {\bibfnamefont {G.~E.}\ \bibnamefont {Volovik}},\ }\href@noop {} {\bibfield  {journal} {\bibinfo  {journal} {Physics--Uspekhi}\ }\textbf {\bibinfo {volume} {61}},\ \bibinfo {pages} {89} (\bibinfo {year} {2018})}\BibitemShut {NoStop}%
\bibitem [{\citenamefont {Lin}\ \emph {et~al.}(2017)\citenamefont {Lin}, \citenamefont {Hu}, \citenamefont {Chen}, \citenamefont {Lee},\ and\ \citenamefont {Zhang}}]{lin2017line}%
  \BibitemOpen
  \bibfield  {author} {\bibinfo {author} {\bibfnamefont {J.~Y.}\ \bibnamefont {Lin}}, \bibinfo {author} {\bibfnamefont {N.~C.}\ \bibnamefont {Hu}}, \bibinfo {author} {\bibfnamefont {Y.~J.}\ \bibnamefont {Chen}}, \bibinfo {author} {\bibfnamefont {C.~H.}\ \bibnamefont {Lee}},\ and\ \bibinfo {author} {\bibfnamefont {X.}~\bibnamefont {Zhang}},\ }\href@noop {} {\bibfield  {journal} {\bibinfo  {journal} {Physical Review B}\ ,\ \bibinfo {pages} {075438}} (\bibinfo {year} {2017})}\BibitemShut {NoStop}%
\bibitem [{\citenamefont {Chang}\ \emph {et~al.}(2017)\citenamefont {Chang}, \citenamefont {Xu}, \citenamefont {Zhou}, \citenamefont {Huang}, \citenamefont {Singh}, \citenamefont {Wang}, \citenamefont {Belopolski}, \citenamefont {Yin}, \citenamefont {Zhang}, \citenamefont {Bansil}, \citenamefont {Lin},\ and\ \citenamefont {Hasan}}]{Chang2017}%
  \BibitemOpen
  \bibfield  {author} {\bibinfo {author} {\bibfnamefont {G.}~\bibnamefont {Chang}}, \bibinfo {author} {\bibfnamefont {S.-Y.}\ \bibnamefont {Xu}}, \bibinfo {author} {\bibfnamefont {X.}~\bibnamefont {Zhou}}, \bibinfo {author} {\bibfnamefont {S.-M.}\ \bibnamefont {Huang}}, \bibinfo {author} {\bibfnamefont {B.}~\bibnamefont {Singh}}, \bibinfo {author} {\bibfnamefont {B.}~\bibnamefont {Wang}}, \bibinfo {author} {\bibfnamefont {I.}~\bibnamefont {Belopolski}}, \bibinfo {author} {\bibfnamefont {J.}~\bibnamefont {Yin}}, \bibinfo {author} {\bibfnamefont {S.}~\bibnamefont {Zhang}}, \bibinfo {author} {\bibfnamefont {A.}~\bibnamefont {Bansil}}, \bibinfo {author} {\bibfnamefont {H.}~\bibnamefont {Lin}},\ and\ \bibinfo {author} {\bibfnamefont {M.~Z.}\ \bibnamefont {Hasan}},\ }\href {https://doi.org/10.1103/PhysRevLett.119.156401} {\bibfield  {journal} {\bibinfo  {journal} {Phys. Rev. Lett.}\ }\textbf {\bibinfo {volume} {119}},\ \bibinfo {pages} {156401} (\bibinfo {year} {2017})}\BibitemShut {NoStop}%
\bibitem [{\citenamefont {Feng}\ \emph {et~al.}(2018)\citenamefont {Feng}, \citenamefont {Yue}, \citenamefont {Song}, \citenamefont {Wu},\ and\ \citenamefont {Wen}}]{Feng2018}%
  \BibitemOpen
  \bibfield  {author} {\bibinfo {author} {\bibfnamefont {X.}~\bibnamefont {Feng}}, \bibinfo {author} {\bibfnamefont {C.}~\bibnamefont {Yue}}, \bibinfo {author} {\bibfnamefont {Z.}~\bibnamefont {Song}}, \bibinfo {author} {\bibfnamefont {Q.}~\bibnamefont {Wu}},\ and\ \bibinfo {author} {\bibfnamefont {B.}~\bibnamefont {Wen}},\ }\href {https://doi.org/10.1103/PhysRevMaterials.2.014202} {\bibfield  {journal} {\bibinfo  {journal} {Phys. Rev. Mater.}\ }\textbf {\bibinfo {volume} {2}},\ \bibinfo {pages} {014202} (\bibinfo {year} {2018})}\BibitemShut {NoStop}%
\bibitem [{\citenamefont {Yi}\ \emph {et~al.}(2018)\citenamefont {Yi}, \citenamefont {Lv}, \citenamefont {Wu}, \citenamefont {Fu}, \citenamefont {Gao}, \citenamefont {Yang}, \citenamefont {Peng}, \citenamefont {Li}, \citenamefont {Huang}, \citenamefont {Richard}, \citenamefont {Shi}, \citenamefont {Li}, \citenamefont {Yazyev}, \citenamefont {Shi}, \citenamefont {Qian},\ and\ \citenamefont {Ding}}]{Yi2018}%
  \BibitemOpen
  \bibfield  {author} {\bibinfo {author} {\bibfnamefont {C.-J.}\ \bibnamefont {Yi}}, \bibinfo {author} {\bibfnamefont {B.~Q.}\ \bibnamefont {Lv}}, \bibinfo {author} {\bibfnamefont {Q.~S.}\ \bibnamefont {Wu}}, \bibinfo {author} {\bibfnamefont {B.-B.}\ \bibnamefont {Fu}}, \bibinfo {author} {\bibfnamefont {X.}~\bibnamefont {Gao}}, \bibinfo {author} {\bibfnamefont {M.}~\bibnamefont {Yang}}, \bibinfo {author} {\bibfnamefont {X.-L.}\ \bibnamefont {Peng}}, \bibinfo {author} {\bibfnamefont {M.}~\bibnamefont {Li}}, \bibinfo {author} {\bibfnamefont {Y.-B.}\ \bibnamefont {Huang}}, \bibinfo {author} {\bibfnamefont {P.}~\bibnamefont {Richard}}, \bibinfo {author} {\bibfnamefont {M.}~\bibnamefont {Shi}}, \bibinfo {author} {\bibfnamefont {G.}~\bibnamefont {Li}}, \bibinfo {author} {\bibfnamefont {O.~V.}\ \bibnamefont {Yazyev}}, \bibinfo {author} {\bibfnamefont {Y.-G.}\ \bibnamefont {Shi}}, \bibinfo {author} {\bibfnamefont {T.}~\bibnamefont {Qian}},\ and\ \bibinfo {author} {\bibfnamefont {H.}~\bibnamefont {Ding}},\ }\href {https://doi.org/10.1103/PhysRevB.97.201107} {\bibfield  {journal} {\bibinfo  {journal} {Phys. Rev. B}\ }\textbf {\bibinfo {volume} {97}},\ \bibinfo {pages} {201107} (\bibinfo {year} {2018})}\BibitemShut {NoStop}%
\bibitem [{\citenamefont {Xie}\ \emph {et~al.}(2019)\citenamefont {Xie}, \citenamefont {Cai}, \citenamefont {Kim}, \citenamefont {Chang},\ and\ \citenamefont {Chen}}]{Xie_2019}%
  \BibitemOpen
  \bibfield  {author} {\bibinfo {author} {\bibfnamefont {Y.}~\bibnamefont {Xie}}, \bibinfo {author} {\bibfnamefont {J.}~\bibnamefont {Cai}}, \bibinfo {author} {\bibfnamefont {J.}~\bibnamefont {Kim}}, \bibinfo {author} {\bibfnamefont {P.-Y.}\ \bibnamefont {Chang}},\ and\ \bibinfo {author} {\bibfnamefont {Y.}~\bibnamefont {Chen}},\ }\bibfield  {journal} {\bibinfo  {journal} {Physical Review B}\ }\textbf {\bibinfo {volume} {99}},\ \href {https://doi.org/10.1103/physrevb.99.165147} {10.1103/physrevb.99.165147} (\bibinfo {year} {2019})\BibitemShut {NoStop}%
\bibitem [{\citenamefont {Ding}\ \emph {et~al.}(2022)\citenamefont {Ding}, \citenamefont {Sun},\ and\ \citenamefont {Wang}}]{ding2022ideal}%
  \BibitemOpen
  \bibfield  {author} {\bibinfo {author} {\bibfnamefont {G.}~\bibnamefont {Ding}}, \bibinfo {author} {\bibfnamefont {T.}~\bibnamefont {Sun}},\ and\ \bibinfo {author} {\bibfnamefont {X.}~\bibnamefont {Wang}},\ }\href@noop {} {\bibfield  {journal} {\bibinfo  {journal} {Physical Chemistry Chemical Physics}\ }\textbf {\bibinfo {volume} {24}},\ \bibinfo {pages} {11175} (\bibinfo {year} {2022})}\BibitemShut {NoStop}%
\bibitem [{\citenamefont {Yan}\ \emph {et~al.}(2018)\citenamefont {Yan}, \citenamefont {Liu}, \citenamefont {Yan}, \citenamefont {Liu}, \citenamefont {Chen}, \citenamefont {Wang},\ and\ \citenamefont {Lu}}]{Yan_2018}%
  \BibitemOpen
  \bibfield  {author} {\bibinfo {author} {\bibfnamefont {Q.}~\bibnamefont {Yan}}, \bibinfo {author} {\bibfnamefont {R.}~\bibnamefont {Liu}}, \bibinfo {author} {\bibfnamefont {Z.}~\bibnamefont {Yan}}, \bibinfo {author} {\bibfnamefont {B.}~\bibnamefont {Liu}}, \bibinfo {author} {\bibfnamefont {H.}~\bibnamefont {Chen}}, \bibinfo {author} {\bibfnamefont {Z.}~\bibnamefont {Wang}},\ and\ \bibinfo {author} {\bibfnamefont {L.}~\bibnamefont {Lu}},\ }\href {https://doi.org/10.1038/s41567-017-0041-4} {\bibfield  {journal} {\bibinfo  {journal} {Nature Physics}\ }\textbf {\bibinfo {volume} {14}},\ \bibinfo {pages} {461–464} (\bibinfo {year} {2018})}\BibitemShut {NoStop}%
\bibitem [{\citenamefont {Yang}\ \emph {et~al.}(2020{\natexlab{b}})\citenamefont {Yang}, \citenamefont {Yang}, \citenamefont {You}, \citenamefont {Chan}, \citenamefont {Mao}, \citenamefont {Guo}, \citenamefont {Ma}, \citenamefont {Xia}, \citenamefont {Fan}, \citenamefont {Xiang},\ and\ \citenamefont {Zhang}}]{Yang_2020}%
  \BibitemOpen
  \bibfield  {author} {\bibinfo {author} {\bibfnamefont {E.}~\bibnamefont {Yang}}, \bibinfo {author} {\bibfnamefont {B.}~\bibnamefont {Yang}}, \bibinfo {author} {\bibfnamefont {O.}~\bibnamefont {You}}, \bibinfo {author} {\bibfnamefont {H.-C.}\ \bibnamefont {Chan}}, \bibinfo {author} {\bibfnamefont {P.}~\bibnamefont {Mao}}, \bibinfo {author} {\bibfnamefont {Q.}~\bibnamefont {Guo}}, \bibinfo {author} {\bibfnamefont {S.}~\bibnamefont {Ma}}, \bibinfo {author} {\bibfnamefont {L.}~\bibnamefont {Xia}}, \bibinfo {author} {\bibfnamefont {D.}~\bibnamefont {Fan}}, \bibinfo {author} {\bibfnamefont {Y.}~\bibnamefont {Xiang}},\ and\ \bibinfo {author} {\bibfnamefont {S.}~\bibnamefont {Zhang}},\ }\bibfield  {journal} {\bibinfo  {journal} {Physical Review Letters}\ }\textbf {\bibinfo {volume} {125}},\ \href {https://doi.org/10.1103/physrevlett.125.033901} {10.1103/physrevlett.125.033901} (\bibinfo {year} {2020}{\natexlab{b}})\BibitemShut {NoStop}%
\bibitem [{\citenamefont {Zhang}\ \emph {et~al.}(2018)\citenamefont {Zhang}, \citenamefont {Yu}, \citenamefont {Zhu}, \citenamefont {Wu}, \citenamefont {Wang}, \citenamefont {Sheng},\ and\ \citenamefont {Yang}}]{Zhang2018}%
  \BibitemOpen
  \bibfield  {author} {\bibinfo {author} {\bibfnamefont {X.}~\bibnamefont {Zhang}}, \bibinfo {author} {\bibfnamefont {Z.-M.}\ \bibnamefont {Yu}}, \bibinfo {author} {\bibfnamefont {Z.}~\bibnamefont {Zhu}}, \bibinfo {author} {\bibfnamefont {W.}~\bibnamefont {Wu}}, \bibinfo {author} {\bibfnamefont {S.-S.}\ \bibnamefont {Wang}}, \bibinfo {author} {\bibfnamefont {X.-L.}\ \bibnamefont {Sheng}},\ and\ \bibinfo {author} {\bibfnamefont {S.~A.}\ \bibnamefont {Yang}},\ }\href {https://doi.org/10.1103/PhysRevB.97.235150} {\bibfield  {journal} {\bibinfo  {journal} {Physical Review B}\ }\textbf {\bibinfo {volume} {97}},\ \bibinfo {pages} {235150} (\bibinfo {year} {2018})}\BibitemShut {NoStop}%
\bibitem [{\citenamefont {Liu}\ \emph {et~al.}(2018)\citenamefont {Liu}, \citenamefont {Lou}, \citenamefont {Guo}, \citenamefont {Wang}, \citenamefont {Sun}, \citenamefont {Li}, \citenamefont {Thirupathaiah}, \citenamefont {Fedorov}, \citenamefont {Shen}, \citenamefont {Liu}, \citenamefont {Lei},\ and\ \citenamefont {Wang}}]{Liu2018}%
  \BibitemOpen
  \bibfield  {author} {\bibinfo {author} {\bibfnamefont {Z.}~\bibnamefont {Liu}}, \bibinfo {author} {\bibfnamefont {R.}~\bibnamefont {Lou}}, \bibinfo {author} {\bibfnamefont {P.}~\bibnamefont {Guo}}, \bibinfo {author} {\bibfnamefont {Q.}~\bibnamefont {Wang}}, \bibinfo {author} {\bibfnamefont {S.}~\bibnamefont {Sun}}, \bibinfo {author} {\bibfnamefont {C.}~\bibnamefont {Li}}, \bibinfo {author} {\bibfnamefont {S.}~\bibnamefont {Thirupathaiah}}, \bibinfo {author} {\bibfnamefont {A.}~\bibnamefont {Fedorov}}, \bibinfo {author} {\bibfnamefont {D.}~\bibnamefont {Shen}}, \bibinfo {author} {\bibfnamefont {K.}~\bibnamefont {Liu}}, \bibinfo {author} {\bibfnamefont {H.}~\bibnamefont {Lei}},\ and\ \bibinfo {author} {\bibfnamefont {S.}~\bibnamefont {Wang}},\ }\href {https://doi.org/10.1103/PhysRevX.8.031044} {\bibfield  {journal} {\bibinfo  {journal} {Phys. Rev. X}\ }\textbf {\bibinfo {volume} {8}},\ \bibinfo {pages} {031044} (\bibinfo {year} {2018})}\BibitemShut {NoStop}%
\bibitem [{\citenamefont {Shao}\ \emph {et~al.}(2018)\citenamefont {Shao}, \citenamefont {Chen}, \citenamefont {Gu}, \citenamefont {Guo}, \citenamefont {Lu}, \citenamefont {Sun}, \citenamefont {Sheng},\ and\ \citenamefont {Xing}}]{Shao_2018}%
  \BibitemOpen
  \bibfield  {author} {\bibinfo {author} {\bibfnamefont {D.}~\bibnamefont {Shao}}, \bibinfo {author} {\bibfnamefont {T.}~\bibnamefont {Chen}}, \bibinfo {author} {\bibfnamefont {Q.}~\bibnamefont {Gu}}, \bibinfo {author} {\bibfnamefont {Z.}~\bibnamefont {Guo}}, \bibinfo {author} {\bibfnamefont {P.}~\bibnamefont {Lu}}, \bibinfo {author} {\bibfnamefont {J.}~\bibnamefont {Sun}}, \bibinfo {author} {\bibfnamefont {L.}~\bibnamefont {Sheng}},\ and\ \bibinfo {author} {\bibfnamefont {D.}~\bibnamefont {Xing}},\ }\bibfield  {journal} {\bibinfo  {journal} {Scientific Reports}\ }\textbf {\bibinfo {volume} {8}},\ \href {https://doi.org/10.1038/s41598-018-19870-5} {10.1038/s41598-018-19870-5} (\bibinfo {year} {2018})\BibitemShut {NoStop}%
\bibitem [{\citenamefont {Huiberts}\ \emph {et~al.}(1996)\citenamefont {Huiberts}, \citenamefont {Rector}, \citenamefont {Wijngaarden}, \citenamefont {Jetten}, \citenamefont {De~Groot}, \citenamefont {Dam}, \citenamefont {Koeman}, \citenamefont {Griessen}, \citenamefont {Hj{\"o}rvarsson}, \citenamefont {Olafsson} \emph {et~al.}}]{huiberts1996synthesis}%
  \BibitemOpen
  \bibfield  {author} {\bibinfo {author} {\bibfnamefont {J.}~\bibnamefont {Huiberts}}, \bibinfo {author} {\bibfnamefont {J.}~\bibnamefont {Rector}}, \bibinfo {author} {\bibfnamefont {R.}~\bibnamefont {Wijngaarden}}, \bibinfo {author} {\bibfnamefont {S.}~\bibnamefont {Jetten}}, \bibinfo {author} {\bibfnamefont {D.}~\bibnamefont {De~Groot}}, \bibinfo {author} {\bibfnamefont {B.}~\bibnamefont {Dam}}, \bibinfo {author} {\bibfnamefont {N.}~\bibnamefont {Koeman}}, \bibinfo {author} {\bibfnamefont {R.}~\bibnamefont {Griessen}}, \bibinfo {author} {\bibfnamefont {B.}~\bibnamefont {Hj{\"o}rvarsson}}, \bibinfo {author} {\bibfnamefont {S.}~\bibnamefont {Olafsson}}, \emph {et~al.},\ }\href@noop {} {\bibfield  {journal} {\bibinfo  {journal} {Journal of Alloys and Compounds}\ }\textbf {\bibinfo {volume} {239}},\ \bibinfo {pages} {158} (\bibinfo {year} {1996})}\BibitemShut {NoStop}%
\bibitem [{\citenamefont {Wang}\ and\ \citenamefont {Chou}(1995)}]{wang1995structural}%
  \BibitemOpen
  \bibfield  {author} {\bibinfo {author} {\bibfnamefont {Y.}~\bibnamefont {Wang}}\ and\ \bibinfo {author} {\bibfnamefont {M.}~\bibnamefont {Chou}},\ }\href@noop {} {\bibfield  {journal} {\bibinfo  {journal} {Physical Review B}\ }\textbf {\bibinfo {volume} {51}},\ \bibinfo {pages} {7500} (\bibinfo {year} {1995})}\BibitemShut {NoStop}%
\bibitem [{\citenamefont {Belopolski}\ \emph {et~al.}(2019)\citenamefont {Belopolski}, \citenamefont {Manna}, \citenamefont {Sanchez}, \citenamefont {Chang}, \citenamefont {Ernst}, \citenamefont {Yin}, \citenamefont {Zhang}, \citenamefont {Cochran}, \citenamefont {Shumiya}, \citenamefont {Zheng}, \citenamefont {Singh}, \citenamefont {Bian}, \citenamefont {Multer}, \citenamefont {Litskevich}, \citenamefont {Zhou}, \citenamefont {Huang}, \citenamefont {Wang}, \citenamefont {Chang}, \citenamefont {Xu}, \citenamefont {Bansil}, \citenamefont {Felser}, \citenamefont {Lin},\ and\ \citenamefont {Hasan}}]{Ilya2019}%
  \BibitemOpen
  \bibfield  {author} {\bibinfo {author} {\bibfnamefont {I.}~\bibnamefont {Belopolski}}, \bibinfo {author} {\bibfnamefont {K.}~\bibnamefont {Manna}}, \bibinfo {author} {\bibfnamefont {D.~S.}\ \bibnamefont {Sanchez}}, \bibinfo {author} {\bibfnamefont {G.}~\bibnamefont {Chang}}, \bibinfo {author} {\bibfnamefont {B.}~\bibnamefont {Ernst}}, \bibinfo {author} {\bibfnamefont {J.}~\bibnamefont {Yin}}, \bibinfo {author} {\bibfnamefont {S.~S.}\ \bibnamefont {Zhang}}, \bibinfo {author} {\bibfnamefont {T.}~\bibnamefont {Cochran}}, \bibinfo {author} {\bibfnamefont {N.}~\bibnamefont {Shumiya}}, \bibinfo {author} {\bibfnamefont {H.}~\bibnamefont {Zheng}}, \bibinfo {author} {\bibfnamefont {B.}~\bibnamefont {Singh}}, \bibinfo {author} {\bibfnamefont {G.}~\bibnamefont {Bian}}, \bibinfo {author} {\bibfnamefont {D.}~\bibnamefont {Multer}}, \bibinfo {author} {\bibfnamefont {M.}~\bibnamefont {Litskevich}}, \bibinfo {author} {\bibfnamefont {X.}~\bibnamefont {Zhou}}, \bibinfo {author} {\bibfnamefont {S.-M.}\ \bibnamefont {Huang}}, \bibinfo {author} {\bibfnamefont {B.}~\bibnamefont {Wang}}, \bibinfo {author} {\bibfnamefont {T.-R.}\ \bibnamefont {Chang}}, \bibinfo {author} {\bibfnamefont {S.-Y.}\ \bibnamefont {Xu}}, \bibinfo {author} {\bibfnamefont {A.}~\bibnamefont {Bansil}}, \bibinfo {author} {\bibfnamefont {C.}~\bibnamefont {Felser}}, \bibinfo {author} {\bibfnamefont {H.}~\bibnamefont {Lin}},\ and\ \bibinfo {author} {\bibfnamefont {M.~Z.}\ \bibnamefont {Hasan}},\ }\href {https://doi.org/10.1126/science.aav2327} {\bibfield  {journal} {\bibinfo  {journal} {Science}\ }\textbf {\bibinfo {volume} {365}},\ \bibinfo {pages} {1278} (\bibinfo {year} {2019})},\ \Eprint {https://arxiv.org/abs/https://www.science.org/doi/pdf/10.1126/science.aav2327} {https://www.science.org/doi/pdf/10.1126/science.aav2327} \BibitemShut {NoStop}%
\bibitem [{\citenamefont {Guin}\ \emph {et~al.}(2019)\citenamefont {Guin}, \citenamefont {Manna}, \citenamefont {Noky}, \citenamefont {Watzman}, \citenamefont {Fu}, \citenamefont {Kumar}, \citenamefont {Schnelle}, \citenamefont {Shekhar}, \citenamefont {Sun}, \citenamefont {Gooth},\ and\ \citenamefont {Felser}}]{Guin_2019}%
  \BibitemOpen
  \bibfield  {author} {\bibinfo {author} {\bibfnamefont {S.~N.}\ \bibnamefont {Guin}}, \bibinfo {author} {\bibfnamefont {K.}~\bibnamefont {Manna}}, \bibinfo {author} {\bibfnamefont {J.}~\bibnamefont {Noky}}, \bibinfo {author} {\bibfnamefont {S.~J.}\ \bibnamefont {Watzman}}, \bibinfo {author} {\bibfnamefont {C.}~\bibnamefont {Fu}}, \bibinfo {author} {\bibfnamefont {N.}~\bibnamefont {Kumar}}, \bibinfo {author} {\bibfnamefont {W.}~\bibnamefont {Schnelle}}, \bibinfo {author} {\bibfnamefont {C.}~\bibnamefont {Shekhar}}, \bibinfo {author} {\bibfnamefont {Y.}~\bibnamefont {Sun}}, \bibinfo {author} {\bibfnamefont {J.}~\bibnamefont {Gooth}},\ and\ \bibinfo {author} {\bibfnamefont {C.}~\bibnamefont {Felser}},\ }\bibfield  {journal} {\bibinfo  {journal} {NPG Asia Materials}\ }\textbf {\bibinfo {volume} {11}},\ \href {https://doi.org/10.1038/s41427-019-0116-z} {10.1038/s41427-019-0116-z} (\bibinfo {year} {2019})\BibitemShut {NoStop}%
\bibitem [{\citenamefont {Markou}\ \emph {et~al.}(2019)\citenamefont {Markou}, \citenamefont {Kriegner}, \citenamefont {Gayles}, \citenamefont {Zhang}, \citenamefont {Chen}, \citenamefont {Ernst}, \citenamefont {Lai}, \citenamefont {Schnelle}, \citenamefont {Chu}, \citenamefont {Sun},\ and\ \citenamefont {Felser}}]{Markou_2019}%
  \BibitemOpen
  \bibfield  {author} {\bibinfo {author} {\bibfnamefont {A.}~\bibnamefont {Markou}}, \bibinfo {author} {\bibfnamefont {D.}~\bibnamefont {Kriegner}}, \bibinfo {author} {\bibfnamefont {J.}~\bibnamefont {Gayles}}, \bibinfo {author} {\bibfnamefont {L.}~\bibnamefont {Zhang}}, \bibinfo {author} {\bibfnamefont {Y.-C.}\ \bibnamefont {Chen}}, \bibinfo {author} {\bibfnamefont {B.}~\bibnamefont {Ernst}}, \bibinfo {author} {\bibfnamefont {Y.-H.}\ \bibnamefont {Lai}}, \bibinfo {author} {\bibfnamefont {W.}~\bibnamefont {Schnelle}}, \bibinfo {author} {\bibfnamefont {Y.-H.}\ \bibnamefont {Chu}}, \bibinfo {author} {\bibfnamefont {Y.}~\bibnamefont {Sun}},\ and\ \bibinfo {author} {\bibfnamefont {C.}~\bibnamefont {Felser}},\ }\href {https://doi.org/10.1103/PhysRevB.100.054422} {\bibfield  {journal} {\bibinfo  {journal} {Phys. Rev. B}\ }\textbf {\bibinfo {volume} {100}},\ \bibinfo {pages} {054422} (\bibinfo {year} {2019})}\BibitemShut {NoStop}%
\bibitem [{\citenamefont {Conway}\ and\ \citenamefont {Gordon}(1983)}]{conway1983}%
  \BibitemOpen
  \bibfield  {author} {\bibinfo {author} {\bibfnamefont {J.~H.}\ \bibnamefont {Conway}}\ and\ \bibinfo {author} {\bibfnamefont {C.~M.}\ \bibnamefont {Gordon}},\ }\href@noop {} {\bibfield  {journal} {\bibinfo  {journal} {Journal of Graph Theory}\ }\textbf {\bibinfo {volume} {7}},\ \bibinfo {pages} {445} (\bibinfo {year} {1983})}\BibitemShut {NoStop}%
\bibitem [{\citenamefont {Kauffman}(1989)}]{kauffmanInvariantsGraphsThreeSpace1989}%
  \BibitemOpen
  \bibfield  {author} {\bibinfo {author} {\bibfnamefont {L.~H.}\ \bibnamefont {Kauffman}},\ }\href@noop {} {\bibfield  {journal} {\bibinfo  {journal} {Transactions of the American Mathematical Society}\ }\textbf {\bibinfo {volume} {311}},\ \bibinfo {pages} {697} (\bibinfo {year} {1989})}\BibitemShut {NoStop}%
\bibitem [{\citenamefont {Flapan}\ \emph {et~al.}(2017)\citenamefont {Flapan}, \citenamefont {Mattman}, \citenamefont {Mellor}, \citenamefont {Naimi},\ and\ \citenamefont {Nikkuni}}]{flapan2017}%
  \BibitemOpen
  \bibfield  {author} {\bibinfo {author} {\bibfnamefont {E.}~\bibnamefont {Flapan}}, \bibinfo {author} {\bibfnamefont {T.~W.}\ \bibnamefont {Mattman}}, \bibinfo {author} {\bibfnamefont {B.}~\bibnamefont {Mellor}}, \bibinfo {author} {\bibfnamefont {R.}~\bibnamefont {Naimi}},\ and\ \bibinfo {author} {\bibfnamefont {R.}~\bibnamefont {Nikkuni}},\ }in\ \href {https://doi.org/10.1090/conm/689/13845} {\emph {\bibinfo {booktitle} {Knots, Links, Spatial Graphs, and Algebraic Invariants}}},\ \bibinfo {series} {Contemporary Mathematics}, Vol.\ \bibinfo {volume} {689}\ (\bibinfo  {publisher} {American Mathematical Society},\ \bibinfo {year} {2017})\ pp.\ \bibinfo {pages} {81--102}\BibitemShut {NoStop}%
\bibitem [{Note2()}]{Note2}%
  \BibitemOpen
  \bibinfo {note} {In the mathematics literature, knotted graphs are usually called \protect \textit {spatial graphs}.}\BibitemShut {Stop}%
\bibitem [{\citenamefont {Alexander}(1928)}]{alexander1928topological}%
  \BibitemOpen
  \bibfield  {author} {\bibinfo {author} {\bibfnamefont {J.~W.}\ \bibnamefont {Alexander}},\ }\href {https://doi.org/10.1090/S0002-9947-1928-1501429-1} {\bibfield  {journal} {\bibinfo  {journal} {Transactions of the American Mathematical Society}\ }\textbf {\bibinfo {volume} {30}},\ \bibinfo {pages} {275} (\bibinfo {year} {1928})}\BibitemShut {NoStop}%
\bibitem [{\citenamefont {Jones}(1985)}]{jones1985polynomial}%
  \BibitemOpen
  \bibfield  {author} {\bibinfo {author} {\bibfnamefont {V.~F.}\ \bibnamefont {Jones}},\ }\href@noop {} {\bibfield  {journal} {\bibinfo  {journal} {Bulletin of the American Mathematical Society}\ }\textbf {\bibinfo {volume} {12}},\ \bibinfo {pages} {103} (\bibinfo {year} {1985})}\BibitemShut {NoStop}%
\bibitem [{\citenamefont {Hoste}\ \emph {et~al.}(1998)\citenamefont {Hoste}, \citenamefont {Thistlethwaite},\ and\ \citenamefont {Weeks}}]{hoste1998first}%
  \BibitemOpen
  \bibfield  {author} {\bibinfo {author} {\bibfnamefont {J.}~\bibnamefont {Hoste}}, \bibinfo {author} {\bibfnamefont {M.}~\bibnamefont {Thistlethwaite}},\ and\ \bibinfo {author} {\bibfnamefont {J.}~\bibnamefont {Weeks}},\ }\href {https://doi.org/10.1007/BF03025227} {\bibfield  {journal} {\bibinfo  {journal} {The Mathematical Intelligencer}\ }\textbf {\bibinfo {volume} {20}},\ \bibinfo {pages} {33} (\bibinfo {year} {1998})}\BibitemShut {NoStop}%
\bibitem [{\citenamefont {Rolfsen}(2003)}]{rolfsen2003knots}%
  \BibitemOpen
  \bibfield  {author} {\bibinfo {author} {\bibfnamefont {D.}~\bibnamefont {Rolfsen}},\ }\href@noop {} {\emph {\bibinfo {title} {Knots and links}}},\ \bibinfo {series} {AMS Chelsea Publishing}, Vol.\ \bibinfo {volume} {346}\ (\bibinfo  {publisher} {American Mathematical Society},\ \bibinfo {year} {2003})\BibitemShut {NoStop}%
\bibitem [{\citenamefont {Bergholtz}\ \emph {et~al.}(2021)\citenamefont {Bergholtz}, \citenamefont {Budich},\ and\ \citenamefont {Kunst}}]{bergholtz2021rmp}%
  \BibitemOpen
  \bibfield  {author} {\bibinfo {author} {\bibfnamefont {E.~J.}\ \bibnamefont {Bergholtz}}, \bibinfo {author} {\bibfnamefont {J.~C.}\ \bibnamefont {Budich}},\ and\ \bibinfo {author} {\bibfnamefont {F.~K.}\ \bibnamefont {Kunst}},\ }\href {https://doi.org/10.1103/RevModPhys.93.015005} {\bibfield  {journal} {\bibinfo  {journal} {Reviews of Modern Physics}\ }\textbf {\bibinfo {volume} {93}},\ \bibinfo {pages} {015005} (\bibinfo {year} {2021})}\BibitemShut {NoStop}%
\bibitem [{\citenamefont {Wang}\ \emph {et~al.}(2024)\citenamefont {Wang}, \citenamefont {Jin},\ and\ \citenamefont {Song}}]{wang2024berry}%
  \BibitemOpen
  \bibfield  {author} {\bibinfo {author} {\bibfnamefont {P.}~\bibnamefont {Wang}}, \bibinfo {author} {\bibfnamefont {L.}~\bibnamefont {Jin}},\ and\ \bibinfo {author} {\bibfnamefont {Z.}~\bibnamefont {Song}},\ }\href {https://doi.org/10.1103/PhysRevB.109.115406} {\bibfield  {journal} {\bibinfo  {journal} {Phys. Rev. B}\ }\textbf {\bibinfo {volume} {109}},\ \bibinfo {pages} {115406} (\bibinfo {year} {2024})}\BibitemShut {NoStop}%
\bibitem [{\citenamefont {Bardakov}\ and\ \citenamefont {Fedoseev}(2025)}]{bardakov2025invariantshandlebodylinksspatialgraphs}%
  \BibitemOpen
  \bibfield  {author} {\bibinfo {author} {\bibfnamefont {V.~G.}\ \bibnamefont {Bardakov}}\ and\ \bibinfo {author} {\bibfnamefont {D.~A.}\ \bibnamefont {Fedoseev}},\ }\href {https://arxiv.org/abs/2504.09718} {\bibinfo {title} {Invariants of handlebody-links and spatial graphs}} (\bibinfo {year} {2025}),\ \Eprint {https://arxiv.org/abs/2504.09718} {arXiv:2504.09718 [math.GT]} \BibitemShut {NoStop}%
\bibitem [{\citenamefont {Ichinokura}\ \emph {et~al.}(2022)\citenamefont {Ichinokura}, \citenamefont {Toyoda}, \citenamefont {Hashizume}, \citenamefont {Horii}, \citenamefont {Kusaka}, \citenamefont {Ideta}, \citenamefont {Tanaka}, \citenamefont {Shimizu}, \citenamefont {Hitosugi}, \citenamefont {Saito},\ and\ \citenamefont {Hirahara}}]{dos_ref}%
  \BibitemOpen
  \bibfield  {author} {\bibinfo {author} {\bibfnamefont {S.}~\bibnamefont {Ichinokura}}, \bibinfo {author} {\bibfnamefont {M.}~\bibnamefont {Toyoda}}, \bibinfo {author} {\bibfnamefont {M.}~\bibnamefont {Hashizume}}, \bibinfo {author} {\bibfnamefont {K.}~\bibnamefont {Horii}}, \bibinfo {author} {\bibfnamefont {S.}~\bibnamefont {Kusaka}}, \bibinfo {author} {\bibfnamefont {S.}~\bibnamefont {Ideta}}, \bibinfo {author} {\bibfnamefont {K.}~\bibnamefont {Tanaka}}, \bibinfo {author} {\bibfnamefont {R.}~\bibnamefont {Shimizu}}, \bibinfo {author} {\bibfnamefont {T.}~\bibnamefont {Hitosugi}}, \bibinfo {author} {\bibfnamefont {S.}~\bibnamefont {Saito}},\ and\ \bibinfo {author} {\bibfnamefont {T.}~\bibnamefont {Hirahara}},\ }\href {https://doi.org/10.1103/PhysRevB.105.235307} {\bibfield  {journal} {\bibinfo  {journal} {Phys. Rev. B}\ }\textbf {\bibinfo {volume} {105}},\ \bibinfo {pages} {235307} (\bibinfo {year} {2022})}\BibitemShut {NoStop}%
\bibitem [{Note3()}]{Note3}%
  \BibitemOpen
  \bibinfo {note} {The nodal set at $E=0$ is just the locus $\{\protect \mathbf {k}\in \protect \text {BZ}: f(\protect \mathbf {k})=0\}$.}\BibitemShut {Stop}%
\bibitem [{\citenamefont {Bode}\ and\ \citenamefont {Dennis}(2016)}]{bode2016constructingpolynomialnodalset}%
  \BibitemOpen
  \bibfield  {author} {\bibinfo {author} {\bibfnamefont {B.}~\bibnamefont {Bode}}\ and\ \bibinfo {author} {\bibfnamefont {M.~R.}\ \bibnamefont {Dennis}},\ }\href {https://arxiv.org/abs/1612.06328} {\bibinfo {title} {Constructing a polynomial whose nodal set is any prescribed knot or link}} (\bibinfo {year} {2016}),\ \Eprint {https://arxiv.org/abs/1612.06328} {arXiv:1612.06328 [math.GT]} \BibitemShut {NoStop}%
\bibitem [{\citenamefont {Bode}\ \emph {et~al.}(2017)\citenamefont {Bode}, \citenamefont {Dennis}, \citenamefont {Foster},\ and\ \citenamefont {King}}]{Bode_2017}%
  \BibitemOpen
  \bibfield  {author} {\bibinfo {author} {\bibfnamefont {B.}~\bibnamefont {Bode}}, \bibinfo {author} {\bibfnamefont {M.~R.}\ \bibnamefont {Dennis}}, \bibinfo {author} {\bibfnamefont {D.}~\bibnamefont {Foster}},\ and\ \bibinfo {author} {\bibfnamefont {R.~P.}\ \bibnamefont {King}},\ }\href {https://doi.org/10.1098/rspa.2016.0829} {\bibfield  {journal} {\bibinfo  {journal} {Proceedings of the Royal Society A: Mathematical, Physical and Engineering Sciences}\ }\textbf {\bibinfo {volume} {473}},\ \bibinfo {pages} {20160829} (\bibinfo {year} {2017})}\BibitemShut {NoStop}%
\bibitem [{\citenamefont {Zhang}\ \emph {et~al.}(2021)\citenamefont {Zhang}, \citenamefont {Li}, \citenamefont {Liu}, \citenamefont {Tai}, \citenamefont {Thomale},\ and\ \citenamefont {Lee}}]{lee2021tidal}%
  \BibitemOpen
  \bibfield  {author} {\bibinfo {author} {\bibfnamefont {X.}~\bibnamefont {Zhang}}, \bibinfo {author} {\bibfnamefont {G.}~\bibnamefont {Li}}, \bibinfo {author} {\bibfnamefont {Y.}~\bibnamefont {Liu}}, \bibinfo {author} {\bibfnamefont {T.}~\bibnamefont {Tai}}, \bibinfo {author} {\bibfnamefont {R.}~\bibnamefont {Thomale}},\ and\ \bibinfo {author} {\bibfnamefont {C.~H.}\ \bibnamefont {Lee}},\ }\href {https://doi.org/10.1038/s42005-021-00535-1} {\bibfield  {journal} {\bibinfo  {journal} {Communications Physics}\ }\textbf {\bibinfo {volume} {4}},\ \bibinfo {pages} {47} (\bibinfo {year} {2021})}\BibitemShut {NoStop}%
\bibitem [{\citenamefont {Yamada}(1989)}]{yamada1989invariant}%
  \BibitemOpen
  \bibfield  {author} {\bibinfo {author} {\bibfnamefont {S.}~\bibnamefont {Yamada}},\ }\href {https://doi.org/10.1002/jgt.3190130503} {\bibfield  {journal} {\bibinfo  {journal} {Journal of Graph Theory}\ }\textbf {\bibinfo {volume} {13}},\ \bibinfo {pages} {537} (\bibinfo {year} {1989})}\BibitemShut {NoStop}%
\bibitem [{\citenamefont {Mellor}(2018)}]{mellorInvariantsSpatialGraphs2018}%
  \BibitemOpen
  \bibfield  {author} {\bibinfo {author} {\bibfnamefont {B.}~\bibnamefont {Mellor}},\ }\href {https://doi.org/10.48550/arXiv.1812.08885} {\bibinfo {title} {Invariants of {{Spatial Graphs}}}} (\bibinfo {year} {2018}),\ \Eprint {https://arxiv.org/abs/1812.08885} {arXiv:1812.08885 [math]} \BibitemShut {NoStop}%
\bibitem [{Note4()}]{Note4}%
  \BibitemOpen
  \bibinfo {note} {The unnormalized Yamada polynomial $\Upsilon $ is invariant under moves \protect \textbf {(II)--(IV)}, corresponding to regular rigid-vertex isotopy. Its normalized form $\protect \overline {\Upsilon }$ removes the monomial factors associated with moves \protect \textbf {(I)} and \protect \textbf {(V)} and is therefore invariant under moves \protect \textbf {(I)--(V)}, corresponding to rigid-vertex isotopy. Ambient isotopy additionally permits edge reordering at vertices through move \protect \textbf {(VI)}. If all vertices have degree $\le 3$---as is the case for every example in \protect \cref {fig:gallery}---$\protect \overline {\Upsilon }$ is consequently a full ambient-isotopy invariant. For graphs containing higher-degree vertices, which do arise for the real materials in \protect \cref {appx:materials}, we use the minimum-crossing planar-projection convention detailed in \protect \cref {appx:Yamada}.}\BibitemShut {Stop}%
\bibitem [{\citenamefont {Kohn}\ \emph {et~al.}(1996)\citenamefont {Kohn}, \citenamefont {Becke},\ and\ \citenamefont {Parr}}]{Kohn1996}%
  \BibitemOpen
  \bibfield  {author} {\bibinfo {author} {\bibfnamefont {W.}~\bibnamefont {Kohn}}, \bibinfo {author} {\bibfnamefont {A.~D.}\ \bibnamefont {Becke}},\ and\ \bibinfo {author} {\bibfnamefont {R.~G.}\ \bibnamefont {Parr}},\ }\href {https://doi.org/10.1021/jp960669l} {\bibfield  {journal} {\bibinfo  {journal} {The Journal of Physical Chemistry}\ }\textbf {\bibinfo {volume} {100}},\ \bibinfo {pages} {12974} (\bibinfo {year} {1996})}\BibitemShut {NoStop}%
\bibitem [{\citenamefont {Korm{\'a}nyos}\ \emph {et~al.}(2015)\citenamefont {Korm{\'a}nyos}, \citenamefont {Burkard}, \citenamefont {Gmitra}, \citenamefont {Fabian}, \citenamefont {Z{\'o}lyomi}, \citenamefont {Drummond},\ and\ \citenamefont {Fal’ko}}]{kormanyos2015k}%
  \BibitemOpen
  \bibfield  {author} {\bibinfo {author} {\bibfnamefont {A.}~\bibnamefont {Korm{\'a}nyos}}, \bibinfo {author} {\bibfnamefont {G.}~\bibnamefont {Burkard}}, \bibinfo {author} {\bibfnamefont {M.}~\bibnamefont {Gmitra}}, \bibinfo {author} {\bibfnamefont {J.}~\bibnamefont {Fabian}}, \bibinfo {author} {\bibfnamefont {V.}~\bibnamefont {Z{\'o}lyomi}}, \bibinfo {author} {\bibfnamefont {N.~D.}\ \bibnamefont {Drummond}},\ and\ \bibinfo {author} {\bibfnamefont {V.}~\bibnamefont {Fal’ko}},\ }\href@noop {} {\bibfield  {journal} {\bibinfo  {journal} {2D Materials}\ }\textbf {\bibinfo {volume} {2}},\ \bibinfo {pages} {022001} (\bibinfo {year} {2015})}\BibitemShut {NoStop}%
\bibitem [{\citenamefont {Wu}\ \emph {et~al.}(2015)\citenamefont {Wu}, \citenamefont {Jo}, \citenamefont {Ochi}, \citenamefont {Huang}, \citenamefont {Mou}, \citenamefont {Bud'ko}, \citenamefont {Canfield}, \citenamefont {Trivedi}, \citenamefont {Arita},\ and\ \citenamefont {Kaminski}}]{temp_induced_lifshitz}%
  \BibitemOpen
  \bibfield  {author} {\bibinfo {author} {\bibfnamefont {Y.}~\bibnamefont {Wu}}, \bibinfo {author} {\bibfnamefont {N.~H.}\ \bibnamefont {Jo}}, \bibinfo {author} {\bibfnamefont {M.}~\bibnamefont {Ochi}}, \bibinfo {author} {\bibfnamefont {L.}~\bibnamefont {Huang}}, \bibinfo {author} {\bibfnamefont {D.}~\bibnamefont {Mou}}, \bibinfo {author} {\bibfnamefont {S.~L.}\ \bibnamefont {Bud'ko}}, \bibinfo {author} {\bibfnamefont {P.~C.}\ \bibnamefont {Canfield}}, \bibinfo {author} {\bibfnamefont {N.}~\bibnamefont {Trivedi}}, \bibinfo {author} {\bibfnamefont {R.}~\bibnamefont {Arita}},\ and\ \bibinfo {author} {\bibfnamefont {A.}~\bibnamefont {Kaminski}},\ }\href {https://doi.org/10.1103/PhysRevLett.115.166602} {\bibfield  {journal} {\bibinfo  {journal} {Phys. Rev. Lett.}\ }\textbf {\bibinfo {volume} {115}},\ \bibinfo {pages} {166602} (\bibinfo {year} {2015})}\BibitemShut {NoStop}%
\bibitem [{\citenamefont {Sunko}\ \emph {et~al.}(2019)\citenamefont {Sunko}, \citenamefont {Abarca~Morales}, \citenamefont {Marković}, \citenamefont {Barber}, \citenamefont {Milosavljević}, \citenamefont {Mazzola}, \citenamefont {Sokolov}, \citenamefont {Kikugawa}, \citenamefont {Cacho}, \citenamefont {Dudin}, \citenamefont {Rosner}, \citenamefont {Hicks}, \citenamefont {King},\ and\ \citenamefont {Mackenzie}}]{Sunko2019}%
  \BibitemOpen
  \bibfield  {author} {\bibinfo {author} {\bibfnamefont {V.}~\bibnamefont {Sunko}}, \bibinfo {author} {\bibfnamefont {E.}~\bibnamefont {Abarca~Morales}}, \bibinfo {author} {\bibfnamefont {I.}~\bibnamefont {Marković}}, \bibinfo {author} {\bibfnamefont {M.~E.}\ \bibnamefont {Barber}}, \bibinfo {author} {\bibfnamefont {D.}~\bibnamefont {Milosavljević}}, \bibinfo {author} {\bibfnamefont {F.}~\bibnamefont {Mazzola}}, \bibinfo {author} {\bibfnamefont {D.~A.}\ \bibnamefont {Sokolov}}, \bibinfo {author} {\bibfnamefont {N.}~\bibnamefont {Kikugawa}}, \bibinfo {author} {\bibfnamefont {C.}~\bibnamefont {Cacho}}, \bibinfo {author} {\bibfnamefont {P.}~\bibnamefont {Dudin}}, \bibinfo {author} {\bibfnamefont {H.}~\bibnamefont {Rosner}}, \bibinfo {author} {\bibfnamefont {C.~W.}\ \bibnamefont {Hicks}}, \bibinfo {author} {\bibfnamefont {P.~D.~C.}\ \bibnamefont {King}},\ and\ \bibinfo {author} {\bibfnamefont {A.~P.}\ \bibnamefont {Mackenzie}},\ }\href {https://doi.org/10.1038/s41535-019-0185-9} {\bibfield  {journal} {\bibinfo  {journal} {npj Quantum Materials}\ }\textbf {\bibinfo {volume} {4}},\ \bibinfo {pages} {46} (\bibinfo {year} {2019})}\BibitemShut {NoStop}%
\bibitem [{\citenamefont {Xiang}\ \emph {et~al.}(2015)\citenamefont {Xiang}, \citenamefont {Ye}, \citenamefont {Shang}, \citenamefont {Lei}, \citenamefont {Wang}, \citenamefont {Yang}, \citenamefont {Liu}, \citenamefont {Meng}, \citenamefont {Luo}, \citenamefont {Zou}, \citenamefont {Sun}, \citenamefont {Zhang},\ and\ \citenamefont {Chen}}]{Xiang2015}%
  \BibitemOpen
  \bibfield  {author} {\bibinfo {author} {\bibfnamefont {Z.~J.}\ \bibnamefont {Xiang}}, \bibinfo {author} {\bibfnamefont {G.~J.}\ \bibnamefont {Ye}}, \bibinfo {author} {\bibfnamefont {C.}~\bibnamefont {Shang}}, \bibinfo {author} {\bibfnamefont {B.}~\bibnamefont {Lei}}, \bibinfo {author} {\bibfnamefont {N.~Z.}\ \bibnamefont {Wang}}, \bibinfo {author} {\bibfnamefont {K.~S.}\ \bibnamefont {Yang}}, \bibinfo {author} {\bibfnamefont {D.~Y.}\ \bibnamefont {Liu}}, \bibinfo {author} {\bibfnamefont {F.~B.}\ \bibnamefont {Meng}}, \bibinfo {author} {\bibfnamefont {X.~G.}\ \bibnamefont {Luo}}, \bibinfo {author} {\bibfnamefont {L.~J.}\ \bibnamefont {Zou}}, \bibinfo {author} {\bibfnamefont {Z.}~\bibnamefont {Sun}}, \bibinfo {author} {\bibfnamefont {Y.}~\bibnamefont {Zhang}},\ and\ \bibinfo {author} {\bibfnamefont {X.~H.}\ \bibnamefont {Chen}},\ }\bibfield  {journal} {\bibinfo  {journal} {Physical Review Letters}\ }\textbf {\bibinfo {volume} {115}},\ \href {https://doi.org/10.1103/physrevlett.115.186403} {10.1103/physrevlett.115.186403} (\bibinfo {year} {2015})\BibitemShut {NoStop}%
\bibitem [{\citenamefont {Dziawa}\ \emph {et~al.}(2012)\citenamefont {Dziawa}, \citenamefont {Kowalski}, \citenamefont {Dybko}, \citenamefont {Buczko}, \citenamefont {Szczerbakow}, \citenamefont {Szot}, \citenamefont {Łusakowska}, \citenamefont {Balasubramanian}, \citenamefont {Wojek}, \citenamefont {Berntsen}, \citenamefont {Tjernberg},\ and\ \citenamefont {Story}}]{Dziawa2012}%
  \BibitemOpen
  \bibfield  {author} {\bibinfo {author} {\bibfnamefont {P.}~\bibnamefont {Dziawa}}, \bibinfo {author} {\bibfnamefont {B.~J.}\ \bibnamefont {Kowalski}}, \bibinfo {author} {\bibfnamefont {K.}~\bibnamefont {Dybko}}, \bibinfo {author} {\bibfnamefont {R.}~\bibnamefont {Buczko}}, \bibinfo {author} {\bibfnamefont {A.}~\bibnamefont {Szczerbakow}}, \bibinfo {author} {\bibfnamefont {M.}~\bibnamefont {Szot}}, \bibinfo {author} {\bibfnamefont {E.}~\bibnamefont {Łusakowska}}, \bibinfo {author} {\bibfnamefont {T.}~\bibnamefont {Balasubramanian}}, \bibinfo {author} {\bibfnamefont {B.~M.}\ \bibnamefont {Wojek}}, \bibinfo {author} {\bibfnamefont {M.~H.}\ \bibnamefont {Berntsen}}, \bibinfo {author} {\bibfnamefont {O.}~\bibnamefont {Tjernberg}},\ and\ \bibinfo {author} {\bibfnamefont {T.}~\bibnamefont {Story}},\ }\href {https://doi.org/10.1038/nmat3449} {\bibfield  {journal} {\bibinfo  {journal} {Nature Materials}\ }\textbf {\bibinfo {volume} {11}},\ \bibinfo {pages} {1023–1027} (\bibinfo {year} {2012})}\BibitemShut {NoStop}%
\bibitem [{\citenamefont {Yang}\ \emph {et~al.}(2026{\natexlab{b}})\citenamefont {Yang}, \citenamefont {Han}, \citenamefont {Ning}, \citenamefont {Wu}, \citenamefont {Yang},\ and\ \citenamefont {Zheng}}]{yang2026exceptional}%
  \BibitemOpen
  \bibfield  {author} {\bibinfo {author} {\bibfnamefont {S.-B.}\ \bibnamefont {Yang}}, \bibinfo {author} {\bibfnamefont {P.-R.}\ \bibnamefont {Han}}, \bibinfo {author} {\bibfnamefont {W.}~\bibnamefont {Ning}}, \bibinfo {author} {\bibfnamefont {F.}~\bibnamefont {Wu}}, \bibinfo {author} {\bibfnamefont {Z.-B.}\ \bibnamefont {Yang}},\ and\ \bibinfo {author} {\bibfnamefont {S.-B.}\ \bibnamefont {Zheng}},\ }\href {https://doi.org/10.1007/s11433-025-2851-8} {\bibfield  {journal} {\bibinfo  {journal} {Science China Physics, Mechanics \& Astronomy}\ }\textbf {\bibinfo {volume} {69}},\ \bibinfo {pages} {230313} (\bibinfo {year} {2026}{\natexlab{b}})}\BibitemShut {NoStop}%
\bibitem [{\citenamefont {Stålhammar}\ and\ \citenamefont {Bergholtz}(2021)}]{stalhammar2021prb}%
  \BibitemOpen
  \bibfield  {author} {\bibinfo {author} {\bibfnamefont {M.}~\bibnamefont {Stålhammar}}\ and\ \bibinfo {author} {\bibfnamefont {E.~J.}\ \bibnamefont {Bergholtz}},\ }\href {https://doi.org/10.1103/PhysRevB.104.L201104} {\bibfield  {journal} {\bibinfo  {journal} {Physical Review B}\ }\textbf {\bibinfo {volume} {104}},\ \bibinfo {pages} {L201104} (\bibinfo {year} {2021})}\BibitemShut {NoStop}%
\bibitem [{\citenamefont {Kawabata}\ \emph {et~al.}(2019)\citenamefont {Kawabata}, \citenamefont {Bessho},\ and\ \citenamefont {Sato}}]{kawabata2019prl}%
  \BibitemOpen
  \bibfield  {author} {\bibinfo {author} {\bibfnamefont {K.}~\bibnamefont {Kawabata}}, \bibinfo {author} {\bibfnamefont {T.}~\bibnamefont {Bessho}},\ and\ \bibinfo {author} {\bibfnamefont {M.}~\bibnamefont {Sato}},\ }\href {https://doi.org/10.1103/PhysRevLett.123.066405} {\bibfield  {journal} {\bibinfo  {journal} {Physical Review Letters}\ }\textbf {\bibinfo {volume} {123}},\ \bibinfo {pages} {066405} (\bibinfo {year} {2019})}\BibitemShut {NoStop}%
\bibitem [{Note5()}]{Note5}%
  \BibitemOpen
  \bibinfo {note} {Here, the non-reciprocal $\sigma _y$ term is also the non-Hermitian contribution, but basis-equivalent results can be obtained even if the Pauli matrices are permuted.}\BibitemShut {Stop}%
\bibitem [{\citenamefont {Ishii}\ and\ \citenamefont {Iwakiri}(2012)}]{ishii2012quandle}%
  \BibitemOpen
  \bibfield  {author} {\bibinfo {author} {\bibfnamefont {A.}~\bibnamefont {Ishii}}\ and\ \bibinfo {author} {\bibfnamefont {M.}~\bibnamefont {Iwakiri}},\ }\href@noop {} {\bibfield  {journal} {\bibinfo  {journal} {Canadian Journal of Mathematics}\ }\textbf {\bibinfo {volume} {64}},\ \bibinfo {pages} {102} (\bibinfo {year} {2012})}\BibitemShut {NoStop}%
\bibitem [{\citenamefont {Ricca}\ and\ \citenamefont {Berger}(1996)}]{ricca1996topological}%
  \BibitemOpen
  \bibfield  {author} {\bibinfo {author} {\bibfnamefont {R.~L.}\ \bibnamefont {Ricca}}\ and\ \bibinfo {author} {\bibfnamefont {M.~A.}\ \bibnamefont {Berger}},\ }\href@noop {} {\bibfield  {journal} {\bibinfo  {journal} {Physics Today}\ }\textbf {\bibinfo {volume} {49}},\ \bibinfo {pages} {28} (\bibinfo {year} {1996})}\BibitemShut {NoStop}%
\bibitem [{\citenamefont {Kleckner}\ and\ \citenamefont {Irvine}(2013)}]{Kleckner2013}%
  \BibitemOpen
  \bibfield  {author} {\bibinfo {author} {\bibfnamefont {D.}~\bibnamefont {Kleckner}}\ and\ \bibinfo {author} {\bibfnamefont {W.}~\bibnamefont {Irvine}},\ }\href {https://doi.org/10.1038/nphys2560} {\bibfield  {journal} {\bibinfo  {journal} {Nature Physics}\ }\textbf {\bibinfo {volume} {9}},\ \bibinfo {pages} {253} (\bibinfo {year} {2013})}\BibitemShut {NoStop}%
\bibitem [{\citenamefont {Kleckner}\ \emph {et~al.}(2016)\citenamefont {Kleckner}, \citenamefont {Kauffman},\ and\ \citenamefont {Irvine}}]{kleckner2016superfluid}%
  \BibitemOpen
  \bibfield  {author} {\bibinfo {author} {\bibfnamefont {D.}~\bibnamefont {Kleckner}}, \bibinfo {author} {\bibfnamefont {L.~H.}\ \bibnamefont {Kauffman}},\ and\ \bibinfo {author} {\bibfnamefont {W.~T.}\ \bibnamefont {Irvine}},\ }\href@noop {} {\bibfield  {journal} {\bibinfo  {journal} {Nature Physics}\ }\textbf {\bibinfo {volume} {12}},\ \bibinfo {pages} {650} (\bibinfo {year} {2016})}\BibitemShut {NoStop}%
\bibitem [{\citenamefont {Liu}(2020)}]{Liu_2020}%
  \BibitemOpen
  \bibfield  {author} {\bibinfo {author} {\bibfnamefont {R.}~\bibnamefont {Liu}},\ }\href {https://doi.org/10.1088/1674-4527/20/10/165} {\bibfield  {journal} {\bibinfo  {journal} {Research in Astronomy and Astrophysics}\ }\textbf {\bibinfo {volume} {20}},\ \bibinfo {pages} {165} (\bibinfo {year} {2020})}\BibitemShut {NoStop}%
\bibitem [{\citenamefont {Bellan}(2003)}]{Bellan_2003}%
  \BibitemOpen
  \bibfield  {author} {\bibinfo {author} {\bibfnamefont {P.~M.}\ \bibnamefont {Bellan}},\ }\href {https://doi.org/10.1063/1.1558275} {\bibfield  {journal} {\bibinfo  {journal} {Physics of Plasmas}\ }\textbf {\bibinfo {volume} {10}},\ \bibinfo {pages} {1999–2008} (\bibinfo {year} {2003})}\BibitemShut {NoStop}%
\bibitem [{\citenamefont {Yokota}(1996)}]{yokotaTopologicalInvariantsGraphs1996}%
  \BibitemOpen
  \bibfield  {author} {\bibinfo {author} {\bibfnamefont {Y.}~\bibnamefont {Yokota}},\ }\href {https://doi.org/10.1016/0040-9383(95)00002-X} {\bibfield  {journal} {\bibinfo  {journal} {Topology}\ }\textbf {\bibinfo {volume} {35}},\ \bibinfo {pages} {77} (\bibinfo {year} {1996})}\BibitemShut {NoStop}%
\bibitem [{\citenamefont {Thompson}(1992)}]{thompsonPolynomialInvariantGraphs1992}%
  \BibitemOpen
  \bibfield  {author} {\bibinfo {author} {\bibfnamefont {A.}~\bibnamefont {Thompson}},\ }\href {https://doi.org/10.1016/0040-9383(92)90056-N} {\bibfield  {journal} {\bibinfo  {journal} {Topology}\ }\textbf {\bibinfo {volume} {31}},\ \bibinfo {pages} {657} (\bibinfo {year} {1992})}\BibitemShut {NoStop}%
\bibitem [{\citenamefont {Huh}(2022)}]{huh2022yamadapolynomialassociatedlink}%
  \BibitemOpen
  \bibfield  {author} {\bibinfo {author} {\bibfnamefont {Y.}~\bibnamefont {Huh}},\ }\href {https://arxiv.org/abs/2206.11450} {\bibinfo {title} {Yamada polynomial and associated link of $\theta$-curves}} (\bibinfo {year} {2022}),\ \Eprint {https://arxiv.org/abs/2206.11450} {arXiv:2206.11450 [math.GT]} \BibitemShut {NoStop}%
\bibitem [{\citenamefont {Negami}(1987)}]{negami1987polynomial}%
  \BibitemOpen
  \bibfield  {author} {\bibinfo {author} {\bibfnamefont {S.}~\bibnamefont {Negami}},\ }\href {https://doi.org/10.1090/S0002-9947-1987-0894562-2} {\bibfield  {journal} {\bibinfo  {journal} {Transactions of the American Mathematical Society}\ }\textbf {\bibinfo {volume} {299}},\ \bibinfo {pages} {601} (\bibinfo {year} {1987})}\BibitemShut {NoStop}%
\bibitem [{\citenamefont {Ishii}(2008)}]{ishii2008moves}%
  \BibitemOpen
  \bibfield  {author} {\bibinfo {author} {\bibfnamefont {A.}~\bibnamefont {Ishii}},\ }\href@noop {} {\bibfield  {journal} {\bibinfo  {journal} {Algebraic \& Geometric Topology}\ }\textbf {\bibinfo {volume} {8}},\ \bibinfo {pages} {1403} (\bibinfo {year} {2008})}\BibitemShut {NoStop}%
\bibitem [{\citenamefont {Vennila}\ \emph {et~al.}(2005)\citenamefont {Vennila}, \citenamefont {Porchelvi}, \citenamefont {Joy}, \citenamefont {Arun},\ and\ \citenamefont {Jaya}}]{Vennila2005}%
  \BibitemOpen
  \bibfield  {author} {\bibinfo {author} {\bibfnamefont {R.~S.}\ \bibnamefont {Vennila}}, \bibinfo {author} {\bibfnamefont {E.~E.}\ \bibnamefont {Porchelvi}}, \bibinfo {author} {\bibfnamefont {K.~M.~F.}\ \bibnamefont {Joy}}, \bibinfo {author} {\bibfnamefont {T.~K.~J.}\ \bibnamefont {Arun}},\ and\ \bibinfo {author} {\bibfnamefont {N.~V.}\ \bibnamefont {Jaya}},\ }\href {https://doi.org/10.1016/j.jallcom.2004.09.004} {\bibfield  {journal} {\bibinfo  {journal} {Journal of Alloys and Compounds}\ }\textbf {\bibinfo {volume} {392}},\ \bibinfo {pages} {24} (\bibinfo {year} {2005})}\BibitemShut {NoStop}%
\bibitem [{\citenamefont {Chen}\ \emph {et~al.}(2013)\citenamefont {Chen}, \citenamefont {Chen}, \citenamefont {Zhang}, \citenamefont {Tang}, \citenamefont {Wei},\ and\ \citenamefont {Sun}}]{Chen2013}%
  \BibitemOpen
  \bibfield  {author} {\bibinfo {author} {\bibfnamefont {S.-J.}\ \bibnamefont {Chen}}, \bibinfo {author} {\bibfnamefont {Y.}~\bibnamefont {Chen}}, \bibinfo {author} {\bibfnamefont {H.}~\bibnamefont {Zhang}}, \bibinfo {author} {\bibfnamefont {Y.}~\bibnamefont {Tang}}, \bibinfo {author} {\bibfnamefont {J.-j.}\ \bibnamefont {Wei}},\ and\ \bibinfo {author} {\bibfnamefont {W.-g.}\ \bibnamefont {Sun}},\ }\href {https://doi.org/10.1155/2013/569537} {\bibfield  {journal} {\bibinfo  {journal} {Journal of Nanomaterials}\ }\textbf {\bibinfo {volume} {2013}},\ \bibinfo {pages} {1} (\bibinfo {year} {2013})}\BibitemShut {NoStop}%
\bibitem [{Note6()}]{Note6}%
  \BibitemOpen
  \bibinfo {note} {A single vertex is a bouquet of zero circles, i.e., $\protect \mathcal {B}_0$.}\BibitemShut {Stop}%
\bibitem [{\citenamefont {Lee}\ \emph {et~al.}(2020{\natexlab{b}})\citenamefont {Lee}, \citenamefont {Yap}, \citenamefont {Tai}, \citenamefont {Xu}, \citenamefont {Zhang},\ and\ \citenamefont {Gong}}]{Lee_2020}%
  \BibitemOpen
  \bibfield  {author} {\bibinfo {author} {\bibfnamefont {C.~H.}\ \bibnamefont {Lee}}, \bibinfo {author} {\bibfnamefont {H.~H.}\ \bibnamefont {Yap}}, \bibinfo {author} {\bibfnamefont {T.}~\bibnamefont {Tai}}, \bibinfo {author} {\bibfnamefont {G.}~\bibnamefont {Xu}}, \bibinfo {author} {\bibfnamefont {X.}~\bibnamefont {Zhang}},\ and\ \bibinfo {author} {\bibfnamefont {J.}~\bibnamefont {Gong}},\ }\bibfield  {journal} {\bibinfo  {journal} {Physical Review B}\ }\textbf {\bibinfo {volume} {102}},\ \href {https://doi.org/10.1103/physrevb.102.035138} {10.1103/physrevb.102.035138} (\bibinfo {year} {2020}{\natexlab{b}})\BibitemShut {NoStop}%
\bibitem [{\citenamefont {Tai}\ and\ \citenamefont {Lee}(2021)}]{tai2021anisotropic}%
  \BibitemOpen
  \bibfield  {author} {\bibinfo {author} {\bibfnamefont {T.}~\bibnamefont {Tai}}\ and\ \bibinfo {author} {\bibfnamefont {C.~H.}\ \bibnamefont {Lee}},\ }\href@noop {} {\bibfield  {journal} {\bibinfo  {journal} {Physical Review B}\ }\textbf {\bibinfo {volume} {103}},\ \bibinfo {pages} {195125} (\bibinfo {year} {2021})}\BibitemShut {NoStop}%
\bibitem [{\citenamefont {Kauffman}\ \emph {et~al.}(1993)\citenamefont {Kauffman}, \citenamefont {Simon}, \citenamefont {Wolcott},\ and\ \citenamefont {Zhao}}]{kauffmanInvariantsThetacurvesOther1993}%
  \BibitemOpen
  \bibfield  {author} {\bibinfo {author} {\bibfnamefont {L.}~\bibnamefont {Kauffman}}, \bibinfo {author} {\bibfnamefont {J.}~\bibnamefont {Simon}}, \bibinfo {author} {\bibfnamefont {K.}~\bibnamefont {Wolcott}},\ and\ \bibinfo {author} {\bibfnamefont {P.}~\bibnamefont {Zhao}},\ }\href {https://doi.org/10.1016/0166-8641(93)90110-Y} {\bibfield  {journal} {\bibinfo  {journal} {Topology and its Applications}\ }\textbf {\bibinfo {volume} {49}},\ \bibinfo {pages} {193} (\bibinfo {year} {1993})}\BibitemShut {NoStop}%
\bibitem [{\citenamefont {Kuratowski}(1930)}]{kuratowski1930}%
  \BibitemOpen
  \bibfield  {author} {\bibinfo {author} {\bibfnamefont {K.}~\bibnamefont {Kuratowski}},\ }\href@noop {} {\bibfield  {journal} {\bibinfo  {journal} {Fundamenta Mathematicae}\ }\textbf {\bibinfo {volume} {15}},\ \bibinfo {pages} {271} (\bibinfo {year} {1930})}\BibitemShut {NoStop}%
\bibitem [{\citenamefont {Wagner}(1937)}]{wagner1937}%
  \BibitemOpen
  \bibfield  {author} {\bibinfo {author} {\bibfnamefont {K.}~\bibnamefont {Wagner}},\ }\href {https://doi.org/10.1007/BF01594196} {\bibfield  {journal} {\bibinfo  {journal} {Mathematische Annalen}\ }\textbf {\bibinfo {volume} {114}},\ \bibinfo {pages} {570} (\bibinfo {year} {1937})}\BibitemShut {NoStop}%
\bibitem [{\citenamefont {Fellows}\ and\ \citenamefont {Langston}(1988)}]{fellows1988}%
  \BibitemOpen
  \bibfield  {author} {\bibinfo {author} {\bibfnamefont {M.~R.}\ \bibnamefont {Fellows}}\ and\ \bibinfo {author} {\bibfnamefont {M.~A.}\ \bibnamefont {Langston}},\ }\href@noop {} {\bibfield  {journal} {\bibinfo  {journal} {Journal of the ACM}\ }\textbf {\bibinfo {volume} {35}},\ \bibinfo {pages} {727} (\bibinfo {year} {1988})}\BibitemShut {NoStop}%
\bibitem [{\citenamefont {Ne{\v{s}}et{\v{r}}il}\ and\ \citenamefont {Thomas}(1985)}]{nesetril1985}%
  \BibitemOpen
  \bibfield  {author} {\bibinfo {author} {\bibfnamefont {J.}~\bibnamefont {Ne{\v{s}}et{\v{r}}il}}\ and\ \bibinfo {author} {\bibfnamefont {R.}~\bibnamefont {Thomas}},\ }\href@noop {} {\bibfield  {journal} {\bibinfo  {journal} {Commentationes Mathematicae Universitatis Carolinae}\ }\textbf {\bibinfo {volume} {26}},\ \bibinfo {pages} {655} (\bibinfo {year} {1985})}\BibitemShut {NoStop}%
\bibitem [{\citenamefont {Sachs}(1983)}]{sachs1983}%
  \BibitemOpen
  \bibfield  {author} {\bibinfo {author} {\bibfnamefont {H.}~\bibnamefont {Sachs}},\ }in\ \href@noop {} {\emph {\bibinfo {booktitle} {Graph Theory}}},\ \bibinfo {series} {Lecture Notes in Mathematics}, Vol.\ \bibinfo {volume} {1018},\ \bibinfo {editor} {edited by\ \bibinfo {editor} {\bibfnamefont {M.}~\bibnamefont {Borowiecki}}, \bibinfo {editor} {\bibfnamefont {J.~W.}\ \bibnamefont {Kennedy}},\ and\ \bibinfo {editor} {\bibfnamefont {M.~M.}\ \bibnamefont {Syso}}}\ (\bibinfo  {publisher} {Springer},\ \bibinfo {address} {Berlin, Heidelberg},\ \bibinfo {year} {1983})\BibitemShut {NoStop}%
\bibitem [{\citenamefont {Sachs}(1984)}]{sachs1984}%
  \BibitemOpen
  \bibfield  {author} {\bibinfo {author} {\bibfnamefont {H.}~\bibnamefont {Sachs}},\ }in\ \href@noop {} {\emph {\bibinfo {booktitle} {Finite and Infinite Sets, Vols. I and II}}},\ \bibinfo {series} {Colloq. Math. Soc. Ja{\'n}os Bolyai}, Vol.~\bibinfo {volume} {37},\ \bibinfo {editor} {edited by\ \bibinfo {editor} {\bibfnamefont {A.}~\bibnamefont {Hajnal}}, \bibinfo {editor} {\bibfnamefont {L.}~\bibnamefont {Lov{\'a}sz}},\ and\ \bibinfo {editor} {\bibfnamefont {V.~T.}\ \bibnamefont {So{\'s}}}}\ (\bibinfo  {publisher} {North Holland},\ \bibinfo {address} {Amsterdam, New York},\ \bibinfo {year} {1984})\ pp.\ \bibinfo {pages} {649--662}\BibitemShut {NoStop}%
\bibitem [{\citenamefont {Robertson}\ \emph {et~al.}(1995)\citenamefont {Robertson}, \citenamefont {Seymour},\ and\ \citenamefont {Thomas}}]{robertson1995}%
  \BibitemOpen
  \bibfield  {author} {\bibinfo {author} {\bibfnamefont {N.}~\bibnamefont {Robertson}}, \bibinfo {author} {\bibfnamefont {P.}~\bibnamefont {Seymour}},\ and\ \bibinfo {author} {\bibfnamefont {R.}~\bibnamefont {Thomas}},\ }\href@noop {} {\bibfield  {journal} {\bibinfo  {journal} {Journal of Combinatorial Theory, Series B}\ }\textbf {\bibinfo {volume} {64}},\ \bibinfo {pages} {185} (\bibinfo {year} {1995})}\BibitemShut {NoStop}%
\bibitem [{\citenamefont {Flapan}\ \emph {et~al.}(2001{\natexlab{a}})\citenamefont {Flapan}, \citenamefont {Naimi},\ and\ \citenamefont {Pommersheim}}]{flapan2001triple}%
  \BibitemOpen
  \bibfield  {author} {\bibinfo {author} {\bibfnamefont {E.}~\bibnamefont {Flapan}}, \bibinfo {author} {\bibfnamefont {R.}~\bibnamefont {Naimi}},\ and\ \bibinfo {author} {\bibfnamefont {J.}~\bibnamefont {Pommersheim}},\ }\href@noop {} {\bibfield  {journal} {\bibinfo  {journal} {Topology and its Applications}\ }\textbf {\bibinfo {volume} {115}},\ \bibinfo {pages} {239} (\bibinfo {year} {2001}{\natexlab{a}})}\BibitemShut {NoStop}%
\bibitem [{\citenamefont {Bowlin}\ and\ \citenamefont {Foisy}(2004)}]{bowlin2004}%
  \BibitemOpen
  \bibfield  {author} {\bibinfo {author} {\bibfnamefont {G.}~\bibnamefont {Bowlin}}\ and\ \bibinfo {author} {\bibfnamefont {J.}~\bibnamefont {Foisy}},\ }\href@noop {} {\bibfield  {journal} {\bibinfo  {journal} {Journal of Knot Theory and its Ramifications}\ }\textbf {\bibinfo {volume} {13}},\ \bibinfo {pages} {1021} (\bibinfo {year} {2004})}\BibitemShut {NoStop}%
\bibitem [{\citenamefont {Flapan}\ \emph {et~al.}(2001{\natexlab{b}})\citenamefont {Flapan}, \citenamefont {Pommersheim}, \citenamefont {Foisy},\ and\ \citenamefont {Naimi}}]{flapan2001nlinked}%
  \BibitemOpen
  \bibfield  {author} {\bibinfo {author} {\bibfnamefont {E.}~\bibnamefont {Flapan}}, \bibinfo {author} {\bibfnamefont {J.}~\bibnamefont {Pommersheim}}, \bibinfo {author} {\bibfnamefont {J.}~\bibnamefont {Foisy}},\ and\ \bibinfo {author} {\bibfnamefont {R.}~\bibnamefont {Naimi}},\ }\href {https://doi.org/10.1142/S0218216501001360} {\bibfield  {journal} {\bibinfo  {journal} {Journal of Knot Theory and its Ramifications}\ }\textbf {\bibinfo {volume} {10}},\ \bibinfo {pages} {1143} (\bibinfo {year} {2001}{\natexlab{b}})}\BibitemShut {NoStop}%
\bibitem [{\citenamefont {Naimi}(2021)}]{naimi2021brief}%
  \BibitemOpen
  \bibfield  {author} {\bibinfo {author} {\bibfnamefont {R.}~\bibnamefont {Naimi}},\ }in\ \href@noop {} {\emph {\bibinfo {booktitle} {Encyclopedia of Knot Theory}}}\ (\bibinfo  {publisher} {Chapman and Hall/CRC},\ \bibinfo {year} {2021})\ pp.\ \bibinfo {pages} {467--476}\BibitemShut {NoStop}%
\bibitem [{\citenamefont {Foisy}(2002)}]{foisy2002}%
  \BibitemOpen
  \bibfield  {author} {\bibinfo {author} {\bibfnamefont {J.}~\bibnamefont {Foisy}},\ }\href@noop {} {\bibfield  {journal} {\bibinfo  {journal} {Journal of Graph Theory}\ }\textbf {\bibinfo {volume} {39}},\ \bibinfo {pages} {178} (\bibinfo {year} {2002})}\BibitemShut {NoStop}%
\bibitem [{\citenamefont {Johnson}\ \emph {et~al.}(2010)\citenamefont {Johnson}, \citenamefont {Kidwell},\ and\ \citenamefont {Michael}}]{johnson2010}%
  \BibitemOpen
  \bibfield  {author} {\bibinfo {author} {\bibfnamefont {B.}~\bibnamefont {Johnson}}, \bibinfo {author} {\bibfnamefont {M.~E.}\ \bibnamefont {Kidwell}},\ and\ \bibinfo {author} {\bibfnamefont {T.~S.}\ \bibnamefont {Michael}},\ }\bibfield  {journal} {\bibinfo  {journal} {Journal of Knot Theory and Its Ramifications}\ }\textbf {\bibinfo {volume} {19}},\ \href {https://doi.org/10.1142/S0218216510008455} {10.1142/S0218216510008455} (\bibinfo {year} {2010})\BibitemShut {NoStop}%
\bibitem [{\citenamefont {Barsotti}\ and\ \citenamefont {Mattman}(2016)}]{Barsotti2015GraphsO2}%
  \BibitemOpen
  \bibfield  {author} {\bibinfo {author} {\bibfnamefont {J.}~\bibnamefont {Barsotti}}\ and\ \bibinfo {author} {\bibfnamefont {T.~W.}\ \bibnamefont {Mattman}},\ }\href {https://doi.org/10.2140/involve.2016.9.591} {\bibfield  {journal} {\bibinfo  {journal} {Involve, a Journal of Mathematics}\ }\textbf {\bibinfo {volume} {9}},\ \bibinfo {pages} {591} (\bibinfo {year} {2016})}\BibitemShut {NoStop}%
\bibitem [{\citenamefont {Kim}\ \emph {et~al.}(2018)\citenamefont {Kim}, \citenamefont {Mattman},\ and\ \citenamefont {Oh}}]{Kim2017}%
  \BibitemOpen
  \bibfield  {author} {\bibinfo {author} {\bibfnamefont {H.}~\bibnamefont {Kim}}, \bibinfo {author} {\bibfnamefont {T.}~\bibnamefont {Mattman}},\ and\ \bibinfo {author} {\bibfnamefont {S.}~\bibnamefont {Oh}},\ }\href {https://doi.org/10.1142/S0218216518500591} {\bibfield  {journal} {\bibinfo  {journal} {Journal of Knot Theory and Its Ramifications}\ }\textbf {\bibinfo {volume} {27}},\ \bibinfo {pages} {1850059} (\bibinfo {year} {2018})}\BibitemShut {NoStop}%
\bibitem [{\citenamefont {Blain}\ \emph {et~al.}(2007)\citenamefont {Blain}, \citenamefont {Bowlin}, \citenamefont {Fleming}, \citenamefont {Foisy}, \citenamefont {Hendricks},\ and\ \citenamefont {LaCombe}}]{blain2007}%
  \BibitemOpen
  \bibfield  {author} {\bibinfo {author} {\bibfnamefont {P.}~\bibnamefont {Blain}}, \bibinfo {author} {\bibfnamefont {G.}~\bibnamefont {Bowlin}}, \bibinfo {author} {\bibfnamefont {T.}~\bibnamefont {Fleming}}, \bibinfo {author} {\bibfnamefont {J.}~\bibnamefont {Foisy}}, \bibinfo {author} {\bibfnamefont {J.}~\bibnamefont {Hendricks}},\ and\ \bibinfo {author} {\bibfnamefont {J.}~\bibnamefont {LaCombe}},\ }\href@noop {} {\bibfield  {journal} {\bibinfo  {journal} {Journal of Knot Theory and its Ramifications}\ }\textbf {\bibinfo {volume} {16}},\ \bibinfo {pages} {749} (\bibinfo {year} {2007})}\BibitemShut {NoStop}%
\bibitem [{\citenamefont {Ozawa}\ and\ \citenamefont {Tsutsumi}(2007)}]{ozawa2007}%
  \BibitemOpen
  \bibfield  {author} {\bibinfo {author} {\bibfnamefont {M.}~\bibnamefont {Ozawa}}\ and\ \bibinfo {author} {\bibfnamefont {Y.}~\bibnamefont {Tsutsumi}},\ }\href@noop {} {\bibfield  {journal} {\bibinfo  {journal} {Revista Matemática Complutense}\ }\textbf {\bibinfo {volume} {20}},\ \bibinfo {pages} {391} (\bibinfo {year} {2007})}\BibitemShut {NoStop}%
\bibitem [{\citenamefont {Miller}\ and\ \citenamefont {Naimi}(2014)}]{miller2014}%
  \BibitemOpen
  \bibfield  {author} {\bibinfo {author} {\bibfnamefont {J.}~\bibnamefont {Miller}}\ and\ \bibinfo {author} {\bibfnamefont {R.}~\bibnamefont {Naimi}},\ }\href {https://doi.org/10.1080/10586458.2014.852033} {\bibfield  {journal} {\bibinfo  {journal} {Experimental Mathematics}\ }\textbf {\bibinfo {volume} {23}},\ \bibinfo {pages} {6} (\bibinfo {year} {2014})}\BibitemShut {NoStop}%
\bibitem [{\citenamefont {Choi}\ \emph {et~al.}(2021)\citenamefont {Choi}, \citenamefont {Kim},\ and\ \citenamefont {No}}]{Choi2021}%
  \BibitemOpen
  \bibfield  {author} {\bibinfo {author} {\bibfnamefont {H.}~\bibnamefont {Choi}}, \bibinfo {author} {\bibfnamefont {H.}~\bibnamefont {Kim}},\ and\ \bibinfo {author} {\bibfnamefont {S.}~\bibnamefont {No}},\ }\href {https://doi.org/10.1016/j.dam.2020.12.009} {\bibfield  {journal} {\bibinfo  {journal} {Discrete Applied Mathematics}\ }\textbf {\bibinfo {volume} {291}},\ \bibinfo {pages} {237} (\bibinfo {year} {2021})}\BibitemShut {NoStop}%
\bibitem [{\citenamefont {Choi}\ \emph {et~al.}(2022)\citenamefont {Choi}, \citenamefont {Kim},\ and\ \citenamefont {No}}]{choi2022}%
  \BibitemOpen
  \bibfield  {author} {\bibinfo {author} {\bibfnamefont {H.}~\bibnamefont {Choi}}, \bibinfo {author} {\bibfnamefont {H.}~\bibnamefont {Kim}},\ and\ \bibinfo {author} {\bibfnamefont {S.}~\bibnamefont {No}},\ }\href {https://arxiv.org/abs/2208.09241} {\bibinfo {title} {Chirality for simple graphs of size up to 12}} (\bibinfo {year} {2022}),\ \Eprint {https://arxiv.org/abs/2208.09241} {arXiv:2208.09241 [math.GT]} \BibitemShut {NoStop}%
\bibitem [{\citenamefont {Flapan}\ and\ \citenamefont {Fletcher}(2013)}]{flapan2013}%
  \BibitemOpen
  \bibfield  {author} {\bibinfo {author} {\bibfnamefont {E.}~\bibnamefont {Flapan}}\ and\ \bibinfo {author} {\bibfnamefont {W.}~\bibnamefont {Fletcher}},\ }\href {https://arxiv.org/abs/1303.5131} {\bibinfo {title} {Intrinsic chirality of multipartite graphs}} (\bibinfo {year} {2013}),\ \Eprint {https://arxiv.org/abs/1303.5131} {arXiv:1303.5131 [math.GT]} \BibitemShut {NoStop}%
\bibitem [{\citenamefont {Flapan}\ and\ \citenamefont {Weaver}(1992)}]{flapan1992}%
  \BibitemOpen
  \bibfield  {author} {\bibinfo {author} {\bibfnamefont {E.}~\bibnamefont {Flapan}}\ and\ \bibinfo {author} {\bibfnamefont {N.}~\bibnamefont {Weaver}},\ }\href {https://doi.org/10.2307/2159591} {\bibfield  {journal} {\bibinfo  {journal} {Proceedings of the American Mathematical Society}\ }\textbf {\bibinfo {volume} {115}},\ \bibinfo {pages} {233} (\bibinfo {year} {1992})},\ \bibinfo {note} {accessed 15 Apr. 2025}\BibitemShut {NoStop}%
\bibitem [{\citenamefont {Bode}(2018)}]{Bode_2018}%
  \BibitemOpen
  \bibfield  {author} {\bibinfo {author} {\bibfnamefont {B.}~\bibnamefont {Bode}},\ }\href {https://doi.org/10.1016/j.topol.2018.05.001} {\bibfield  {journal} {\bibinfo  {journal} {Topology and its Applications}\ }\textbf {\bibinfo {volume} {243}},\ \bibinfo {pages} {33–51} (\bibinfo {year} {2018})}\BibitemShut {NoStop}%
\end{thebibliography}%

\clearpage
\appendix
\renewcommand{\thefigure}{S\arabic{figure}}
\renewcommand{\thetable}{S\arabic{table}}
\onecolumngrid

\begin{center}
    \textbf{
    \Large Supplemental Material 
    }
\end{center}

\begin{appendices}
\AppendixSectionPrefixNumbers
\DoToC

\bigskip
The supplemental material collects a glossary overview, computational details, invariant constructions, and extended results that support the main text. They cover the skeletonization pipeline, Yamada polynomial evaluation, multi-boundary compression-body generalization and Yamada set,  additional Yamada sequences for real materials, and other pure graph-theoretic signatures such as planarity and intrinsic linkedness.

\section{Glossary}\label[smsection]{appx:glossary}

\begin{figure}[!h]
    \centering
    \includegraphics[width=0.8\linewidth]{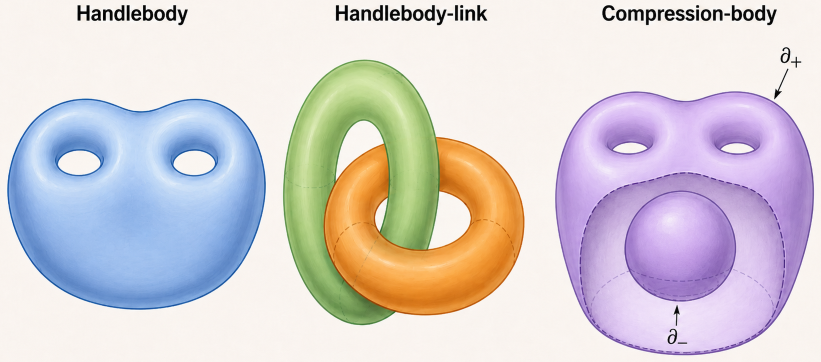}
    \caption{\textbf{Illustration of handlebody, handlebody-link, and compression body.}
    The compression-body's outer boundary is denoted by $\partial_+$, which is topologically equivalent to the boundary of the handlebody on the left.
    The inner boundary of the compression body is denoted by $\partial_-$, which comes from a void inside the volume.
    \textbf{Compression-body links} (not illustrated) would be linked unions of such compression bodies.}
    \label{fig:compression-body}
\end{figure}

\begin{itemize}[itemsep=0ex]
    \item \textbf{Brillouin zone (BZ).} Momentum space of a 3D lattice. Under periodic boundary condition it is topologically a 3-torus, $T^3$.

    \item \textbf{Constant-energy surface / Fermi surface (level set).} For a chosen band $n$ with dispersion $\varepsilon_n(\mathbf{k})$, the constant-energy surface at Fermi energy $E$ is
    $\Sigma_E^{(n)}:=\{\mathbf{k}\in\mathrm{BZ}:\varepsilon_n(\mathbf{k})=E\}$.
    This is equivalently the Fermi surface at Fermi energy $E$.
    When $E$ is a regular value of $\varepsilon_n$, $\Sigma_E^{(n)}$ is a smooth embedded closed surface, possibly with several connected components and nested inner boundaries.
    In the main text, after choosing one representative energy in the $i$-th window whose two ends are Lifshitz transitions, we denote the corresponding surface topology by $\Sigma_{[i]}$ (or in general $\Sigma_{[i]}^{(n)}$ with explicit band label).

    \item \textbf{Fermi sea / Fermi volume (sublevel set).} We use ``Fermi volume'' as shorthand for the enclosed sublevel region
    $\mathcal{F}_E^{(n)}:=\{\mathbf{k}\in\mathrm{BZ}:\varepsilon_n(\mathbf{k})\le E\}$.
    For regular $E$, it is a compact 3-manifold with boundary $\partial\mathcal{F}_E^{(n)}=\Sigma_E^{(n)}$.
    A generic Fermi volume may have multiple disjoint components and each possibly with nested inner boundaries, in which case we analyze the volume boundary-wise.
    Once a representative energy is selected in the $i$-th window, we write $\mathcal{F}_{[i]}$ or $\mathcal{F}_{[i]}^{(n)}$ to represent the topological equivalent (homeomorphic) Fermi volume.

    \item \textbf{Handlebody.} A compact, connected, orientable three-manifold obtained by attaching 1-handles to a 3-ball. Equivalently, it is the regular neighborhood of an embedded graph.
    Fermi volumes in \cref{fig:hopfsequence} are precisely handlebodies.
    A handlebody up to homeomorphism is in one-to-one correspondence with its embedded knotted graph up to ambient isotopy and contraction moves (\cref{appx:compression-body}).

    \item \textbf{Compression body.} A multiply-connected generalization of a handlebody with a distinguished outer boundary $\partial_+$ and possibly several inner boundaries (voids) $\partial_-$ (\cref{fig:compression-body}).
    A generic Fermi volume is a compression body having multiple boundaries, and thus generic Fermi surface is multi-component. This extension beyond handlebodies is necessary to characterize the realistic Fermi surfaces that arise in the materials considered in this study.

    \item \textbf{Knotted graph (Spatial graph).} An embedding of an abstract graph into a three-manifold (here the Brillouin zone $BZ \cong T^3$). Two such embeddings are considered the same when they can be connected by ambient isotopy and contraction moves, unless stated otherwise. In the math literature~\cite{ishii2012quandle,bardakov2025invariantshandlebodylinksspatialgraphs}, the embedded graph onto which a handlebody deformation-retracts is also called a \textit{spine}; different spines of the same handlebody are related by ambient isotopy together with contraction/expansion moves (Whitehead moves).

    \item \textbf{Skeleton graph.} The specific knotted graph obtained by our numerical skeletonization procedure (\cref{appx:skeletonization})---a ``canonical'' representative of the equivalence class of knotted graphs associated with a given Fermi volume.
    We denote it by $\mathcal{G}_{[i]}$ or $\mathcal{G}_{[i]}^{(n)}$ for the $i$-th window and $n$-th band.
    In practice, it is obtained by topology-preserving thinning and pruning of a voxelized Fermi volume.
    For compression-body components with inner boundaries, the corresponding invariants are recorded boundary-wise through the Yamada-set construction.

    \item \textbf{Planar diagram (Spatial-graph diagram).} A generic projection of a spatial graph to $\mathbb{R}^2$, in which vertices project to distinct points, edge intersections away from vertices are \textit{crossing}s (in math term, \textit{transverse double point}s), and each crossing carries over/under information. 
    Thus a planar diagram is a \textit{diagrammatic encoding} of a spatial embedding (not the same thing as a crossing-free planar embedding of the abstract graph).
    Generalized Reidemeister moves and Yamada polynomial are defined in terms of planar diagrams (2D projections), not directly in terms of 3D embeddings.
    \textbf{Throughout this manuscript, we denote the \textit{diagram} of the spatial graph $\mathcal{G}$ by the same notation $\mathcal{G}$ for brevity.}

    \item \textbf{Generalized Reidemeister moves.} The local moves \textbf{(I)--(VI)} on planar diagrams of spatial graphs. Moves \textbf{(I)--(III)} are the usual knot/link Reidemeister moves, while moves \textbf{(IV)--(VI)} involve graph vertices. Moves \textbf{(I)--(V)} generate rigid (flat) vertex isotopy; adding move \textbf{(VI)} gives pliable vertex isotopy (i.e. ambient isotopy).

    \item \textbf{Ambient isotopy / Pliable vertex isotopy.} A continuous deformation of the ambient three-manifold (here $BZ \cong T^3$) carrying one embedded graph to another. In diagrammatic terms, this is the equivalence generated by generalized Reidemeister moves \textbf{(I)--(VI)}.

    \item \textbf{Rigid-vertex isotopy.} An isotopy in which the cyclic order of the half-edges at each vertex is fixed. Diagrammatically, it is generated by moves \textbf{(I)--(V)}; move \textbf{(VI)} is excluded because it can change that cyclic order.

    \item \textbf{Yamada polynomial and its normalization.} The Yamada polynomial $\Upsilon(\mathcal{G};Y)$ is a one-variable \textit{Laurent} polynomial defined by the axioms in \cref{tab:YamadaTable}, with $\sigma := Y + 1 + Y^{-1}$. The unnormalized polynomial is invariant under \textit{regular} rigid-vertex isotopy, i.e.\ under moves \textbf{(II)--(IV)}. We use the normalized form $\overline{\Upsilon}(\mathcal{G};Y)$, which removes the monomial factors from moves \textbf{(I)} and \textbf{(V)} so that $\overline{\Upsilon}$ is invariant under moves \textbf{(I)--(V)}. In general it is not invariant under move \textbf{(VI)}; however, if all vertices have degree at most $3$, then move \textbf{(VI)} is generated by moves \textbf{(I,IV,V)}, hence $\overline{\Upsilon}$ is also an ambient-isotopy invariant in this case.

    \item \textbf{Yamada set (multi-boundary invariant).} For a generic Fermi surface with multiple boundaries in the $i$-th energy window of band $n$, we record the multiset of normalized Yamada polynomials associated with each boundary,
    $\overline{\Upsilon}_{\partial\mathcal{F}_{[i]}^{(n)}}:=\{\overline{\Upsilon}(\mathcal{G}_{[i],\alpha}^{(n)};Y)\}_{\alpha=1}^{m_{[i]}^{(n)}}$,
    where $\alpha$ runs over all boundary components of $\mathcal{F}_{[i]}^{(n)}$.
    This boundary-wise encoding generalizes the single Yamada polynomial that is appropriate only for handlebody.

    \item \textbf{Yamada sequence.} For a Hamiltonian $H$, the Yamada sequence is the ordered list of invariants encountered across successive energy windows. If $E_1<\cdots<E_{m+1}$ are consecutive Lifshitz energies and one representative energy is chosen in each window $E\in(E_i,E_{i+1})$, then in the handlebody case $\boldsymbol{\overline{\Upsilon}}(H)=\bigl[\overline{\Upsilon}(\mathcal{G}_{[1]}),\ldots,\overline{\Upsilon}(\mathcal{G}_{[m]})\bigr]$. For generic multi-boundary Fermi surface (Fermi volume being compression-body), the $i$-th entry is the corresponding multiset-valued Yamada set $\overline{\Upsilon}_{\partial\mathcal{F}_{[i]}}$. In other words, a Yamada sequence is a sequence of Yamada sets, and reduces to a sequence of Yamada polynomials in the handlebody case.

    \item \textbf{Lifshitz transition.} The change of Fermi surface topology at a critical energy. I.e., as $E$ crosses a critical value of $\varepsilon_n$, and the Fermi volume changes by a standard Morse-theoretic move such as handle attachment/removal or component merger/splitting. Meanwhile it implies that the Fermi surface crosses at least one \textbf{van Hove singularity} $\mathbf{k}: \nabla_{\mathbf{k}}\varepsilon_n=0$.
    In semiclassical language, the Fermi velocity is the flow field \textit{on} the Fermi surface, van Hove singularities are its zeros, and a Lifshitz transition means crossing a van Hove point. 
    Thus, Lifshitz transition is associated with the global reorganization of Fermi velocity field on the Fermi surface.

    \item \textbf{Exceptional surface.} For the perturbed non-Hermitian two-band model $H_E(\mathbf{k})=H(\mathbf{k})+iE\,\sigma_y$ (\cref{eq:NonhermitianThickening}) (note $E$ here is the perturbation strength parameter, not the constant-energy), the exceptional surface is the momentum set where $H_E$ becomes defective.

    \item \textbf{Biorthogonal Berry connection and curvature.} Let the right/left eigenvectors for a non-Hermitian band be $|u_n^R(\mathbf{k})\rangle$ and $\langle u_n^L(\mathbf{k})|$, normalized biorthogonally by $\langle u_n^L|u_n^R\rangle=1$; $n$ is the band index. The Berry connection and curvature are
    $\mathbf{A}_n(\mathbf{k})=i\langle u_n^L|\nabla_{\mathbf{k}}|u_n^R\rangle$ and $\boldsymbol{\Omega}_n(\mathbf{k})=\nabla_{\mathbf{k}}\times\mathbf{A}_n(\mathbf{k})$.

    \item \textbf{Abelian edge flow.} An assignment $\phi$ of values in an Abelian group $A$ to oriented edges of a graph such that reversing the orientation reverses the value, $\phi(\bar e)=-\phi(e)$, and the signed sum of incident edge values vanishes at each vertex (Kirchhoff junction rule). In our exceptional surface example, the flow on a knotted-graph edge is defined by the Berry flux $\Phi_B \in \mathbb{R}$ flowing through the tube cross-section, and thus the Abelian group is the additive real numbers $A=(\mathbb{R},+)$.
\end{itemize}

\section{Ansatzes of Nodal Hamiltonians and their Fermi Surfaces}
\label[smsection]{appx:ansatz}

Many of the illustrative knotted graph and Yamada polynomial fingerprints in the main text are based on the Fermi surface of the two-band nodal-knot Hamiltonian ansatz \cref{eq:two_band_hamiltonian} \cite{Bi_2017,Li2019}:
\begin{equation}
H(\mathbf{k})=[\mathrm{Re}\,f(\mathbf{k})]\,\sigma_x+[\mathrm{Im}\,f(\mathbf{k})]\,\sigma_z,
\end{equation}
whose dispersion relation is
\begin{equation}
\varepsilon_\pm(\mathbf{k})=\pm |f(\mathbf{k})|,
\end{equation}
with $f(\mathbf{k})$ constructed from the two building blocks $z(\mathbf{k}),w(\mathbf{k})$ defined in \cref{eq:z_w}:
\begin{align}
z(\mathbf{k})&=\cos(2k_z)+\tfrac12+i\bigl(\cos k_x+\cos k_y+\cos k_z-2\bigr) \notag\\
w(\mathbf{k})&=\sin k_x+i\,\sin k_y
\end{align}
A key advantage of this ansatz is that desired knot topologies can be straightforwardly designed: by construction, it generates a family of Hamiltonians with nodal sets at $E=0$ forming knots, links, and knotted graphs---namely, the \textit{nodal knot} / \textit{nodal link} / \textit{nodal graph}. 

\begin{table}[h!]
\centering
\begin{tabular}{@{} l c @{}}
\toprule
\textbf{Nodal set type} & \makecell[c]{\textbf{Ansatz $f(z,w)$} \\ \textbf{in Hamiltonian [\cref{eq:two_band_hamiltonian}]}} \\
\midrule
Unknot & $z$ \\[2pt] 
Hopf link & $z^2-w^2$ \\[2pt]
Trefoil knot & $z^2-w^3$ \\[2pt]
Solomon's knot & $z^2-w^4$ \\[2pt]
Cinquefoil knot & $z^2-w^5$ \\[2pt]
Three-link & $z(z^2-w^2)$ \\[2pt]
$(p,q)$-torus knot & $z^p-w^q$ \\[2pt]
\makecell[l]{\textit{Awesome} knotted graph \\ (a fine-tuned nodal graph)} & $z(z^2-w^4+w)$ \\
\bottomrule
\end{tabular}
\caption{\textbf{Ansatzes of nodal knots/links/graphs used in this work.}
The nodal set at $E=0$ is given by $f(z(\mathbf{k}),w(\mathbf{k}))=0$, while Fermi  surfaces are $|f(\mathbf{k})|=E$.}
\label{tab:nodal-knots}
\end{table}

\begin{figure*}[h!]
    \centering
    \includegraphics[width=\linewidth]{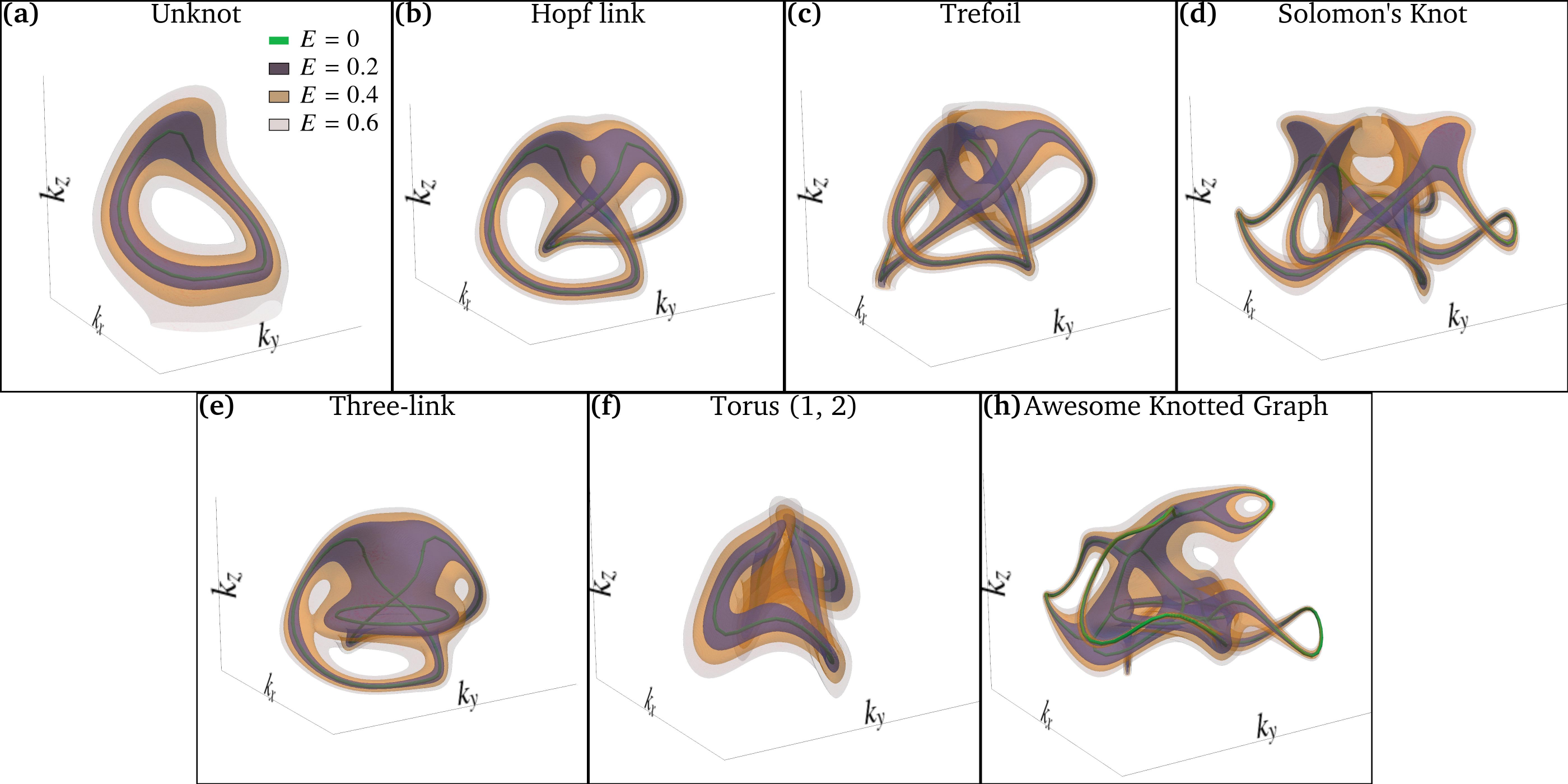}
    \caption{\textbf{Nodal knots/links/graphs and their Fermi surfaces.}
    Representative level sets $|f(\mathbf{k})|=E$ for the models in \cref{tab:nodal-knots} at $E=0$ (green), $E=0.2$ (dark purple), $E=0.4$ (brown), and $E=0.6$ (grey).
    As the Fermi energy $E$ increases, the Fermi surfaces thicken around the nodal set and can undergo Lifshitz transitions (topology of Fermi surface changes), which are captured by changes in the associated knotted graphs.}
    \label{fig:energyisosurface}
\end{figure*}

\cref{tab:nodal-knots} lists some functional forms of $f(z,w)$ and the corresponding nodal set types. \cref{fig:energyisosurface} shows some of the nodal sets and overlays their Fermi surfaces at various Fermi energies.

For these models, the Fermi surfaces of the positive band at $E$ are the level sets
\begin{equation}
\Sigma_{E}=\{\mathbf{k}\in\mathrm{BZ}:|f(\mathbf{k})|=E\},
\end{equation}
and in the nodal limit $E=0$ the knotted graph reduces to $\mathcal{G}_{[1]}\cong \Sigma_{(E=0)}$---the nodal knot/link/graph itself.

\section{Computational Skeletonization Pipeline}
\label[smsection]{appx:skeletonization}

This section explains our numerical method that maps a Fermi surface to its knotted graph used for topological classification.

\subsection{From Fermi sea to skeleton graph}

\begin{figure*}
    \centering
    \includegraphics[width=\linewidth]{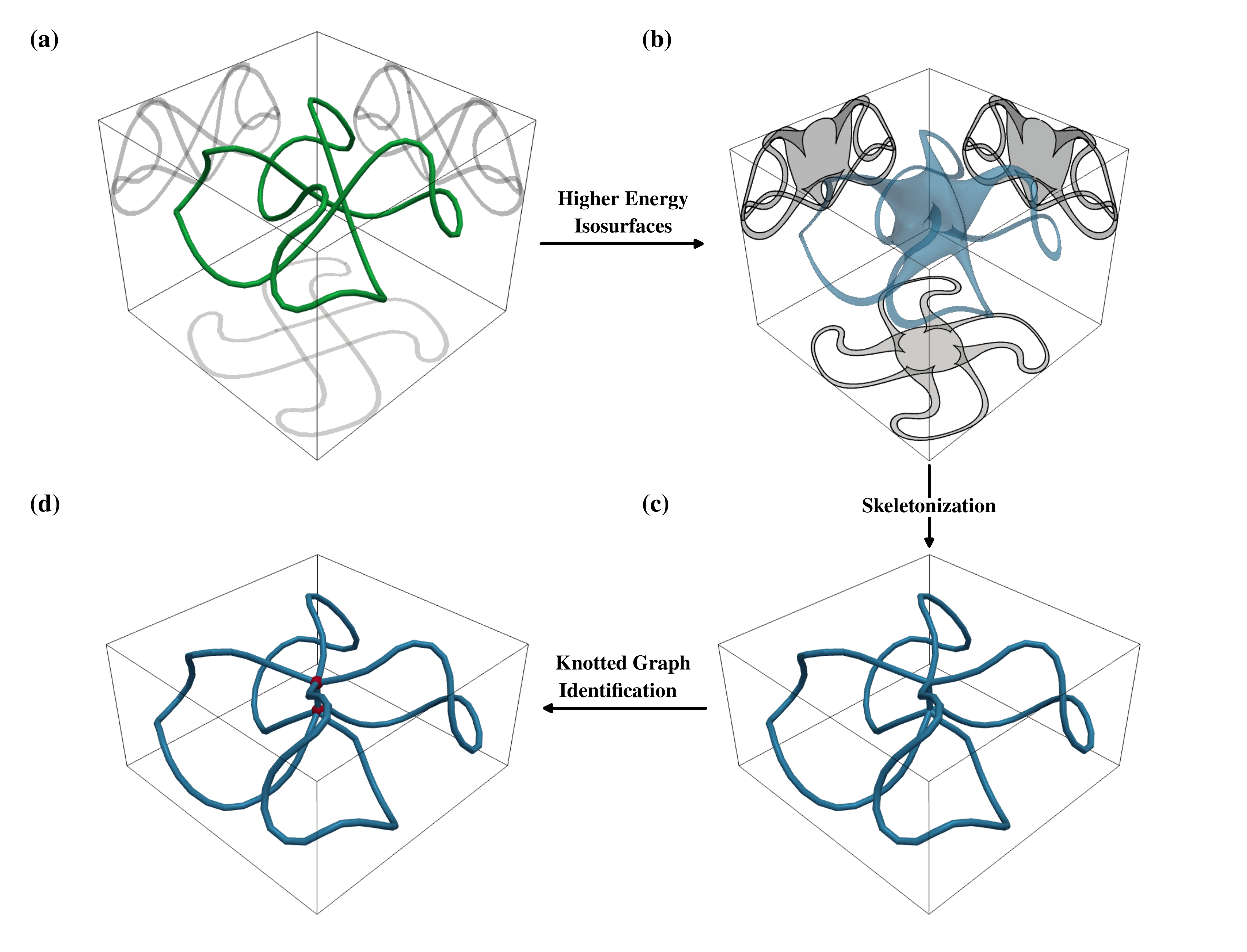}
    \caption{\textbf{Nodal knot, Fermi surface, and knotted graph.}
    \textbf{(a)} The nodal knot (collapsed Fermi surface at $E=0$) of the Solomon's knot model [\cref{tab:nodal-knots}].
    \textbf{(b)} A Fermi surface $\Sigma_E$ at finite Fermi energy $E>0$ of the nodal knot. As the Fermi energy is varied, Lifshitz transitions can create or remove necks and thereby change the connectivity of the Fermi surface. The Fermi volume $\mathcal{F}_E$ is the sublevel set enclosed by the Fermi surface $\Sigma_E = \partial \mathcal{F}_E$. 
    \textbf{(c)} Skeletonization (topology-preserving thinning) of the voxelized Fermi volume yields a raw 1-voxel-thick skeleton. 
    A voxel is the 3D analogue of a pixel, i.e., a small cube in a regular 3D grid.
    \textbf{(d)} Parsing the skeleton voxels, pruning short spurs, and contracting degree-two chains converts this raw skeleton into the final knotted graph, whose vertices track branching or merging of the topological skeleton.
    }
    \label{fig:knot_to_spatialgraph}
\end{figure*}

For a chosen band $n$ and Fermi energy $E$, we start from the Fermi surface $\Sigma_E^{(n)}=\{\mathbf{k}\in\mathrm{BZ}:\varepsilon_n(\mathbf{k})=E\}$ in the Brillouin zone.
After selecting one representative energy in the $i$-th interval between Lifshitz transitions, we denote this surface by $\Sigma_{[i]}$ or $\Sigma_{[i]}^{(n)}$.
For regular $E$, the surface is smooth and embedded.
Topology changes occur only at critical energies (Lifshitz transitions), where the level set becomes singular and components merge, split, or change genus through handle attachment/removal.

For our purposes, the surface topology is most conveniently represented through its enclosed region.
We therefore take the sublevel set, i.e., the Fermi sea or \textbf{Fermi volume}, as input:
\begin{equation}
\mathcal{F}_E^{(n)}=\{\mathbf{k}\in\mathrm{BZ}:\varepsilon_n(\mathbf{k})\le E\},
\end{equation}
and deterministically extract a 1D skeleton graph $\mathcal{G}_{[i]}^{(n)}$ from it.

This volume-based method is numerically more stable than trying to infer graph connectivity directly from a surface triangulation, and it also matches the conceptual picture of \textit{handlebody} used throughout the paper.

This is the progression illustrated in \cref{fig:knot_to_spatialgraph}.
At $E=0$, the Fermi surface collapses to the nodal knot itself.
At finite Fermi energy, the level set thickens into a surface bounding a Fermi volume, and Lifshitz transitions can create additional necks and handles.
The computational task is then to collapse that volume to a graph that captures its topology.

The pipeline summarized in \cref{fig:knot_to_spatialgraph}(b--d) is: (i) voxelize the enclosed volume on a 3D grid; (ii) apply topology-preserving thinning to obtain a 1-voxel-thick skeleton; (iii) convert the remaining skeleton voxels into vertices and edges; and (iv) prune artifacts that do not affect the Fermi topology.
The result is the embedded graph $\mathcal{G}_{[i]}^{(n)}$ used in the subsequent Yamada-polynomial computation.

\subsection{Medial axis as an ``innermost'' skeleton}

\begin{figure}[!h]
    \centering
    \includegraphics[width=0.47\linewidth]{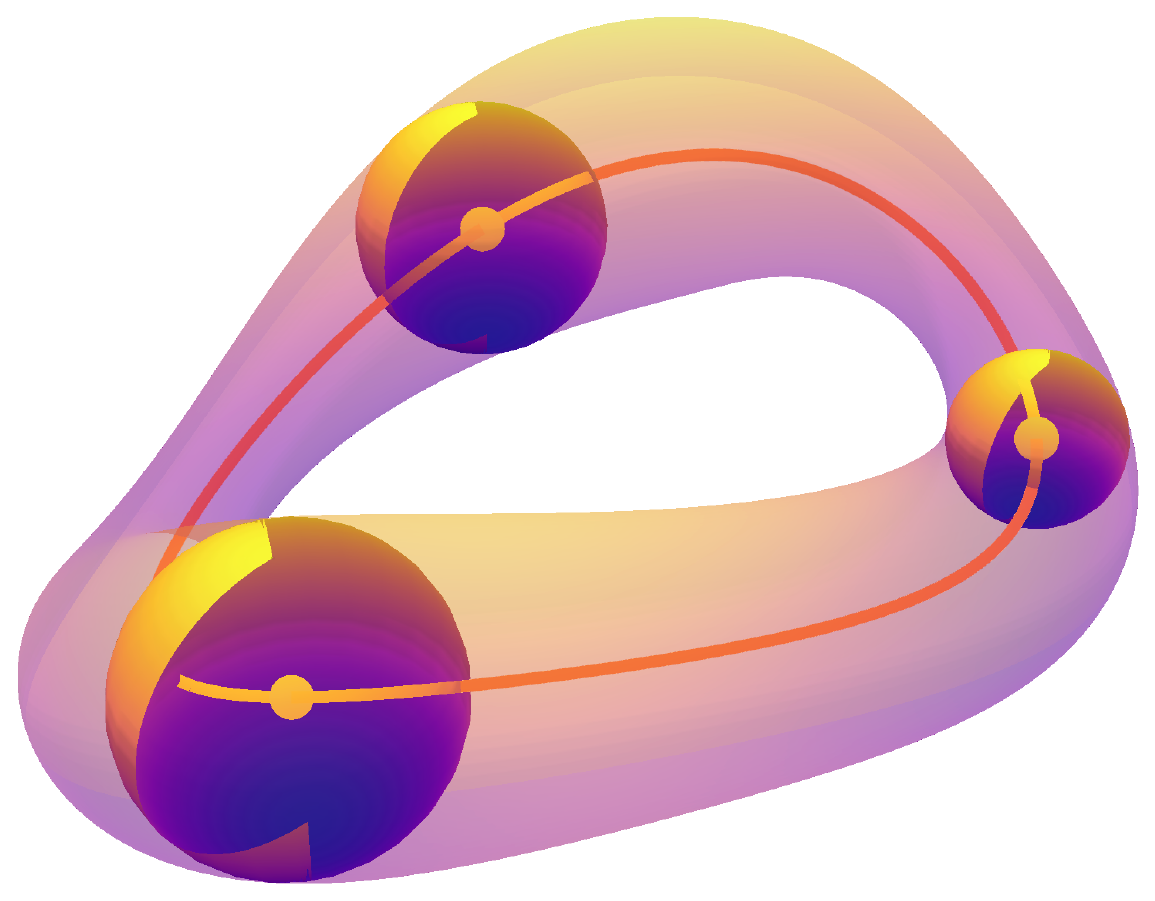}
    \caption{\textbf{Medial-axis picture behind skeletonization.}
    An example Fermi surface for an unknot model (a distorted torus) is shown together with representative maximal inscribed balls within.
    The centers of these balls form the \textit{medial axis}.
    For tubular, handlebody-like Fermi volumes, this set can be continuously collapsed to a 1D graph, which motivates the extraction of a knotted-graph skeleton.
    Numerically, we compute a topology-preserving 1-voxel-thick skeleton of the voxelized Fermi volume and then prune numeric artifacts to obtain the clean graph output.}
    \label{fig:medialaxisunknot}
\end{figure}

The thinning step is motivated by the ``medial-axis'' conceptual picture in \cref{fig:medialaxisunknot}.
For a solid region $V\subset\mathbb{R}^3$, the \textbf{medial axis is the set of points that have more than one closest point on the closed boundary $\partial V$}. Equivalently, it is \textbf{the set of centers of maximal inscribed balls in $V$}. Informally, it is the most ``interior'' center set of the volume.

For a generic 3D object, the medial axis can contain 2D sheets (usually when geometry is symmetric), so it is not necessarily always 1D.
These 2D medial-axis sheets are geometrically meaningful, but they are more detailed than what is needed for the topological invariant used in our analysis.
For topological classification, we merely need a skeleton that preserves the connected components and tunnels of the volume.
Thus, more precisely, our thinning procedure is a topology-preserving collapse inspired by the medial axis: in practice, we perform iterative thinning that also collapses sheet-like parts of the medial-axis to a 1D skeleton.

\subsection{Practical pipeline: iterative morphological thinning and graph pruning}

We begin by voxelizing the Fermi volume on a regular 3D grid: first discretize the Brillouin zone into a 3D grid of small cubes (voxels), then mark each voxel as inside or outside $\mathcal{F}_E^{(n)}$.

We then iteratively remove boundary voxels that are \textit{simple} in the sense of digital topology.
Deleting a simple voxel leaves both the occupied voxel set and its complement with the same connectivity, so no components, tunnels, or cavities are created or destroyed.
The iteration stops when no further simple boundary voxels remain.
The result is a topology-preserving \textit{1-voxel-thick} skeleton.

This ``raw'' skeleton is not yet the graph used for invariants.
It typically contains short spurious spurs caused by discretization noise, as well as topology-wise redundant degree-two vertices scattered along otherwise smooth edges.

To obtain a clean spatial graph, we propose and implement two pruning steps:
\begin{itemize}
    \item \textbf{Trim short degree-one spurs (``bristles'').}
    Remove short degree-one branches up to a certain length threshold.
    These \textit{short} spurs are artifacts of numeric noise, and they are irrelevant to the cycle structure that encodes the underlying knot/link/graph topology.
    In geometric topology, this equivalence-preserving operation is known as \textbf{contraction}; the inverse operation of this is \textbf{expansion} which adds leaf edges.
    However, we do keep \textit{long} leaf edges in the output, as they carry physical information rather than being voxel-scale noise.

    \item \textbf{Remove degree-two vertices.}
    Degree-two vertices are also irrelevant to the spatial graph topology; they only appear when a spatial edge has sharp twists, which is identified by our skeleton-to-graph program as a turning vertex.
    Removing degree-two vertices is equivalent to merging consecutive edge segments into a single edge.
    In geometric topology, this equivalence-preserving operation is known as \textbf{un-subdivision}, whose inverse is \textbf{subdivision} that adds degree-two vertices on edges.
    This yields a minimal embedded graph of interest and significantly reduces numeric overhead for invariant computation (e.g., the Yamada polynomial) without changing the ambient-isotopy class of the knotted graph.
\end{itemize}

After these topological simplification steps, the resulting graph, denoted by $\mathcal{G}_{[i]}$, is the knotted graph fingerprint that is used throughout this work.

\section{Yamada Polynomial --- Definition, Property, and Efficient evaluation}
\label[smsection]{appx:Yamada}

\begin{table*}[t]
\centering
\includegraphics[width=\linewidth]{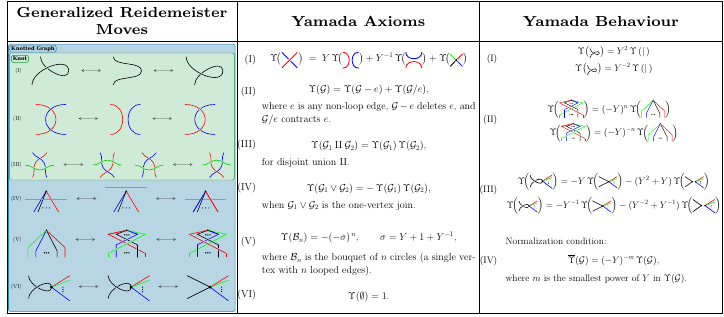}
\caption{\textbf{Generalized Reidemeister moves and Yamada polynomial.}
\textbf{(left)} generalized Reidemeister moves for spatial graph diagrams. The usual Reidemeister moves (I--III) govern knot ambient isotopy (highlighted with a green box); the additional vertex moves (IV--VI) generalize the conventional moves, and in conjuntion with moves (I--III), they govern the ambient isotopy of spatial graph (highlighted with a blue box).
\textbf{(middle)} defining axioms (including skein relations) of Yamada polynomial $\Upsilon(\mathcal{G};Y)$. These axioms define a ``raw'' Yamada polynomial that is invariant up to moves (II,III,IV), namely, \textit{regular rigid vertex isotopy}.
\textbf{(right)} additional formulae that constrains $\Upsilon$ and normalizes $\Upsilon\to \overline{\Upsilon}$, such that it is invariant up to moves (I--V), namely, \textit{rigid vertex isotopy}.
If the maximum degree of the vertices is 3 or less, e.g. conventional knots/links, and all examples in \cref{fig:gallery}(a--f), $\overline{\Upsilon}(\mathcal{G};Y)$ becomes invariant up to all moves (I--VI) and thus it's an invariant of \textit{ambient isotopy} (or \textit{pliable vertex isotopy}, or simply \textit{isotopy}). 
The normalized $\overline{\Upsilon}$ is reported consistently throughout this work, if not stated otherwise.}
\label{tab:YamadaTable}
\end{table*}

In conventional knot theory, knots and links are studied as a \textit{planar diagram}, i.e. a direct projection from 3D embedding to 2D plane $\mathbb{R}^2$ (see an example in \cref{fig:YamadaResolutions}). 
Knot's ambient isotopy is governed by the conventional Reidemeister moves (I--III) [\cref{tab:YamadaTable} (left, green box)], which are defined and operate on the planar diagram. 
Topological invariants e.g. Alexander polynomial can effectively distinguish non-isotopic knots.

Knotted-graphs (spatial-graphs), which are much less commonly known, similarly have \textit{graph diagrams}, which in additional allow vertices (of arbitrary degree). 
Knotted-graph's ambient isotopy is in turn governed by the \textit{generalized} Reidemeister moves (I--VI) \cite{mellorInvariantsSpatialGraphs2018,kauffmanInvariantsGraphsThreeSpace1989} [\cref{tab:YamadaTable} (left, blue box)].

To account for the additional vertex moves (IV--VI), knot invariants are generalized to invariants of knotted graphs that capture both knot-theoretic and graph-theoretic structure. The best-known such invariant is the \textit{Yamada polynomial} \cite{yamada1989invariant}.
The definition and evaluation reviewed in this section are standard \cite{yamada1989invariant,mellorInvariantsSpatialGraphs2018,kauffmanInvariantsGraphsThreeSpace1989}; our extensions beyond existing literature---the compression-body framework, Yamada set, and Yamada sequence---are developed in \cref{appx:compression-body} and applied to real materials in \cref{appx:materials}.

\begin{figure*}[t]
    \centering
    \includegraphics[width=\linewidth]{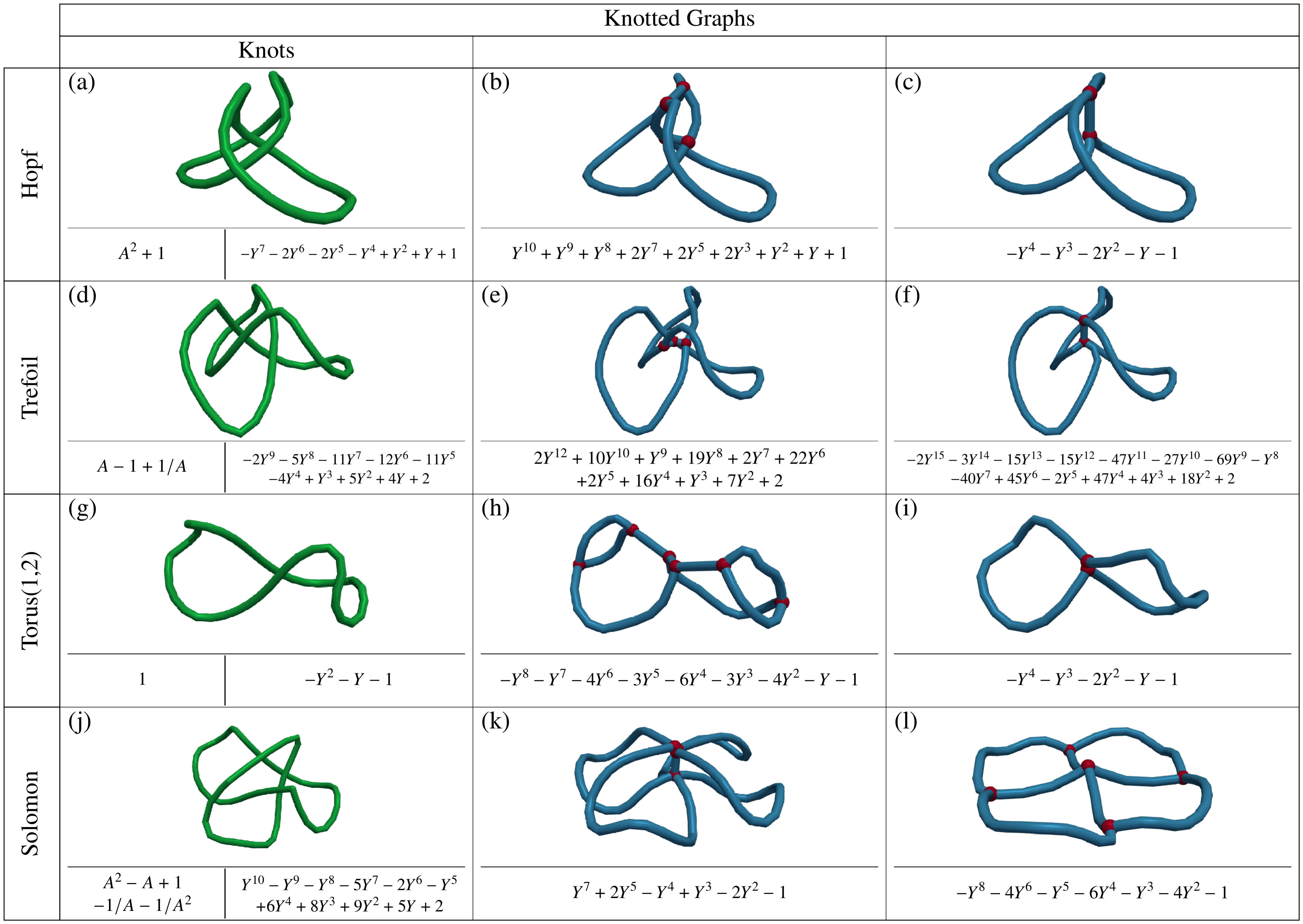}
    \caption{\textbf{Knots/links versus knotted graphs.}
    Examples of nodal knots/links (green, panels (\textbf{a,d,g,j})) and knotted graphs from Fermi surfaces at higher Fermi energies (blue, panels (\textbf{b,c,e,f,h,i,k,l})).
    For knots/links (no vertices), standard knot invariants such as the Alexander polynomial $\mathcal{A}$ apply; the Yamada polynomial $\overline{\Upsilon}$ also applies in this special case.
    Once genuine graph vertices appear, knot-only invariants cease to apply, whereas $\overline{\Upsilon}$ continues to furnish an invariant.}
    \label{fig:YamadaSpatialGraphExamples}
\end{figure*}

For readers familiar with knot theory, the Yamada polynomial $\Upsilon(Y)$ for knotted graphs plays a role analogous to the Alexander polynomial $\mathcal{A}(A)$ for ordinary knots and links. In the nodal regime $E\gtrsim 0$ prior to the first Lifshitz transition, the skeleton of the Fermi surface forms an ordinary knot or link, which can be classified using conventional knot invariants; examples of Alexander polynomials are shown in \cref{fig:YamadaSpatialGraphExamples}(a,d,g,j). 
Once Lifshitz transitions occur and vertices appear on the skeleton graph, however, it leaves the class of knots and enters the broader class of knotted graphs, illustrated in \cref{fig:YamadaSpatialGraphExamples}(b,c,e,f,h,i,k,l). Knot-only invariants are then no longer applicable, whereas the Yamada polynomial continues to apply.

Equivalently, knots can be viewed as a special subset of knotted graphs. In that sense, the Yamada polynomial naturally extends the scope of knot invariants: it applies to ordinary knots/links when no vertices are present, and it remains applicable after vertices appear. 
This is the main reason it is the appropriate invariant for the family of spines studied in this work, as it covers both the nodal knot limit and all the connectivity changes due to Lifshitz transitions in a unified way.

\subsection{Definition, normalization, and properties}

The Yamada polynomial is defined purely by local \textit{skein relations} \cref{tab:YamadaTable}(middle)---recursive relations on graph diagrams.
An efficient algorithmic way to compute the Yamada polynomial is through a \textit{state-sum}, namely by resolving crossings into planar \textit{states}.
We discuss the definition, properties, and state-sum evaluation of the Yamada polynomial in order next.

The unnormalized Yamada polynomial can be defined as the unique Laurent polynomial $\Upsilon(\mathcal{G};Y)$ that satisfies the axioms in \cref{tab:YamadaTable}(middle).

Starting from these axioms, it can be shown that the Laurent polynomial is unchanged under moves (II), (III), and the rigid-vertex move (IV). 
In mathematical language unnormalized Yamada polynomial is an invariant of spatial graphs up to \textit{regular rigid vertex isotopy}. 

Generally, the moves (I), (V) or (VI) would change the unnormalized $\Upsilon(\mathcal{G};Y)$.
In particular, moves (I) and (V) change the polynomial by multiplying a monomial of $(-Y)$; the behavior is summarized as the \textit{first two formulae} in \cref{tab:YamadaTable}(right). Following this convention, we can remove the dependence on moves (I,V) by \textit{normalizing} the polynomial:
\begin{equation}
\boxed{
\overline{\Upsilon}(\mathcal{G};Y):=(-Y)^{-m}\,\Upsilon(\mathcal{G};Y),\qquad m:=\min\{\text{exponents of }Y\text{ appearing in }\Upsilon(\mathcal{G};Y)\}
}
\label{eq:yamada_normalization}
\end{equation}
This \textbf{normalized} $\overline{\Upsilon}(\mathcal{G};Y)$ is an invariant of \textit{rigid vertex isotopy}, i.e., invariance under moves (I--V).

An important special scenario is that, for graphs with maximum vertex degree $\le 3$, move (VI) can be decomposed into moves (V)$\to$(IV,I). Hence, $\overline{\Upsilon}$ is an \textit{ambient isotopy} invariant in these cases \cite{mellorInvariantsSpatialGraphs2018}. 
E.g., for conventional knots/links, and all examples in \cref{fig:gallery}(a--f), $\overline{\Upsilon}$ is an ambient isotopy invariant.
Ambient isotopy is also known as \textit{pliable vertex isotopy}, or simply \textit{isotopy}.

\paragraph*{Convention for graphs with higher-degree vertices.}
When $\mathcal{G}$ has a vertex of degree $\ge 4$, move (VI) is no longer generated by moves (I--V), and $\overline{\Upsilon}$ is only a \textit{rigid vertex isotopy} invariant: two diagrams of the same embedded graph that differ by a cyclic reordering of the edges around such a vertex may yield different $\overline{\Upsilon}$.
Such vertices do occur in this work---the real-material skeletons of \cref{appx:materials} contain vertices of degree up to $6$ (in $\theta_6$) and $12$ (in $\theta_{12,P}$).
To fix a definite diagram in these cases, we adopt a \textbf{minimum-crossing planar-projection convention}: among the generic planar projections of the embedded graph, we retain one whose diagram realizes the smallest crossing number, and quote the $\overline{\Upsilon}$ of that diagram.
In practice we keep the projection plane fixed to be the $xy$ plane and rotate the embedded graph instead.
A generic rotation carries three Euler angles, but the third one merely rotates the resulting diagram within the projection plane, and reversing the viewing direction (a flip of the $z$ axis) only shows the same diagram from behind.
Since neither operation changes the diagram obtained, the genuinely distinct projections are labelled by a viewing direction taken up to antipodal identification, i.e.\ by one half of the direction sphere, over which we sample uniformly.
We draw $N_{\mathrm{rot}}$ such rotations ($N_{\mathrm{rot}}=10$ by default), project each rotated graph onto the $xy$ plane, and evaluate $\overline{\Upsilon}$ on whichever of the resulting diagrams carries the fewest crossings.
We do not prove that this sampling is guaranteed to reach a globally unique minimum-crossing polynomial for an arbitrary knotted graph; we only observe that the attained minimum already saturates at $N_{\mathrm{rot}}=10$ for every knotted graph appearing in this manuscript.
A selection procedure with an accompanying guarantee is left for future work.
Whenever every vertex has degree $\le 3$, no such choice is needed and the projection may be taken arbitrarily.

Relatedly, other ambient-isotopy invariants also exist (e.g., the Yokota polynomial \cite{yokotaTopologicalInvariantsGraphs1996} and Thompson polynomial \cite{thompsonPolynomialInvariantGraphs1992}), but they are typically much more expensive to evaluate.

\subsection{State-sum evaluation via the specialized Negami polynomial}

A practical way to compute the Yamada polynomial is through a \textit{state-sum} expansion \cite{huh2022yamadapolynomialassociatedlink}.
One starts from a planar diagram of a spatial graph $\mathcal{G}$, namely a 2D projection together with the over/under-crossing information.
At each crossing $\scalebox{0.7}{\CrossPic}$, the skein relation in \cref{tab:YamadaTable}(middle) yields three local resolutions: the two \textit{smoothing}s $\scalebox{0.7}{\SidePic}$ and $\scalebox{0.7}{\TopPic}$, and the \textit{vertex resolution} $\scalebox{0.7}{\VertexPic}$.
After resolving every crossing, one obtains a crossing-free planar graph, called a \textit{state}.
Here ``planar graph'' means that the resulting graph can be drawn in the plane without any remaining edge intersections; do not conflate it with ``planar diagram''.
If the original diagram has $c$ crossings, then there are $3^c$ possible states.

Let $S$ be the set of all such states.
For a given state $s \in S$, denote by $\mathcal{G}_s$ the planar graph obtained after all crossings are resolved.
\textbf{Let $p(s)$ and $m(s)$ be the numbers of $\scalebox{0.7}{\SidePic}$ and $\scalebox{0.7}{\TopPic}$ smoothings used in the construction of $s$, respectively}; the remaining crossings are resolved by $\scalebox{0.7}{\VertexPic}$.
The Yamada polynomial is then obtained as the weighted sum over all states:
\begin{equation}
\boxed{
\Upsilon(\mathcal{G};Y)
=
\sum_{s\in S}
Y^{p(s)-m(s)}
\, h\!\left(\mathcal{G}_s;\,-1,\,-Y-2-Y^{-1}\right)
}
\label{eq:yamada_negami}
\end{equation}
This formula separates the problem into two steps: first resolve the crossings, and then evaluate a purely graph-theoretic quantity on the resulting planar graphs.

For a crossing-free graph $\mathcal{G}_s$, the required two-variable polynomial $h(\mathcal{G}_s;x,y)$ is the specialization of the Negami polynomial used in Yamada's construction \cite{negami1987polynomial}:
\begin{equation}
h(\mathcal{G}_s;x,y)
=
\sum_{F\subseteq E(\mathcal{G}_s)}
(-x)^{-|F|}
\,x^{\mu(\mathcal{G}_s-F)}
\,y^{\beta(\mathcal{G}_s-F)} ,
\label{eq:negami}
\end{equation}
where $F$ is a subset of edges to be deleted, $\mathcal{G}_s-F$ is the graph obtained after deleting those edges, and $\sum_{F\subseteq E(\mathcal{G}_s)}$ sums over all possible edge subsets.
$\mu(\mathcal{G}_s-F)$ is the number of connected components of the remaining graph, and $\beta(\mathcal{G}_s-F)$ is its first Betti number, also called the \textit{cycle rank}, which counts the number of independent cycles.
In the Yamada specialization $x=-1$, the prefactor $(-x)^{-|F|}$ becomes unity, so the contribution of each deleted-edge configuration is controlled only by the topology of the remaining graph through $\mu$ and $\beta$.

For graphs, the first Betti number can be computed from the Euler--Poincar\'e relation,
\begin{equation}
\beta(\mathcal{G}_s-F)
=
|E(\mathcal{G}_s-F)|
-
|V(\mathcal{G}_s)|
+
\mu(\mathcal{G}_s-F) ,
\label{eq:betti1}
\end{equation}
where $|E(\mathcal{G}_s-F)|$ and $|V(\mathcal{G}_s)| = |V(\mathcal{G}_s-F)|$ are the numbers of edges and vertices, respectively.
Thus, once a state has been produced, its contribution can be computed entirely from standard combinatorial data of the planar graph, and thereby \textbf{the Yamada polynomial can be expressed in terms of planar graphs and their standard combinatorial data}:
\begin{equation}
\boxed{
\Upsilon(\mathcal{G};Y)
=
\sum_{s\in S}
Y^{p(s)-m(s)}
\,\sum_{F\subseteq E(\mathcal{G}_s)}
\,(-1)^{\mu(\mathcal{G}_s-F)}
\,(-Y-2-Y^{-1})^{|E(\mathcal{G}_s-F)|-|V(\mathcal{G}_s)|+\mu(\mathcal{G}_s-F)}
}
\label{eq:yamada_negami_expanded}
\end{equation}

In summary, the state-sum method converts the 3D topological problem into a sum over crossing resolutions, with each resolved state evaluated by a specialized graph polynomial, although the state space scales exponentially with the crossing number.

Notably, this state-sum procedure enables us to construct an algorithmic pipeline to compute the Yamada polynomial, significantly expanding our investigation scope of our topological fingerprinting framework beyond the limits of manual computation.

\begin{figure*}
    \centering
    \includegraphics[width=\linewidth]{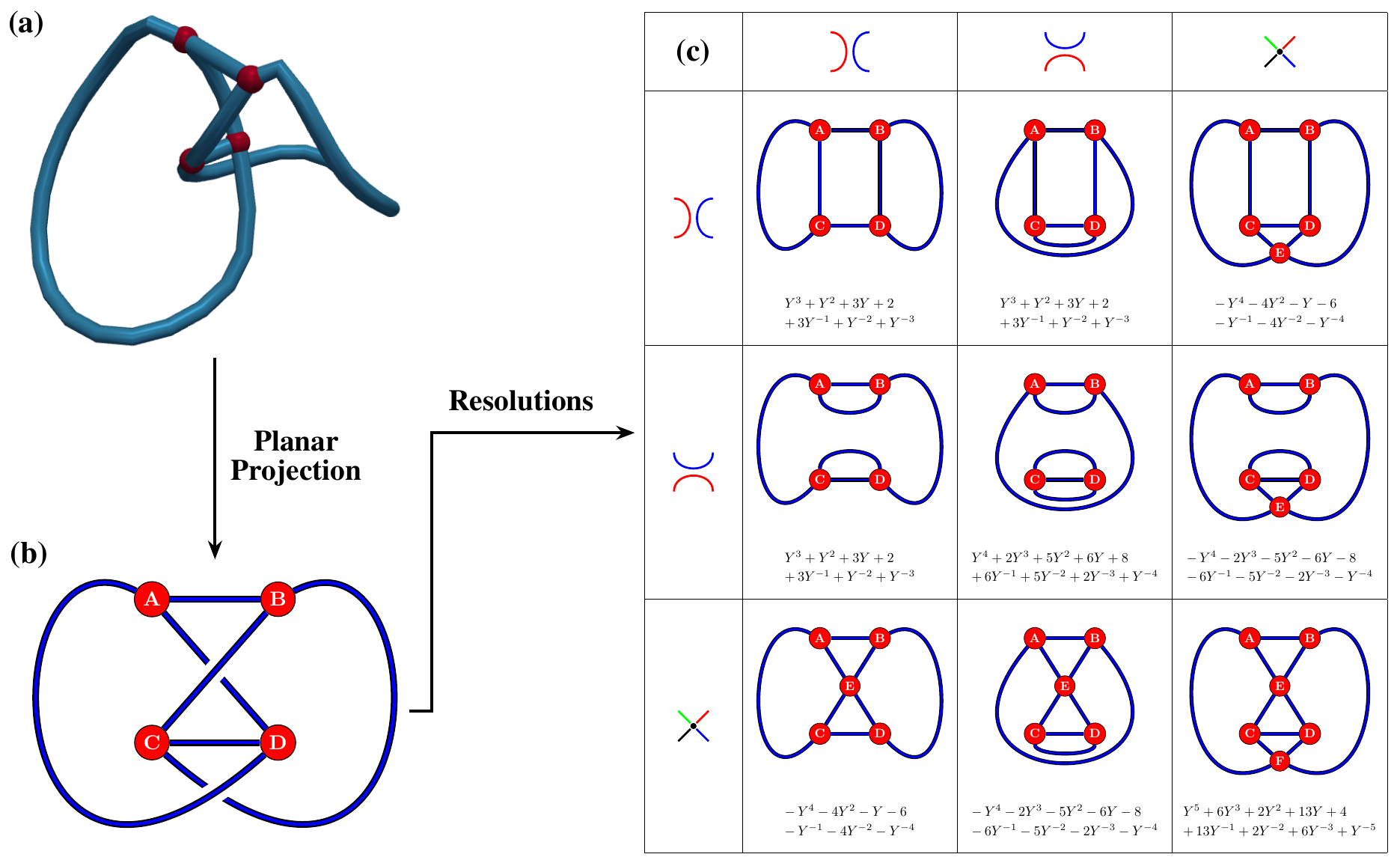}
    \caption{\textbf{Example Yamada polynomial evaluation by state-sum expansion on a two-crossing diagram.}
    \textbf{(a)} The 3D knotted graph extracted from the constant-energy surface $\Sigma_{[2]}$ at $E=0.3$ in \cref{fig:hopfsequence}(c).
    \textbf{(b)} The \textit{graph diagram} after a chosen planar projection. The \textit{four} red nodes are true graph \textit{vertices}, while the \textit{two} self-intersections are \textit{crossings}.%
    \textbf{(c)} The $3^2=9$ planar \textit{states} obtained by resolving each crossing in three ways.
    Each state $s$ contributes 
    $Y^{p(s)-m(s)}\,h\!\left(\mathcal{G}_s;-1,-Y-2-Y^{-1}\right)$ [\cref{eq:negami}] to the un-normalized Yamada polynomial, and summing all nine contributions yields $\Upsilon(\mathcal{G};Y)$ [\cref{eq:yamada_negami}].}
    \label{fig:YamadaResolutions}
\end{figure*}

\subsection{Worked example: a two-crossing diagram}

We now illustrate the state-sum procedure on a minimal nontrivial example. 
We consider the knotted graph extracted from the Fermi surface $\Sigma_{[2]}$ of the Hopf-link Hamiltonian at $E=0.3$ shown in \cref{fig:hopfsequence}(c), whose knotted graph is now displayed separately in \cref{fig:YamadaResolutions}(a). 
Because this graph is trivalent, the normalized polynomial $\overline{\Upsilon}$ is an ambient-isotopy invariant, so any projection angle may be used. 
We choose the projection shown in \cref{fig:YamadaResolutions}(b).

The projected diagram has 4 vertices, 6 edges, and 2 crossings. 
Note that the crossings are \textit{not} graph vertices; in math language, they are called \textit{transverse double points}, the point where two edges overlap in the projection and which one goes over the other is important information.

Since each crossing admits three local resolutions, the diagram generates $3^2=9$ planar states, displayed in \cref{fig:YamadaResolutions}(c). 
For each state $s$, the contribution to the un-normalized Yamada polynomial is [\cref{eq:yamada_negami_expanded}]:
$$
Y^{p(s)-m(s)}\,h\!\left(\mathcal{G}_s;-1,-Y-2-Y^{-1}\right)
=
Y^{p(s)-m(s)}
\sum_{F\subseteq E(\mathcal{G}_s)}
(-1)^{\mu(\mathcal{G}_s-F)}
(-Y-2-Y^{-1})^{|E(\mathcal{G}_s-F)|-|V(\mathcal{G}_s)|+\mu(\mathcal{G}_s-F)},
$$
Once a state is fixed, the problem is purely graph-theoretic: one only needs the number of connected components $\mu$ and the Betti-1 number $\beta=|E|-|V|+\mu$ for the graphs obtained by deleting subsets of edges.

As an explicit example of how to compute one \textit{state}, consider the state obtained by choosing the $\bigl(\scalebox{0.7}{\TopPic},\scalebox{0.7}{\TopPic}\bigr)$ resolution pair in \cref{fig:YamadaResolutions}(c), i.e. the center one in the diagram matrix.

The resulting planar, crossing-free graph is a disjoint union of two $\theta_3$-graph (two vertices connected by three parallel edges constitute a $\theta_3$-graph).
Similar to the axiom (III) in \cref{tab:YamadaTable}(middle), Negami polynomial has a similar skein relation for disjoint unions of graphs:
\begin{equation}
h\bigl(
\scalebox{0.4}{\TopPic},\scalebox{0.4}{\TopPic}
\bigr)
=
h\bigl(
\raisebox{-1.3ex}{\scalebox{0.4}{\ThreeGraphThreeEdgeAB}} 
\amalg 
\raisebox{-1.3ex}{\scalebox{0.4}{\ThreeGraphThreeEdgeCD}}
\bigr)
=
\left[h\bigl(
\raisebox{-1.3ex}{\scalebox{0.4}{\ThreeGraphThreeEdgeCD}}
\bigr)\right]^2
\end{equation}
This is straightforward to understand: the planar $\theta_3$-graph is crossing-free, so the Yamada polynomial is given by the Negami specialization.
Therefore, we only need to evaluate the specialized Negami polynomial $h(\theta_3; -1, -Y-2-Y^{-1})$ of a single $\theta_3$-graph.

For one $\theta_3$ component, since the graph is very ``symmetric'', the edge-deletion subsets $F\subseteq E(\theta_3)$ can be grouped by $|F|=0,1,2,3$.
The four groups are 
$\binom{3}{0} \times \left(\raisebox{-1.3ex}{\scalebox{0.4}{\ThreeGraphThreeEdgeCD}}\right)$, 
$\binom{3}{1} \times \left(\raisebox{-0.8ex}{\scalebox{0.4}{\ThreeGraphTwoEdge}}\right)$, 
$\binom{3}{2} \times \left(\raisebox{-0.4ex}{\scalebox{0.4}{\ThreeGraphOneEdge}}\right)$, 
and $\binom{3}{3} \times \left(\raisebox{-0.4ex}{\scalebox{0.4}{\ThreeGraphZeroEdge}}\right)$, 
where the first factor counts the number of ways to delete $|F|$ edges, and the second factor is the identical resulting graph after deleting $|F|$ edges.
Their corresponding values of $(\mu, |E|, |V|, \beta=\mu+|E|-|V|)$ are $(1,3,2,2)$, $(1,2,2,1)$, $(1,1,2,0)$, and $(2,0,2,0)$, respectively.

\begin{center}
\renewcommand{\arraystretch}{2}
\setlength{\tabcolsep}{15pt}
\begin{tabular}{|c|c|c|cc|c|}
\toprule[1.5pt]
$|F|$ & remaining graph & $\mu$ & $|E|$ & $|V|$ & $\beta=\mu+|E|-|V|$ \\
\midrule
0 & \raisebox{-1.6ex}{\scalebox{0.6}{\ThreeGraphThreeEdgeCD}} & 1 & 3 & 2 & 2 \\
1 & \raisebox{-1.4ex}{\scalebox{0.6}{\ThreeGraphTwoEdge}}    & 1 & 2 & 2 & 1 \\
2 & \raisebox{-0.7ex}{\scalebox{0.6}{\ThreeGraphOneEdge}}    & 1 & 1 & 2 & 0 \\
3 & \raisebox{-0.7ex}{\scalebox{0.6}{\ThreeGraphZeroEdge}}   & 2 & 0 & 2 & 0 \\
\bottomrule[1.5pt]
\end{tabular}
\end{center}

Substituting these data into \cref{eq:negami} gives
\begin{equation}
\begin{split}\label{eq:combinatorictheta3}
h(\theta_3;-1,-Y-2-Y^{-1})
& = \binom{3}{0} (-1)^1 (-Y-2-Y^{-1})^2
+ \binom{3}{1} (-1)^1 (-Y-2-Y^{-1})^1 \\
& \quad
+ \binom{3}{2} (-1)^1 (-Y-2-Y^{-1})^0
+ \binom{3}{3} (-1)^2 (-Y-2-Y^{-1})^0 \\
& = -Y^2-Y-2-Y^{-1}-Y^{-2}.
\end{split}
\end{equation}

As a useful consistency check, this agrees with the standard closed form of the Yamada polynomial for a planar $\theta_s$ graph,
\begin{equation}\label{eq:stheta}
\Upsilon(\theta_s)=\frac{\sigma+(-\sigma)^s}{\sigma+1},
\qquad
\sigma:=Y+1+Y^{-1},
\end{equation}
evaluated at $s=3$.

Therefore the $\bigl(\scalebox{0.4}{\TopPic},\scalebox{0.4}{\TopPic}\bigr)$ state contributes
\begin{align}\label{eq:Y_TopPic_squared}
h(\scalebox{0.4}{\TopPic},\scalebox{0.4}{\TopPic})
&=\Bigl[h(\theta_3;-1,-Y-2-Y^{-1})\Bigr]^2 
=\left(-Y^2-Y-2-Y^{-1}-Y^{-2}\right)^2 \notag\\
&=Y^4+2Y^3+5Y^2+6Y+8+6Y^{-1}+5Y^{-2}+2Y^{-3}+Y^{-4}.
\end{align}

The remaining eight states are handled in exactly the same way. 
Summing all nine weighted contributions shown in \cref{fig:YamadaResolutions}(c) yields the un-normalized Yamada polynomial of the knotted graph:
\begin{equation}\label{eq:overallyamada}
\begin{split}
\Upsilon\Bigl(\raisebox{-3.5ex}{\scalebox{0.25}{\Overallgraph}}\Bigr)
&=
\sum_{s\in S}
Y^{p(s)-m(s)}\,h\!\left(\mathcal{G}_s;-1,-Y-2-Y^{-1}\right) \\
&=
Y^{2-0}h\Bigl(\raisebox{0.5ex}{\scalebox{0.4}{\SidePic},\scalebox{0.4}{\SidePic}}\Bigr)
+ Y^{1-1}h\Bigl(\raisebox{0.5ex}{\scalebox{0.4}{\SidePic},\scalebox{0.4}{\TopPic}}\Bigr) 
+ Y^{1-0}h\Bigl(\raisebox{0.5ex}{\scalebox{0.4}{\SidePic},\scalebox{0.4}{\VertexPic}}\Bigr)
+ Y^{1-1}h\Bigl(\raisebox{0.5ex}{\scalebox{0.4}{\TopPic},\scalebox{0.4}{\SidePic}}\Bigr)
+ \dots
+ Y^{0-0}h\Bigl(\raisebox{0.5ex}{\scalebox{0.4}{\VertexPic},\scalebox{0.4}{\VertexPic}}\Bigr) \\
&=
Y^{4}+Y^{3}+Y^{2}+2Y+2Y^{-1}+2Y^{-3}+Y^{-4}+Y^{-5}+Y^{-6}.
\end{split}
\end{equation}

Finally, normalizing according to \cref{eq:yamada_normalization} gives
\begin{equation}\label{eq:overallyamadanorm}
\begin{split}
\overline{\Upsilon}\Bigl(\raisebox{-3.5ex}{\scalebox{0.25}{\Overallgraph}}\Bigr)
&=
(-Y)^6\,
\Upsilon\Bigl(\raisebox{-3.5ex}{\scalebox{0.25}{\Overallgraph}}\Bigr) \\
&=
Y^{10}+Y^9+Y^8+2Y^7+2Y^5+2Y^3+Y^2+Y+1.
\end{split}
\end{equation}

This example demonstrates the computational workflow used throughout this work: resolve all crossings, evaluate each planar state from ordinary graph data, and then sum the weighted contributions. For more complicated knotted graphs the bookkeeping becomes longer, but the logic is unchanged. 

We use this same state-sum pipeline to compute the normalized Yamada polynomials reported throughout this work.

\section{Application in Other Scenarios: \\ Non-Hermitian Perturbed Nodal Hamiltonians and Fingerprinting their Exceptional Surfaces}
\label[smsection]{appx:ES}

In this section, we show how the knotted-graph framework extends to a simple class of non-Hermitian band structures, for which the relevant geometric object is an \textit{exceptional surface} rather than a Fermi surface.

The section is organized as follows. We first apply a non-Hermitian perturbation to the nodal-Hamiltonian ansatz and describe the resulting exceptional surface. This surface marks a $\mathcal{PT}$-symmetry-breaking transition: its interior is the $\mathcal{PT}$-broken region with purely imaginary eigenvalues, whereas its exterior is $\mathcal{PT}$-unbroken and has purely real eigenvalues. Skeletonizing the broken region then produces a knotted-graph fingerprint of the exceptional surface.

We next compute the Berry curvature in the $\mathcal{PT}$-broken region and show that its Berry flux through transverse cross-sections of the exceptional-volume tubes naturally \textit{orients} and \textit{weights} the skeleton edges. Because there is no Berry-curvature monopole in the smooth interior, the fluxes incident at each vertex obey \textit{Kirchhoff's law} and define an \textit{Abelian edge flow} \cite{ishii2012quandle}.

Finally, we construct two Hamiltonians with identical complex spectra and exceptional surfaces but opposite biorthogonal Berry curvatures. This pair makes explicit the \textit{eigenstate} information carried by the flow beyond the unflowed, dispersion-based skeleton. More broadly, the construction points to a way of characterizing geometric objects endowed with vector-field structure.

\subsection{Non-Hermitian perturbation, exceptional surface, and $\mathcal{PT}$-symmetry-broken volume}

In the previously shown Hermitian nodal systems, a nodal knot is a one-dimensional locus of band degeneracies: the eigenvalues coincide along the knot, but the Hamiltonian remains diagonalizable. 
In the non-Hermitian setting considered here, the exceptional surface is instead a two-dimensional locus of exceptional points, where both eigenvalues and eigenvectors coalesce and the Hamiltonian becomes defective.
The volume enclosed by the exceptional surface in the 3D Brillouin zone can be skeletonized in the same way as the Fermi volume discussed earlier.

To make the connection with the Hermitian discussion explicit, recall the two-band nodal Hamiltonian
\begin{equation}
H(\mathbf{k})=[\Re \,f(\mathbf{k})]\,\sigma_x+[\Im \,f(\mathbf{k})]\,\sigma_z
\end{equation}
The non-Hermitian ``thickening'' introduced in the main text is
\begin{equation}
H_E(\mathbf{k})
=H(\mathbf{k})+iE\,\sigma_y
=[\Re \,f(\mathbf{k})]\,\sigma_x+[\Im \,f(\mathbf{k})]\,\sigma_z+iE\,\sigma_y
\label{eqS:nonHermHamiltonian}
\end{equation}
as in \cref{eq:NonhermitianThickening}, whose dispersion is
\begin{equation}
E_\pm(\mathbf{k})=\pm \sqrt{[\Re f(\mathbf{k})]^2+[\Im f(\mathbf{k})]^2+(i E)^2}=\pm \sqrt{|f(\mathbf{k})|^2-E^2}
\end{equation}

Here the notation deserves one remark: throughout this section, the symbol $E$ denotes the \textit{strength of the non-Hermitian perturbation}; it is \textit{not} the Fermi energy used in the Hermitian Fermi-surface discussion. Without loss of generality, we assume $E$ is a \textit{positive real} number.

The exceptional surface of $H_E(\mathbf{k})$ can be found by solving the band-closing condition:
\begin{gather}
\sqrt{|f(\mathbf{k})|^2-E^2}=0
\qquad \Rightarrow \qquad
|f(\mathbf{k})|=E \\
\therefore \text{Exceptional surface of } H_E(\mathbf{k}) = \{\mathbf{k}\in \mathrm{BZ}: |f(\mathbf{k})|=E\}
\end{gather}

Note that it is the same equation as the Fermi-surface condition for the Hermitian Hamiltonian $H(\mathbf{k})$ at Fermi energy $E$.

Therefore, the exceptional surface of $H_E(\mathbf{k})$ has exactly the \textit{same geometry} as the Hermitian level set, but a \textbf{\it different physical meaning}: it is now the locus where $H_E(\mathbf{k})$ becomes defective.

The physical meaning of the enclosed interior region is also different. For a Fermi surface, the two sides are distinguished by their energy relative to the Fermi energy. Here, {\bfseries the exceptional surface marks a \textit{$\mathcal{PT}$-symmetry-breaking transition}: its interior is the $\mathcal{PT}$-broken region with \textit{purely imaginary eigenvalues}, whereas its exterior is the $\mathcal{PT}$-unbroken region with real eigenvalues}.

In other words, the ``exceptional volume'' to be skeletonized is the region bounded by the exceptional surface in which the dispersion is purely imaginary:
\begin{align}
\text{Exceptional volume of } H_E(\mathbf{k})
:=& \{\mathbf{k}\in \mathrm{BZ}: |f(\mathbf{k})|<E\} \notag \\
=& \{\mathbf{k}\in \mathrm{BZ}: \Re(E_\pm(\mathbf{k}))=0\}
\end{align}

\begin{figure}[!ht]
    \centering
    \includegraphics[width=\linewidth]{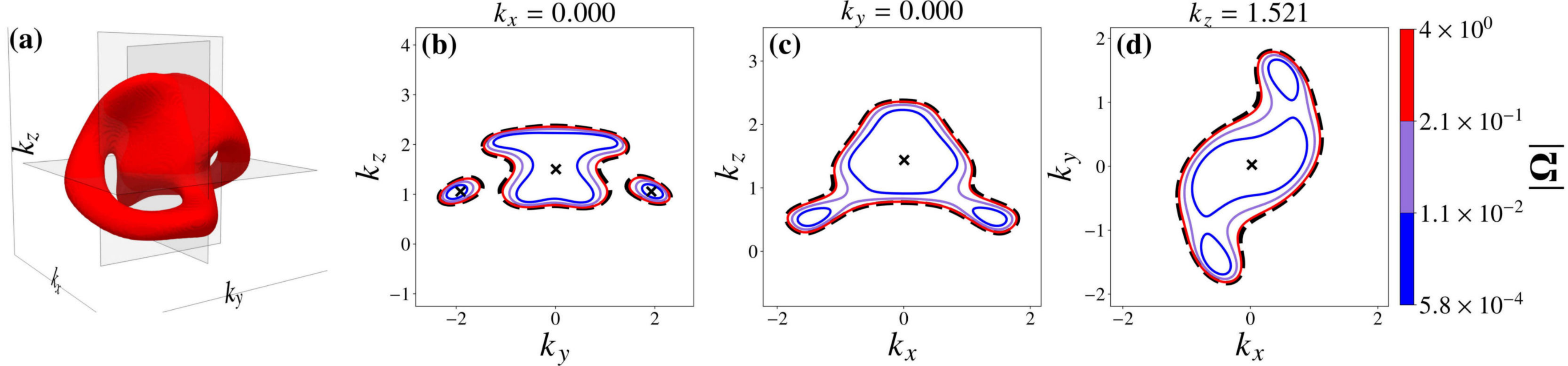}
    \caption{\textbf{Berry-curvature profile of an exceptional surface (Hopf-link example).}
    \textbf{(a)} Exceptional surface (red), which bounds the $\mathcal{PT}$-broken region that is skeletonized.
    \textbf{(b)--(d)} Three planar slices of the 3D Brillouin zone at $k_x=0$, $k_y=0$, and $k_z=1.521$, respectively, showing the biorthogonal Berry-curvature magnitude $|\boldsymbol{\Omega}(\mathbf{k})|$. The ``$\times$'' markers indicate the intersections of the slice with the knotted graph. In agreement with \cref{eq:berry_curvature_inside}, the magnitude of $\boldsymbol{\Omega}$ diverges on the exceptional surface (the black dashed boundary) and decays towards the interior.}
    \label{figLBerryprofile}
\end{figure}

As a concrete example, the Hopf-link model is obtained by choosing $f(\mathbf{k})=z(\mathbf{k})^2-w(\mathbf{k})^2$ in \cref{tab:nodal-knots}. For a representative perturbation strength, e.g. $E=0.8$, the resulting exceptional surface and its Berry-curvature profile are shown in \cref{figLBerryprofile}.

\subsection{Berry curvature in the $\mathcal{PT}$-broken region}

Because $H_E(\mathbf{k})$ is non-Hermitian, its right and left eigenvectors are different. Denoting them by $|u_n^R(\mathbf{k})\rangle$ and $\langle u_n^L(\mathbf{k})|$ (from the lower band), they satisfy
\begin{equation}
H_E(\mathbf{k})|u_n^R(\mathbf{k})\rangle=\varepsilon_n(\mathbf{k})|u_n^R(\mathbf{k})\rangle,\qquad
\langle u_n^L(\mathbf{k})|H_E(\mathbf{k})=\varepsilon_n(\mathbf{k})\langle u_n^L(\mathbf{k})|,
\end{equation}
and are normalized biorthogonally as $\langle u_m^L|u_n^R\rangle=\delta_{mn}$. On the exceptional surface, the Hamiltonian becomes defective and thus non-diagonalizable, and this biorthogonal normalization condition no longer holds.

The \textit{Berry connection} and \textit{Berry curvature} generalized to a non-Hermitian band are then biorthogonally defined \cite{wang2024berry} by
\begin{equation}
\boxed{
\mathbf{A}_n(\mathbf{k})=i\langle u_n^L(\mathbf{k})|\nabla_{\mathbf{k}}|u_n^R(\mathbf{k})\rangle,
\qquad
\boldsymbol{\Omega}_n(\mathbf{k})=\nabla_{\mathbf{k}}\times \mathbf{A}_n(\mathbf{k})
}
\end{equation}
where $n$ is the band index, and in our two-band case, $n \in \{+,-\}$. These are the quantities that enter the discussion below.

For the two-band class of models in \cref{eqS:nonHermHamiltonian}, the eigenvectors and Berry curvature can be obtained analytically \cite{wang2024berry}.

For compactness, write
\begin{gather}
f_R:=\Re \,f(\mathbf{k}),\qquad
f_I:=\Im \,f(\mathbf{k}),\qquad
|f|:=\sqrt{f_R^2+f_I^2},\qquad
\varepsilon:=\sqrt{E^2-|f|^2}.
\end{gather}

The biorthogonal Berry connection $\mathbf{A}(\mathbf{k})$ and curvature $\boldsymbol{\Omega}(\mathbf{k})=\nabla_{\mathbf{k}}\times\mathbf{A}(\mathbf{k})$ behave qualitatively differently on the two sides of the exceptional surface.

\paragraph{Inside the exceptional surface ($|f|<E$).}
\begin{gather}
\mathbf{A}(\mathbf{k})=
\frac{(\varepsilon-E)(f_I\nabla_{\mathbf{k}}f_R-f_R\nabla_{\mathbf{k}}f_I)}
{2\varepsilon |f|^2}
\label{eq:berry_connection_inside} \\
\boldsymbol{\Omega}(\mathbf{k})
=\frac{E}{2\varepsilon^3}
(\nabla_{\mathbf{k}}f_R\times\nabla_{\mathbf{k}}f_I)
\label{eq:berry_curvature_inside}
\end{gather}
The main physical message of \cref{eq:berry_curvature_inside} is simple: inside the exceptional surface the Berry curvature is nonzero, and its magnitude diverges as $\varepsilon\to 0$, i.e. as one approaches the exceptional surface from the interior. This is exactly what is seen in \cref{figLBerryprofile}.

\paragraph{Outside the exceptional surface ($|f|>E$).}
In the exterior, which is the $\mathcal{PT}$-unbroken region of this model, the biorthogonal Berry curvature vanishes,
\begin{equation}
\boldsymbol{\Omega}(\mathbf{k})=0,
\label{eq:berry_curvature_outside}
\end{equation}
and the Berry connection is a curl-free field. Thus, for this particular class of models, the Berry curvature is supported only inside the exceptional surface and becomes singular on the exceptional surface itself.

\subsection{Hamiltonian pairs with same dispersion but opposite Berry curvature}
\label[smsection]{appx:isospectral-opposite-flow}

The preceding expressions reveal a distinction that is invisible to the complex spectrum alone. For any Hamiltonian of the form in \cref{eqS:nonHermHamiltonian}, consider the pair
\begin{align}
H_E^{(1)}(\mathbf{k})
&=f_R(\mathbf{k})\,\sigma_x+f_I(\mathbf{k})\,\sigma_z+iE\,\sigma_y, \notag\\
H_E^{(2)}(\mathbf{k})
&=f_R(\mathbf{k})\,\sigma_x-f_I(\mathbf{k})\,\sigma_z+iE\,\sigma_y
=\sigma_x\bigl[H_E^{(1)}(\mathbf{k})\bigr]^T\sigma_x .
\label{eq:isospectral_pair}
\end{align}
Equivalently, the second Hamiltonian is obtained by replacing $f$ with $f^*$ while leaving the non-Hermitian perturbation unchanged. Since
\begin{equation}
\bigl[H_E^{(a)}(\mathbf{k})\bigr]^2
=\bigl(|f(\mathbf{k})|^2-E^2\bigr)\mathbb{I},
\qquad a=1,2,
\end{equation}
the two characteristic polynomials, and hence both complex bands, coincide pointwise:
\begin{equation}
\varepsilon_{\pm}^{(1)}(\mathbf{k})
=\varepsilon_{\pm}^{(2)}(\mathbf{k})
=\pm\sqrt{|f(\mathbf{k})|^2-E^2}.
\label{eq:isospectral_pair_spectrum}
\end{equation}
They therefore have the same exceptional surface $|f|=E$, the same $\mathcal{PT}$-broken volume $|f|<E$, and the same unflowed knotted-graph skeleton.
Their Hermitian parents, obtained by removing the $iE\sigma_y$ term, likewise share the dispersion $\pm|f(\mathbf{k})|$. Thus, all Fermi surfaces and knotted-graph fermiology derived solely from their eigenvalues are also identical.

Their biorthogonal eigenstates, however, are related by an exchange of left and right eigenvectors. Let $|u_n^R\rangle$ and $\langle u_n^L|$ be a biorthonormal pair for $H_E^{(1)}$ away from the exceptional surface. A corresponding pair for $H_E^{(2)}$ is
\begin{equation}
|\widetilde u_n^R\rangle
=\sigma_x(\langle u_n^L|)^T,
\qquad
\langle\widetilde u_n^L|
=(|u_n^R\rangle)^T\sigma_x.
\end{equation}
Because $\sigma_x$ is independent of momentum, the associated Berry connection obeys
\begin{align}
\widetilde{\mathbf A}_n
&=i(|u_n^R\rangle)^T\nabla_{\mathbf{k}}(\langle u_n^L|)^T \notag\\
&=i(\nabla_{\mathbf{k}}\langle u_n^L|)|u_n^R\rangle
=-i\langle u_n^L|\nabla_{\mathbf{k}}|u_n^R\rangle
=-\mathbf A_n,
\end{align}
where the penultimate equality uses $\nabla_{\mathbf{k}}\langle u_n^L|u_n^R\rangle=0$, so that $(\nabla_{\mathbf{k}}\langle u_n^L|)|u_n^R\rangle=-\langle u_n^L|\nabla_{\mathbf{k}}|u_n^R\rangle$.
Consequently,
\begin{equation}
\widetilde{\boldsymbol{\Omega}}_n(\mathbf{k})
=-\boldsymbol{\Omega}_n(\mathbf{k}).
\label{eq:isospectral_pair_curvature}
\end{equation}
This is the general transpose construction specialized to our model: for \textit{any diagonalizable non-Hermitian} $H(\mathbf{k})$, transposition preserves its eigenvalues, exchanges its left and right eigenvectors, and reverses the biorthogonal Berry curvature of each nondegenerate band. A momentum-independent similarity transformation, such as the conjugation by $\sigma_x$ in \cref{eq:isospectral_pair}, does not alter that conclusion.

The same reversal follows directly from \cref{eq:berry_curvature_inside}: the transformation $f_I\mapsto-f_I$ leaves $\varepsilon$ unchanged but reverses $\nabla_{\mathbf{k}}f_R\times\nabla_{\mathbf{k}}f_I$. Thus, identical dispersion and exceptional-surface geometry do not fix the direction of the biorthogonal Berry-curvature field.

\subsection{Berry flux orients the knotted graph edges}
\label[smsection]{appx:oriantability_of_knottedgraphs}

With the Berry-curvature field in hand, we can define the Berry flux and Berry phase associated with a closed loop in the 3D Brillouin zone.

Given a closed loop $\mathcal{L}$ that does not cross singularities in the Brillouin zone and a surface $\mathcal{S}$ whose boundary is the loop $\partial\mathcal{S}=\mathcal{L}$, the corresponding Berry phase and Berry flux are:
\begin{gather}
\gamma_B = \oint_{\mathcal{L}}\mathbf{A}(\mathbf{k})\cdot d\mathbf{k},
\label{eq:berryphase} \\
\Phi_B = \iint_{\mathcal{S}}\boldsymbol{\Omega}(\mathbf{k})\cdot d\mathbf{S},
\label{eq:berryflux}
\end{gather}
and for most of the scenarios considered in this work, the Berry phase and Berry flux are related by the usual Stokes' theorem, modulo $2\pi$:
\begin{equation}
\oint_{\mathcal{L}}\mathbf{A}(\mathbf{k})\cdot d\mathbf{k}
= \iint_{\mathcal{S}}\boldsymbol{\Omega}(\mathbf{k})\cdot d\mathbf{S}
\quad (\text{modulo}\ 2\pi).
\end{equation}

For the discussion in this section, the key object is not the Berry phase (which will play a role in \cref{appx:fluxpaths}) but how the Berry-curvature flux is routed through the tubular neighborhood of the skeleton.
Once the exceptional volume is coarse-grained to its knotted-graph skeleton, that flux turns a previously unoriented spatial graph into an oriented one.

Consider a segment of an edge away from the vertices, and choose the cross-section disk $D_e$ transverse to a local flux tube, such that it covers the total flux through the tube and only the particular edge of interest passes through it.
The Berry flux through that cross section,
\begin{equation}
\Phi_e
:=
\iint_{D_e}\boldsymbol{\Omega}(\mathbf{k})\cdot \hat{\mathbf{n}}\,dS,
\label{eq:edge_flux_def}
\end{equation}
is the natural quantity associated with the edge.
Its sign selects a preferred orientation, while its magnitude gives the corresponding edge weight. Moreover, $\Phi_e$ does not depend on the precise choice of transverse section along the same smooth tube segment. Thus each edge inherits a well-defined signed flux.

\bigskip

The next issue is what happens at a vertex, where several such tubes meet. For the class of models in \cref{eqS:nonHermHamiltonian}, the Berry curvature inside the exceptional volume has the form
\begin{equation}
\boldsymbol{\Omega}(\mathbf{k})
=g\bigl(f_R(\mathbf{k}),f_I(\mathbf{k})\bigr)\,
\bigl(\nabla_{\mathbf{k}}f_R\times\nabla_{\mathbf{k}}f_I\bigr),
\end{equation}
with $g=E/[2(E^2-f_R^2-f_I^2)^{3/2}]$. Since $g$ depends only on $f_R$ and $f_I$, its gradient is a linear combination of $\nabla f_R$ and $\nabla f_I$, and therefore
\begin{equation}
\nabla_{\mathbf{k}}\cdot \boldsymbol{\Omega}(\mathbf{k}) = 0
\label{eq:continuity}
\end{equation}
everywhere in the smooth interior of the exceptional volume. Equivalently, there is no bulk monopole source of Berry curvature; the only singular behavior is the divergence of $|\boldsymbol{\Omega}|$ on the exceptional surface itself.

Now choose a volume $V_v$ with boundary consisting of transverse disks $D_i$ on the incident tubes together with side walls on (infinitesimally close to) the exceptional surface:
\begin{equation}
\partial V_v=\left(\bigcup_i D_i\right)\cup S_{\mathrm{side}},
\end{equation}
where $S_{\mathrm{side}}$ is the part of the boundary running along the exceptional surface. Let $\hat{\mathbf n}_{V}$ denote the outward normal of this control volume. Applying the divergence theorem using \cref{eq:continuity} gives
\begin{align}
0
&=\iiint_{V_v}\nabla_{\mathbf{k}}\cdot\boldsymbol{\Omega}(\mathbf{k})\,d^3k \notag\\
&=\oiint_{\partial V_v}\boldsymbol{\Omega}(\mathbf{k})\cdot\hat{\mathbf n}_{V}\,dS \notag\\
&=\sum_i \iint_{D_i}\boldsymbol{\Omega}(\mathbf{k})\cdot\hat{\mathbf n}_{V}\,dS
  + \iint_{S_{\mathrm{side}}}\boldsymbol{\Omega}(\mathbf{k})\cdot\hat{\mathbf n}_{V}\,dS \notag\\
&=\sum_i \Phi_i^{(v)} + 0
\end{align}
The side-wall term vanishes because, on the exceptional surface, the Berry curvature is tangent to the surface and hence has no normal component through $S_{\mathrm{side}}$.
The signed edge flux at the vertex is defined by the outward-normal convention, where outgoing tubes contribute with positive sign and incoming tubes with negative sign.

Take a closer look at the vertex constraint:
\begin{equation}
\boxed{\sum_{i\in E(v)} \Phi_i^{(v)}=0.}
\label{eq:kirchhoff_stationary}
\end{equation}
\cref{eq:kirchhoff_stationary} is just \textbf{Kirchhoff's law} in continuum form: the total outgoing Berry flux equals the total incoming Berry flux at a source-free junction.

\bigskip

This conservation law has two immediate consequences. First, not every local orientation pattern at a vertex is allowed; the edge directions must be compatible with flux balance. Second, the skeleton is richer than an unoriented knotted graph: each edge carries a signed real number, and those numbers obey a vertex conservation rule. In graph-theoretic language, the Berry-curvature field equips the knotted graph edge with an $\mathbb{R}$-valued \textit{Abelian flow}.

\subsection{Abelian edge flow and flowed knotted graph invariants}\label[smsection]{appx:Aflow}

The previous subsection established the continuum input needed for a graph-theoretic description: the Berry flux through a transverse section gives a signed edge weight, and \cref{eq:kirchhoff_stationary} imposes a zero-sum rule at every junction. The purpose of this subsection is to translate that statement into the standard language of \textit{flowed} spatial graphs. This translation is useful because it distinguishes a genuinely ``flowed'' skeleton from an ``arbitrarily oriented'' graph, and because it identifies which diagrammatic moves preserve the thickened exceptional volume represented by the knotted graph.

We proceed in three steps.
First, we recall the definition of an $A$-flow and identify the Berry-flux skeleton as an $\mathbb{R}$-flowed spatial graph.
Second, we explain the equivalence relation for flowed graphs and the additional contraction moves required when the graph represents a thickened volume (i.e., is used as its spine in the sense of~\cite{ishii2012quandle}).
Finally, we comment on the finite-group invariants of Ishii and Iwakiri and why applying them directly to real-valued Berry flux requires an extra discretization step which we do not attempt, but we remark that this is an interesting direction for future work.

\paragraph{$A$-flow and $A$-flowed spatial graphs.}
In graph-theoretic terms, \cref{eq:kirchhoff_stationary} is exactly the vertex condition in an Abelian edge flow.

\begin{boxeddefinition}{$A$-flow and $A$-flowed spatial graph}{A_flow}
Let $A$ be an Abelian group. Fix a reference orientation on each edge of a knotted graph $\mathcal{G}$. An \textit{$A$-flow} is an assignment of an element of $A$ to each oriented edge such that reversing the orientation reverses the value, and such that at every vertex $v$ the sum of the values on the incident edges vanishes:
\begin{equation}
\boxed{
\varphi(\bar e)=-\varphi(e), \qquad
\sum_{e\in E_v} \varphi(e)=0,
}
\label{eq:A_flow}
\end{equation}
where $E_v$ is the set of oriented edges incident on $v$ and taken to point outward.
For Berry-curvature flux, the natural choice of the Abelian group is the real numbers with addition, i.e. $A=(\mathbb{R}, +)$, with $\varphi(e)$ given by the integrated flux $\Phi_e$ through a transverse cross section of the corresponding tube.

\tcblower

An \textit{$A$-flowed spatial graph} is the pair $(\mathcal{G},\varphi)$, where $\mathcal{G}$ is a spatial graph and $\varphi$ is an $A$-flow on $\mathcal{G}$.
\end{boxeddefinition}

Thus the Berry-curvature skeleton is naturally an $\mathbb{R}$-flowed spatial graph: the graph is the skeleton of the exceptional volume, and the flow value on each edge is the integrated Berry flux through a transverse section of the corresponding tube.

For the isospectral pair in \cref{eq:isospectral_pair}, hold the common Brillouin-zone embedding fixed and use the same reference orientation and transverse disk for corresponding edges. Then \cref{eq:edge_flux_def,eq:isospectral_pair_curvature} give
\begin{equation}
\Phi_e^{(2)}=-\Phi_e^{(1)},
\qquad
\varphi^{(2)}(e)=-\varphi^{(1)}(e).
\label{eq:isospectral_pair_flow}
\end{equation}
Hence the two models produce the same unflowed spatial graph $\mathcal G$ but opposite $\mathbb{R}$-flows, $(\mathcal G,\varphi)$ and $(\mathcal G,-\varphi)$. \textbf{Both assignments satisfy Kirchhoff's law, so the distinction cannot be recovered from the spectrum or from an invariant of the unflowed graph alone.}

\paragraph{Equivalence and contraction moves.}
For ordinary flowed spatial graphs, equivalence means ambient isotopy that transports the flow labels. On diagrams, this equivalence is generated by the $A$-flowed Reidemeister moves \cite[Lemma~2.2]{ishii2012quandle}.

\begin{figure}[!ht]
    \centering
    \includegraphics[width=0.8\linewidth]{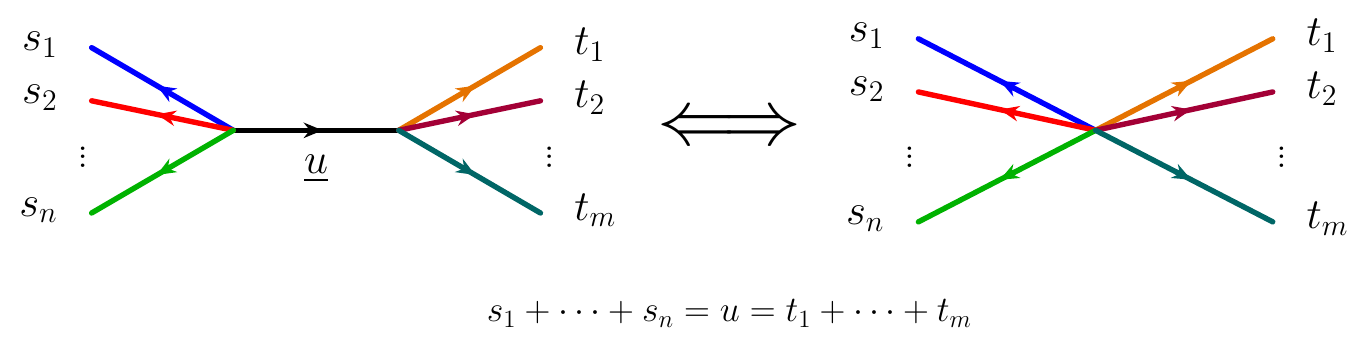}
    \caption{\textbf{($A$-flowed) Contraction move.}
    The left and right graphs are related by a ($A$-flowed) contraction move, which shrinks the middle edge and merges the two vertices into one.
    \textbf{Such contractions, together with ambient isotopy}, define \textit{an equivalence class of spines (knotted graphs)} that maps \textbf{one-to-one} to the \textit{handlebody-link (Fermi volume / exceptional volume)} up to homeomorphism.
    The illustration here is $A$-flowed and requires Kirchhoff's law to be satisfied, and it applies to the $A$-flow knotted graph setting.
    The \textbf{``unflowed''} version of this move applies to the unflowed knotted graph setting in the previous Fermi-surface context.}
    \label{fig:AflowContraction}
\end{figure}

For our application, however, the graph is not just any embedded spatial graph: it is a spine of a thickened exceptional volume. One must therefore also allow \textit{contraction moves} (see \cref{fig:AflowContraction}), because different spines can represent the same thickened object.
In the unflowed setting, \cite[Theorem~2.3]{ishii2012quandle} states that two spatial graphs represent the same handlebody-link if and only if they are related by contraction moves and ambient isotopies.
Correspondingly, \cite[Proposition~2.4]{ishii2012quandle} shows that any invariant of $A$-flowed spatial graphs that is also invariant under $A$-flowed contraction moves descends to an invariant of the underlying handlebody-link.

Here a \textit{handlebody-link} means a disjoint union of embedded handlebodies, i.e. thickened neighborhoods of spatial graphs. In the present non-Hermitian setting, this thickened object is the exceptional volume; in the Hermitian setting, it is the Fermi volume. The unflowed and multiboundary version of this viewpoint is developed in \cref{appx:compression-body}.

\paragraph{Quandle coloring and cocycle state-sum invariants.}
The finite-flow invariant literature gives a natural direction beyond polynomial invariants such as the Yamada polynomial.

Ishii and Iwakiri construct quandle-coloring and cocycle state-sum invariants for flowed spatial graphs and the handlebody-links they represent \cite{ishii2012quandle}. Their standard setup uses \textit{finite} flow groups, typically $A=\mathbb{Z}_n$, because the flow label enters the coloring rule as a discrete iterate of the quandle operation. By contrast, the Berry-curvature fluxes in our three-parameter setting are naturally real-valued and generally not quantized. Applying those invariants directly would therefore require an extra discretization step, for instance reducing the flux modulo $2\pi$ and then mapping it into $\mathbb{Z}_n$. We do not carry out this discretization here, but it provides a concrete route for future flowed-graph refinements of exceptional-surface fingerprints.

The isospectral construction also clarifies what the flow contributes to the classification. The unflowed skeleton $\mathcal G$ records the topology and embedding of the exceptional volume determined by the eigenvalues, and is therefore identical for $H_E^{(1)}$ and $H_E^{(2)}$. In contrast, the signed flow $\varphi$ is derived from the biorthogonal Berry curvature and reverses when the left--right eigenstate structure is exchanged, even though the spectrum is unchanged. The Berry-flowed spatial graph $(\mathcal G,\varphi)$ is therefore not a repackaging of dispersion data. Taken together, its two components form \textit{a joint spectral--eigenstate fingerprint}: $\mathcal G$ captures exceptional-volume topology and embedding, while $\varphi$ captures the Berry-curvature geometry of the biorthogonal eigenstates.

\FloatBarrier

\section{Handlebodies, Compression-bodies, and Fermi Surfaces with Multiple Boundary Components}
\label[smsection]{appx:compression-body}

The preceding \cref{appx:ansatz,appx:Yamada,appx:ES} use \textit{one} knotted graph as the topological fingerprint of a regular constant-energy surface with \textit{one} boundary component.
In the cleanest case, the Fermi volume bounded by the surface is connected and has no internal boundary components; topologically, the volume is a \textit{handlebody}, and a single knotted graph records its topology. (In the handlebody-link literature~\cite{ishii2012quandle,bardakov2025invariantshandlebodylinksspatialgraphs}, such an embedded graph is called a \textit{spine}; we adopt this term in the present section when discussing the corresponding mathematical theorems.)
\textbf{This section supplies the geometric topology justification for the framework and then explains how it must be generalized for realistic Fermi volumes with \textit{multiple boundary components}}.
The handlebody-link equivalence theorem reviewed below is due to Ishii et al.~\cite{ishii2012quandle}; the extension to compression bodies, the boundary-wise knotted graph formulation, the compression-body equivalence theorem (\cref{thm:compression_body_equiv}), and the Yamada set are all \textit{newly} developed in this work.

There are two points to separate:
\begin{itemize}
	\item First, as long as the energy has no critical value in an interval, the Fermi surface does not topologically change, so any knotted graph (spine) remains equivalent up to ambient isotopy and contraction moves. 
    At a Lifshitz transition, the Fermi-volume topology changes; for a handlebody component, this must change the neighborhood-equivalence class of its spine.
    
	\item Second, a single spine is sufficient only for a handlebody (i.e. with one boundary component). If the Fermi volume contains multiple boundary components, \textit{the knotted graph data must be recorded boundary-wise}.
\end{itemize}

The section is organized as follows. 
We first recall the handlebody-link $\cong$ spine correspondence theorem (which naturally applies to the handlebody case). 
We then explain why \textbf{a Lifshitz transition forces a transition of the neighborhood-equivalence class of knotted graph}, and why raw numerical graph changes should be interpreted with care. 
Finally, we extend from handlebody to the more generic \textbf{compression body}, define the \textbf{boundary-wise graph data}, and introduce the notion \textbf{Yamada set} that packages the corresponding normalized Yamada polynomials.

\subsection{Handlebody(-link), spine, and contraction moves.}

\begin{boxeddefinition}{Handlebody, Spine, and Handlebody-Link}{handlebody_spine}
A \textit{handlebody} is a compact, connected, orientable 3-manifold obtained from a 3-ball by attaching 1-handles; equivalently, it is a \textit{regular neighborhood} of a spatial graph.

In other words, any handlebody $\mathcal{H}$ admits a \textit{spine}: an embedded graph $\mathcal{G} \subset \mathcal{H}$ such that $\mathcal{H}$ deformation-retracts onto $\mathcal{G}$.

Spines are not unique as embedded graphs, but different choices of spine for the same handlebody are related by isotopy and \textit{contraction/expansion} operations (sometimes called Whitehead moves, \cref{fig:AflowContraction}) that do not change the homeomorphism type of the regular neighborhood.

A \textit{handlebody-link} is a disjoint union of handlebodies, and its spine has multiple connected components, each of which is a spine of the corresponding handlebody component.
\end{boxeddefinition}

This viewpoint is rigorously studied in geometric topology. The relevant equivalence theorem \cite{ishii2008moves,ishii2012quandle} is:
\begin{boxedtheorem}{Handlebody-link equivalence \cite[Theorem~2.3]{ishii2012quandle}}{handlebody_equiv}
Two knotted graphs represent equivalent handlebody-links if and only if they are related by a finite sequence of contraction moves and ambient isotopies.
\end{boxedtheorem}

This lays the mathematical foundation for our approach, with one caveat that the numerically extracted knotted graph is just one ``canonical'' representative of the neighborhood-equivalent class of spines.

\subsection{Lifshitz transitions force knotted-graph transitions.}

Following the above equivalence theorem, Lifshitz transitions are reflected in the change of knotted-graph equivalence classes, and thus the transition in the numerical representatives.

To break it down, let $n$ be the band index of the dispersion $\varepsilon_n(\mathbf{k})$, and the bracketed index $[i]$ labels energy window. 
It is worth noting that this band index was not important in all previous symmetric two-band model examples, where the Fermi surface evolution and Yamada sequences are exactly the same for the two bands. 
However, the band index becomes important in multiband models, e.g. real-material effective Hamiltonians, as different bands can contribute distinct Fermi-volume evolutions and distinct Lifshitz events.

Consider a particular band $n$ and an energy interval $(E_-,E_+)$ containing no critical value of $\varepsilon_n$, then all level sets $\Sigma_{[i]}^{(n)}$ for $E\in(E_-,E_+)$ are smooth, and the corresponding sublevel sets $\mathcal{F}_{[i]}^{(n)}$ vary by homeomorphism in the Brillouin zone. 
Put differently, the homeomorphism type of each connected component of $\mathcal{F}_{[i]}^{(n)}$ is constant throughout the interval, and thus by the theorem, the knotted graph up to neighborhood-equivalence is also constant.

On the other hand, at a Lifshitz transition energy $E_c$, the topology of the Fermi surface changes, which implies that at least one component of the Fermi volume changes its homeomorphism type, and thus the neighborhood-equivalence class of its spine must change as well. Different spine classes are mutually exclusive, so our numerically obtained knotted graph must be topologically different across the transition. 

\begin{boxedproposition}{Lifshitz transition forces knotted graph transition}{lifshitz_spine_transition}
A Lifshitz transition forces a change in the neighborhood-equivalence class of the spine graph. Consequently, any representative spine (including the numerical skeleton graph) in the class must be topologically different across the transition.
\end{boxedproposition}

\bigskip

The converse---\textit{whether a transition in the knotted graph fingerprint sufficiently indicates a Lifshitz transition}---has to be treated with care.

At the level of neighborhood-equivalence classes, a genuine change of the spine class does imply a change in the represented handlebody topology, and therefore signals a Lifshitz transition. 
However, the graph produced by our numerical skeletonization is a particular representative, not the equivalence class itself. Different representatives may transition to one another by contraction/expansion moves without any Lifshitz event happening. 
Such knotted graph transitions do not sufficiently determine Lifshitz transitions unless they survive the quotient by ambient isotopy and contraction moves.
In accordance, the Yamada polynomial signatures should be interpreted with similar care. 

\begin{figure*}[!ht]
    \centering
    \includegraphics[width=0.65\linewidth]{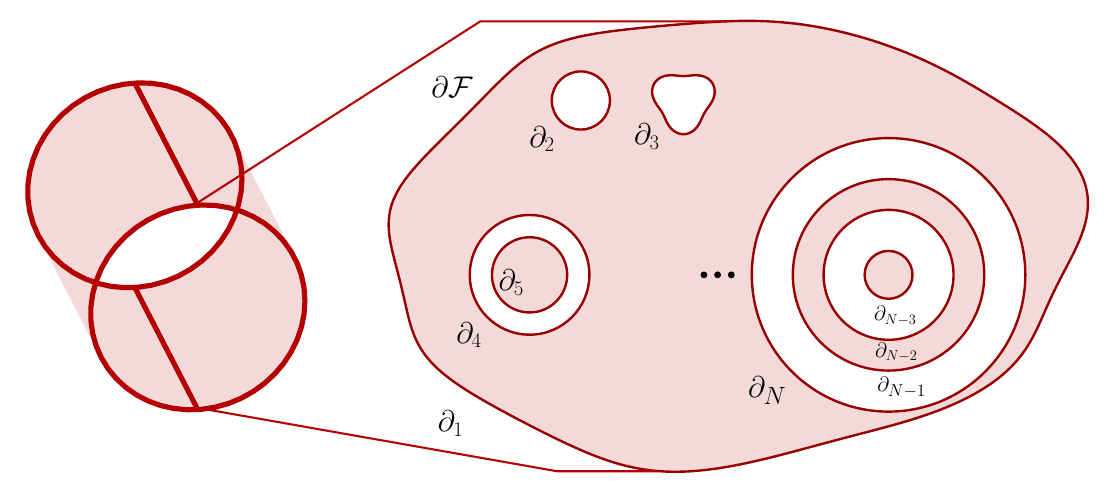}
    \caption{\textbf{Compression-body and Yamada-set encoding.}
    A schematic cross section of a Fermi-volume component with an outer boundary and (nested) inner boundaries. In the handlebody case, a single skeleton graph is sufficient. In the compression-body case, the topology is recorded boundary-wise by the collection of knotted graphs $\{\mathcal{G}_{[i],\alpha}^{(n)}\}_{\alpha=1}^{m_{[i]}^{(n)}}$ and their corresponding Yamada set $\overline{\Upsilon}_{\partial\mathcal{F}_{[i]}^{(n)}}$, where $n$ is the band index, $[i]$ is the energy window index, and $\alpha$ labels the boundary components.}
    \label{fig:YamadaMultiSetGenus}
\end{figure*}

\subsection{Compression bodies and the encoding of multiple boundaries.}
\label{subsec:compression_body_encoding}

The handlebody discussion above assumes that each relevant Fermi volume component has one outer boundary and no inner boundaries. 
Real-material Fermi volumes can be more intricate. A connected component of $\mathcal{F}_{[i]}^{(n)}$ may contain nested voids, so its boundary has one outer component together with one or more inner components. 
This is naturally expected from the fact that Lifshitz transitions can create or annihilate inner pockets without changing the outer boundary. Therefore, a single skeleton graph is not sufficient in such cases.

To describe this structure, we use the language of \textbf{compression bodies}.
For our purposes, a compression body can be viewed as a three-manifold with one distinguished outer boundary $\partial_+$ and, possibly, several inner boundary components collectively denoted by $\partial_-$. 

Now we formulate the boundary-wise graph data and the corresponding equivalence theorem for compression bodies, which generalizes the handlebody-link equivalence theorem quoted above.

\begin{boxeddefinition}{Boundary-wise knotted graph data}{boundarywise_graph_data}
Let
\begin{equation}
\partial\mathcal{F}_{[i]}^{(n)}
=
\bigsqcup_{\alpha=1}^{m_{[i]}^{(n)}}
\partial_\alpha \mathcal{F}_{[i]}^{(n)}
\end{equation}
be the decomposition of the Fermi-volume boundary (Fermi surface) into its connected boundary components. Here $m_{[i]}^{(n)}$ is the number of boundary components in the $i$-th energy window of band $n$, and the index $\alpha$ labels both outer and inner boundary components. \cref{fig:YamadaMultiSetGenus} illustrates a compression-body region where a tubular component encloses multiple inner boundary components. We associate each boundary component $\partial_\alpha \mathcal{F}_{[i]}^{(n)}$ a knotted graph
\begin{equation}
\mathcal{G}_{[i],\alpha}^{(n)},
\qquad
\alpha=1,\ldots,m_{[i]}^{(n)} .
\end{equation}
Thus the relevant object is not, in general, a single knotted graph, but the \textbf{multiset} of knotted graphs
\begin{equation}
\left\{
\mathcal{G}_{[i],\alpha}^{(n)}
\right\}_{\alpha=1}^{m_{[i]}^{(n)}} ,
\end{equation}
together with the nesting relation between the corresponding boundary components.
\end{boxeddefinition}

The boundary-wise data should still be treated with the same care as in the previous subsection.

The following statement is the compression-body analogue of the handlebody-link equivalence theorem quoted above.

\begin{boxedtheorem}{Equivalence between compression-body link and boundary-wise knotted graph}{compression_body_equiv}
Two compression-body links are equivalent \textit{if and only if} their boundary-wise knotted graphs (spines) are related by a finite sequence of contraction moves and ambient isotopy of the full embedded graph collection, while preserving the nesting structure.

\tcblower

\noindent\textit{Proof.}
The statement follows by reducing the compression-body case to the handlebody case. Consider one connected compression-body component $\mathcal{C}_{[i]}^{(n)}\subset \mathcal{F}_{[i]}^{(n)}$, with outer boundary $\partial_+\mathcal{C}_{[i]}^{(n)}$ and inner boundary components $\partial_-\mathcal{C}_{[i]}^{(n)}$. We orient $\mathcal{C}_{[i]}^{(n)}$ as the original Fermi-volume component. With this orientation, $\partial_+\mathcal{C}_{[i]}^{(n)}$ is oriented by the outward normal of the Fermi volume. By contrast, each inner void boundary component is oriented oppositely, because it bounds a void inside $\mathcal{C}_{[i]}^{(n)}$: the outward normal of $\mathcal{C}_{[i]}^{(n)}$ along an inner boundary points into the void, whereas the outward normal of the void points back into $\mathcal{C}_{[i]}^{(n)}$.

This opposite orientation allows us to attach a compatible auxiliary filling along each inner boundary. Write
\begin{equation}
\partial_-\mathcal{C}_{[i]}^{(n)}
=
\bigsqcup_{\alpha=1}^{r_{[i]}^{(n)}}
\partial_{-,\alpha}\mathcal{C}_{[i]}^{(n)} .
\end{equation}
For each $\partial_{-,\alpha}\mathcal{C}_{[i]}^{(n)}$, choose an auxiliary filling handlebody $\mathcal{V}_{[i],\alpha}^{(n)}$, attached with the opposite boundary orientation and identified with $\partial_{-,\alpha}\mathcal{C}_{[i]}^{(n)}$. Attaching these fillings to $\mathcal{C}_{[i]}^{(n)}$ caps off all inner voids:
\begin{equation}
\widehat{\mathcal{C}}_{[i]}^{(n)}
=
\mathcal{C}_{[i]}^{(n)}
\cup_{\partial_-\mathcal{C}_{[i]}^{(n)}}
\left(
\bigsqcup_{\alpha=1}^{r_{[i]}^{(n)}}
\mathcal{V}_{[i],\alpha}^{(n)}
\right).
\end{equation}
By the definition of a compression body, the resulting region $\widehat{\mathcal{C}}_{[i]}^{(n)}$ has only the outer boundary $\partial_+\mathcal{C}_{[i]}^{(n)}$ and is an ordinary handlebody.

The same filling procedure can be applied componentwise to every disconnected component of the original Fermi volume. Therefore, after all inner boundaries are filled, the original multicomponent compression-body link is converted into a handlebody-link.

The key observation is that, the filled volume handlebodies together with the auxiliary reverse-oriented fillings give a one-to-one encoding of the compression-body data, provided that the nesting structure and the distinction between original volume components and auxiliary fillings are kept. The filled volume handlebodies record the original volume components, while the reverse-oriented auxiliary handlebodies record the voids. Conversely, the original compression body is recovered by placing each reverse-oriented auxiliary handlebody inside the appropriate filled volume handlebody and deleting its interior. Thus the filling procedure is reversible and preserves the full compression-body information.

Since the original volume components and the reverse-oriented auxiliary fillings are now represented as a collection of handlebodies, we can apply the handlebody-link equivalence theorem \cite[Theorem~2.3]{ishii2012quandle}.

Since the handlebody collections described above uniquely encode the corresponding compression bodies, two compression-body links are equivalent if and only if their boundary-wise knotted graph collections are related by componentwise contraction/expansion moves and ambient isotopies of the full embedded collection, with the inner-boundary data and nesting structure preserved.
\end{boxedtheorem}

This equivalence explains and justifies why the graph data must be kept boundary-wise. For an ordinary handlebody, one knotted graph is enough. For a compression body, the inner boundary components are part of the topology, so the graph representatives associated with different boundary components must be retained separately, together with their mutual nesting.

\paragraph{From boundary-wise graphs to Yamada sets.}
The remaining question is how to package this boundary-wise graph collection into the invariant sequence that forms a dispersion-level fingerprint.

For example, one could combine the boundary-wise graphs into a single disjoint-union (no knotting or linking) graph,
\(
\mathcal{G}_{[i],\mathrm{tot}}^{(n)}
:=
\coprod_{\alpha=1}^{m_{[i]}^{(n)}}
\mathcal{G}_{[i],\alpha}^{(n)} .
\)
By the disjoint-union property of the Yamada polynomial
(\cref{tab:YamadaTable}), this gives
\(
\overline{\Upsilon}
\!\left(
\mathcal{G}_{[i],\mathrm{tot}}^{(n)}
\right)
=
\prod_{\alpha=1}^{m_{[i]}^{(n)}}
\overline{\Upsilon}
\!\left(
\mathcal{G}_{[i],\alpha}^{(n)}
\right).
\label{eq:YamadaProductCompressionBody}
\)
However, this product hides component-resolved information: it does not record which boundary component was created, annihilated, or reconnected across a Lifshitz transition.

We therefore keep the boundary-wise invariants separately and define the \textbf{Yamada set}:
\begin{boxeddefinition}{Yamada set}{yamada_set}
\begin{equation}
\overline{\Upsilon}_{\partial\mathcal{F}_{[i]}^{(n)}}
:=
\left\{
\overline{\Upsilon}
\!\left(
\mathcal{G}_{[i],\alpha}^{(n)};Y
\right)
\right\}_{\alpha=1}^{m_{[i]}^{(n)}} .
\label{eq:YamadaSET}
\end{equation}
Here the braces denote a \textbf{multiset}: repeated polynomial values are retained, while the ordering of boundary components is not physically meaningful.
\end{boxeddefinition}
When nested boundaries are present, this multiset is understood together with the nesting structure of the corresponding boundary components.

The Yamada set reduces to a single normalized Yamada polynomial when $\mathcal{F}_{[i]}^{(n)}$ is a handlebody with one boundary component. 
In the general compression-body case, it provides a boundary-resolved fingerprint of the Fermi-surface topology.

Accordingly, the Yamada sequence \cref{eq:yamadaseq_def} should be more generically understood as \textbf{a sequence whose entries are multiset-valued} Yamada sets instead of single polynomials:
\begin{align}
\boldsymbol{\overline{\Upsilon}}(H)
:=&
\left[
\overline{\Upsilon}_{\partial\mathcal{F}_{[1]}^{(n)}},
\overline{\Upsilon}_{\partial\mathcal{F}_{[2]}^{(n)}},
\ldots,
\overline{\Upsilon}_{\partial\mathcal{F}_{[m]}^{(n)}}
\right] \\
=&
\left[
\left\{\overline{\Upsilon}
\!\left(\mathcal{G}_{[1],\alpha}^{(n)};Y
\right)\right\}_{\alpha=1}^{m_{[1]}^{(n)}},
\left\{\overline{\Upsilon}
\!\left(\mathcal{G}_{[2],\alpha}^{(n)};Y
\right)\right\}_{\alpha=1}^{m_{[2]}^{(n)}},
\ldots,
\left\{\overline{\Upsilon}
\!\left(\mathcal{G}_{[m],\alpha}^{(n)};Y
\right)\right\}_{\alpha=1}^{m_{[m]}^{(n)}}
\right]
\label{eq:yamada_Set_seq_def}
\end{align}

The nontrivial content of this construction is that the boundary-wise data is not merely a convenient bookkeeping device: the compression-body equivalence theorem (\cref{thm:compression_body_equiv}) guarantees it is a mathematically complete characterization of the Fermi-volume topology.

This extension lays the foundation for the more generic Fermi-surface topology characterization that applies to real-material examples, which we present next.

\section{Fingerprinting Real Materials --- Full Yamada Sequences across All Energy Windows}
\label[smsection]{appx:materials}
In this section, we examine the nontrivial topologies realized in
representative real materials and show how these features naturally manifest
as knotted graphs. For each material (Ti$_3$Al, YH$_3$, TiB$_2$, and
Co$_2$MnGa), we track the topology of finite-thickness momentum-space
surfaces as the interband-gap scale $E$ is varied, and compute the
corresponding Yamada polynomials to obtain compact, topology-sensitive
characterizations. Here and throughout this section, $E$ denotes the interband-gap threshold
used to define the finite-thickness momentum-space region,
$\{\mathbf{k}:|\varepsilon_1(\mathbf{k})-\varepsilon_0(\mathbf{k})|\le E\}$;
increasing $E$ therefore thickens the region surrounding the nodal structure. Notably, the
point-group symmetries of the underlying crystals are frequently reflected
in the associated knotted graphs.

Unlike the ansatz examples of \cref{fig:gallery}, the skeleton graphs encountered below contain vertices of degree greater than three: a $\theta_s$ graph has two vertices of degree $s$ (here up to $s=6$), and its periodic variant $\theta_{s,P}$ has two vertices of degree $s$ together with $s$ degree-four midpoints, so that $\theta_{12,P}$ carries vertices of degree twelve.
For such graphs, move (VI) is not generated by moves (I--V), so $\overline{\Upsilon}$ is a \textit{rigid vertex isotopy} invariant rather than a full ambient-isotopy invariant.
All polynomials quoted in this section are therefore evaluated on the diagram fixed by the minimum-crossing planar-projection convention introduced in \cref{appx:Yamada}.

\begin{figure}
    \centering
    \includegraphics[width=0.6\linewidth]{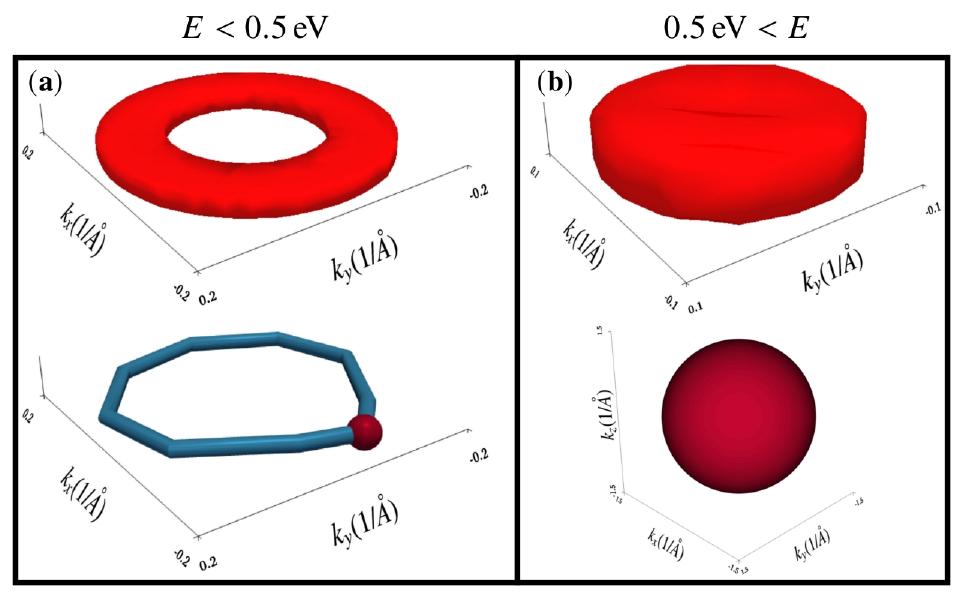}
     \caption{\textbf{Fermi-surface and knotted-graph evolution for Ti$_3$Al.}
    Fermi surfaces computed from the DFT-fitted dispersion \cref{eq:ti3al_dispersion}.
    \textbf{(a)} At lower $E$, the Fermi surface is a torus, corresponding to the bouquet graph $\mathcal{B}_1$ (unknot).\textbf{(b)} Upon increasing $E$, a gap-closing event changes the surface topology to a sphere, represented by a single vertex $\mathcal{B}_0$.}
    \label{fig:Ti3Al_seq}
\end{figure}

\subsection{Ti$_3$Al}

Employing the DFT-fitted low-energy dispersion \cite{Zhang2018} for Ti$_3$Al, we use
\begin{equation}
E_{\mathrm{Ti_3Al}}(\mathbf{k})
=\frac{1}{2}\left( h_1+h_2+\sqrt{(h_1-h_2)^2+h^2}\right),
\label{eq:ti3al_dispersion}
\end{equation}
where
\begin{equation}
h_i=A_i\left(k_x^2+k_y^2\right)+B_i k_z^2+M_i,
\qquad
h=2Ck_z,
\end{equation}
with parameter values
\begin{align}
\begin{array}{@{}c@{\hspace{2em}}c@{\hspace{2em}}c@{\hspace{2em}}c@{}}
A_1=-9.66~\mathrm{eV\AA^2},\quad
A_2=11.37~\mathrm{eV\AA^2},\quad
B_1=36.22~\mathrm{eV\AA^2},\quad
B_2=-25.71~\mathrm{eV\AA^2}. \\[0.8em]
\multicolumn{4}{c}{
M_1=0.12~\mathrm{eV},\qquad
M_2=-0.52~\mathrm{eV},\qquad
C=22.34~\mathrm{eV\AA}
}
\end{array}
\end{align}

Using this parametrization, we compute the Fermi surfaces of experimentally realized Ti$_3$Al  \cite{Vennila2005,Chen2013} and present two representative regimes of the Fermi energy $E$ in \cref{fig:Ti3Al_seq}. Ti$_3$Al exhibits the simplest nontrivial topological evolution of the Fermi surface: a torus (\cref{fig:Ti3Al_seq}(a)) deforms into a sphere (\cref{fig:Ti3Al_seq}(b)) through a gap-closing event as $E$ increases. In our graph-based encoding, this corresponds to a knotted-graph transition from the bouquet $\mathcal{B}_1$ (an unknot) to a single vertex,\footnote{A single vertex is a bouquet of zero circles, i.e., $\mathcal{B}_0$.} i.e., $\mathcal{B}_1 \to \mathcal{B}_0$. By Yamada axiom (V) in \cref{tab:YamadaTable}, the associated Yamada sequence is
\begin{equation}
\boldsymbol{\overline{\Upsilon}}_{\mathrm{Ti_3Al}}
=\bigl[\overline{\Upsilon}(\mathcal{B}_1),\overline{\Upsilon}(\mathcal{B}_0)\bigr]
=\bigl[-(Y^2+Y+1), -1\bigr].
\end{equation}

\subsection{YH$_3$}

Using the DFT-fitted low-energy dispersion for YH$_3$ \cite{Shao_2018}, which has been synthesized and experimentally characterized \cite{huiberts1996synthesis,wang1995structural}, we use
\begin{equation}
E_{\mathrm{YH_3}}(\mathbf{k})
=\sqrt{\bigl(g_1^2+h_1^2\bigr)\bigl(g_2^2+h_2^2\bigr)\bigl(g_3^2+h_3^2\bigr)},
\label{eq:yh3_dispersion}
\end{equation}
where each factor $g_i^2+h_i^2$ yields an individual nodal line, and
\begin{align}
g_1 &= \sin(k_z), &
h_1 &= a_1\!\left[r_1\bigl(\cos k_x\bigr)^{n_1}+s_1\bigl(\cos k_y\bigr)^{n_1}+t_1\bigl(\cos k_z\bigr)^{n_1}-m_1\right], \\
g_2 &= \sin(k_x), &
h_2 &= a_2\!\left(\cos k_x+\cos k_y+\cos k_z-m_2\right), \\
g_3 &= \sin(k_y), &
h_3 &= a_3\!\left(\cos k_x+\cos k_y+\cos k_z-m_3\right).
\end{align}

\begin{figure}
    \centering
    \includegraphics[width=\linewidth]{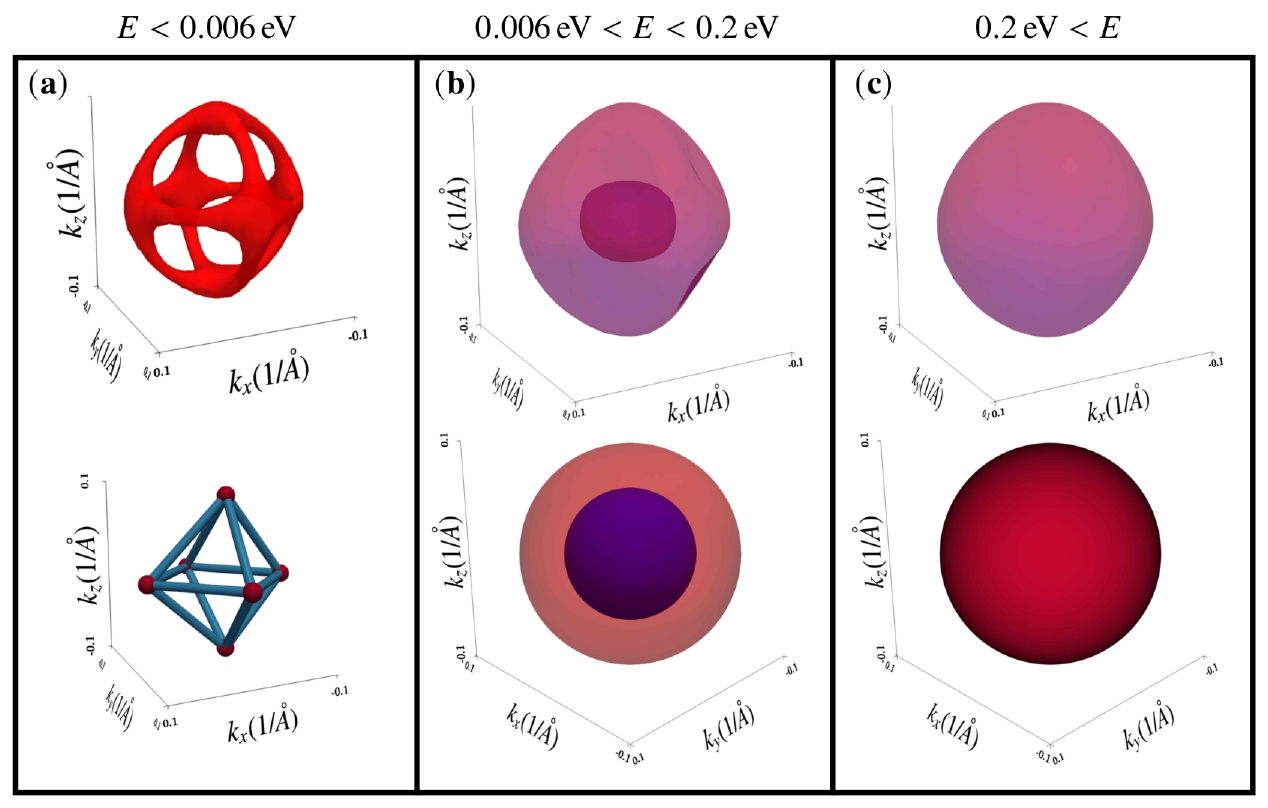}
    \caption{\textbf{Fermi-surface and multiset knotted-graph evolution for YH$_3$.}
    Fermi surfaces computed from the DFT-fitted dispersion \cref{eq:yh3_dispersion}.
    \textbf{(a)} At low $E$, the Fermi surface forms a single connected component whose associated knotted graph has an octahedral geometry.
    \textbf{(b)} Increasing $E$ closes all outer gaps and generates an additional interior genus, yielding two disconnected components (inner/outer), which are encoded separately as $\partial_-,\partial_+$.
    \textbf{(c)} At higher $E$, the inner component also closes, leaving a single spherical component represented by a single vertex.}
    \label{fig:YH3}
\end{figure}

The fitted parameters are $m_1=2.99$, $a_1=2$, $r_1=s_1=t_1=1.032$, $n_1=3$, $m_2=m_3=2.96$, and $a_2=a_3=4$. Using this parametrization, we plot the Fermi surfaces of YH$_3$ as a function of the Fermi energy $E$ in \cref{fig:YH3}.

In the low-energy regime (\cref{fig:YH3}(a)), the knotted graph exhibits an octahedral geometry whose associated knotted graph is a variant of the $\theta$-graph. This structure can be generated by periodically connecting adjacent edges of the $\theta$, hence we denote it by $\theta_{s,P}$ (here, $\theta_{4,P}$). The Yamada polynomial for this knotted graph is:
\begin{equation}\label{eq:Lyamada1Yh3}
\begin{aligned}
\overline{\Upsilon}(\theta_{4,P})
&= -[Y^{14} + 2Y^{13} + 13Y^{12} + 18Y^{11} + 60Y^{10} + 64Y^{9} + 125Y^{8} \\&+ 97Y^{7} + 125Y^{6} + 64Y^{5} + 60Y^{4} + 18Y^{3} + 13Y^{2} + 2Y + 1] \\
&= -Y^{7}\left(\sigma^{7} - 5\sigma^{6} + 15\sigma^{5} - 29\sigma^{4} + 40\sigma^{3} - 32\sigma^{2} + 11\sigma\right)= -Y^{7}\sum_{i=1}^{7} c_i\,\sigma^{i}.
\end{aligned}
\end{equation}

where $\sigma$ is used to write the expression more compactly and is defined in Yamada axiom~(V).

Increasing energy leads to the simultaneous closure of all outer gaps in \cref{fig:YH3}(a), while generating an additional genus in the interior region. This produces two distinct surfaces, i.e., two connected components, as highlighted by the different colors in \cref{fig:YH3}(b). Since a single knotted graph can only represent the topology of a single connected component, we instead use the multiset Yamada invariant $\Upsilon_{\partial\mathcal{F}}$ defined in \cref{appx:compression-body}. Labeling the inner/outer components as $\partial_-/\partial_+$, we assign a knotted graph to each component separately. In the present case, both components are genus-zero spheres, so each corresponds to a single node; we depict the outer component by a red node and the inner component by a purple node in the second row of \cref{fig:YH3}(b). These nodes have $\overline{\Upsilon}=-1$, and the multiset Yamada invariant is
\begin{equation} \overline{\Upsilon}_{\partial\mathcal{F}}=\!\{\Upsilon(\mathcal{G})\,:\, \mathcal{G}\in\pi_0(\partial\mathcal{F})\}\! =\Big\{\overline{\Upsilon}(\mathcal{G}_{\partial_-}),\,\overline{\Upsilon}(\mathcal{G}_{\partial_+})\Big\}=\Big\{-1,-1\Big\} \end{equation}
with $\pi_0(\partial\mathcal{F})={\partial_-,\partial_+}$.

Even at higher energies, the inner surface gap closes as well, yielding a sphere topology whose associated knotted graph is again a single node, as shown in \cref{fig:YH3}(c). Collectively, the three energy regimes can be summarized by the multiset Yamada invariants $\overline{\Upsilon}_{\partial\mathcal{F}}$ as follows:
\begin{subequations}
\begin{align}
\overline{\Upsilon}_{(E<0.006,\partial\mathcal{F})}
&=\Big\{\overline{\Upsilon}(\mathcal{G}_{\partial_-}),\,\overline{\Upsilon}(\mathcal{G}_{\partial_+})\Big\}
=\Big\{\varnothing,\, -Y^{7}\sum_{i=1}^{7} c_i\,\sigma^{i}\Big\},
\label{eq:octohedronyamada}
\\[2pt]
\overline{\Upsilon}_{(0.006<E<0.2,\partial\mathcal{F})}
&=\Big\{\overline{\Upsilon}(\mathcal{G}_{\partial_-}),\,\overline{\Upsilon}(\mathcal{G}_{\partial_+})\Big\}
=\Big\{-1,-1\Big\},
\label{eq:2sphere}
\\[2pt]
\overline{\Upsilon}_{(E>0.2,\partial\mathcal{F})}
&=\Big\{\overline{\Upsilon}(\mathcal{G}_{\partial_-}),\,\overline{\Upsilon}(\mathcal{G}_{\partial_+})\Big\}
=\Big\{\varnothing,-1\Big\}.
\label{eq:1sphere}
\end{align}
\end{subequations}

where $\varnothing$ denotes the empty set, indicating the absence of a corresponding connected component, i.e., the associated surface does not exist in that energy regime.

When comparing two multiset Yamada invariants, a topological change in any connected component implies a topological change of the overall structure. Consequently, a difference in any element of the multiset signals a topological transition. Using these multiset Yamada invariants, we define the overall Yamada sequence associated with the Hamiltonian as
\begin{equation}
\boldsymbol{\overline{\Upsilon}}(H_{\mathrm{YH_3}})
=\bigl[\quad
\Big\{\varnothing, -Y^{7}\sum_{i=1}^{7} c_i\,\sigma^{i}\Big\},\quad
\Big\{-1,-1\Big\},\quad
\Big\{\varnothing,-1\Big\}\quad
\bigr].
\end{equation}
where $c_i$ are given in \cref{eq:Lyamada1Yh3}.

\subsection{TiB$_2$}
\label[smsection]{appx:real-TiB2}

\begin{figure}
    \centering
    \includegraphics[width=\linewidth]{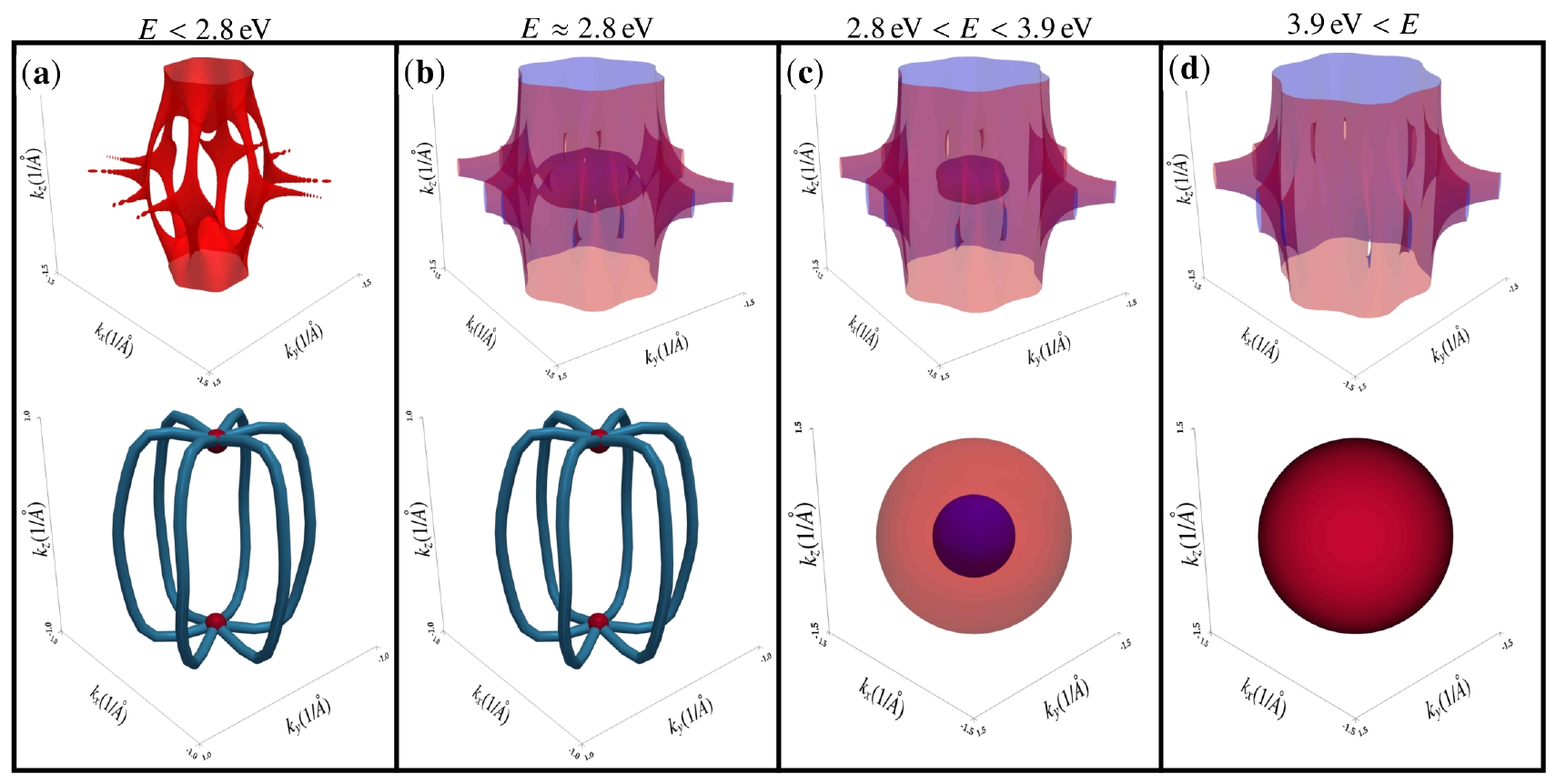}
    \caption{\textbf{Fermi-surface evolution and multiset knotted-graph encoding for TiB$_2$ near $\Gamma$.}
    Fermi surfaces computed for TiB$_2$ using the effective description near $\Gamma$ with parameters \cref{eq:TiB2_params_a}.
    \textbf{(a)} At low $E$, the Fermi surface is encoded by the $\theta_6$ graph.
    \textbf{(b)} Increasing $E$ toward $E\simeq 2.8$ progressively closes the outer-surface gaps while generating an interior gap.
    \textbf{(c)} With a further increase in $E$, the outer surface closes simultaneously, producing an interior gap and yielding genus-zero (single-vertex) topology for both the inner ($\partial_-$) and outer ($\partial_+$) components. \textbf{(d)} Upon increasing $E$ further, the remaining interior gap associated with $\partial_-$ also closes, leaving a single connected spherical component. }
    \label{fig:TiB2_a}
\end{figure}

Using the DFT-fitted low-energy $k\!\cdot\!p$ model for TiB$_2$ from Ref. \cite{Feng2018}, an experimentally established material \cite{Liu2018,Yi2018}, we adopt the three-band Hamiltonian
\begin{equation}
H_{\mathrm{TiB_2}}(\mathbf{k})=
\begin{pmatrix}
Q_1(\mathbf{k}) & h_{12}(\mathbf{k}) & h_{13}(\mathbf{k})\\
h_{12}^{\dagger}(\mathbf{k}) & Q_1(\mathbf{k}) & h_{23}(\mathbf{k})\\
h_{13}^{\dagger}(\mathbf{k}) & h_{23}^{\dagger}(\mathbf{k}) & Q_2(\mathbf{k})
\end{pmatrix},
\label{eq:tib2_H}
\end{equation}
where $k_{\pm}=k_x\pm i k_y$ and $k_\perp^2=k_x^2+k_y^2$. The diagonal terms are
\begin{align}
Q_1(\mathbf{k}) &= F_1 + A_1 k_\perp^2 + B_1 k_z^2,\\
Q_2(\mathbf{k}) &= F_2 + A_2 k_\perp^2 + B_2 k_z^2
+ L\,k_\perp^4 + M\,(k_+^6+k_-^6),
\end{align}
and the off-diagonal couplings are
\begin{align}
h_{12}(\mathbf{k}) &= C\,k_-^2 + F\,k_+^4,\\
h_{13}(\mathbf{k}) &= D\,k_- k_z,\\
h_{23}(\mathbf{k}) &= D\,k_+ k_z,
\end{align}
with the parameter set:
\begin{equation}
A_1=2.6,\quad B_1=-3.8,\quad F_1=1.787,\quad
A_2=1.63,\quad B_2=5.1,\quad F_2=-2.12,\quad
L=1.3,\quad C=3.55,
\label{eq:tib2_base_params}
\end{equation}
and consider three fitted parameter panels (corresponding to \cref{fig:TiB2_a}, \cref{fig:TiB2_b} and \cref{fig:TiB2_c}):
\begin{align}
\text{(\cref{fig:TiB2_a})}\ \qquad& M=0.65,\qquad F=1.83,\qquad D=5.1,\label{eq:TiB2_params_a}\\
\text{(\cref{fig:TiB2_b})}\ \qquad & M=0.0,\qquad\ \ F=1.18,\qquad D=2.53,\label{eq:TiB2_params_b}\\
\text{(\cref{fig:TiB2_c})}\ \qquad& M=0.65,\qquad F=7.6,\qquad\ \ D=5.1.\label{eq:TiB2_params_c}
\end{align}
Here, \cref{fig:TiB2_a} uses the baseline parameter set obtained by fitting the $\Gamma$-centered $k\!\cdot\!p$ model to the DFT bands near $\Gamma$ point  \cite{Feng2018}. Since this effective description is local, it captures the near-$\Gamma$ nodal-net geometry but may not reproduce features tied to the Brillouin-zone boundary, such as the nexus point $A$ and the nodal line beyond it. Panels \cref{fig:TiB2_b} and \cref{fig:TiB2_c} therefore serve as symmetry-consistent, representative parameter variations to isolate how specific terms in the $k\!\cdot\!p$ model affect the nodal net: \cref{fig:TiB2_b} demonstrates that a mirror-plane nodal line becomes visible when the fourth- and sixth-order contributions are reduced, whereas \cref{fig:TiB2_c} shows that the nexus point disappears when $F$ is taken to be sufficiently large.
In the following sections, we analyze these three cases in detail within the effective $k\!\cdot\!p$ framework introduced in Ref.~\cite{Feng2018}.

\subsubsection{Effective Description Near $\Gamma$ (Parameter Set \cref{eq:TiB2_params_a})}

Plotting the Fermi-surface evolution in \cref{fig:TiB2_a}, we find that at low energies the system first exhibits a surface topology whose associated knotted graph is a $\theta$-graph, as shown in \cref{fig:TiB2_a}(a). More specifically, this configuration corresponds to a $\theta_6$ graph. Using \cref{eq:stheta}, the associated normalized Yamada polynomial can be written compactly as
\begin{equation}\label{eq:6theta}
    \overline{\Upsilon}
    =-Y^5 \frac{\sigma+(-\sigma)^6}{\sigma+1}
    =-Y^5 \sigma(\sigma^4 - \sigma^3 + \sigma^2 - \sigma + 1)=-Y^5\sum_{i=1}^{5} c_i\,\sigma^{i},
\end{equation}
where $\sigma$ is defined by Yamada axiom~(V).

As the Fermi energy is increased toward $E \approx 2.8$, we observe that the gaps of the outer surface progressively close in a manner that generates an inner gap, as illustrated in \cref{fig:TiB2_a}(b). With a further increase in $E$, the outer surface undergoes complete closure [\cref{fig:TiB2_a}(c–d)], exhibiting an evolution analogous to that shown in \cref{fig:YH3}(b–c). Using previous results (\cref{eq:1sphere} and \cref{eq:2sphere}), we can express the Yamada sequence associated with TiB$_2$ near $\Gamma$ for the parameter set \cref{eq:TiB2_params_a} as:
\begin{equation}
\boldsymbol{\overline{\Upsilon}}(H_{\mathrm{TiB_2},\,\ref{eq:TiB2_params_a}})
=\bigl[\quad
\Big\{\varnothing,\,-Y^5\sum_{i=1}^{5} c_i\,\sigma^{i}\Big\},\quad
\Big\{-1,-1\Big\},\quad
\Big\{\varnothing,-1\Big\}\quad
\bigr].
\end{equation}
where $c_i$ are given in \cref{eq:6theta}.

\subsubsection{Reduced Higher Order Contributions (Parameter Set \cref{eq:TiB2_params_b})}

\begin{figure}
    \centering
    \includegraphics[width=\linewidth]{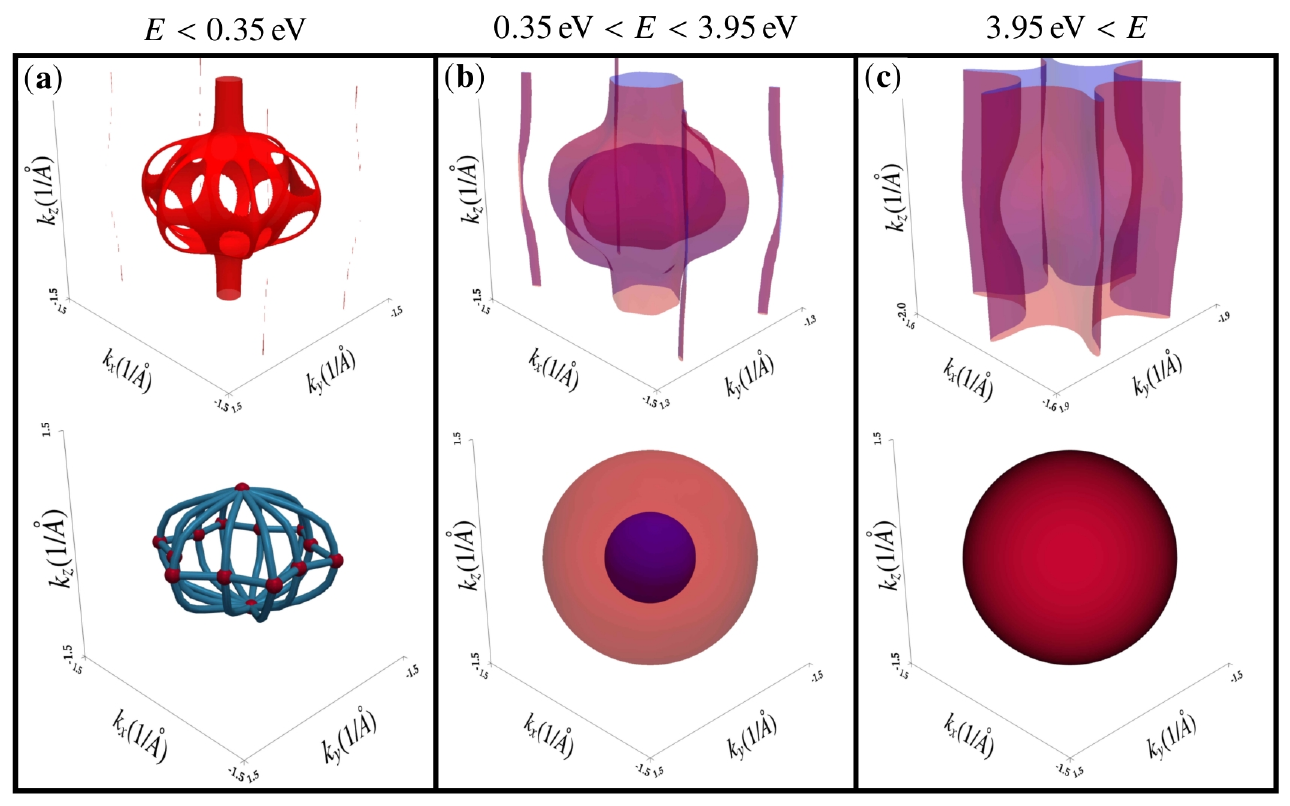}
    \caption{\textbf{Fermi-surface evolution and multiset knotted-graph encoding for TiB$_2$ with reduced higher order contributions.}
    Fermi surfaces computed for TiB$_2$ using the effective description near $\Gamma$ with parameters \cref{eq:TiB2_params_b}.
    \textbf{(a)} At low $E$, the Fermi surface is encoded by the periodically connected $\theta$-graph $\theta_{12,P}$.
    \textbf{(b)} Owing to symmetry, the outer-surface gaps close simultaneously, generating an interior surface and yielding two sphere surface component.
    \textbf{(c)} With further increase in $E$, the inner gap closes, leaving a single connected spherical component represented by a single vertex.}
    \label{fig:TiB2_b}
\end{figure}

Plotting the Fermi-surface evolution in \cref{fig:TiB2_b}, we find that at low energies the system first realizes a surface topology whose associated knotted graph is the previously introduced $\theta_{s,P}$, as shown in \cref{fig:TiB2_b}(a). Its normalized Yamada polynomial is:
\begin{equation}
\begin{aligned}\label{eq:theta_pbc}
\overline{\Upsilon}(\theta_{12,P})
&= -Y^{49}\Bigl[
105\sigma^{49}-105\sigma^{48}-148\sigma^{46}+562\sigma^{45}-414\sigma^{44}+581\sigma^{43}-821\sigma^{42}+1582\sigma^{41}\\
&\qquad-3036\sigma^{40}+1694\sigma^{39}-1978\sigma^{38}+7298\sigma^{37}-5800\sigma^{36}+4392\sigma^{35}-11573\sigma^{34}\\
&\qquad+12471\sigma^{33}-12002\sigma^{32}+19177\sigma^{31}-20004\sigma^{30}+26708\sigma^{29}-34599\sigma^{28}\\
&\qquad+32589\sigma^{27}-43601\sigma^{26}+58566\sigma^{25}-55133\sigma^{24}+63543\sigma^{23}-81481\sigma^{22}\\
&\qquad+84874\sigma^{21}-92790\sigma^{20}+106858\sigma^{19}-114419\sigma^{18}+129524\sigma^{17}-143256\sigma^{16}\\
&\qquad+147629\sigma^{15}-168991\sigma^{14}+185918\sigma^{13}-195710\sigma^{12}+202948\sigma^{11}-230406\sigma^{10}\\
&\qquad+227939\sigma^{9}-239477\sigma^{8}+235754\sigma^{7}-230303\sigma^{6}+206225\sigma^{5}-158742\sigma^{4}\\
&\qquad+85688\sigma^{3}-28580\sigma^{2}+4083\sigma
\Bigr] \\
&= -Y^{49}\sum_{i=1}^{49} c_i\,\sigma^{i}.
\end{aligned}
\end{equation}
where, as before, $\sigma$ is defined in Yamada axiom (V). It is worth emphasizing that evaluating this polynomial via \cref{eq:negami} is computationally prohibitive: the sum runs over all edge subsets, so the cost scales exponentially with the edge set size, i.e., as $2^{\lvert E(\mathcal{G})\rvert}$. Therefore, exploiting the previously noted recursive structure of the $\theta_{s,P}$ graphs was essential for this calculation.

We observe that, owing to symmetry, the gap closures occur simultaneously (as in the previous cases). This process first generates an inner surface (\cref{fig:TiB2_b}(b)), and subsequently leads to complete closure (\cref{fig:TiB2_b}(c)). The corresponding multiset Yamada invariants are given in \cref{eq:1sphere} and \cref{eq:2sphere}. Collectively Yamada sequence becomes:

\begin{equation}
\boldsymbol{\overline{\Upsilon}}(H_{\mathrm{TiB_2},\,\ref{eq:TiB2_params_b}})
=\bigl[\quad
\Big\{\varnothing,\,-Y^{49}\sum_{i=1}^{49} c_i\,\sigma^{i}\Big\},\quad
\Big\{-1,-1\Big\},\quad
\Big\{\varnothing,-1\Big\}\quad
\bigr].
\end{equation}
where $c_i$ are given in \cref{eq:theta_pbc}.

\begin{figure}
    \centering
    \includegraphics[width=\linewidth]{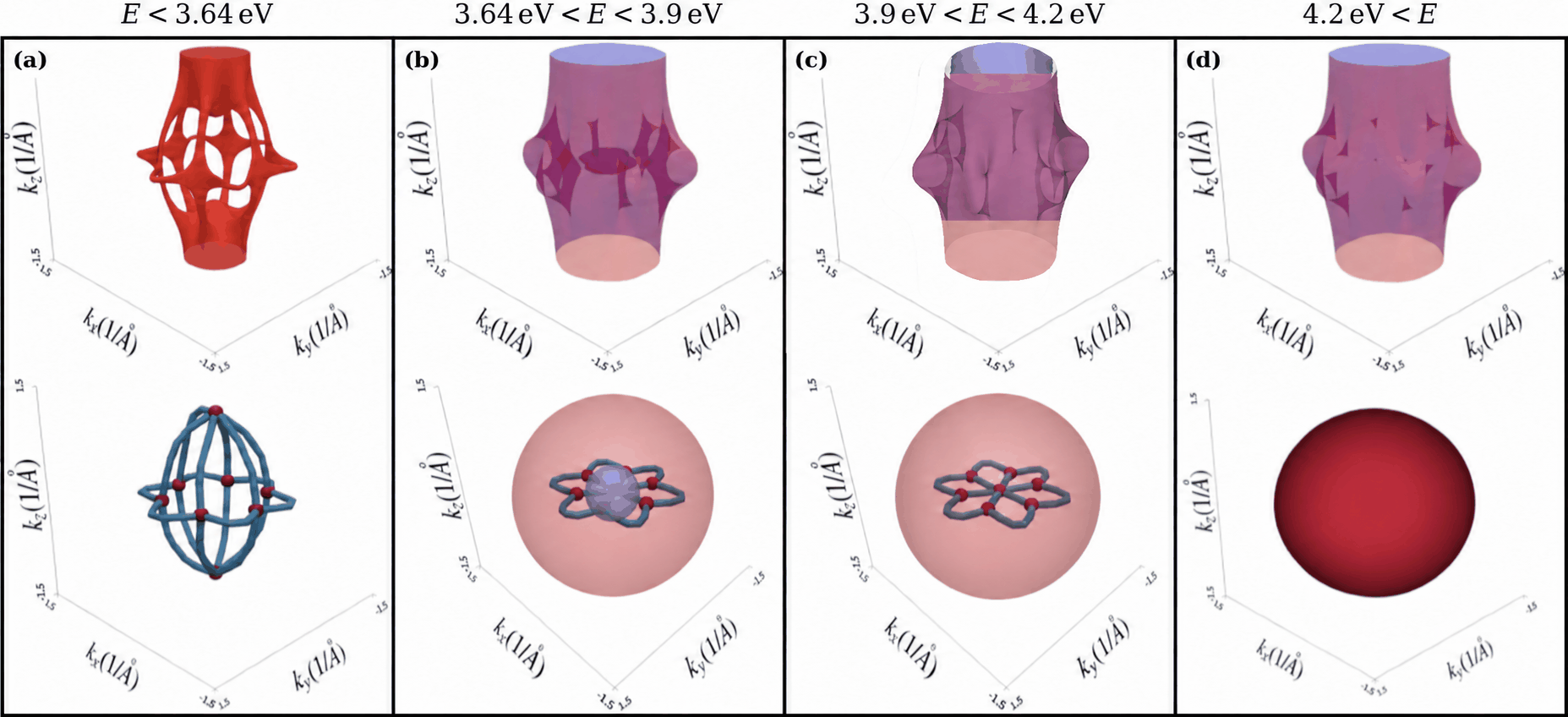}
    \caption{
    \textbf{Fermi-surface evolution and multiset knotted-graph encoding for TiB$_2$ in the large-$F$ limit.} Fermi surfaces computed using the effective description near $\Gamma$ with parameters \cref{eq:TiB2_params_c}.
    \textbf{(a)} For $E<3.64\,\mathrm{eV}$, the surface is encoded by the periodically connected $\theta$-graph $\theta_{6,P}$.
    \textbf{(b)} For $3.64\,\mathrm{eV}<E<3.9\,\mathrm{eV}$, the topology contains three boundary components: an inner spherical boundary, a flower-like middle boundary, and an outer spherical boundary.
    \textbf{(c)} For $3.9\,\mathrm{eV}<E<4.2\,\mathrm{eV}$, the innermost gap closes and its spherical boundary disappears, leaving the same flower-like boundary together with the outer spherical boundary.
    \textbf{(d)} For $E>4.2\,\mathrm{eV}$, the remaining interior structure closes, leaving a single spherical component represented by a single vertex.
    }
    \label{fig:TiB2_c}
\end{figure}
\subsubsection{Large $F$ Limit (Parameter Set \cref{eq:TiB2_params_c})}

For the large-$F$ parameter set \cref{eq:TiB2_params_c}, the low-energy surface topology is captured by the periodically connected graph $\theta_{6,P}$ shown in \cref{fig:TiB2_c}(a). Its normalized Yamada polynomial is
\begin{equation}
\begin{aligned}\label{eq:theta_pbc_s6}
\overline{\Upsilon}(\theta_{6,P}) &= -Y^{13}\Bigl[ 9\sigma^{13}-9\sigma^{12}+19\sigma^{11}-39\sigma^{10} +74\sigma^{9}-120\sigma^{8}+187\sigma^{7}-281\sigma^{6}\\
&\qquad +364\sigma^{5}-415\sigma^{4}+382\sigma^{3} -227\sigma^{2}+57\sigma \Bigr] \\
&= -Y^{13}\sum_{i=1}^{13} c_i\,\sigma^{i}.
\end{aligned}
\end{equation}

For $E<3.64\,\mathrm{eV}$, the surface therefore carries the $\theta_{6,P}$ topology. As $E$ increases through $E\simeq3.64\,\mathrm{eV}$, the outer surface closes while nontrivial interior structure remains. The resulting configuration in \cref{fig:TiB2_c}(b) contains three boundary components: an inner spherical boundary, a middle boundary with flower-like topology, and the outer spherical boundary. The flower-like component has normalized Yamada polynomial
\begin{equation}
\begin{aligned}\label{eq:TiB2_c_c}
\overline{\Upsilon} &= Y^{6}\sigma\Bigl[ -\sigma^{5} +6\sigma^{4} -15\sigma^{3} +20\sigma^{2} -15\sigma +5 \Bigr] \\
&= Y^{6}\sum_{i=1}^{6} d_i\,\sigma^{i}.
\end{aligned}
\end{equation}
Since each spherical boundary is represented by a single vertex with $\overline{\Upsilon}=-1$, the corresponding multiset Yamada invariant for $3.64\,\mathrm{eV}<E<3.9\,\mathrm{eV}$ is
\begin{equation}
\overline{\Upsilon}_{\partial\mathcal{F}} = \Bigg\{ -1,\, Y^{6}\sum_{i=1}^{6} d_i\,\sigma^{i},\, -1 \Bigg\}.
\label{eq:tib2_largeF_three_boundaries}
\end{equation}

At $E\simeq3.9\,\mathrm{eV}$, the innermost gap closes. The corresponding inner spherical boundary therefore disappears, while the flower-like middle boundary retains its topology. The configuration in \cref{fig:TiB2_c}(c) consequently contains two boundary components: the flower-like boundary and the outer spherical boundary. Its multiset Yamada invariant is
\begin{equation}
\overline{\Upsilon}_{\partial\mathcal{F}} = \Bigg\{\varnothing, Y^{6}\sum_{i=1}^{6} d_i\,\sigma^{i},\, -1 \Bigg\}, \qquad 3.9\,\mathrm{eV}<E<4.2\,\mathrm{eV}.
\label{eq:tib2_largeF_two_boundaries}
\end{equation}

Upon increasing $E$ beyond $4.2\,\mathrm{eV}$, the remaining interior structure also closes. Only the outer spherical boundary remains, as shown in \cref{fig:TiB2_c}(d), and the topology is therefore represented by a single vertex with $\overline{\Upsilon}=-1$.

Collecting the four regimes, the corrected multiset Yamada sequence for the large-$F$ Hamiltonian is
\begin{equation}
\begin{aligned}
\boldsymbol{\overline{\Upsilon}} (H_{\mathrm{TiB_2},\,\ref{eq:TiB2_params_c}}) = \biggl[\, & \Bigg\{ \varnothing,\, -Y^{13}\sum_{i=1}^{13} c_i\,\sigma^{i},\varnothing \Bigg\}, \quad \Bigg\{ -1,\, Y^{6}\sum_{i=1}^{6} d_i\,\sigma^{i},\, -1 \Bigg\}, \\
& \Bigg\{ \varnothing,Y^{6}\sum_{i=1}^{6} d_i\,\sigma^{i},\, -1 \Bigg\}, \quad \Bigg\{ \varnothing,\varnothing,\,-1 \Bigg\} \,\biggr].
\label{eq:tib2_largeF_corrected_sequence}
\end{aligned}
\end{equation}
Here $c_i$ and $d_i$ are defined in \cref{eq:theta_pbc_s6,eq:TiB2_c_c}, respectively.

\subsection{Co$_2$MnGa}

\begin{figure}
    \centering
    \includegraphics[width=\linewidth]
    {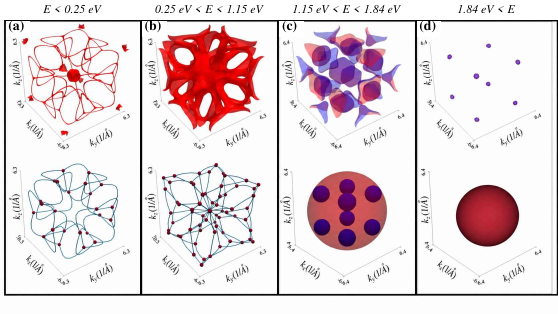}
    \caption{
    \textbf{Finite-thickness momentum-space evolution in Co$_2$MnGa on an enlarged Brillouin-zone cube.}
    Surfaces obtained by diagonalizing the six-band Hamiltonian
    $H_{\mathrm{Co_2MnGa}}(\mathbf{k})$.
    For visualization, the momentum-space structures are rendered on a cube
    whose side length is twice that of the first Brillouin zone.
    \textbf{(a)} For $E<0.25\,\mathrm{eV}$, the low-threshold structure
    consists of weakly connected components with a sparse knotted-graph
    skeleton.
    \textbf{(b)} For $0.25<E<1.15\,\mathrm{eV}$, the surfaces thicken and
    reconnect into a highly intricate multiply connected topology.
    \textbf{(c)} For $1.15<E<1.84\,\mathrm{eV}$, the surrounding region
    closes while a nontrivial internal cavity remains. In the enlarged
    rendering, the corresponding internal structure appears through
    periodically repeated copies.
    \textbf{(d)} For $E>1.84\,\mathrm{eV}$, the remaining internal cavity
    closes, leaving only the outer spherical surface.
    }
    \label{fig:Co2mnGa2cellsize}
\end{figure}
Using the DFT-fitted tight-binding model for Co$_2$MnGa from Ref. \cite{Chang2017}, an experimentally established magnetic Heusler compound \cite{Ilya2019,Guin_2019,Markou_2019}, we consider a six-band Hamiltonian built from three Mn $d$ orbitals and three Ga $p$ orbitals. In reciprocal space, with the basis $(d_{xz},d_{yz},d_{xy},p_x,p_y,p_z)$, the model reads
\begin{equation}
H_{\mathrm{Co_2MnGa}}(\mathbf{k})=
\begin{pmatrix}
\xi^{d}_1 & 0 & 0 & \xi^{dp}_{11} & 0 & \xi^{dp}_{13} \\
0 & \xi^{d}_2 & 0 & 0 & \xi^{dp}_{22} & \xi^{dp}_{23} \\
0 & 0 & \xi^{d}_3 & \xi^{dp}_{31} & \xi^{dp}_{32} & 0 \\
\xi^{dp}_{11} & 0 & \xi^{dp}_{31} & \xi^{p}_1 & \xi^{p}_{12} & \xi^{p}_{31} \\
0 & \xi^{dp}_{22} & \xi^{dp}_{32} & \xi^{p}_{12} & \xi^{p}_2 & \xi^{p}_{23} \\
\xi^{dp}_{13} & \xi^{dp}_{23} & 0 & \xi^{p}_{31} & \xi^{p}_{23} & \xi^{p}_3
\end{pmatrix}\!(\mathbf{k}),
\label{eq:co2mnga_H}
\end{equation}
with matrix elements

\begin{equation}\label{eq:co2mnga_xi}
\begin{aligned}
\begin{array}{@{}c@{\qquad}c@{}}
\begin{aligned}[t]
\xi^{d}_1(\mathbf{k}) &= 4t_1\cos\frac{k_x}{2}\cos\frac{k_z}{2}
+2t_2(\cos k_x+\cos k_z)\\
&\quad +2t_3\cos k_y+\epsilon_d,
\end{aligned}
&
\begin{aligned}[t]
\xi^{d}_2(\mathbf{k}) &= 4t_1\cos\frac{k_y}{2}\cos\frac{k_z}{2}
+2t_2(\cos k_y+\cos k_z)\\
&\quad +2t_3\cos k_x+\epsilon_d,
\end{aligned}
\\[4pt]
\begin{aligned}[t]
\xi^{d}_3(\mathbf{k}) &= 4t_1\cos\frac{k_x}{2}\cos\frac{k_y}{2}
+2t_2(\cos k_x+\cos k_y)\\
&\quad +2t_3\cos k_z+\epsilon_d,
\end{aligned}
&
\begin{aligned}[t]
\xi^{p}_1(\mathbf{k}) &= 4t_4\cos\frac{k_y}{2}\cos\frac{k_z}{2}
+2t_5(\cos k_y+\cos k_z)\\
&\quad +2t_6\cos k_x+\epsilon_p,
\end{aligned}
\\[4pt]
\begin{aligned}[t]
\xi^{p}_2(\mathbf{k}) &= 4t_4\cos\frac{k_x}{2}\cos\frac{k_z}{2}
+2t_5(\cos k_x+\cos k_z)\\
&\quad +2t_6\cos k_y+\epsilon_p,
\end{aligned}
&
\begin{aligned}[t]
\xi^{p}_3(\mathbf{k}) &= 4t_4\cos\frac{k_x}{2}\cos\frac{k_y}{2}
+2t_5(\cos k_x+\cos k_y)\\
&\quad +2t_6\cos k_z+\epsilon_p,
\end{aligned}
\end{array}
\\[6pt]
\begin{array}{@{}l@{\hspace{2em}}l@{\hspace{2em}}l@{}}
\xi^{p}_{12}(\mathbf{k}) = -4t_7\sin\frac{k_x}{2}\sin\frac{k_y}{2}, &
\xi^{p}_{23}(\mathbf{k}) = -4t_7\sin\frac{k_y}{2}\sin\frac{k_z}{2}, &
\xi^{p}_{31}(\mathbf{k}) = -4t_7\sin\frac{k_x}{2}\sin\frac{k_z}{2}, \\
\xi^{dp}_{11}(\mathbf{k}) = \xi^{dp}_{22}(\mathbf{k}) = 2t_8\sin\frac{k_z}{2}, &
\xi^{dp}_{13}(\mathbf{k}) = \xi^{dp}_{32}(\mathbf{k}) = 2t_8\sin\frac{k_x}{2}, &
\xi^{dp}_{23}(\mathbf{k}) = \xi^{dp}_{31}(\mathbf{k}) = 2t_8\sin\frac{k_y}{2}.
\end{array}
\end{aligned}
\end{equation}

\begin{figure}
    \centering
    \includegraphics[width=\linewidth]
    {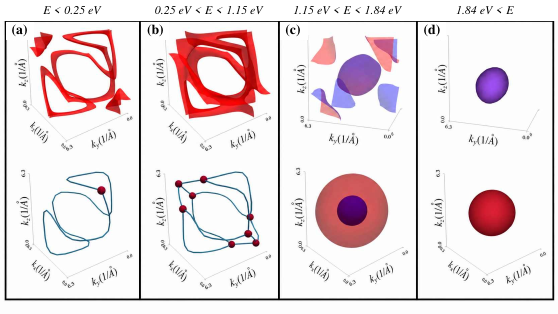}
    \caption{
    \textbf{Single-unit-cell momentum-space surfaces used for
    knotted-graph extraction and Yamada-polynomial evaluation in
    Co$_2$MnGa.}
    The invariant calculation is performed within a single
    Brillouin-zone unit cell.
    \textbf{(a)} For $E<0.25\,\mathrm{eV}$, three connected components
    are present, consisting of two spherical components and a central
    knotted component; only the knotted graph of the latter is shown.
    \textbf{(b)} For $0.25<E<1.15\,\mathrm{eV}$, the components expand
    and merge into a single highly connected surface.
    \textbf{(c)} For $1.15<E<1.84\,\mathrm{eV}$, the outer region has
    closed while a distinct internal cavity remains, producing separate
    outer and inner boundary components.
    \textbf{(d)} For $E>1.84\,\mathrm{eV}$, further thickening closes
    the remaining internal cavity, leaving only a single outer spherical
    boundary.
    }
    \label{fig:Co2mnGa2singlecellsize}
\end{figure}

The fitted parameters are
$t_1=-0.31$, $t_2=-0.018$, $t_3=-0.01$, $t_4=0.2$, $t_5=-0.02$, $t_6=0.04$, $t_7=0.28$, $t_8=-0.34$,
$\epsilon_d=-0.6$, and $\epsilon_p=0.6$ \cite{Chang2017}.
Diagonalizing $H_{\mathrm{Co_2MnGa}}(\mathbf{k})$ yields the six-band dispersion used to compute the finite-thickness momentum-space regions. For clarity, we visualize all nodal structures in \cref{fig:Co2mnGa2cellsize} on a cube whose side length is twice that of the first Brillouin zone. However, for computational tractability, we restrict the analysis to a single unit cell when extracting the corresponding knotted graphs and evaluating the Yamada polynomials; the single-cell configurations used in the calculations are shown in \cref{fig:Co2mnGa2singlecellsize}.

As the Fermi energy $E$ is increased, the initial momentum-space structure in
\cref{fig:Co2mnGa2cellsize}(a) progressively thickens and reconnects,
producing the highly intricate topology in
\cref{fig:Co2mnGa2cellsize}(b).
At $E\simeq1.15\,\mathrm{eV}$, the surrounding region closes while a
nontrivial internal cavity remains, yielding the distinct outer and
inner-boundary structure shown in
\cref{fig:Co2mnGa2cellsize}(c).
In the enlarged momentum-space rendering, this internal structure appears
through periodic copies of the corresponding single-cell boundary.
Upon increasing $E$ beyond approximately $1.84\,\mathrm{eV}$, the remaining
internal cavity also closes, leaving only the outer spherical surface shown
in \cref{fig:Co2mnGa2cellsize}(d). Notably,  \cite{Lee_2020,tai2021anisotropic} demonstrated that the number of Fermi-surface touchings---associated with multiple velocity turning points---can enhance the nonlinearity of the intraband optical response, and Co$_2$MnGa was found to exhibit one of the strongest such responses. In this context, knotted-graph representations provide a natural framework to quantify the expected nonlinear strength: the degree of a node directly counts the number of nodal touchings, and thus serves as a proxy for the density of velocity turning points. We believe that establishing a quantitative connection between these graph-theoretic descriptors and the resulting nonlinear optical response is a promising direction for future research.

As can be seen, the knotted-graph and surface topologies in the second row of \cref{fig:Co2mnGa2cellsize} are highly intricate, making the evaluation of the Yamada polynomial via \cref{eq:yamada_negami} challenging both analytically and computationally. Accordingly, as stated above, we restrict to single-unit-cell Fermi surfaces when extracting the knotted graphs and evaluating the Yamada polynomial; the resulting single-cell configurations are shown in \cref{fig:Co2mnGa2singlecellsize}. In the viewpoint of \cref{fig:Co2mnGa2singlecellsize}(a), the surface consists of three connected components: the closed surface in the upper-left (topologically a sphere), the knotted component in the center, and the closed surface in the lower-right (again topologically a sphere). The resulting multiset Yamada invariant is:
\begin{equation}\label{eq:yamadaco2mnga_lowE}
\begin{aligned}
\overline{\Upsilon}_{(E<0.25,\partial\mathcal{F})}
&=\Big\{
\overline{\Upsilon}(\mathcal{G}_{\partial_1}),\,
\overline{\Upsilon}(\mathcal{G}_{\partial_2}),\,
\overline{\Upsilon}(\mathcal{G}_{\partial_3})
\Big\}
=\Big\{
-1,\,
-\bigl(Y^{4}+3Y^{3}+5Y^{2}+4Y+2\bigr),\,
-1
\Big\} \\
&=\Big\{
-1,\,
\sum_{i=0}^{4} c_i\,Y^{i},\,
-1
\Big\}.
\end{aligned}
\end{equation}

In \cref{fig:Co2mnGa2singlecellsize}(a) we display only the knotted graph associated with the interior component, since the remaining two components are topologically apparent. At intermediate energies, the surfaces expand and merge into a single connected component, as shown in \cref{fig:Co2mnGa2singlecellsize}(b). In this regime, the Yamada invariant becomes:
\begin{equation}\label{eq:yamadaco2mnga_middleE}
\begin{aligned}
\overline{\Upsilon}_{(0.25<E<1.15,\partial\mathcal{F})}
&= -Y^{10} + 2Y^{9} - 7Y^{8} + 5Y^{7} - 14Y^{6} + 6Y^{5} - 14Y^{4} + 5Y^{3} - 7Y^{2} + 2Y - 1=\sum_{i=0}^{10} d_i\,Y^{i}.
\end{aligned}
\end{equation}

For $1.15<E<1.84\,\mathrm{eV}$, the finite-thickness region contains
distinct spherical outer and inner boundary components, as shown in
\cref{fig:Co2mnGa2singlecellsize}(c). The corresponding boundary-resolved
Yamada invariant is
\begin{equation}
\overline{\Upsilon}_{(1.15<E<1.84,\partial\mathcal{F})}
=
\Big\{\varnothing,-1,-1\Big\}.
\end{equation}

For $E>1.84\,\mathrm{eV}$, the remaining internal cavity closes, leaving
only the outer spherical boundary shown in
\cref{fig:Co2mnGa2singlecellsize}(d). The corresponding boundary-resolved
Yamada invariant is
\begin{equation}
\overline{\Upsilon}_{(E>1.84,\partial\mathcal{F})}
=
\Big\{\varnothing,-1,\varnothing\Big\}.
\end{equation}

Collecting all regimes, the Yamada sequence for Co$_2$MnGa is therefore:

\begin{equation}
\boldsymbol{\overline{\Upsilon}}(H_{\mathrm{Co_2MnGa}})
=\bigl[\quad
\Big\{
-1,\,
\sum_{i=0}^{4} c_i\,Y^{i},\,
-1
\Big\},\quad
\Big\{
\varnothing,\,
\sum_{i=0}^{10} d_i\,Y^{i},\,
\varnothing
\Big\},\quad
\Big\{\varnothing,-1,-1\Big\},\quad\Big\{\varnothing,-1,\varnothing\Big\}\quad
\bigr].
\end{equation}
where $c_i$ and $d_i$ are given in \cref{eq:yamadaco2mnga_lowE,eq:yamadaco2mnga_middleE}, respectively.

\bigskip

These material examples illustrate several aspects that go beyond the scope of existing Yamada polynomial methods. First, three of the four materials---YH$_3$, TiB$_2$, and Co$_2$MnGa---develop multi-boundary Fermi volumes that \textit{require} the Yamada set formalism developed in \cref{appx:compression-body}; a single Yamada polynomial cannot encode their topology. Second, the algorithmic pipeline (\cref{appx:Yamada}) was essential for evaluating polynomials of high complexity, such as the degree-49 invariant of $\theta_{12,P}$ in TiB$_2$. Together, these results constitute the first systematic Yamada-polynomial fingerprinting of realistic multiband Fermi surfaces.

\subsection{Parameter-resolved topology maps of material Hamiltonians}
\label[smsection]{appx:material_parameter_maps}

The fixed-parameter analyses above establish the topology sequences of TiB$_2$ and Co$_2$MnGa for representative Hamiltonians. We now promote one microscopic coefficient in each model to a continuous deformation parameter, thereby resolving how these established topology classes connect through the larger parameter space shown in \cref{fig:topophasespace}(b,c).

\begin{figure}[t]
    \centering
    \includegraphics[width=\linewidth]{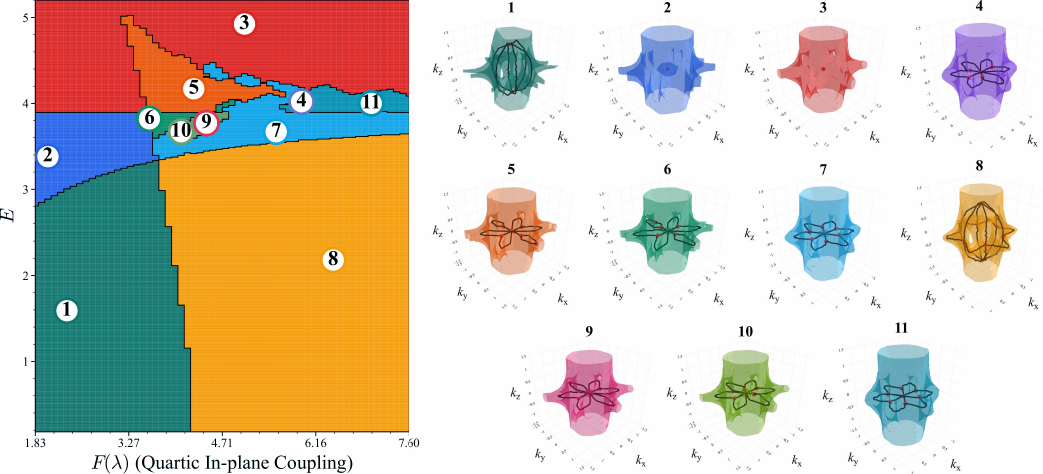}
    \caption{
    \textbf{Detailed parameter-resolved topology map of the TiB$_2$ Hamiltonian underlying \cref{fig:topophasespace}(b).} The horizontal axis follows the quartic in-plane coupling $F(\lambda)$ in \cref{eq:tib2_parameter_path}. Colored regions denote eleven stabilized topology-signature sectors, with numbered markers identifying the representative geometries and knotted-graph skeletons shown at right. Surface colors match the corresponding regions; black curves denote graph edges and red points denote graph vertices. The $F=1.83$ boundary yields the sequence $1\rightarrow2\rightarrow3$, reproducing the topology progression in \cref{fig:TiB2_a}. At the opposite boundary, $F=7.60$, the sequence $8\rightarrow7\rightarrow11\rightarrow3$ reproduces the four regimes in \cref{fig:TiB2_c}: the periodically connected $\theta_{6,P}$ topology, the three-boundary configuration containing an inner sphere, a flower-like middle boundary, and an outer sphere, the subsequent two-boundary configuration obtained after closure of the innermost gap, and finally the single spherical component. The representative geometries retain the expected sixfold organization throughout the deformation.
    }
    \label{fig:tib2_parameter_phase_map}
\end{figure}


\subsubsection{TiB$_2$: deformation of the quartic in-plane coupling}

For TiB$_2$, we use the three-band $D_6$ $k\!\cdot\!p$ Hamiltonian introduced in \cref{eq:tib2_H} and continuously vary the quartic in-plane coupling $F$ entering
\begin{equation}
h_{12}(\mathbf{k}) = Ck_-^2+Fk_+^4,
\label{eq:tib2_h12_parameter_map}
\end{equation}
while keeping $M=0.65$, $D=5.1$, and all remaining coefficients fixed. The deformation is
\begin{equation}
F(\lambda) = (1-\lambda)\,1.83+\lambda\,7.60, \qquad 0\le\lambda\le1.
\label{eq:tib2_parameter_path}
\end{equation}
Thus, the left boundary $F=1.83$ is precisely the DFT-fitted $\Gamma$-centered Hamiltonian studied in \cref{fig:TiB2_a} \cite{Feng2018}, whereas the right boundary $F=7.60$ is the symmetry-preserving large-$F$ Hamiltonian studied independently in \cref{fig:TiB2_c}. The parameter sweep therefore connects two previously characterized limiting Hamiltonians while isolating the role of the quartic hybridization channel.

The microscopic origin of the resulting phase structure follows directly from the competition between the two terms in \cref{eq:tib2_h12_parameter_map}. Writing $k_\pm=k_\perp e^{\pm i\phi}$ gives
\begin{equation}
|h_{12}|^2 = k_\perp^4 \left[ C^2+F^2k_\perp^4 +2CFk_\perp^2\cos(6\phi) \right].
\end{equation}
The quadratic channel dominates near $\Gamma$, whereas the quartic channel grows more rapidly with $k_\perp$; their characteristic competition occurs at $k_\perp\sim\sqrt{C/F}$. Increasing $F$ therefore moves this competition toward smaller momenta. The relative phase winds as $e^{6i\phi}$, producing six symmetry-related directions of destructive interference, with exact cancellation at
\begin{equation}
k_\perp^2=C/F, \qquad \phi=(2m+1)\pi/6, \qquad m=0,\ldots,5 .
\end{equation}
The deformation consequently moves a sixfold-organized set of hybridization-suppressed loci through momentum space, providing a natural mechanism for successive reconnections and handle reorganizations of the corresponding momentum-space geometry. It is important to note that varying $F$ preserves this sixfold symmetry, hence the associated knotted graphs remain $C_6$ symmetric throughout the trajectory.

The resulting phase space is shown in \cref{fig:tib2_parameter_phase_map}. Eleven stabilized topology-signature sectors are resolved. Much of the parameter plane is occupied by a few broad phases, while sectors 4--7 and 9--10 form a sequence of narrower regions in the crossover between the two limiting Hamiltonians. These intermediate sectors correspond to successive reconnections that cannot be inferred by examining only \cref{fig:TiB2_a} and \cref{fig:TiB2_c} separately.

The two parameter limits of the phase diagram provide direct consistency checks against those independently computed fixed-parameter results. At the left boundary $F_{\min}=1.83$, the vertical topology sequence is
\begin{equation}
1\longrightarrow2\longrightarrow3.
\end{equation}
Sector 1 recovers the $\theta_6$ topology of \cref{fig:TiB2_a}(a). Sector 2 corresponds to the subsequent multi-boundary genus-zero configuration of \cref{fig:TiB2_a}(c), while sector 3 recovers the final single spherical component of \cref{fig:TiB2_a}(d). Thus, the $F_{\min}$ boundary of the two-dimensional map reproduces the previously established topology sequence of the fitted Hamiltonian. At the opposite boundary $F_{\max}=7.60$, the corresponding sequence is
\begin{equation}
8\longrightarrow7\longrightarrow11\longrightarrow3.
\end{equation}
Sector 8 reproduces the periodically connected $\theta_{6,P}$ topology of \cref{fig:TiB2_c}(a). Sector 7 corresponds to \cref{fig:TiB2_c}(b), where the outer surface has closed and the topology contains three boundary components: an inner spherical boundary, a flower-like middle boundary, and an outer spherical boundary. At the next transition, the innermost gap closes while the flower-like boundary retains its topology, giving sector 11 and the two-boundary configuration shown in \cref{fig:TiB2_c}(c). Sector 3 then recovers the final single spherical component of \cref{fig:TiB2_c}(d).

The agreement at both $F_{\min}$ and $F_{\max}$ constitutes an independent limiting-case validation of the parameter map. In particular, the $F_{\max}$ boundary resolves the complete four-stage sequence of \cref{fig:TiB2_c}, including the distinction between sectors 7 and 11 that arises from closure of the innermost boundary while the flower-like component persists. Together with the exact $C_6$ symmetry constraint, these limiting cases provide stringent checks on the intermediate sectors: their representative geometries must connect the independently established endpoint topology classes while retaining the symmetry required by the Hamiltonian. The additional sectors in the interior of \cref{fig:tib2_parameter_phase_map} therefore resolve topology changes along the continuous deformation that are absent from either endpoint analysis alone.

\begin{figure}[t]
    \centering
    \includegraphics[width=\linewidth]{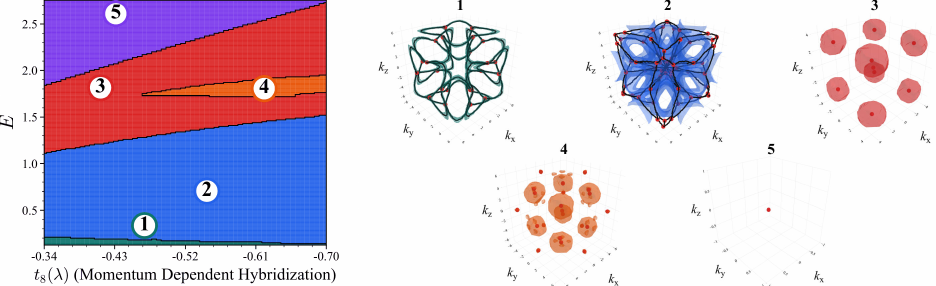}
    \caption{
    \textbf{Detailed parameter-resolved topology map of the Co$_2$MnGa Hamiltonian underlying \cref{fig:topophasespace}(c).} The horizontal axis follows the momentum-dependent $d$--$p$ hybridization coefficient $t_8(\lambda)$ in \cref{eq:co2mnga_parameter_path}. The parameter plane separates into five stabilized topology-signature sectors. Sector 1 represents the initial topology, sector 2 the extended
    multiply connected network produced after branch merger, sector 3 the
    internal-boundary configuration obtained after closure of the surrounding
    region, sector 4 an additional intermediate topology that appears over part
    of the deformation range, and sector 5 the final spherical topology after
    closure of the remaining internal cavity. Numbered markers identify
    representative geometries shown at right; surface colors match their
    corresponding regions, black curves denote graph edges, and red points denote
    graph vertices. The $t_8=-0.34$ boundary reproduces the sequence
    $1\rightarrow2\rightarrow3\rightarrow5$ identified independently in
    \cref{fig:Co2mnGa2singlecellsize}(a--d). At $t_8=-0.70$, the corresponding
    sequence is
    $2\rightarrow3\rightarrow4\rightarrow3\rightarrow5$.
    }
    \label{fig:co2mnga_parameter_phase_map}
\end{figure}

\subsubsection{Co$_2$MnGa: deformation of momentum-dependent $d$--$p$ hybridization}

For Co$_2$MnGa, we start from the six-band Hamiltonian of \cref{eq:co2mnga_H} derived in Ref.~\cite{Chang2017} and vary the coefficient $t_8$ appearing in \cref{eq:co2mnga_xi}. Its contribution is
\begin{equation}
\begin{aligned}
\xi^{dp}_{11}(\mathbf{k}) &= \xi^{dp}_{22}(\mathbf{k}) = 2t_8\sin\frac{k_z}{2},\\
\xi^{dp}_{13}(\mathbf{k}) &= \xi^{dp}_{32}(\mathbf{k}) = 2t_8\sin\frac{k_x}{2},\\
\xi^{dp}_{23}(\mathbf{k}) &= \xi^{dp}_{31}(\mathbf{k}) = 2t_8\sin\frac{k_y}{2}.
\end{aligned}
\label{eq:co2mnga_t8_terms}
\end{equation}
All other Hamiltonian coefficients are held fixed. We deform
\begin{equation}
t_8(\lambda) = (1-\lambda)(-0.34)+\lambda(-0.70), \qquad 0\le\lambda\le1,
\label{eq:co2mnga_parameter_path}
\end{equation}
so that $t_8=-0.34$ reproduces the DFT-fitted Hamiltonian used in \cref{fig:Co2mnGa2cellsize,fig:Co2mnGa2singlecellsize} \cite{Chang2017}, while $t_8=-0.70$ provides a controlled deformation of the same microscopic hybridization channel.

The physical role of $t_8$ is particularly transparent near the Brillouin-zone center, where
\begin{equation}
\sin\frac{k_i}{2} \simeq \frac{k_i}{2}.
\end{equation}
Thus, $t_8$ controls the leading momentum-linear $d$--$p$ mixing along the three crystallographic directions. Increasing $|t_8|$ strengthens the corresponding hybridization away from the symmetry planes on which the relevant sine factor vanishes, while leaving those symmetry-enforced zero loci fixed. The deformation therefore changes the band repulsion around the nodal structure and continuously shifts where neighboring low-gap branches expand, meet, and reconnect.

In the topology map of \cref{fig:co2mnga_parameter_phase_map}, the
finite-thickness geometries organize into five topology sectors. At fixed $t_8$, increasing the threshold $E$ drives the connectivity
evolution: sector 1 corresponds to the initial low-threshold topology,
sector 2 to the extended multiply connected network formed after branch
merger, sector 3 to the internal-boundary configuration that remains after
closure of the surrounding region, sector 4 to an additional intermediate
topology appearing under deformation, and sector 5 to the final spherical
configuration after closure of the remaining internal cavity.

The fitted limit
\begin{equation}
t_{8,\max}=-0.34
\end{equation}
provides a direct consistency check with the fixed-parameter Co$_2$MnGa analysis. Along this boundary the phase map follows
\begin{equation}
1\longrightarrow2\longrightarrow3\longrightarrow5.
\end{equation}
matching the progression in \cref{fig:Co2mnGa2singlecellsize}(a--d): the initially separated components evolve into the highly connected
intermediate structure, followed by closure of the surrounding region while
an internal boundary remains, and finally closure of the remaining internal
cavity, leaving the single spherical outer boundary.

At the opposite deformation limit,
\begin{equation}
t_{8,\min}=-0.70,
\end{equation}
the vertical topology sequence is
\begin{equation}
2\longrightarrow3\longrightarrow4\longrightarrow3\longrightarrow5.
\end{equation}
Sector 1 lies below the sampled low-$E$ range at this boundary. In addition,
sector 4 appears between two sector-3 regions, so that increasing $E$ drives
the sequence $3\rightarrow4\rightarrow3$ before the final transition into
sector 5. Thus varying $t_8$ changes both the transition boundaries and the
set of topology sectors encountered along a fixed-$t_8$ energy sweep.

The calculations underlying \cref{fig:topophasespace}(b,c) and \cref{fig:tib2_parameter_phase_map,fig:co2mnga_parameter_phase_map} use $60$ uniformly spaced deformation values over $\lambda\in[0,1]$ and a $140^3$ momentum grid. For TiB$_2$ we sample $\mathbf{k}\in[-1.5,1.5]^3$ over $0.18\le E\le5.20$, while for Co$_2$MnGa we use
$\mathbf{k}\in[-2.05\pi,2.05\pi]^3$ over
$0.125\le E\le 2.75$.

\section{Additional Graph-theoretic Signatures and Combinatorics of Flowed Cycles}
\label[smsection]{appx:combinatorics}

The previous sections use Yamada polynomials and Yamada sets as the main fingerprints of Fermi-surface and exceptional-surface topology.
This final section records two complementary directions.
First, we describe a possible way to use the cycles of a Berry-oriented knotted graph to organize Berry-flux integration paths; this is only an outlook and is not developed into a result here.
Second, we show that standard graph-theoretic properties, illustrated by \textbf{planarity} and \textbf{intrinsic linkedness}, provide additional diagnostics of the extracted knotted graphs.

\subsection{Combinatorics of the cycles in flowed knotted graphs (an outlook).}
\label[smsection]{appx:fluxpaths}

The flowed-graph construction in \cref{appx:ES} suggests a natural question: \textbf{can the cycles and routed edge paths of a Berry-oriented skeleton organize inequivalent Berry-flux integration contours?}

For a tubular edge of an exceptional volume, one may distinguish a \textit{meridional} contour, where one encircles the knotted graph in a plane normal to the local edge direction, from a longitudinal or \textit{equatorial} contour, where one traces a closed loop that runs parallel to and follows the graph itself.
\textbf{The latter viewpoint is appealing because different graph cycles could label different candidate integration paths} through the same Berry-curvature field.

In this tentative organization, these meridional and equatorial paths would form representative closed loops for Berry-flux integration.
This gives a possible bridge between the topology of the knotted graph and the space of distinct Berry-flux integrals.
For example, starting from the red nodes in \cref{fig:directed}(b), one can identify three distinct equatorial loops on the $\theta_3$ graph that return to the same point; each corresponds to an inequivalent path and can contribute differently to the Berry flux.
However, for exceptional surfaces arising from Hamiltonians defined over a 3D parameter space, the Berry curvature is generally non-quantized \cite{wang2024berry}.
As a result, this construction would distinguish inequivalent integration paths rather than uniquely quantized Berry fluxes for $H_{3D}(\mathbf{k})$.

The situation could become more topological in higher-parameter settings.
A recent work \cite{yang2026exceptional} shows that 2D exceptional surfaces arising from \textit{four-dimensional} parameter spaces can exhibit \textit{quantized} Berry curvature and is associated with \textit{DD invariants}.
In particular, \cite{yang2026exceptional} shows that a parameter loop encircling an exceptional ring yields a quantized Berry flux, $\Phi_{\mathbf B}=0$ or $\Phi_{\mathbf B}\neq 0$, depending on whether the loop links the ring.
In the same spirit, if analogous higher-parameter exceptional surfaces can be realized for the topologically intricate geometries considered here, then the number of distinct quantized Berry-flux values that can be accumulated would be naturally organized by the set of closed loops supported by the associated knotted graph.

To the best of our knowledge, this combinatorial exploration of Berry-flux paths has not yet been developed in the literature.
We remark that this is an interesting future direction: cycles of the knotted-graph may provide an organizing tool for Berry-flux integration paths, and potentially would become much more interesting in higher-dimensional settings.

\bigskip

The rest of this section focuses on two concrete graph-theoretic diagnostics that require no additional Berry-flux construction: planarity and intrinsic linkedness.

\begin{figure}[ht!]
    \centering
    \includegraphics[width=\linewidth]{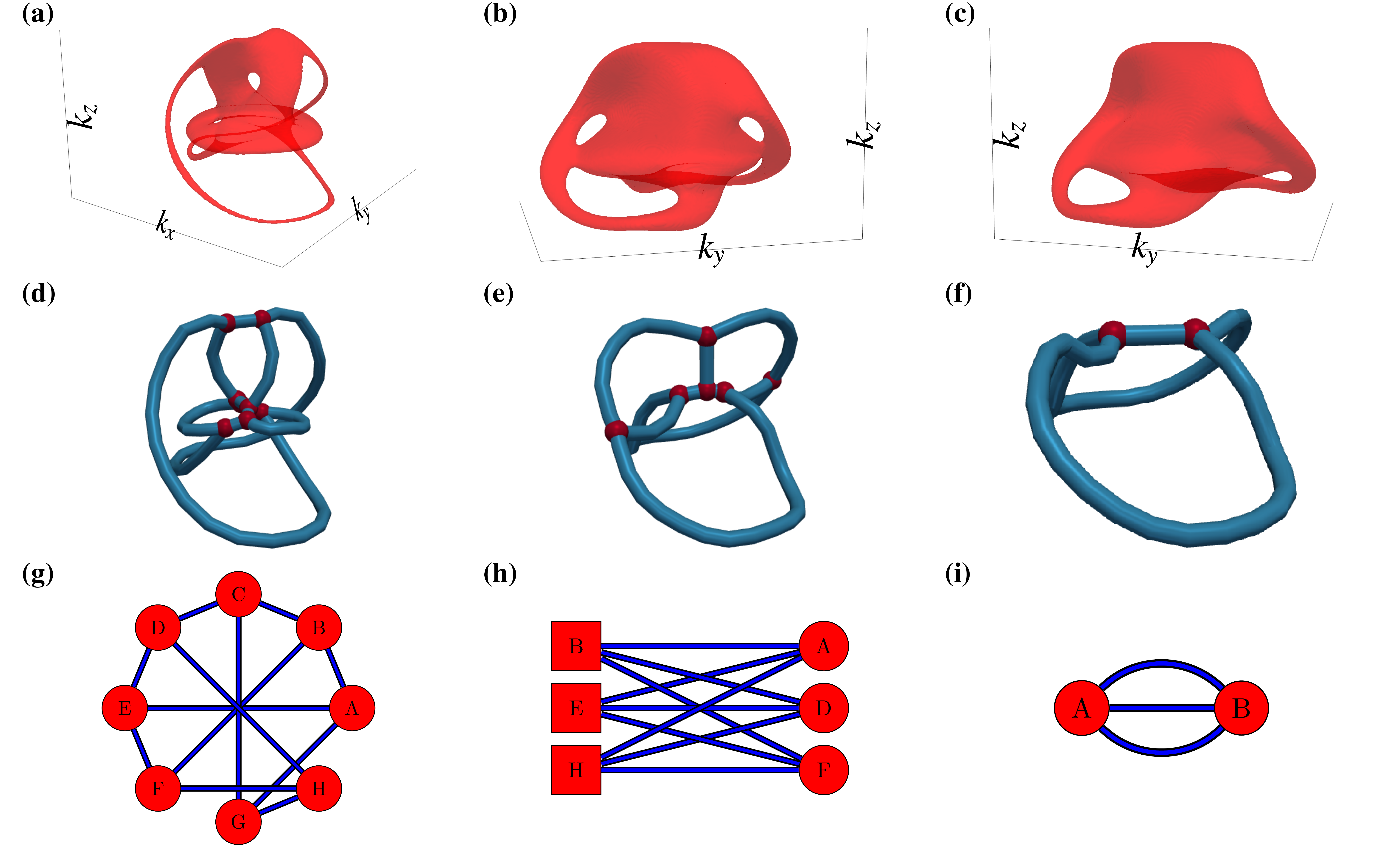}
    \caption{\textbf{Planarity as a graph-theoretic signature of rich topology in knotted exceptional graphs.}
    We show the evolution of the exceptional surfaces, the associated knotted graphs, and their corresponding 2D graph embeddings (drawn without knotting information, since we focus on intrinsic planarity) for the 3-link Hamiltonian at $E=0.115,\,0.4,\,0.5$.
    \textbf{(a,d,g)} At low energy, the extracted graph contains a $\mathcal{K}_{3,3}$ subgraph and is therefore intrinsically non-planar.
    \textbf{(b,e,h)} Upon increasing the energy, the $\mathcal{K}_{3,3}$ structure directly appears in the knotted-graph representation and its planar projection.
    \textbf{(c,f,i)} As the surface enlarges and handles close, the resulting graph admits a planar embedding, indicating an intrinsically planar graph.}
    \label{fig:planarity}
\end{figure}

\subsection{Planarity}
\label[smsection]{appx:Planarity}

We begin with a fundamental signature from graph theory: planarity. A graph is planar if it can be drawn in the plane without any edge crossings. This property provides the simplest bridge between graph theory and knot theory, since any planar graph can be embedded in $\mathbb{R}^3$ without introducing crossings and hence without forming any knottings \cite{kauffmanInvariantsThetacurvesOther1993}. Kuratowski’s theorem states that a finite graph is planar if and only if it contains no subgraph homeomorphic to either the complete graph $\mathcal{K}_5$ or the complete bipartite graph $\mathcal{K}_{3,3}$ \cite{kuratowski1930,wagner1937}.

Considering the energy isosurfaces of the 3-link Hamiltonian in \cref{tab:nodal-knots} at $E=0.115,\,0.4,\,0.5$, we obtain the surfaces shown in \cref{fig:planarity}(a--c) and their associated knotted graphs in \cref{fig:planarity}(d--f). For $E=0.115$, the embedding in \cref{fig:planarity}(g) contains a $\mathcal{K}_{3,3}$ subgraph, implying intrinsic non-planarity. As the energy increases to $E=0.4$, the canonical $\mathcal{K}_{3,3}$ structure appears directly in \cref{fig:planarity}(e), with a representative embedding shown in \cref{fig:planarity}(h). Finally, upon further thickening to $E=0.5$, handles close and the resulting graph in \cref{fig:planarity}(f) admits a planar embedding \cref{fig:planarity}(i). Altogether, this sequence illustrates how intrinsic planarity versus non-planarity emerges directly from the topology of the Fermi surfaces.

\begin{figure}[ht!]
    \centering
    \includegraphics[width=0.7\textwidth]{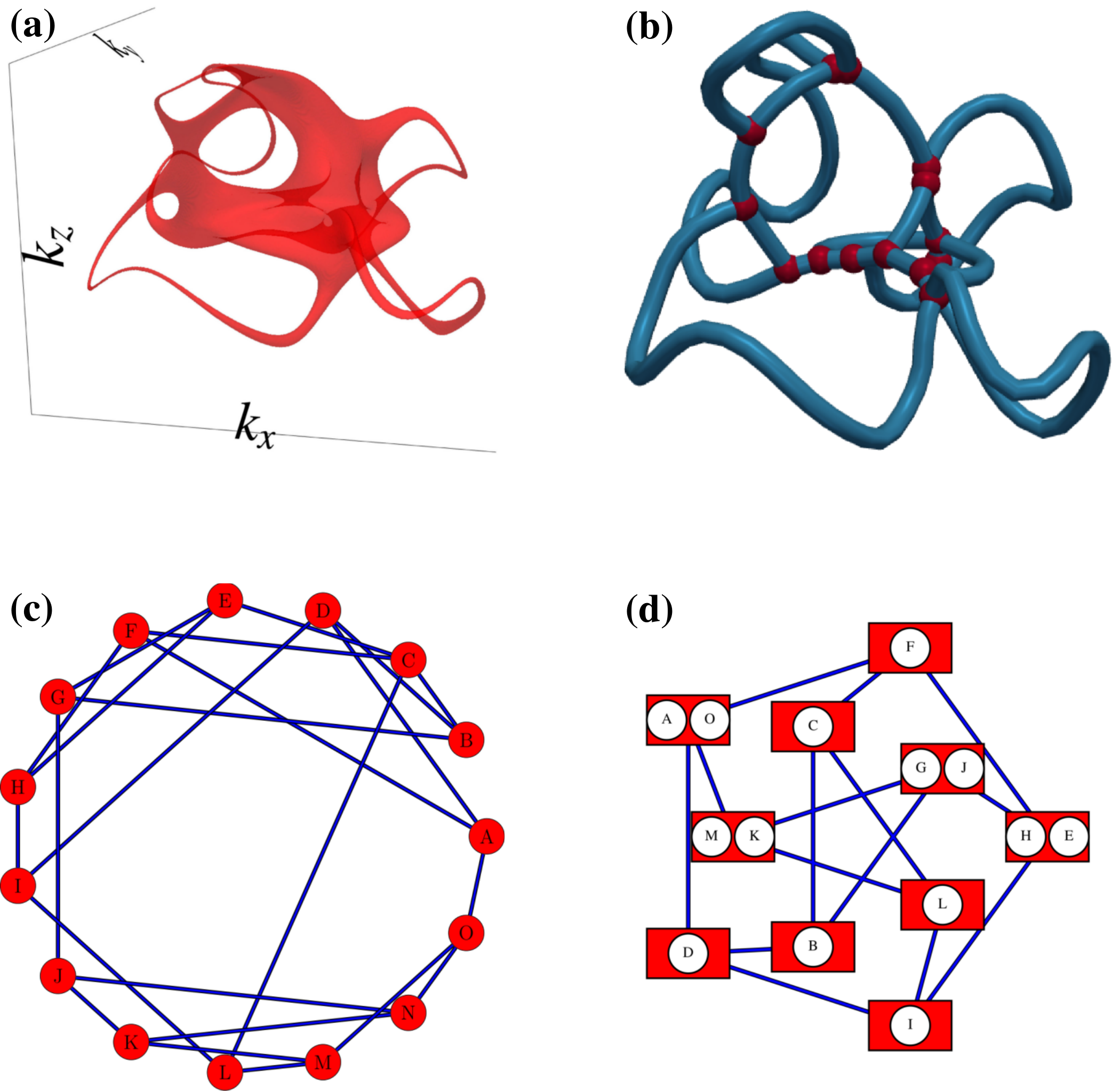}
    \caption{\textbf{An intrinsically linked exceptional surface.} \textbf{(a)} A thickened exceptional surface. \textbf{(b)} The corresponding knotted skeleton graph. \textbf{(c)} A convenient embedding of the graph. \textbf{(d)} After contracting edges, the Petersen graph appears as a minor (boxed), implying intrinsic linkedness.}
    \label{fig:intrinsiclinkedness}
\end{figure}

\subsection{Intrinsic linkedness}
\label[smsection]{appx:IntrinsicLinkedness}

A graph is \textit{intrinsically linked} (IL) if every spatial embedding of the graph contains a nontrivial link. A convenient way to characterize IL uses the notion of a \textit{minor}: a graph $\mathcal{H}$ is a minor of $\mathcal{G}$ if it can be obtained from $\mathcal{G}$ by contracting and deleting edges. The property of admitting a linkless embedding is minor-closed (hereditary): if $\mathcal{G}$ has a linkless embedding, then every minor of $\mathcal{G}$ also has a linkless embedding \cite{fellows1988,nesetril1985}. Sachs showed that every graph in the Petersen family is intrinsically linked \cite{sachs1983,sachs1984}, and Robertson--Seymour--Thomas proved that a graph is intrinsically linked if and only if it contains a Petersen-family graph as a minor \cite{robertson1995}. This theorem gives a practical criterion for testing IL: check whether one of the Petersen-family graphs appears as a minor of $\mathcal{G}$.

We find that such knotted-graph signatures arise naturally from our energy isosurfaces. As an illustrative example, we consider the awesome-knotted-graph Hamiltonian in \cref{tab:nodal-knots}. At $E=0.5$, the corresponding surface is shown in \cref{fig:intrinsiclinkedness}(a), and the associated skeleton graph is shown in \cref{fig:intrinsiclinkedness}(b). Reorganizing the embedding reveals a Petersen-graph minor via edge contractions, as indicated in \cref{fig:intrinsiclinkedness}(d), and therefore establishes intrinsic linkedness.

One can also consider stronger ``intrinsic'' properties. For instance, a graph is \textit{intrinsically $n$-linked} (InL) if every spatial embedding contains an $n$-component link. Examples of InL graphs are known \cite{flapan2001triple,bowlin2004,flapan2001nlinked}, but the classification of minor-minimal InL graphs is not fully settled \cite{naimi2021brief}. More broadly, the same framework extends to intrinsic knottedness \cite{conway1983,foisy2002,johnson2010,Barsotti2015GraphsO2,Kim2017,blain2007,ozawa2007,miller2014}, intrinsic chirality \cite{Choi2021,choi2022,flapan2013,flapan1992}, and quantitative invariants such as linking and crossing numbers \cite{Bode_2018,conway1983}, among others.

\end{appendices}

\end{document}